\documentclass[10pt,a4paper]{article}
\usepackage[english]{babel}

\usepackage{amsmath}
\usepackage{amssymb}
\usepackage{graphicx,epsfig}
\usepackage[mathscr]{eucal}
\usepackage{wrapfig}
\usepackage{lscape}
\usepackage{rotating}
\usepackage{epstopdf}
\usepackage{placeins}

\usepackage{color}
\usepackage[numbers,sort&compress]{natbib}
\usepackage{multirow}
\usepackage{tabularray}
\usepackage{tabularx}
\UseTblrLibrary{booktabs}
\usepackage{float}

\usepackage{slashed}
\usepackage{braket}

\usepackage{xspace}
\usepackage{setspace}
\usepackage{xstring}
\usepackage{mathrsfs}
\usepackage{amsbsy}

\usepackage{subfig}

\newcommand{\OpenLoops}{{\scshape OpenLoops2}\xspace}
\newcommand{\Collier}{{\scshape Collier}\xspace}

\newcommand{\Recola}{{\scshape Recola2}\xspace}
\newcommand{\Matrix}{{\scshape Matrix}\xspace}
\newcommand\TOPplusplus{{\scshape TOP++}\xspace}

\newcommand{\alphasNf}{\alpha_s^{(n_f)}}
\newcommand{\alphasNl}{\alpha_s^{(n_l)}}
\newcommand{\alphas}{\alpha_s}
\newcommand{\MSbar}{\overline{\mathrm{MS} }}

\newcommand{\M}{\mathcal{M}}
\newcommand{\J}{\mathcal{J}}

\def\muF{{\mu_F}}
\def\muR{{\mu_R}}
\def\muIR{{\mu_{\mathrm{IR}}}}
\def\rcut{r_{\mathrm{cut}}}
\def\rcuti{r^{i}_{\mathrm{cut}}}

\newcommand{\PA}{\mathrm{DPA}}
\newcommand{\fact}{\mathrm{fact}}
\newcommand{\nonfact}{\mathrm{non-fact}}

\newcommand{\textfact}{\textit{factorisable}\xspace}

\newcommand{\LO}{\mathrm{LO}}
\newcommand{\NLO}{\mathrm{NLO}}
\newcommand{\NNLO}{\mathrm{NNLO}}
\newcommand{\proddec}{\mathrm{prod} \times \mathrm{dec}}
\newcommand{\offshell}{\mathrm{off-shell}}

\newcommand{\Gatphys}{\Gamma_t^{\rm phys}}

\newcommand{\raisedhat}[1]{\hat{\vphantom{\raisebox{0.3ex}{#1}}#1}}

\usepackage{scalefnt,pstricks}
\usepackage{cancel}

\usepackage{array}
\newcolumntype{L}[1]{>{\raggedright\let\newline\\\arraybackslash\hspace{0pt}}m{#1}}
\newcolumntype{C}[1]{>{\centering\let\newline\\\arraybackslash\hspace{0pt}}m{#1}}
\newcolumntype{R}[1]{>{\raggedleft\let\newline\\\arraybackslash\hspace{0pt}}m{#1}}

\usepackage[font=normal, labelfont=bf]{caption}

\usepackage{lipsum}
\usepackage[colorlinks=true,allcolors={blue!70!black}]{hyperref}

\begin{document}
\begin{titlepage}
\begin{flushright}
ZU-TH-47/25 \\
TUM-HEP-1565/25 \\
CERN-TH-2025-130
\end{flushright}

\vspace*{0.5cm}

\begin{center}
  {\Large \bf Towards NNLO QCD predictions \\ \vspace{0.2cm}
   for off-shell top-quark pair production and decays}
\end{center}

\par \vspace{2mm}
\begin{center}
    {\bf Luca Buonocore${}^{(a)}$}, {\bf Massimiliano Grazzini${}^{(b)}$},\\[0.25cm]
    {\bf Stefan Kallweit${}^{(b)}$}, {\bf Jonas M. Lindert${}^{(c)}$} and {\bf Chiara Savoini${}^{(d)}$}

\vspace{5mm}
 
${}^{(a)}$Theoretical Physics Department, CERN, CH-1211 Geneva 23, Switzerland\\[0.25cm]

${}^{(b)}$Physik Institut, Universit\"at Z\"urich, 8057 Z\"urich, Switzerland\\[0.25cm]

${}^{(c)}$Department of Physics and Astronomy, University of Sussex, Brighton BN1 9QH, UK\\[0.25cm]

${}^{(d)}$Technical University of Munich, TUM School of Natural Sciences, Physics Department, James-Franck-Stra{\ss}e 1, 85748 Garching, Germany

\vspace{5mm}

\end{center}

\par \vspace{2mm}
\begin{center} {\large \bf Abstract}

\end{center}
\begin{quote}
\pretolerance 10000

We consider QCD radiative corrections to $W^+W^-b {\bar b}$ production with leptonic decays and massive bottom quarks at the LHC. We perform an exact next-to-leading order (NLO) calculation within the $q_T$-subtraction formalism and validate it against an independent computation in the dipole subtraction scheme. 
Non-resonant and off-shell effects related to the top quarks and the leptonic decays of the $W^\pm$ bosons are consistently included. We also consider the approximation in which the real-emission contribution is computed exactly while the virtual is evaluated in the double-pole approximation (DPA), which formally requires the inclusion of both factorisable and non-factorisable corrections. 
We evaluate such contributions and show that the DPA performs remarkably well at both the inclusive and differential levels.  
We then extend our calculation to the next-to-next-to-leading order (NNLO). All tree-level and one-loop amplitudes are evaluated exactly, while the missing two-loop virtual contribution is estimated using the DPA.
The factorisable two-loop corrections are explicitly computed by relying on available results for the polarised two-loop on-shell top-quark pair production amplitudes and the corresponding top-quark decays.
The non-factorisable contributions are inferred by exploiting the cancellation of logarithmic singularities in the \mbox{$\Gamma_t\to 0$} limit through an on-shell matching procedure. 
The NNLO corrections for the inclusive cross section are found to increase the NLO prediction by approximately $11\%$, with a numerical uncertainty that is conservatively estimated to be below the $2\%$ level -- significantly smaller than the $5\%$ residual perturbative uncertainties.

\end{quote}

\vspace*{\fill}
\begin{flushleft}
July 2025
\end{flushleft}

\end{titlepage}

\tableofcontents

\section{Introduction}
\label{sec:intro}


The production of top quarks at high-energy colliders is a crucially important process, both
in testing the validity of the Standard Model (SM) and in the quest for New Physics. 
Within the SM, the main source of top-quark events in hadronic collisions is top-quark pair ($t \bar t$) production.

At energies of the Large Hadron Collider (LHC), the main inclusive $t \bar t$
production mechanism is gluon fusion (approximately $90\%$ of the events), while
only $10\%$ of the $t \bar t$ pairs are produced via $q \bar q$ annihilation.
The production is mediated dominantly by QCD, while electroweak production modes
play a negligible role. With the huge amount of top-quark pairs produced,
accurate experimental measurements of cross sections and differential
distributions have been made possible. This enabled the precise
determination of the top-quark mass ($m_t$) along with tests of QCD predictions at one
of the highest accessible energy scales. At the same time, these studies have a
wider relevance given that top-quark events constitute a crucial background in
many new-physics searches. The sensitivity of such analyses, currently limited
by systematics in the modelling of the $t \bar t$ final state, critically relies
on the precision of the theoretical predictions.

The state-of-the-art of theoretical predictions for stable $t\bar t$ production is represented by NNLO QCD calculations for the total cross section~\cite{Czakon:2013goa,Catani:2019iny} as well as for differential distributions~\cite{Czakon:2015owf,Catani:2019hip}, combined with NLO EW corrections~\cite{Czakon:2017wor}. Fixed-order predictions have been further matched with a parton shower in Refs.~\cite{Mazzitelli:2020jio, Mazzitelli:2021mmm}, where tree-level decays of the top quarks as well as spin-correlation effects have been taken into account.
However, a more realistic treatment should also include quantum corrections to the top-quark decay. 

Within the SM, the top quark is an unstable particle and decays almost exclusively into a $W$ boson plus a bottom quark, thus making the final-state signature dependent on the (hadronic or leptonic) decay mode of the $W$ boson. 
The ``golden" channel, from the experimental point of view, is represented by the so-called \textit{dilepton channel}
\begin{equation}
	p p \to t \bar t \to l^+ \nu b \, l^- \bar \nu \bar b + X \,,
	\label{eq:ttx_dilepton}
\end{equation}
resulting from the decay of the top quarks into states containing two leptons. 
In order to suppress SM backgrounds, exactly two isolated leptons with opposite charge ($e^+e^-$, $\mu^+\mu^-$ or $e^{\pm}\mu^{\mp}$) are usually selected. Overall, the dilepton channel has the smallest branching fraction but the most favourable signal-to-background ratio, since it does not suffer from a large QCD multi-jet background. The clear experimental signature of this channel allowed the ATLAS and CMS collaborations to obtain accurate cross-section measurements at \mbox{$\sqrt{s} = 7,8$\,TeV}~\cite{Aad:2014kva} and $13$\,TeV~\cite{CMS:2018fks,CMS:2019esx,Aad:2019hzw,ATLAS:2023gsl}. Dedicated measurements for the off-shell $W^+W^-b {\bar b}$ process in the leptonic decay channel at \mbox{$13$\,TeV} have been presented in Refs.~\cite{ATLAS:2018ivx,ATLAS:2025doc}, where phase-space regions sensitive to the quantum interference between $t\bar t$ and $tW$ production have also been probed.

From the theoretical point of view, the process~\eqref{eq:ttx_dilepton} has been extensively studied at NLO in QCD. 
Initially, the calculations~\cite{Bernreuther:2004jv,Melnikov:2009dn,Bernreuther:2010ny,Biswas:2010sa,Melnikov:2011qx,Campbell:2012uf} have been performed in the narrow-width approximation~(NWA), which treats the top quarks on-shell without losing spin information. This approximation is well motivated by the small width ($\Gamma_t$) of the top quark, and its accuracy is parametrised by the ratio \mbox{$\mathcal{O}(\Gamma_t/m_t) \approx 1\%$}.  
Dedicated phenomenological studies have confirmed the theoretically expected $\mathcal{O}(\Gamma_t/m_t)$ suppression~\cite{Melnikov:1993np,Fadin:1993kt} of finite top-quark width effects for the total cross section as well as for many differential observables that are insensitive to the off-shellness of the top quark. This indicates that the approximation is reliable as long as one does not look into phase-space regions that are particularly sensitive to off-shell effects.
The narrow-width approach has been extended to approximate NNLO in top-quark production and full NNLO in the decay~\cite{Gao:2017goi}.
Full NNLO accuracy has been achieved in Ref.~\cite{Behring:2019iiv} for certain spin-correlation observables, and in Ref.~\cite{Czakon:2020qbd} for a complete set of inclusive and fiducial one- and two-dimensional differential distributions of leptons, bottom jets and top quarks.

Even though off-shell effects can be negligible at the level of the total cross section, the investigation of finite top-quark width effects in exclusive measurements is even more important since their magnitude is not known a priori. 
A first systematic study of these effects was performed in Ref.~\cite{SM:2012sed} by comparing against the computation in NWA. This study showed that the corrections to phenomenologically important observables can range from a few per mille to tens of per cent.

The expected amount of collected data and sensitivity of future experimental analyses require, therefore, more precise theoretical predictions at the level of the complete \mbox{$2 \to 6$} process
\begin{equation}
	p p \to e^+ \nu_e b \, \mu^- \bar{\nu}_{\mu} \bar b + X \,,
	\label{eq:off-shell_ttx_dilepton}
\end{equation}
including all relevant off-shell effects, irreducible backgrounds and interferences. 
As already mentioned, the total cross section for the process~\eqref{eq:off-shell_ttx_dilepton} is dominated by the double-resonant topology~\eqref{eq:ttx_dilepton}, but the same final-state signature also comes from single-top production in association with a $W$ boson ($t W$) as well as non-resonant diagrams (i.e.\ dilepton final states not mediated by top quarks). 

The separation of the $tW$ and $t\bar t$ contributions has been extensively discussed in the literature, and different prescriptions
have been proposed~\cite{Frixione:2008yi,White:2009yt,Demartin:2016axk}.
However, such methods are subject to a certain degree of arbitrariness that is either due to ad-hoc prescriptions, violations of gauge invariance, or the treatment of interference and off-shell effects.
As a consequence, the modelling of the $t \bar t$ -- $tW$ interference represents a significant source of systematic uncertainty (see Refs.~\cite{ATLAS:2018ivx,ATLAS:2021pyq}) for many beyond SM searches~\cite{ATLAS:2016seq,ATLAS:2017www,ATLAS:2015zwy,ATLAS:2016qyi,ATLAS:2017drc,ATLAS:2016maz}. 
A unified description of double-, single- and non-resonant topologies together with their quantum interferences is therefore of paramount importance.

More than ten years ago, advances and automation in one-loop calculations made it possible to account for these effects at NLO QCD~\cite{Denner:2010jp,Bevilacqua:2010qb,Denner:2012yc,Frederix:2013gra,Cascioli:2013wga,Heinrich:2017bqp} and later at NLO EW~\cite{Denner:2016jyo}.
The computations in Refs.~\cite{Denner:2010jp,Bevilacqua:2010qb,Denner:2012yc,Heinrich:2017bqp,Denner:2016jyo} were performed in the 5 flavour scheme (FS) with massless bottom quarks, while Refs.~\cite{Frederix:2013gra,Cascioli:2013wga} considered massive bottom quarks in the 4FS. The latter allowed for a unified description of $t\bar t$ and $tW$ production, including NLO interference and off-shell effects.
Fixed-order computations for the process~\eqref{eq:off-shell_ttx_dilepton} in association with an additional jet have been presented in Refs.~\cite{Bevilacqua:2015qha,Bevilacqua:2016jfk} for massless bottom quarks.
The matching of off-shell NLO QCD calculations to parton showers was enabled by a non-trivial resonance-aware extension of standard matching techniques~\cite{Jezo:2015aia}. It was presented for the first time in Ref.~\cite{Jezo:2016ujg} for the leptonic final states, and recently extended in Ref.~\cite{Jezo:2023rht} to account for semileptonic channels. 

Extending off-shell NLO calculations to the next perturbative order is an extremely demanding task.
The NNLO corrections to the full process~(\ref{eq:off-shell_ttx_dilepton}) require the combination of \mbox{$2\to 6$}, \mbox{$2\to7$} and \mbox{$2\to 8$} topologies.
The corresponding contributions are all infrared (IR) divergent, and a method to handle and cancel IR singularities is needed.
As long as the bottom quarks are massive, the $q_T$-subtraction formalism \cite{Catani:2007vq} can be used, which is now fully developed for heavy-quark pair production \cite{Bonciani:2015sha,Catani:2019iny,Catani:2019hip,Catani:2020kkl} and related processes \cite{Catani:2022mfv,Buonocore:2022pqq,Buonocore:2023ljm,Devoto:2024nhl}.
In particular the complete process~(\ref{eq:off-shell_ttx_dilepton}) belongs to the class of $Q{\bar Q}F$ processes (with \mbox{$Q=t,b$} and $F$ any colourless final state) for
which the perturbative ingredients required to implement the method are fully available \cite{Catani:2021cbl,Catani:2023tby,inprep}.
Nevertheless, the high multiplicity and the smallness of the bottom-quark mass ($m_b$) challenge the numerical integration of the double-real contribution as well as the applicability of the method, due to huge cancellations between the real contributions and the non-local counterterms.
The major bottleneck towards a complete off-shell NNLO calculation is, however, represented by the double-virtual contribution. Despite the progress of the last few years, the two-loop amplitudes for such a complicated multi-parton and multi-scale process are clearly out of reach.

In this paper we perform a first off-shell $W^+W^-b{\bar b}$ computation at NNLO.
The problem of the missing two-loop amplitudes is addressed 
by using the so-called \textit{double-pole approximation} (DPA) (see Ref.~\cite{Denner:2019vbn} and references therein).
We first carry out a complete NLO computation in the 4FS,
consistently including non-resonant and off-shell contributions. We study the
quality of the DPA for the virtual contribution at NLO by comparing the exact
results with those obtained via this approximation, at both fiducial and
differential levels. We then move to NNLO, where we consider a fully inclusive
scenario. We exactly include all the double-real and real--virtual contributions
and the corresponding subtraction counterterms. As for the double-virtual
contributions, the one-loop squared terms are included exactly, while the
genuine two-loop terms are computed in DPA. More precisely, only the {\it
  factorisable} contributions are explicitly computed by combining available
results for the polarised two-loop $t {\bar t}$ production amplitudes
\cite{Chen:2017jvi} with the corresponding polarised amplitudes for the
top-quark decay. The latter are evaluated in the massless bottom-quark limit
\cite{Bonciani:2008wf,Beneke:2008ei,Asatrian:2008uk,Bell:2008ws}, and logarithmic mass corrections are effectively
restored through a {\it massification} procedure
\cite{Penin:2005eh,Mitov:2006xs,Becher:2007cu,Engel:2018fsb,Wang:2023qbf}. We
then study the on-shell limit and exploit the cancellation of logarithmic
singularities for \mbox{$\Gamma_t\to 0$} to obtain, through a properly defined
on-shell matching procedure, the missing \textit{non-factorisable} contribution.
As we will show, this procedure, which we fully validate at NLO and for the
off-diagonal channels at NNLO, allows us to obtain an excellent approximation of
the inclusive off-shell NNLO cross section. In passing, we would like
  to highlight that the contribution to the newly computed NNLO cross section
  due to a resolved additional jet represents, to the best of our knowledge, the
  first ever NLO computation of the process~\eqref{eq:off-shell_ttx_dilepton} in
  association with a jet, performed in the 4FS with massive bottom quarks.

The paper is organised as follows. 
In Sec.~\ref{sec:DPA} we introduce the DPA. 
We first start from the one-loop order in Sec.~\ref{sec:DPA_NLO}, where we discuss both the factorisable (Sec.~\ref{sec:fact_oneloop}) and non-factorisable (Sec.~\ref{sec:nonfact_oneloop}) contributions. 
We then move to the two-loop order in Sec.~\ref{sec:DPA_NNLO}. Our construction of the factorisable corrections is described in Sec.~\ref{sec:fact_twoloop}, and the relevant ingredients such as the polarised two-loop amplitudes for the top-quark decay (Sec.~\ref{sec:heavy-to-light_FF}) and the on-shell $t{\bar t}$ production (Sec.~\ref{sec:onshell_twoloop_amplitudes}) are presented. The massive bottom-quark contribution to the two-loop production amplitudes is discussed in Sec.~\ref{sec:mb-effects_onshell_ttx_amplitudes}. 
In Sec.~\ref{sec:framework} we comment on the computational framework and the numerical challenges we have to face to carry out the calculation. In Sec.~\ref{sec:ttxoffshell_NLO_results} we present our NLO results, at both fiducial (Sec.~\ref{sec:totalXS_ttx_off-shell}) and differential (Sec.~\ref{sec:distributions_ttx_off-shell}) levels, and we perform a detailed validation of the DPA at NLO in Sec.~\ref{sec:DPAvalidation_ttx_off-shell}.
In Sec.~\ref{sec:ttxoffshell_NNLO_results} we describe our computation of the NNLO cross section. We first introduce our on-shell matching procedure in Sec.~\ref{sec:Delta-term}, we discuss the dependence on the slicing parameter in Sec.~\ref{sec:rcut_dependence} and then move to the numerical extrapolation procedures in Secs.~\ref{sec:rcut_extrapolation} and \ref{sec:Gammat_extrapolation}.
We finally present our predictions for the inclusive NNLO cross section in Sec.~\ref{sec:NNLO_results}.
Our findings are summarised in Sec.~\ref{sec:summa}.
Additional technical details are provided in four Appendices.

\section{Double-pole approximation at one- and two-loop level}
\label{sec:DPA}
%
The pole approximation enables a systematic and gauge-invariant separation of scattering amplitudes into leading contributions, with a maximum number of resonant $s$-channel topologies, and additional subleading and non-resonant contributions~\cite{Denner:2019vbn}. For the process~\eqref{eq:off-shell_ttx_dilepton}, the leading-pole contribution involves double-resonant diagrams, while single-resonant and non-resonant contributions are neglected, i.e.\ in the expansion
\begin{align}
	\mathcal{M_{\rm off-shell}} =\frac{\mathcal{R}_{t\bar t}(p_t,p_{\bar t})}{(p_t^2-\mu_t)(p_{\bar t}^2-\mu_t)} + \frac{\mathcal{R}_t(p_t,p_{\bar t})}{(p_t^2-\mu_t)} + \frac{\mathcal{R}_{\bar t}(p_t,p_{\bar t})}{(p_{\bar t}^2-\mu_t)} + \mathcal{N}(p_t,p_{\bar t})
\end{align}
only the first term is kept. Here, $p_t, p_{\bar t}$ are the off-shell top and anti-top quark momenta, and \mbox{$\mu_t^2 = m_t^2 - i \Gamma_t m_t$} is the complex top-quark mass in the complex-mass scheme~\cite{Denner:2005fg}. In order to ensure gauge invariance of this approximation, the corresponding double-pole residue has to be evaluated on-shell, i.e.\ \mbox{$p_t \to \raisedhat{p}_t$}, \mbox{$p_{\bar t} \to \raisedhat{p}_{\bar t}$} with \mbox{$\raisedhat{p}_t^2 = \raisedhat{p}_{\bar t}^2 = m_t^2$}.

Our ultimate aim is to apply such a double-pole approximation only at the level of the two-loop virtual amplitude in the NNLO computation of $W^+W^-b {\bar b}$ production, whose exact evaluation is clearly out of reach at present.
All other perturbative ingredients, including double-real and real--virtual contributions, are available as provided by \OpenLoops~\cite{Cascioli:2011va,Buccioni:2017yxi,Buccioni:2019sur} and will be included exactly. 
This procedure is motivated by the dominance, at the cross-section level, of double-resonant diagrams and the suppression of off-shell effects for sufficiently inclusive observables. 

In the following two subsections, we present the construction of the (virtual) DPA at NLO and NNLO.  
While the DPA can also be applied to real-emission amplitudes, in our discussion, we will only treat virtual (one-loop at NLO and two-loop at NNLO) corrections in the DPA.
The detailed NLO construction and numerical investigations, which will be presented in Sec.~\ref{sec:ttxoffshell_NLO_results}, enable us to assess the reliability of this approximation. As we will show in Sec.~\ref{sec:ttxoffshell_NLO_results}, the DPA turns out to work very well at both fiducial and differential cross-section levels.
Additionally, this detailed NLO study will provide insights into the potential efficacy of the approximation at NNLO.

Based on the DPA, the process~\eqref{eq:off-shell_ttx_dilepton} naturally factorises into production and decay subprocesses, connected by the intermediate top and anti-top quark resonances. At higher orders in perturbation theory, the different subprocesses receive independent radiative corrections as part of the \textit{factorisable} corrections, which must be combined in a way that preserves spin correlations and finite-width effects. However, additional \textit{non-factorisable} corrections account for interferences between the different hard subprocesses. These contributions cannot be separated into production and decay stages. 
It is worth stressing that the distinction between factorisable and non-factorisable corrections is not made at the level of individual Feynman diagrams, but at the level of scattering amplitudes. 
For example, the self-energy correction to the intermediate resonance contains both factorisable and non-factorisable terms, and the two have to be carefully disentangled (via partial fractioning of the resonant propagators) to avoid double counting.

Non-factorisable one-loop corrections for top-quark pair production and decays have
been known for a long time~\cite{Beenakker:1999ya,Falgari:2013gwa} and have been
studied in detail, e.g. for four-fermion production at LEP~\cite{Denner:1997ia}.
Here, we recompute these one-loop contributions following the strategy of
Ref.~\cite{Dittmaier:2015bfe}, which is based on a Feynman-diagrammatic approach
using an extended soft-gluon approximation.%
\footnote{Alternatively, non-factorisable corrections can also be derived from
  dedicated effective field theories as formulated in
  Refs.~\cite{Chapovsky:2001zt,Beneke:2003xh,Beneke:2004km,Falgari:2013gwa}.
  This approach, based on a well-defined power counting, enables a more
  systematic definition of factorisable and non-factorisable corrections.
  Moreover, it could be more easily extended to higher perturbative orders as
  well as to the computation of power-suppressed terms in $\Gamma_t/m_t$.}
Corresponding two-loop results for the non-factorisable corrections are not
available at present.

\subsection{Double-pole approximation at one-loop level}
\label{sec:DPA_NLO}
\begin{figure}[tb]
	\centering
	\includegraphics[width=\textwidth]{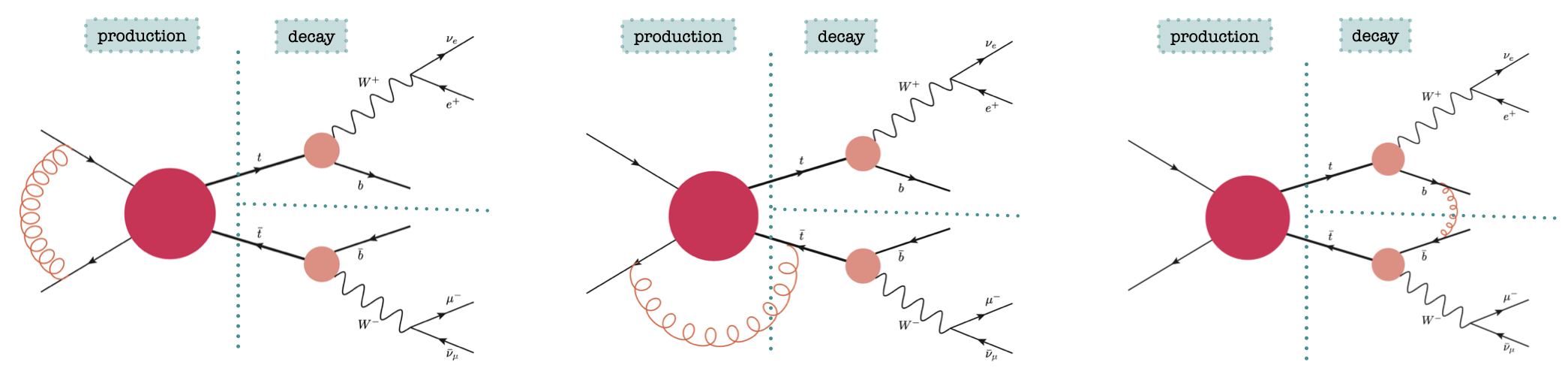}
	\caption[]{Three diagrams contributing to the factorisable (left) and non-factorisable corrections (right), respectively. The diagram in the centre contributes to both factorisable and non-factorisable corrections. }
	\label{fig:fact_nonfact_cartoon}
\end{figure}

In the following, we describe the construction of the DPA at one-loop level. To the
purpose of computing the factorisable corrections, $t \bar t$ production and
top/anti-top decays factorise and can be considered as independent subprocesses.
The different subprocesses are connected only by the kinematics and the helicity
state of the resonant $t \bar t$ pair. On the contrary, the non-factorisable
corrections involve resonant diagrams in which a soft gluon is exchanged between
production and decay stages, or between the different decay stages. Since they
originate from soft wide-angle radiation, the non-factorisable corrections do
not generate any logarithmic dependence on the masses of the external particles,
and can therefore be computed in the limit of massless bottom quarks up to power
corrections in the bottom-quark mass. In
Fig.~\ref{fig:fact_nonfact_cartoon} we display examples of factorisable (left)
and non-factorisable (right) corrections for the process at hand. The diagram in
the centre of Fig.~\ref{fig:fact_nonfact_cartoon} contributes to both
factorisable and non-factorisable corrections. The separation starting from the
full off-shell amplitude requires a partial fractioning of the intermediate
resonant propagator.

\subsubsection{Factorisable one-loop corrections }\label{sec:fact_oneloop}
As described above, the factorisable one-loop corrections involve production and decay stages separately. 
The only connection between the subprocesses lies in the kinematics and the helicity states of the resonant top and anti-top quarks. 
For the process~\eqref{eq:off-shell_ttx_dilepton}, the \textfact contribution to the one-loop amplitude is given by%
\footnote{We denote the perturbative expansion of QCD amplitudes as \mbox{$|{\cal M}\rangle=\sum_{l=0}^{\infty}\left(\alphas/(2\pi)\right)^l |{\cal M}^{(l)}\rangle$} where \mbox{$l=0$} corresponds to the tree-level contribution, \mbox{$l=1$} to the one-loop correction, and so forth.}

\begingroup
\allowdisplaybreaks
\begin{align}
	\hspace{-0.7cm} \ket{ \M^{(1)}_{\fact} } &= 
	\frac{1}{(p_t^2 -\mu_t^2)(p_{\bar{t}}^2 -\mu_t^2)} 
	\sum_{\{ \vec{\mathfrak{c}} \}}
	 \sum_{\{ \vec{h}_I \}} \sum_{\{\vec{h}_{R_t}\}}  \sum_{\{\vec{h}_{R_{\bar{t}} }\}}  
	\sum_{\lambda_t,\lambda_{\bar{t}} } \biggl\{ \M_{t \bar t}^{(1), \vec{\mathfrak{c}} }(\vec{h}_I, \lambda_t,\lambda_{\bar{t}})\,\M_{t}^{(0)}(\vec{h}_{R_t}, \lambda_t)\,\M_{\bar{t}}^{(0)}(\vec{h}_{R_{\bar{t}} }, \lambda_{\bar{t}}) \notag \\
	& 
	+ \M_{t \bar t}^{(0), \vec{\mathfrak{c}} }(\vec{h}_I, \lambda_t,\lambda_{\bar{t}})\,
	\left( \M_{t}^{(1)}(\vec{h}_{R_t}, \lambda_t)\,\M_{\bar{t}}^{(0)}(\vec{h}_{R_{\bar{t}} }, \lambda_{\bar{t}})  +
	\M_{t}^{(0)}(\vec{h}_{R_t},\lambda_t)\,\mathcal{M}_{\bar{t}}^{(1)}(\vec{h}_{R_{\bar{t}} }, \lambda_{\bar{t}}) \right)
	\biggr\} \ket{ \mathfrak{C}_{\vec{\mathfrak{c}}} } \,,
\label{eq:fact_oneloop_ttx}
\end{align}
\endgroup
where we specify explicit sums over the colour \mbox{$\vec{\mathfrak{c}} = \{\mathfrak{c}_1,\mathfrak{c}_2, \mathfrak{c}_t, \mathfrak{c}_{\bar t} \}$} of the initial-~(IS) and final-state~(FS) particles in $t \bar t$ production, the helicity states \mbox{$\vec{h}_I = \{h_1,h_2\}$} of the IS partons and over the intermediate helicities $\lambda_t, \lambda_{\bar t}$ of the top and anti-top resonances, respectively.  Production and decay matrix elements are evaluated on the same helicities, thus inducing spin correlations between the different production and decay subprocesses. The remaining two nested sums refer to the helicity configurations of the decay products of the two resonances: $R_{t}$ and $R_{\bar{t}}$ are collective indices for 
\begin{equation}
	R_{t} = \{ e^+, \nu_e, b \} ~,~~ R_{\bar{t}} = \{\mu^-, \bar{\nu}_{\mu}, \bar b \} \,.
	\label{eq:indices_topdecay_products}
\end{equation}
The tree-level and one-loop amplitudes appearing in Eq.~\eqref{eq:fact_oneloop_ttx} represent  the matrix elements associated with the $l$-loop production $\M_{t \bar t}^{(l)}$ and decay $\M_{t}^{(l)}, \M_{\bar t}^{(l)}$ subprocesses, respectively.
The abstract vector $\ket{ \mathfrak{C}_{\vec{\mathfrak{c}}} }$ is an element of the chosen colour basis.
In Eq.~\eqref{eq:fact_oneloop_ttx}, we do not write a colour decomposition for the matrix elements associated with the decay subprocesses since only one colour structure contributes. 
In order to ensure gauge invariance, all matrix elements appearing in Eq.~\eqref{eq:fact_oneloop_ttx} have to be evaluated on the set of on-shell projected momenta $\{ \raisedhat{p}_i \}$. The employed on-shell projection is described in Appendix~\ref{app:Appendix_mapping}.

Analogously, the Born matrix element in DPA is given by
\begin{equation}
	\ket{ \M^{(0)}_{\PA} } = \frac{1}{(p_t^2 -\mu_t^2)(p_{\bar{t}}^2 -\mu_t^2)} \sum_{\{ \vec{\mathfrak{c}} \}} \sum_{\{ \vec{h}_I \}} \sum_{\{\vec{h}_{R_t}\}}  \sum_{\{\vec{h}_{R_{\bar{t}} }\}}  
	\sum_{\lambda_t,\lambda_{\bar{t}} } \M_{t \bar t}^{(0), \vec{\mathfrak{c}} }(\vec{h}_I, \lambda_t,\lambda_{\bar{t}})\,\M_{t}^{(0)}(\vec{h}_{R_t}, \lambda_t)\,\M_{\bar{t}}^{(0)}(\vec{h}_{R_{\bar{t}} }, \lambda_{\bar{t}}) \ket{ \mathfrak{C}_{\vec{\mathfrak{c}}} } \,.
\label{eq:LO_PA_ttx}
\end{equation}
The loop--tree interference of Eqs.~\eqref{eq:fact_oneloop_ttx} and~\eqref{eq:LO_PA_ttx} yields the factorisable corrections at NLO,
\begingroup
\allowdisplaybreaks
\begin{align}
	2 \mathrm{Re} &\left \{\braket{\M^{(0)}_{\PA} | \M^{(1)}_{\fact} }  \right\} = \frac{1}{N_{c \bar c}} 
	\sum_{\{ \vec{h}_I \}} \sum_{\{ \vec{\mathfrak{c}} \}} \sum_{\{ \vec{\mathfrak{d}} \}}  \sum_{\{\vec{h}_{R_t}\}}  \sum_{\{\vec{h}_{R_{\bar{t}} }\}} \sum_{\lambda_t,\lambda_{\bar{t}} } \sum_{\tilde{\lambda}_t,\tilde{\lambda}_{\bar{t}} } \frac{\braket{  \mathfrak{C}_{\vec{\mathfrak{d}}} |  \mathfrak{C}_{\vec{\mathfrak{c}}} } }{| p_t^2 -\mu_t^2|^2 | p_{\bar{t}}^2 -\mu_t^2 |^2}   \nonumber \\
	&\hspace{-0.2cm}\times \biggl\{  
	\left( \M_{t \bar t}^{(0), \vec{\mathfrak{d}} }(\vec{h}_I, \tilde{\lambda}_t, \tilde{\lambda}_{\bar{t}}) \right)^* \M_{t \bar t}^{(1), \vec{\mathfrak{c}} }(\vec{h}_I, \lambda_t,\lambda_{\bar{t}}) \notag 
	 \left( \M_{t}^{(0)}(\vec{h}_{R_t}, \tilde{\lambda}_t) \right)^* \M_{t}^{(0)}(\vec{h}_{R_t}, \lambda_t)\,\left( \M_{\bar{t}}^{(0)}(\vec{h}_{R_{\bar{t}} }, \tilde{\lambda}_{\bar{t}}) \right)^* \M_{\bar{t}}^{(0)}(\vec{h}_{R_{\bar{t}} }, \lambda_{\bar{t}})   \notag \\
	&+ \left( \M_{t \bar t}^{(0), \vec{\mathfrak{d}} }(\vec{h}_I, \tilde{\lambda}_t, \tilde{\lambda}_{\bar{t}}) \right)^* \M_{t \bar t}^{(0), \vec{\mathfrak{c}} }(\vec{h}_I, \lambda_t,\lambda_{\bar{t}}) \notag  \biggl[ \left( \M_{t}^{(0)}(\vec{h}_{R_t}, \tilde{\lambda}_t) \right)^* \M_{t}^{(1)}(\vec{h}_{R_t}, \lambda_t)\,\left( \M_{\bar{t}}^{(0)}(\vec{h}_{R_{\bar{t}} }, \tilde{\lambda}_{\bar{t}}) \right)^* \M_{\bar{t}}^{(0)}(\vec{h}_{R_{\bar{t}} }, \lambda_{\bar{t}})  \notag \\
	&+ \left( \M_{t}^{(0)}(\vec{h}_{R_t}, \tilde{\lambda}_t) \right)^* \M_{t}^{(0)}(\vec{h}_{R_t}, \lambda_t)\,\left( \M_{\bar{t}}^{(0)}(\vec{h}_{R_{\bar{t}} }, \tilde{\lambda}_{\bar{t}}) \right)^* \M_{\bar{t}}^{(1)}(\vec{h}_{R_{\bar{t}} }, \lambda_{\bar{t}})
	\biggr]
	\biggr\} + {\rm c.c.}
	\label{eq:fact_M1M0_ttx}
\end{align}
\endgroup
Here, we have exploited the orthogonality of helicity amplitudes with respect to the external states, resulting in a single sum over $\{ \vec{h}_I \}, \{\vec{h}_{R_{t} }\}$ and $\{\vec{h}_{R_{\bar{t}} }\}$. 
The double sum over the colour configurations $\vec{\mathfrak{c}}, \vec{\mathfrak{d}}$ is due to the fact that the chosen colour basis is not necessarily orthonormal, i.e.\ $\braket{  \mathfrak{C}_{\vec{\mathfrak{d}}} |  \mathfrak{C}_{\vec{\mathfrak{c}}} }$ can be different from zero for \mbox{$\vec{\mathfrak{c}} \ne \vec{\mathfrak{d}}$}.
The overall factor $N_{c \bar c}$ depends on the partonic subprocess and takes into account the average over the IS colours and spins, i.e.\
\vspace{-0.15cm}
\begin{equation}
	N_{c \bar c} = \begin{cases}
		(N_c^2 -1)^2 \times 4 ~~~~\text{if}~~ c = g \,, \\
		N_c^2 \times 4 ~~~~~~~~~~~~~\text{if}~~ c = q \,.
		\end{cases}
	\label{eq:Ncc_average-factor}
\end{equation}

We have implemented these factorisable one-loop corrections as specified in Eq.~\eqref{eq:fact_M1M0_ttx} in the \Matrix framework~\cite{Grazzini:2017mhc}, relying on \Recola~\cite{Denner:2017wsf} for the evaluation of the colour--helicity amplitudes associated with the production and decay subprocesses. This implementation has been validated against a corresponding dedicated DPA evaluation mode provided by \Recola. 
Moreover, we have verified that the contributions from interferences between matrix elements with \mbox{$\lambda_t \ne \tilde{\lambda}_t$} and/or \mbox{$\lambda_{\bar t} \ne \tilde{\lambda}_{\bar t}$} are numerically suppressed.%
\footnote{These contributions count at the per mille level of the entire factorisable corrections.}
Therefore, the construction of the factorisable virtual corrections can alternatively be simplified as
\begingroup
\allowdisplaybreaks
\begin{align}
	2 \mathrm{Re} \left \{\braket{\M^{(0)}_{\PA} | \M^{(1)}_{\fact} }  \right\} \approx \frac{1}{| p_t^2 -\mu_t^2|^2 | p_{\bar{t}}^2 -\mu_t^2 |^2} 
	\sum_{\lambda_t,\lambda_{\bar{t}} } &\biggl\{  2 \mathrm{Re}\left( \M_{t \bar t}^{(0) *}(\lambda_t, \lambda_{\bar{t}}) \M_{t \bar t}^{(1)}(\lambda_t, \lambda_{\bar{t}}) \right) \left| \M_t^{(0)}( \lambda_t) \right|^2 \left| \M_{\bar{t}}^{(0)}(\lambda_{\bar{t}}) \right|^2 \notag \\
	&+ \left| \M_{t \bar t}^{(0)}(\lambda_t, \lambda_{\bar{t}}) \right|^2  \biggl[ 2\mathrm{Re}\left( \M_t^{(0) *}( \lambda_t) \M_t^{(1)}( \lambda_t) \right) \left| \M_{\bar{t}}^{(0)}(\lambda_{\bar{t}}) \right|^2  \notag \\
	&+ \left| \M_t^{(0)}(\lambda_t) \right|^2 2\mathrm{Re}\left(\M_{\bar{t}}^{(0) *}(\lambda_{\bar{t}}) \M_{\bar{t}}^{(1)}(\lambda_{\bar{t}})  \right)  \biggr] \biggr\} \,,
	\label{eq:approx_fact_M1M0_ttx}
\end{align}
\endgroup
where the polarised squared Born matrix elements and the loop--tree interferences can be directly evaluated with the standard routines of \Recola or \OpenLoops, without requiring explicit colour--helicity amplitudes. 
Here, the average over colours and spins of the IS partons is implicitly understood. 
This approximation, which is diagonal in helicity space, will be crucial for constructing parts of the factorisable corrections at two-loop order, as will be discussed in Sec.~\ref{sec:DPA_NNLO}.

\subsubsection{Non-factorisable one-loop corrections}\label{sec:nonfact_oneloop}
Non-factorisable one-loop corrections for the process~\eqref{eq:off-shell_ttx_dilepton} arise from the exchange of a soft gluon between production and decay subprocesses 
or between the two decay subprocesses. Due to their soft origin, these corrections can be written in terms of extended soft currents \cite{Dittmaier:2015bfe}, whose expression is given by
\begin{align}
	\boldsymbol{\J}^{\mu}_{\mathrm{prod},t}(q)
	&= - 2 g_s \biggl[ \frac{\mathbf{T}_t p^{\mu}_t}{q^2 - 2 p_t \cdot q} - \frac{\mathbf{T}_{\bar t} p^{\mu}_{\bar t}}{q^2 + 2 p_{\bar t} \cdot q} + \frac{\mathbf{T}_1 p^{\mu}_1}{q^2 - 2 p_1 \cdot q} +  \frac{\mathbf{T}_2 p^{\mu}_2}{q^2 - 2 p_2 \cdot q} \biggr] \,,
	\label{eq:ESPA_prod_top-quark}\\
	\boldsymbol{\J}^{\mu}_{\mathrm{dec},t}(q)
	&= 2 g_s \biggl[ -\frac{\mathbf{T}_t p^{\mu}_t}{q^2 + 2 p_t \cdot q} + \frac{\mathbf{T}_b p^{\mu}_b}{q^2 + 2 p_b \cdot q}  \biggr] \frac{p_t^2 - \mu_t^2}{(p_t + q)^2 -\mu_t^2} \,,
	\label{eq:ESPA_dec_top-quark}
\end{align}
for the production and decay of the resonant top quark, respectively. 
Analogous currents can be written for the anti-top quark, by simply replacing \mbox{$t \to \bar t$}, \mbox{$b \to \bar b$} in Eqs.~\eqref{eq:ESPA_prod_top-quark} and \eqref{eq:ESPA_dec_top-quark}. Here, $\mathbf{T}_t$ ($\mathbf{T}_{\bar t}$) is the colour operator \cite{Catani:1996vz} of the resonant top (anti-top) quark, considered as outgoing from the production to the decay subprocess, and $\mathbf{T}_1 (\mathbf{T}_2)$ and $\mathbf{T}_{b} (\mathbf{T}_{\bar b})$ are the colour operators of the IS partons and the FS bottom (anti-bottom) quark, respectively.
In our convention, the IS momenta $p_1$ and $p_2$ are incoming, while all other momenta are considered outgoing.
The reconstructed top-quark momenta are defined by the on-shell projection detailed in Appendix~\ref{app:Appendix_mapping}.

The complete non-factorisable one-loop corrections are obtained via the loop--tree interference
\begin{equation}
	2\times \frac{\alphas}{2\pi}\, \mathrm{Re} \left\{ \braket{ \M^{(0)}_{\PA} | \M^{(1)}_{\nonfact}} \right\} = \bra{\M^{(0)}_{\PA}} \boldsymbol{\delta}_{\nonfact} \ket{ \M^{(0)}_{\PA} } \,,
	 \label{eq:nonfact_M1M0_ttx}
\end{equation}
where $\boldsymbol{\delta}_{\nonfact}$ is an operator acting on colour space and defined by
\begin{equation}
	 \boldsymbol{\delta}_{\nonfact} = 2\, \mathrm{Re} \biggl\{  \int\!\! \frac{d^d q}{(2\pi)^d} \frac{-i}{q^2 + i0} 
	 \biggl[ \boldsymbol{\J}_{\! \mathrm{prod},t}(q) \cdot \boldsymbol{\J}_{\! \mathrm{dec},t}(-q) 
	 + \boldsymbol{\J}_{\! \mathrm{prod},\bar t}(q) \cdot \boldsymbol{\J}_{\! \mathrm{dec},\bar t}(-q)  
	 + \boldsymbol{\J}_{\! \mathrm{dec},t }(q) \cdot \boldsymbol{\J}_{\! \mathrm{dec},\bar t}(-q) \biggr]  \biggr\} \,.
	 \label{eq:deltanonfact_M1M0_ttx_currents}
\end{equation}
Based on Eq.~\eqref{eq:deltanonfact_M1M0_ttx_currents} we can translate all individual contributions to the correction factor $\boldsymbol{\delta}_{\nonfact}$ into a form expressed in terms of standard scalar one-loop integrals.
We follow a notation similar to Appendix B of Ref.~\cite{Dittmaier:2015bfe} (which considers QED non-factorisable corrections) and rewrite $\boldsymbol{\delta}_{\nonfact}$ as
\begingroup
\allowdisplaybreaks
\begin{align}
	\boldsymbol{\delta}_{\nonfact} = - \left(\frac{\alphas}{2\pi}\right)
	&2\mathrm{Re} \biggl\{ -C_F \left( \Delta_{mm}(t) + \Delta_{mm}(\bar{t}) + \Delta_{mf}(t,b) +\Delta_{mf}(\bar{t},\bar{b})  \right)  \notag \\
&\hspace{0.8cm} + \mathbf{T}_b \cdot \mathbf{T}_{\bar b} \left( \Delta_{mm'}(t;\bar{t}) + \Delta_{ff'}(t,b;\bar{t},\bar{b}) + \Delta_{mf'}(t;\bar{t},\bar{b}) + \Delta_{mf'}(\bar{t}; t, b) \right)        \notag \\
&\hspace{0.8cm} +  \sum_{j=1,2} \!\left[ \mathbf{T}_j \cdot \mathbf{T}_b \left( \Delta_{if}(t,b;j) + \Delta_{im}(t;j) \right)  + 
\mathbf{T}_j \cdot \mathbf{T}_{\bar b}  \left( \Delta_{if}(\bar{t},\bar{b};j) + \Delta_{im}(\bar{t};j) \right)  \right]   \biggr\}  \,.
 	\label{eq:deltanonfact_M1M0_ttx}
\end{align}
\endgroup
Each $\Delta$ function appearing in Eq.~\eqref{eq:deltanonfact_M1M0_ttx} corresponds to a specific class of double-resonant Feynman diagrams.

\noindent More precisely, the different terms can be classified as:
\begin{enumerate}
	\item \textit{manifestly non-factorisable} corrections (for example diagrams in Fig.~\ref{fig:manifestly-nonfact_diagrams}):
		\begin{itemize}
			\item $\Delta_{ff'}(i,a; j,b)$ refers to the exchange of a soft gluon between two FS partons $a$ and $b$, coming from the decay of the resonances $i$ and $j$ (\mbox{$i \ne j$} and \mbox{$i,j \in \overline{R}=\{ t, \bar t \}$}), respectively;
			\item $\Delta_{mf'}(i; j, b)$ results from the emission of a soft gluon by a FS parton \mbox{$b \in R_j$}~\footnote{The sets $R_t$ and $R_{\bar t}$ contain the labels associated with the decay products of the top and anti-top quarks, respectively, as defined in Eq.~\eqref{eq:indices_topdecay_products}.}, \mbox{$j \in \overline{R}$}, which is then reabsorbed by another resonance \mbox{$i \ne j$};
			\item $\Delta_{if}(i,a;b)$ takes into account a soft gluon exchanged between an IS parton \mbox{$b \in I = \{1,2\}$} and a FS coloured particle \mbox{$a \in R_i$}; 
		\end{itemize}
	\vspace{-0.15cm}
	\begin{figure}[h]
		\begin{center}
		\includegraphics[width=0.85\textwidth]{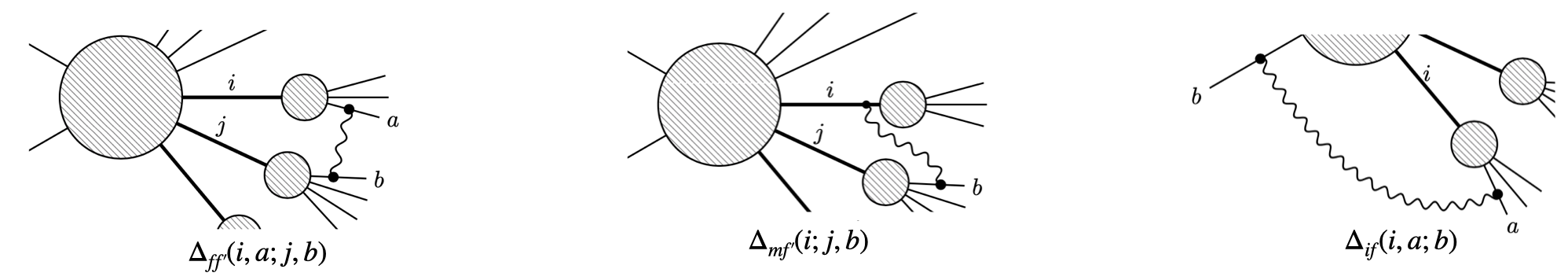}
		\end{center}
		\vspace{-0.5cm}
	\caption[]{Manifestly non-factorisable diagrams contributing to off-shell $t \bar t$ production.} 
	\label{fig:manifestly-nonfact_diagrams}
	\end{figure}
	\item \textit{non-manifestly non-factorisable} corrections (see Fig.~\ref{fig:non-manifestly-nonfact_diagrams}):
		\begin{itemize}
			\item $\Delta_{mm}(i)$ and $\Delta_{mf}(i,a)$ result from the emission of a soft gluon by the resonance $i$, which is then reabsorbed by the same resonance or by one of its decay products \mbox{$a \in R_i$}, respectively;
			\item $\Delta_{mm'}(i;j)$ refers to the exchange of a soft gluon between two resonances \mbox{$i \ne j$};
			\item $\Delta_{im}(i;b)$ takes into account a soft gluon exchanged between an IS parton \mbox{$b \in I$} and the resonance \mbox{$i \in \overline{R}$}.
		\end{itemize}
	\vspace{-0.15cm}
	\begin{figure}[h]
		\begin{center}
		\includegraphics[width=\textwidth]{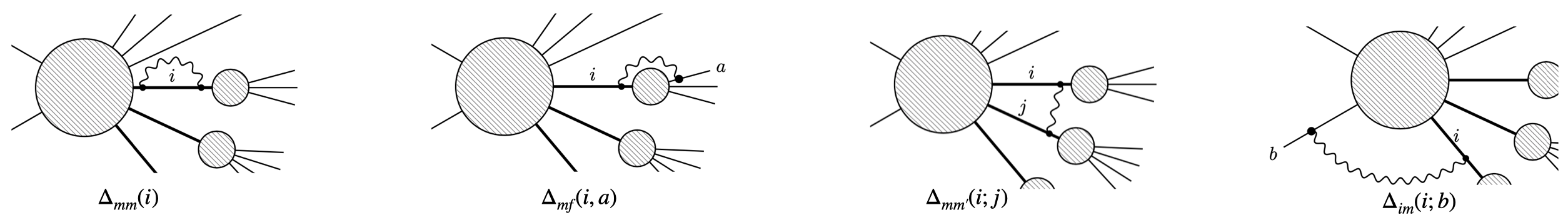}
		\end{center}
		\vspace{-0.5cm}
	\caption[]{Non-manifestly non-factorisable diagrams contributing to off-shell $t \bar t$ production.}
	\label{fig:non-manifestly-nonfact_diagrams}
	\end{figure}
\end{enumerate}
Explicit expressions for these $\Delta$ functions, written in terms of scalar one-loop integrals, are given in Appendix~\ref{app:Appendix0}.
In our \Matrix implementation of the non-factorisable corrections according to Eq.~\eqref{eq:deltanonfact_M1M0_ttx}, we rely on \Collier~\cite{Denner:2016kdg} for the evaluation of the required one-loop scalar integrals.

Considering only the contributions from IR-divergent one-loop scalar integrals, we can analytically (exploiting the one-loop integral representation of Ref.~\cite{Ellis:2007qk} together with {\scshape Package-X}~\cite{Patel:2015tea}) reconstruct the IR $1/\epsilon$ poles of the non-factorisable corrections,
\begingroup
\allowdisplaybreaks
\begin{align}
  \hspace{-0.5cm} 2 \mathrm{Re} \left\{\! \braket{ \M^{(0)}_{\PA} | \M^{(1)}_{\nonfact} } \!\right\}\! \biggl|_{\mathrm{poles}} \!\!\!\!\!&= -
  \frac{(4\pi)^{\epsilon}}{\Gamma(1-\epsilon)} 	
	\biggl\{ 
	- \frac{C_F}{\epsilon} \! \left[  4 + \frac{1}{v_{tb}} \log\left( \frac{1-v_{tb}}{1+v_{tb}} \right) +\frac{1}{v_{\bar{t}\bar{b}}} \log\left( \frac{1-v_{\bar{t}\bar{b}}}{1+v_{\bar{t}\bar{b}}} \right) \right]  |\M^{(0)}_{\PA}|^2   \notag \\
	& +  \left[ -  \frac{1}{v_{b\bar{b}}} \log\left( \frac{1-v_{b\bar{b}}}{1+v_{b\bar{b}}} \right) \textcolor{brown}{\frac{1}{\epsilon}}  + \frac{1}{v_{t\bar{t}}} \log\left( \frac{1-v_{t\bar{t}}}{1+v_{t\bar{t}}} \right)\frac{1}{\epsilon}   \right] \bra{\M^{(0)}_{\PA}} \mathbf{T}_b \cdot \mathbf{T}_{\bar b} \ket{ \M^{(0)}_{\PA} }        \notag \\
	& +  2 \sum_{i=1,2} \left[  \log\left( \frac{ 2\raisedhat{p}_b \cdot \raisedhat{p}_i}{m_b \mu} \right)\textcolor{brown}{\frac{1}{\epsilon}} - \log\left(\frac{2\raisedhat{p}_t \cdot \raisedhat{p}_i}{m_t \mu} \right)  \frac{1}{\epsilon} \right] \bra{\M^{(0)}_{\PA}} \mathbf{T}_b \cdot \mathbf{T}_i \ket{ \M^{(0)}_{\PA} }    \notag \\
	& + 2 \sum_{i=1,2} \left[  \log\left( \frac{ 2\raisedhat{p}_{\bar{b}} \cdot \raisedhat{p}_i}{m_b \mu} \right)\textcolor{brown}{\frac{1}{\epsilon}} - \log\left(\frac{2\raisedhat{p}_{\bar t} \cdot \raisedhat{p}_i}{m_t \mu} \right) \frac{1}{\epsilon}  \right] \bra{\M^{(0)}_{\PA}} \mathbf{T}_{\bar b} \cdot \mathbf{T}_i \ket{ \M^{(0)}_{\PA} }
	 \biggr\}   \,, \label{eq:nonfact-poles}
\end{align}
where $\mu$ is the renormalisation scale and $v_{ij}$ the relative
velocity of two heavy quarks $i$ and $j$ defined as
\mbox{$v_{ij} = \sqrt{1-  m_i^2m_j^2/(\hat{p}_i \cdot \hat{p}_j)^2}$}. All poles
highlighted in \textcolor{brown}{brown} are those we would expect from the IR
subtraction operator describing the singular structure of the full one-loop
matrix element. They arise from the exchange of a soft gluon between an
IS parton and the FS massive bottom (or anti-bottom) quark, or between the two FS bottom quarks. The remaining poles are
spurious in the sense that they are cancelled by analogous poles generated in
the factorisable one-loop corrections. The absence of $1/\epsilon^2$ poles
confirms the pure soft origin of the non-factorisable corrections. As expected,
no dependence on the top-quark width $\Gamma_t$ appears in the singularity structure
detailed in~Eq.~\eqref{eq:nonfact-poles}.

Here we want to highlight two important features of the non-factorisable corrections:
\begin{enumerate}
	\item the absence of logarithmic contributions in the mass of the external partons;
	\item a logarithmic dependence on the resonant off-shell propagators, which implies the presence of logarithms of the top-quark width $\Gamma_t$ in the finite remainder.
\end{enumerate}
The first property is clearly pointed out already in Refs.~\cite{Falgari:2013gwa,Dittmaier:2015bfe}. 
We analytically verified that the correction factor $\boldsymbol{\delta}_{\nonfact}$ in Eq.~\eqref{eq:deltanonfact_M1M0_ttx} is free of mass singularities of the bottom quarks, in the limit where $m_b^2$ is much smaller than $m_t^2$ and the other invariants $s_{ij}$.
In particular for the $m_b$-dependent poles, the dependence on $m_b$ identically cancels, by applying colour conservation, up to power corrections in $m_b$.

The required cancellation is non-trivial since each individual contribution that depends on $m_b$, i.e.\ $\Delta_{if}$, $\Delta_{mf}$, $\Delta_{mf'}$, $\Delta_{ff'}$, gives rise to $\log(m_b)$ terms. 
In order not to spoil the cancellation of mass singularities, the on-shell limit \mbox{$p_t^2, p_{\bar t}^2 \to m_t^2$} has to be carefully taken. Indeed, residual $\log(m_b)$ contributions can appear at higher orders in the small-$\Gamma_t$ expansion. 
We checked the cancellation also numerically, by fitting the dependence of $\boldsymbol{\delta}_{\nonfact}$ in the small-$m_b$ limit: it turns out that the mass dependence, in the on-shell top-quark limit, is at most quadratic in $m_b$. 

The fact that the non-factorisable corrections do not exhibit a logarithmically enhanced behaviour, in the small-mass limit
of an external particle, can be understood since the role of the mass is to screen collinear singularities, while the non-factorisable
corrections originate from soft wide-angle radiation.
Thus, for small $m_b$, it would be possible to simplify the
computation of the finite remainder of the non-factorisable one-loop 
corrections by setting the mass to zero and neglecting power corrections in $m_b$.
This will change individual singular
loop integrals, but not the final result for 
$\boldsymbol{\delta}_{\nonfact}$. The possibility to set the external masses
to zero could be a great advantage for the computation of the non-factorisable
corrections at higher perturbative orders, since the number of kinematic
scales is reduced. Its feasibility has to be further investigated.

The logarithmic dependence on $\Gamma_t$ is expected since the factorisable corrections only carry an overall resonant factor $1/(K_t K_{\bar t})$, where \mbox{$K_t = p_t^2 - \mu_t^2$} and \mbox{$K_{\bar t} = p_{\bar t}^2 - \mu_t^2$}, while real corrections are enhanced by large logarithms of $\Gamma_t/m_t$ originating from soft-gluon emissions off the resonances. 
These logarithms must cancel against analogous logarithmic terms from the non-factorisable virtual corrections. 
Indeed, it is well-known that, for observables sufficiently inclusive over the resonances, the cross section is finite in the limit \mbox{$\Gamma_t \to 0$} and that the remaining finite-width effects yield rather small contributions of order $\Gamma_t/m_t$~\cite{Melnikov:1993np, Fadin:1993kt}. 
We anticipate that, in Sec.~\ref{sec:ttxoffshell_NNLO_results}, we will exploit this property, which holds at any loop order, to numerically determine the non-factorisable two-loop corrections.

Starting from the expressions of the soft scalar integrals $\Delta$ given in Appendix~\ref{app:Appendix0}, we can analytically extract the explicit logarithmic dependence of the non-factorisable corrections in Eq.~\eqref{eq:deltanonfact_M1M0_ttx} on $\Gamma_t$. As expected, only a single power of $\log(\Gamma_t)$ can arise per loop order, as observed in Ref.~\cite{Falgari:2013gwa}, and we can write the explicit  $\Gamma_t$ dependence as
\begin{equation}
\boldsymbol{\delta}_{\nonfact}  = \boldsymbol{\mathcal{C}}^{(1)}_{\log(\Gamma_t)} \log\left( \frac{\Gamma_t}{Q_{h}} \right) + \boldsymbol{\mathcal{C}}^{(1)}_0 +\mathcal{O}\left ( \frac{\Gamma_t}{Q_{h}}\right)
	\label{eq:oneloop-nonfact_Gtdep} \,,
\end{equation}
where the superscript of the coefficients refers to the one-loop order, and $\boldsymbol{\mathcal{C}}_0^{(1)}$ does not depend on $\Gamma_t$.
The explicit dependence on $\Gamma_t$ originates from the overall scaling factor 
\begin{equation}
	\left( \frac{\mu^2}{\Gamma_t^2}\right)^{\epsilon} \left(-1 + i o^+ \right)^{\epsilon} \,,
	\label{eq:overall-factor_Gammat}
\end{equation}
with $i o^+$ an infinitesimal and positive imaginary part.
This suggests that the natural scale choice $\mu$ for the non-factorisable corrections is \mbox{$\mu_{\mathrm{soft}} \sim \Gamma_t$}, in order to avoid potentially large logarithms. 
In Eq.~\eqref{eq:oneloop-nonfact_Gtdep}, spurious terms of $\mathcal{O}(\Gamma_t/Q_{h})$ arise from the numerical evaluation of the scalar integrals appearing in the $\Delta$ functions within Eq.~\eqref{eq:deltanonfact_M1M0_ttx}.

Based on the obtained explicit logarithmic dependence on $\Gamma_t$, we observe that the coefficient $\boldsymbol{\mathcal{C}}_{\log(\Gamma_t)}^{(1)}$ of $\log(\Gamma_t)$ is fully determined by the pole structure in Eq.~\eqref{eq:nonfact-poles}, namely
\begingroup
\allowdisplaybreaks
\begin{align}
	\boldsymbol{\mathcal{C}}_{\log(\Gamma_t)}^{(1)} = \frac{\alphas}{2 \pi} 2 \biggl\{ &- C_F \left( 2 \mathcal{C}_{mm}^{(-\epsilon)} + \mathcal{C}_{mf}^{(-\epsilon)}(t,b) + \mathcal{C}_{mf}^{(-\epsilon)}(\bar t, \bar b) \right) 
	+ \mathbf{T}_{b} \cdot \mathbf{T}_{\bar b}\left( \mathcal{C}_{mm'}^{(-\epsilon)} + \mathcal{C}_{ff'}^{(-\epsilon)} \right)  \notag \\
	&+ \sum_{k=1,2} \!\!\left[ \mathbf{T}_b \cdot \mathbf{T}_k\, \left( \mathcal{C}_{im}^{(-\epsilon)} + \mathcal{C}_{if}^{(-\epsilon)}(k,b) \right) + (b \leftrightarrow \bar b) \right]  \biggr\} \,,
	\label{eq:coefficient_logGt}
\end{align}
where the coefficients $\mathcal{C}_{ij}^{(-\epsilon)}$ of the $1/\epsilon$ poles are given by:
\begin{align}
	&\mathcal{C}_{mm}^{(-\epsilon)} = 1 ~,~~~~ \hspace{3.6cm} \mathcal{C}_{mf}^{(-\epsilon)}(t,b) = \frac{1}{2 v_{tb}}\log\left( \frac{1-v_{tb}}{1+ v_{tb}} \right)  \,, \notag \\
	&\mathcal{C}_{mm'}^{(-\epsilon)} = \frac{1}{2 v_{t\bar t}}\log\left( \frac{1-v_{t\bar t}}{1+ v_{t\bar t}} \right) + \frac{i \pi}{v_{t\bar t}} ~,~~~~  \mathcal{C}_{ff'}^{(-\epsilon)} =  -\frac{1}{2 v_{b\bar b}}\log\left( \frac{1-v_{b\bar b}}{1+ v_{b\bar b}} \right) - \frac{i \pi}{v_{b\bar b}} \,, \notag \\
	&\mathcal{C}_{im}^{(-\epsilon)} = \frac{i \pi}{2} ~,~~~~\hspace{3.4cm}   \mathcal{C}_{if}^{(-\epsilon)}(k,b) = \log\left(\frac{ \raisedhat{p}_{b} \cdot \raisedhat{p}_k}{ \raisedhat{p}_{t} \cdot \raisedhat{p}_k} \frac{m_t}{m_b} \right) -\frac{i \pi}{2} ~~\forall k= 1,2 \,.
\end{align}
\endgroup

\subsection{Double-pole approximation at two-loop level}
\label{sec:DPA_NNLO}
The separation into factorisable and non-factorisable corrections as part of the construction of the DPA at two-loop order in QCD is significantly more involved than in the one-loop case.
First, there is a factorisable two-loop contribution given by diagrams where the insertion of a two-loop correction leads to a complete factorisation into virtual amplitudes involving the production or decay subprocesses separately. 
Second, there are contributions that can be attributed to the interference between factorisable and non-factorisable one-loop corrections. The latter could be regarded as factorisable one-loop corrections ``dressed" by the exchange of a soft gluon between production and decay stages. 
Third, there are genuinely non-factorisable two-loop corrections.

A first attempt in organising multi-leg two-loop corrections in pole approximation was performed more than twenty years ago in Ref.~\cite{Chapovsky:2001zt}, for QED corrections to production and decay of a neutral resonance. 
The consistent extension to QCD corrections is more involved due to the complicated gauge structure arising from the gluon self-coupling and has, to the best of our knowledge, not been presented to date.

Given that off-shell $t \bar t$ production is effectively a \mbox{$2\to 4$} process (from the QCD viewpoint), two-loop six-point functions with internal and external masses are expected to appear in the calculation of the non-factorisable corrections.
Certain simplifications may occur, as only soft-gluon exchanges contribute to the double-pole expansion at leading power. Nonetheless, the explicit evaluation of the non-factorisable corrections is beyond the scope of the present paper.

In this subsection, we will concentrate on the detailed construction of the genuinely factorisable two-loop corrections for the process~\eqref{eq:off-shell_ttx_dilepton}, as a first step towards the evaluation of the complete two-loop QCD amplitudes in DPA. In Sec.~\ref{sec:Gammat_extrapolation} we will present a method to infer the remaining non-factorisable corrections based on a numerical fit of the \mbox{$\Gamma_t \to 0$} behaviour, combined with an on-shell matching procedure.

\subsubsection[Factorisable two-loop corrections for off-shell \texorpdfstring{$t \bar t$}{ttx} production]{Factorisable two-loop corrections for off-shell \texorpdfstring{$\boldsymbol{t \bar t}$}{ttx} production}
\label{sec:fact_twoloop}
Analogously to the one-loop construction, the factorisable two-loop corrections involve production and decay subprocesses separately.
The only connection between the different stages is through the momentum flow and helicity states of the resonant top quarks. 
The factorisable two-loop corrections can schematically be decomposed into three building blocks: 
\begin{itemize}
	\item[(1)] two-loop corrections to $t \bar{t}$ production times tree-level decays ($\M^{(2;0)}$);
	\item[(2)] one-loop corrections to $t \bar{t}$ production times one-loop corrections to either the top \textit{or} the anti-top quark decay, and tree-level production times one-loop corrections to the top \textit{and} the anti-top quark decays ($\M^{(1;1)}$);
	\item[(3)] tree-level production times two-loop corrections to the top \textit{or} the anti-top quark decay ($\M^{(0;2)}$).
\end{itemize}
Thus, for a fixed helicity configuration $\{ \vec{h}_I, \vec{h}_{R_t}, \vec{h}_{R_{\bar{t}}} \}$ of the external particles, the factorisable two-loop corrections can be written as
\begin{align}
	\ket{ \M^{(2)}_{\mathrm{fact}} (\vec{h}_I, \vec{h}_{R_t}, \vec{h}_{R_{\bar{t}}}) } \equiv \ket{ \M^{(2;0)}(\vec{h}_I, \vec{h}_{R_t}, \vec{h}_{R_{\bar{t}}}) } + \ket{ \M^{(1;1)}(\vec{h}_I, \vec{h}_{R_t}, \vec{h}_{R_{\bar{t}}}) } + \ket{ \M^{(0;2)}(\vec{h}_I, \vec{h}_{R_t}, \vec{h}_{R_{\bar{t}}}) } \,,
	\label{eq:fact_twoloop_ttx}
\end{align}
with 
\begin{align}
	\hspace{-.18cm}\ket{ \M^{(2;0)}(\vec{h}_I, \vec{h}_{R_t}, \vec{h}_{R_{\bar{t}}}) } = 
	\frac{1}{(p_t^2 -\mu_t^2)(p_{\bar{t}}^2 -\mu_t^2)}
	 \sum_{\{ \vec{\mathfrak{c}} \}} \sum_{\lambda_t,\lambda_{\bar{t}} }  \biggl\{ \!
\M_{t \bar t}^{(2), \vec{\mathfrak{c}}}(\vec{h}_I, \lambda_t,\lambda_{\bar{t}})\,\M_{t}^{(0)}(\vec{h}_{R_t},\lambda_t)\,\M_{\bar{t}}^{(0)}(\vec{h}_{R_{\bar{t}} },\lambda_{\bar{t}}) \!\biggr\} \ket{ \mathfrak{C}_{\vec{\mathfrak{c}}} } \,,
	\label{eq:fact_twoloop_2x0}
\end{align}
\begin{align}
	\hspace{-.1cm}\ket{ \M^{(1;1)}(\vec{h}_I, \vec{h}_{R_t}, \vec{h}_{R_{\bar{t}}}) } =&
	\frac{1}{(p_t^2 -\mu_t^2)(p_{\bar{t}}^2 -\mu_t^2)} \,\times \notag \\
	\sum_{\{ \vec{\mathfrak{c}} \}}& \sum_{\lambda_t,\lambda_{\bar{t}} } 
	 \biggl\{
	\M_{t \bar t}^{(1), \vec{\mathfrak{c}}}(\vec{h}_I, \lambda_t,\lambda_{\bar{t}})  
		\biggl( \M_{t}^{(1)}(\vec{h}_{R_t},\lambda_t)\,\M_{\bar{t}}^{(0)}(\vec{h}_{R_{\bar{t}} },\lambda_{\bar{t}})  
	+ \M_{t}^{(0)}(\vec{h}_{R_t},\lambda_t)\,\M_{\bar{t}}^{(1)}(\vec{h}_{R_{\bar{t}} },\lambda_{\bar{t}}) \biggr)  \notag \\
	&\hspace{0.5cm}
	+ \M_{t \bar t}^{(0), \vec{\mathfrak{c}}}(\vec{h}_I, \lambda_t,\lambda_{\bar{t}})\,\M_{t}^{(1)}(\vec{h}_{R_t},\lambda_t)\,\M_{\bar{t}}^{(1)}(\vec{h}_{R_{\bar{t}} },\lambda_{\bar{t}}) \biggr\} \ket{ \mathfrak{C}_{\vec{\mathfrak{c}}} }\,,
	\label{eq:fact_twoloop_1x1}
\end{align}
\begin{align}
	\hspace{-1.5cm}\ket{ \M^{(0;2)}(\vec{h}_I, \vec{h}_{R_t}, \vec{h}_{R_{\bar{t}}}) }  &=
	\frac{1}{(p_t^2 -\mu_t^2)(p_{\bar{t}}^2 -\mu_t^2)} 
	 \sum_{\{ \vec{\mathfrak{c}} \}} \sum_{\lambda_t,\lambda_{\bar{t}} } \biggl\{
	\M_{t \bar t}^{(0), \vec{\mathfrak{c}}}(\vec{h}_I, \lambda_t,\lambda_{\bar{t}}) \biggl( \M_{t}^{(2)}(\vec{h}_{R_t},\lambda_t)\,\M_{\bar{t}}^{(0)}(\vec{h}_{R_{\bar{t}} },\lambda_{\bar{t}})  \notag \\
	&\hspace{4.5cm}
	+ \M_{t}^{(0)}(\vec{h}_{R_t},\lambda_t)\,\M_{\bar{t}}^{(2)}(\vec{h}_{R_{\bar{t}} },\lambda_{\bar{t}}) \biggr)  \biggr\} \ket{ \mathfrak{C}_{\vec{\mathfrak{c}}} } \,,
	\label{eq:fact_twoloop_0x2}
\end{align}
with colour and helicity sums as in Eq.~\eqref{eq:fact_oneloop_ttx}.
As in the one-loop case the abstract vector $\ket{ \mathfrak{C}_{\vec{\mathfrak{c}}} }$ is an element of the chosen colour basis, while $\M_{t \bar t}^{(l),\vec{\mathfrak{c}} }$ stands for the corresponding $l$-loop colour-stripped amplitude. 
As already noted in Sec.~\ref{sec:fact_oneloop}, all matrix elements appearing in the r.h.s.\ of Eqs.~\eqref{eq:fact_twoloop_2x0}--\eqref{eq:fact_twoloop_0x2} have to be evaluated on the set of on-shell momenta $\{ \raisedhat{p}_i \}$ obtained by applying the on-shell projection described in Appendix~\ref{app:Appendix_mapping}. Off-shell effects are kept in the top-quark propagators.
After summing over the helicities of the external particles, the complete factorisable two-loop corrections in Eq.~\eqref{eq:fact_twoloop_ttx} have to be interfered with the Born matrix element in DPA~\eqref{eq:LO_PA_ttx}, thus obtaining the factorisable part of the double-virtual contribution denoted by $2\mathrm{Re}\{ \braket{ \M^{(0)}_{\PA} | \M^{(2)}_{\mathrm{fact}}} \}$. 

As discussed in Sec.~\ref{sec:fact_oneloop}, interferences between different helicity states 
\mbox{$\lambda_t \ne \lambda'_t$} and/or \mbox{$\lambda_{\bar t} \ne \lambda'_{\bar t}$} are strongly suppressed in the loop--tree interference $\braket{\M^{(0)}_{\PA} | \M^{(1)}_{\fact}}$, numerically accounting for less than a per mille of the entire factorisable corrections. Therefore, we neglect such interference effects in the contributions $\braket{\M^{(0)}_{\PA} | \M^{(2;0)} }$ and $\braket{\M^{(0)}_{\PA} | \M^{(0;2)} }$.
 By doing so, our construction only requires the knowledge of the squared Born matrix elements and loop--tree interferences for the production of a polarised $t \bar t$ pair, and for the decay of polarised (anti-)top quarks.
At the level of the two-loop finite remainders, we can directly exploit the results of Ref.~\cite{Chen:2017jvi} for the on-shell $t \bar t$ production, and the analytic results reported in Appendix~\ref{app:Appendix1} for the (anti-)top-quark decay.

To implement the factorisable two-loop corrections, tree-level, one-loop and two-loop helicity amplitudes for $t\bar{t}$ production and (anti-)top-quark decays are required.
For the evaluation of the tree-level and one-loop amplitudes, we exploit the automated amplitude generators \Recola or \OpenLoops. 
This allows us to easily construct the interference $\braket{\M^{(0)}_{\PA} | \M^{(1;1)}}$, starting from amplitudes decomposed on a colour $\otimes$ spin basis. The procedure is analogous to that described in Sec.~\ref{sec:fact_oneloop} for the loop--tree interference $\braket{\M^{(0)}_{\PA} | \M^{(1)}_{\fact}}$.

Regarding the polarised two-loop amplitudes for the (anti-)top-quark decay, we construct them starting from the heavy-to-light two-loop form factors~\cite{Bonciani:2008wf}. We rely on the so-called \textit{massification} procedure~\cite{Penin:2005eh,Mitov:2006xs,Becher:2007cu,Engel:2018fsb,Wang:2023qbf}, which is valid in the high-energy limit and allows us to recover the dependence on the bottom-quark mass up to power corrections. 
Details on this construction are given in Sec.~\ref{sec:heavy-to-light_FF}, while analytic results for the two-loop finite remainders are presented in Appendix~\ref{app:Appendix1}.

The polarised two-loop amplitudes for on-shell $t\bar{t}$ production have been computed in Ref.~\cite{Chen:2017jvi}. 
More details on their implementation in our \Matrix framework will be given in Sec.~\ref{sec:onshell_twoloop_amplitudes}.
We also observe that effects due to massive bottom-quark loops are not included in Ref.~\cite{Chen:2017jvi}, where a single mass scale (the top-quark mass) is kept in the two-loop amplitudes. Since we are performing the off-shell computation in the 4FS, by retaining the full dependence on the bottom-quark mass in the tree-level and one-loop amplitudes, it is appropriate to restore $m_b$ effects also in the two-loop on-shell $t \bar t$ amplitudes. This can be achieved by applying again the aforementioned mass factorisation formula, recently extended in Ref.~\cite{Wang:2023qbf} to account for effects from the presence of several heavy flavours in the theory. More details will be given in Sec.~\ref{sec:mb-effects_onshell_ttx_amplitudes}.

We have performed several consistency checks of our construction of the factorisable two-loop corrections. 
Besides the tests of the IR pole cancellation and correct scale dependence, a stringent check of the contribution $2 \mathrm{Re}\{ \braket{\M^{(0)}_{\PA} | \M^{(2;0)} } \}$ was perfomed: we verified that, in the \mbox{$\Gamma_t \to 0$} limit, this term numerically coincides with the double-virtual contribution to on-shell $t \bar t$ production multiplied by the branching ratio for the top and anti-top quark decays into leptons. 

\subsubsection{Heavy-to-light quark form factors}\label{sec:heavy-to-light_FF}
In this section, we discuss the construction of the two-loop QCD amplitudes for the (anti-)top-quark decay, at fixed helicity of the decaying quark. 
The required full set of two-loop master integrals for the heavy-to-light form factors has, in principle, been computed in Ref.~\cite{Chen:2018dpt}, for arbitrary momentum transfer, and expressed in terms of generalised polylogarithms (GPLs). These integrals would allow us to determine, in a fully analytic way and without any approximation, the two-loop QCD corrections to the decay of a heavy fermion into a massive lighter fermion. Indeed, in Ref.~\cite{Engel:2018fsb} these master integrals were exploited to construct the two-loop amplitudes by retaining the full dependence on the masses of both the heavy and light fermions. However, in Ref.~\cite{Engel:2018fsb} it was noticed that the numerical evaluation of the appearing GPLs is time-consuming and therefore difficult to implement directly in an NNLO Monte Carlo event generator.

Here, we are interested in a process whose typical energy scale $Q$ is of the same order as the top-quark mass. Since the mass ratio $m_b/ m_t$ with $m_b$ denoting the mass of the bottom quark is small enough, an expansion in the bottom-quark mass is fully justified. In other words, the mass factorisation formula%
\footnote{For a thorough discussion and derivation of the mass factorisation formula, we refer the reader to Sec.2.2 of Ref.~\cite{Devoto:2024nhl}.} provides a very good approximation of the exact two-loop amplitudes, as observed in Ref.~\cite{Engel:2018fsb}. Therefore, we construct the two-loop amplitudes for the top-quark decay via a massification of the corresponding heavy-to-light form factor with massless FS quarks as obtained in Ref.~\cite{Bonciani:2008wf}.
\\
\\
In the following, we consider the decay of a top quark into a massive bottom quark and an off-shell $W$ boson
\begin{equation}
	t(p_t, m_t) \rightarrow b(p_b, m_b) + W^+(\rightarrow l^+ \nu_l) \,.
\end{equation}
We label the momentum of the off-shell $W$ boson with \mbox{$q=p_3+p_4$}, where $p_3$ and $p_4$ refer to the momenta of the neutrino and the (massless) charged lepton, respectively. 

In order to build the massified two-loop amplitudes for the top-quark decay, we start from the results of Ref.~\cite{Bonciani:2008wf} for the heavy-to-light quark form factors, where the light quark is considered massless (\mbox{$p_b^2=0$}). 
In Ref.~\cite{Bonciani:2008wf}, the calculation has been carried out in dimensional regularisation with a suitable prescription to handle $\gamma_5$. More specifically, a $\gamma_5$ that anticommutes with all Dirac matrices $\gamma_{\mu}$ in \mbox{$d= 4 -2 \epsilon$} dimensions was chosen. 
The UV renormalisation has been performed in a mixed scheme: on-shell scheme for the heavy- and light-quark wave functions and the heavy-quark mass, and $\MSbar$ scheme for the strong coupling. 

The most general QCD vertex correction in the SM can be described in terms of six form factors $F_i$ and $G_i$ (\mbox{$i = 1,2,3$}), associated with vector and axial-vector structures, respectively. However, since the bottom quark is treated as massless, only three of the above form factors are independent, and the vertex structure can be expressed as
\begin{equation}
	V^{\nu}(q, \tilde{q}) = -i\frac{g_W}{2\sqrt 2} V_{bt}\, \biggl[ G_1(q^2)\gamma^{\nu}(1-\gamma_5) +  
	\frac{1}{2 m_t} \biggl( G_2(q^2) q^{\nu} + G_3(q^2) \tilde{q}^{\nu} \biggr) (1+\gamma_5)  \biggr] \,,
	\label{eq:heavy-to-light_QCDvertex}
\end{equation}
where $\nu$ is the Lorentz index carried by the off-shell $W$ boson, \mbox{$g_W = e/\sin\theta_w$} is the weak-interaction coupling constant and $V_{bt}$ represents the CKM matrix element. The $W$-boson momentum transfer is \mbox{$q = p_t - p_b = p_3 + p_4$} while \mbox{$\tilde{q} = p_t + p_b$}. 

The bare form factors are expanded in powers of the bare strong coupling $\alphas^0$ as
\begin{equation}
	G_i(q^2) = \sum_{n=0}^2 \biggl( \frac{\alphas^0}{\pi} \biggr)^{\!n} G_i^{(n)}(q^2) + \mathcal{O}(\alphas^3) \,~~~~~~~\forall \,i=1,2,3\,,
\end{equation}
where, at tree-level, \mbox{$G_1^{(0)} = 1$} and \mbox{$G_2^{(0)} = G_3^{(0)} = 0$}.

By taking into account the $W$-boson decay into leptons, the bare matrix element (up to two-loop order) for the top-quark decay into a massless bottom quark is given by
\begin{align}
	\hspace{-0.7cm} \M^{\mathrm{bare}} &= \frac{g_W^2}{8} V_{bt} \,\bar{u}(p_b) \sum_{n=0}^2 
	\biggl(\! \frac{\alphas^0}{\pi} \! \biggr)^{\!n} \biggl[ G_1^{(n)}(q^2)\gamma^{\nu}(1-\gamma_5) +  
	\frac{1}{2 m_t} \biggl(\! G_2^{(n)}(q^2) q^{\nu} + G_3^{(n)}(q^2) \tilde{q}^{\nu} \! \biggr) (1+\gamma_5)  \biggr] u(p_t)  \notag \\
	&\hspace{4cm} \times  \frac{i}{q^2 - \mu_W^2}  \bar{u}(p_3) \gamma_{\nu} (1-\gamma_5) v(p_4) \,,
\end{align}
where \mbox{$\mu^2_W = m_W^2 - i \Gamma_W m_W $} is the complex $W$-boson mass. The dependence of $\M^{\mathrm{bare}}$ on the external momenta and masses is left understood. 
It is worth mentioning that $G_2^{(n)}$ does not contribute at any perturbative order since \mbox{$\bar{u}(p_3) (\slashed{p}_3+ \slashed{p}_4)(1-\gamma_5) v(p_4) = 0$} due to the Dirac equation (for lepton and neutrino).

For our application, we want to select a specific helicity state of the top quark. 
To this end, we define the spin projectors 
\begin{equation}
	P_{\Sigma}(\pm s_t) \equiv \frac{1 \pm \gamma_5 \slashed{s}_t}{2} \,,
	\label{eq:spin_projectors}
\end{equation}
with $s_t$ being the spin four-vector obtained by boosting the rest-frame spin four-vector $s_t'$ along the direction of the top-quark four-momentum $p_t$. 
The explicit expression of $s_t$ is 
\begin{equation}
	s_t = \frac{1}{m_t} \left( |\vec{p}_t|, E_t \frac{\vec{p}_t}{|\vec{p}_t|} \right)
	\label{eq:top_spin-four-vector}
\end{equation}
with \mbox{$s_t \cdot p_t = 0$} and \mbox{$s_t^2 = -1$}. 
The spin projectors are idempotent and orthogonal, i.e.\
\begingroup
\allowdisplaybreaks
\begin{equation}
	P_{\Sigma}(\pm s_t) P_{\Sigma}(\pm s_t) = P_{\Sigma}(\pm s_t) ~~~~~\text{and}~~~~~ P_{\Sigma}(\pm s_t) P_{\Sigma}(\mp s_t) = 0 \,.
\end{equation}	
\endgroup
Thus, the polarised bare matrix element (up to two-loop order) describing the decay of a top quark with polarisation \mbox{$\pm s_t$} reads
\begingroup
\allowdisplaybreaks
\begin{align}
	\M^{\mathrm{bare}}_{(\pm s_t)} &= \frac{g_W^2}{8} V_{bt} \,\bar{u}(p_b) \sum_{n=0}^2 \biggl( \frac{\alphas^0}{\pi} \biggr)^{\!n} \biggl[ G_1^{(n)}(q^2)\gamma^{\nu}(1-\gamma_5) +  
	\frac{1}{m_t} G_3^{(n)}(q^2) p_b^{\nu} (1+\gamma_5)  \biggr] P_{\Sigma}(\pm s_t) u(p_t)  \notag \\
	&\hspace{4cm} \times  \frac{i}{q^2 - \mu_W^2}  \bar{u}(p_3) \gamma_{\nu} (1-\gamma_5) v(p_4) \notag \\
	&= g_W^2 V_{bt} \, \frac{i}{q^2 - \mu_W^2} \,\sum_{n=0}^2 \biggl( \frac{\alphas^0}{\pi} \biggr)^{\!n} \biggl[ G_1^{(n)}(q^2) M_{(\pm s_t)}^{(G_1)} + G_3^{(n)}(q^2) M_{(\pm s_t)}^{(G_3)} \biggr] \,,
\end{align}
\endgroup
where, for brevity, we introduce the notation $M_{(\pm s_t)}^{(G_i)}$ to refer to the Lorentz structure corresponding to the form factor $G_i$.

Having the previous discussion in mind, we decompose the polarised UV-renormalised amplitude (in the hybrid scheme mentioned above) as
\begin{align}
	\M_{(\pm s_t)}^{\mathrm{ren}}(\mu, \epsilon) = g_W^2 V_{bt} \, \frac{i}{q^2 - \mu_W^2} \, \left[ \M_{(\pm s_t)}^{(0)} +  \frac{\alphasNf}{\pi} \M_{(\pm s_t)}^{(1), \mathrm{ren}}(\mu, \epsilon) +  \left( \frac{\alphasNf}{\pi} \right)^{\!2} \!\! \M_{(\pm s_t)}^{(2), \mathrm{ren}}(\mu, \epsilon) \right] + \mathcal{O}(\alphas^3) \,,
	\label{eq:top_decay_ampdecomp}
\end{align}
where $\mu$ is the renormalisation scale.
The strong coupling \mbox{$\alphasNf \equiv \alphasNf(\mu)$} is renormalised in a theory with \mbox{$n_f = n_l +  n_h$} flavours, where $n_l$ denotes the number of massless quarks and $n_h$ is the number of heavy quarks with mass $m_t$, according to the conventions of Ref.~\cite{Bonciani:2008wf}.
In our specific case, \mbox{$n_l=4$} and \mbox{$n_h=1$}.
The dependence of the renormalised scattering amplitude $\M_{(\pm s_t)}^{\mathrm{ren}}$ on the kinematical variables has been suppressed.

The one-loop renormalised amplitude (stripped off the couplings and the $W$-boson propagator) is given by
\begin{align}
	\M_{(\pm s_t)}^{(1),\mathrm{ren}}(\mu, \epsilon) =  \mathcal{N} \left\{ \biggl( \sum_{i=-2}^{1} \,G_1^{(1l),i} \,\epsilon^i \biggr) M_{(\pm s_t)}^{(G_1)}+ 
	 \biggl(\sum_{i=0}^{1} \,G_3^{(1l),i} \,\epsilon^i \biggr) M_{(\pm s_t)}^{(G_3)} \right\} \,,
	 \label{top_decay_1Lmassless}
\end{align}
where the normalisation factor \mbox{$\mathcal{N} = \Gamma(1+\epsilon) e^{\gamma_E \epsilon} \left( \mu^2/m_t^2 \right)^{\epsilon} $} takes into account the different convention in the overall amplitude normalisation \mbox{$S_\epsilon = (4\pi)^\epsilon e^{-\epsilon \gamma_E}$} with respect to \mbox{$C(\epsilon) = (4\pi)^\epsilon \Gamma(1+ \epsilon)$} used in Ref.~\cite{Bonciani:2008wf}.
The UV-renormalised one-loop form factors $G_1^{(1l)}$ and $G_3^{(1l)}$ are expanded in Laurent series up to $\mathcal{O}(\epsilon)$, and the coefficients of this expansion, namely $G_1^{(1l),i}$ and $G_3^{(1l),i}$, are functions of \mbox{$x \equiv q^2/m_t^2 = 1- 2 p_t \cdot p_b/m_t^2 $} and \mbox{$L_{\mu} \equiv \log\left( \mu^2/m_t^2 \right)$}. Their analytic expressions can be directly extracted from Ref.~\cite{Bonciani:2008wf}.\footnote{Note that, differently from Ref.~\cite{Bonciani:2008wf}, we have included the colour factor $C_F$ in the definition of the Laurent coefficients $G_1^{(1l),i}$ and $G_3^{(1l),i}$. The same comment also holds at the two-loop order.}

In order to construct the two-loop massified finite remainder starting from the UV-renormalised massless amplitudes, it is crucial to know also the $\mathcal{O}(\epsilon^2)$ terms of the massless one-loop amplitude. These contributions are partially provided in Ref.~\cite{Bonciani:2008wf} if one expands the prefactor $\mathcal{N}$ in $\epsilon$. 
The entire $\mathcal{O}(\epsilon^2)$ contribution can be derived 
from Ref.~\cite{Beneke:2008ei} by identifying
\begingroup
\allowdisplaybreaks
\begin{align}
	\tilde{G}_1^{(1l),2}(x) &= \frac{1}{4}F_1^{(1),2}(1-x) \,,\\
	\tilde{G}_3^{(1l),2}(x) &= \frac{1}{4}\biggl( F_2^{(1),2}(1-x) + \frac{2}{1-x}F_3^{(1),2}(1-x) \biggr) \,,
\end{align}
\endgroup	
where $F_i^{(1),2}$ stands for the $\mathcal{O}(\epsilon^2)$ coefficient of the functions $F_i^{(1)}$ defined in Ref.~\cite{Beneke:2008ei}.
Therefore, we can rewrite Eq.~\eqref{top_decay_1Lmassless} as
\begingroup
\allowdisplaybreaks
\begin{align}
\M_{(\pm s_t)}^{(1),\mathrm{ren}}(\mu, \epsilon) =  \biggl( \sum_{i=-2}^{2} \,\tilde{G}_1^{(1l),i} \,\epsilon^i \biggr) M_{(\pm s_t)}^{(G_1)}+ 
	 \biggl(\sum_{i=0}^{2} \,\tilde{G}_3^{(1l),i} \,\epsilon^i \biggr) M_{(\pm s_t)}^{(G_3)}  + \mathcal{O}(\epsilon^3)\,,
	 \label{top_decay_1Lmassless_complete}
\end{align}
\endgroup	
where, for \mbox{$i < 2$}, $\tilde{G}_1^{(1l),i}$ and $\tilde{G}_3^{(1l),i}$ coincide with $G_1^{(1l),i}$ and $G_3^{(1l),i}$ except for terms arising from the $\epsilon$ expansion of $\mathcal{N}$ up to  $\mathcal{O}(\epsilon)$.

Analogously, the renormalised massless two-loop amplitude reads
\begingroup
\allowdisplaybreaks
\begin{align}
	\M_{(\pm s_t)}^{(2),\mathrm{ren}}(\mu, \epsilon)   &= \mathcal{N}^{\,2} \biggl\{ \biggl( \sum_{i=-4}^{0} \,G_1^{(2l),i}\epsilon^i \biggr) M_{(\pm s_t)}^{(G_1)} + 
	  \biggl( \sum_{i=-2}^{0} \,G_3^{(2l),i}\epsilon^i \biggr) M_{(\pm s_t)}^{(G_3)}\biggr\} \notag \\
	 &=  \biggl( \sum_{i=-4}^{0} \,\tilde{G}_1^{(2l),i}\epsilon^i \biggr) M_{(\pm s_t)}^{(G_1)} + 
	  \biggl( \sum_{i=-2}^{0} \,\tilde{G}_3^{(2l),i}\epsilon^i \biggr) M_{(\pm s_t)}^{(G_3)} + \mathcal{O}(\epsilon) \,,
	 \label{top_decay_2Lmassless}
\end{align}
\endgroup
where the two-loop form factors $G_1^{(2l)}$ and $G_3^{(2l)}$ are, as at one-loop level, functions of $x$ and $L_{\mu}$, and they have been computed in Ref.~\cite{Bonciani:2008wf}.  

Before applying the massification procedure, we need to decouple the top quark from the running of $\alphas$. This can be achieved by expanding Eq.~\eqref{eq:top_decay_ampdecomp} in $\alphasNl$ through a finite renormalisation shift (see e.g.\ Ref.~\cite{Grozin:2005yg}) given by
\begingroup
\allowdisplaybreaks
\begin{equation}
	\alphasNf(\mu) = \alphasNl(\mu) \biggl[ 1 +  \frac{\alphasNl(\mu)}{\pi}  \frac{T_R}{3} n_h\biggl( \Gamma(\epsilon) e^{\gamma_E \epsilon} \biggl( \frac{\mu^2}{m_t^2} \biggr)^{\epsilon} -\frac{1}{\epsilon} \biggl) +\mathcal{O}(\alphas^2)   \biggr]  \,.
	\label{oneloop_decoupling}
\end{equation}
\endgroup
It follows that, in the decoupling scheme, Eq.~\eqref{eq:top_decay_ampdecomp} becomes
\begin{align}
	\M_{(\pm s_t)}^{\mathrm{ren}}(\mu, \epsilon) = \frac{i g_W^2 V_{bt} }{q^2 - \mu_W^2} \, \left[ \M_{(\pm s_t)}^{(0)} +  \frac{\alphasNl}{\pi} \M_{(\pm s_t)}^{(1), \mathrm{ren, dec}}(\mu, \epsilon) + \biggl( \frac{\alphasNl}{\pi} \biggr)^2 \!\! \M_{(\pm s_t)}^{(2), \mathrm{ren, dec}}(\mu, \epsilon) \right] + \mathcal{O}(\alphas^3) \,,
	\label{eq:top_decay_ampdecomp_decoupled}
\end{align}
where \mbox{$\alphasNl \equiv \alphasNl(\mu)$} and
\begin{align}
	\M_{(\pm s_t)}^{(1),\mathrm{ren, dec}}(\mu, \epsilon) &=  \M_{(\pm s_t)}^{(1),\mathrm{ren}}(\mu, \epsilon) \,,  \\
	 \M_{(\pm s_t)}^{(2),\mathrm{ren, dec}}(\mu, \epsilon) &=  \M_{(\pm s_t)}^{(2),\mathrm{ren}}(\mu, \epsilon) 
	 + \frac{T_R n_h}{3}\biggl( \frac{\Gamma(\epsilon)}{\Gamma(1+\epsilon)} \mathcal{N}-\frac{1}{\epsilon} \biggl) \M_{(\pm s_t)}^{(1),\mathrm{ren}}(\mu, \epsilon) \,.
	 \label{top_decay_masslessAmp_decoupled}
\end{align}
In order to simplify the notation, we will drop the superscript ``$\mathrm{ren, dec}$" in the remaining formulae of this section, and we will use $\M_{(\pm s_t)}^{(1)}$ and $\M_{(\pm s_t)}^{(2)}$ to refer to the UV-renormalised (in the decoupling scheme) one-loop and two-loop amplitudes with massless bottom quarks.

We start the construction of the massified top-quark decay amplitudes by neglecting, in first approximation, the contributions from massive bottom-quark loops.
Therefore, according to Ref.~\cite{Mitov:2006xs}, in the small bottom-mass limit (\mbox{$m_b^2 \ll m_t^2 \sim Q^2$}), the dependence on $m_b$ can be restored (up to power corrections) as 
\begin{equation}
	\M_{(\pm s_t)}^{(m_b)}(\mu,\epsilon) = \bigl(  Z_{[Q]}^{(m_b|0)} \bigr)^{1/2} \M_{(\pm s_t)}(\mu,\epsilon) \,,
	\label{top_decay_massified}
\end{equation}
since we have a single external parton that we want to promote from massless to massive. 
$\M_{(\pm s_t)}^{(m_b)}$ follows a decomposition analogous to Eq.~\eqref{eq:top_decay_ampdecomp_decoupled}, where the weak coupling and the $W$-boson propagator have been factorised. 

The factor $Z_{[Q]}^{(m_b|0)}$ admits the following perturbative expansion in $\alphas$,
\begingroup
\allowdisplaybreaks
\begin{equation}
	Z_{[Q]}^{(m_b|0)}\biggl( \frac{\mu^2}{m_b^2}, \alphasNl; \epsilon \biggr) = 1+ \frac{\alphasNl}{\pi} Z_{[Q]}^{(1)}\biggl( \frac{\mu^2}{m_b^2}; \epsilon \biggr) + \biggl( \frac{\alphasNl}{\pi} \biggr)^2 Z_{[Q]}^{(2)}\biggl( \frac{\mu^2}{m_b^2}; \epsilon \biggr) + \mathcal{O}(\alphas^3) \,,
\end{equation}
\endgroup
where $Z_{[Q]}^{(k)}$ \mbox{$(k=1,2)$} coincide with the functions computed in Ref.~\cite{Mitov:2006xs} up to an overall factor $\left( 1/4 \right)^k$ due to the different $\alphas$ expansion.
By expanding Eq.~\eqref{top_decay_massified} in $\alphas$, we can extract the massified one-loop and two-loop amplitudes as
\begingroup
\allowdisplaybreaks
\begin{align}
	\M_{(\pm s_t)}^{(1),(m_b)}(\mu, \epsilon) &=  \frac{1}{2}Z_{[Q]}^{(1)}\,\M_{\pm s_t}^{(0)} + \M_{(\pm s_t)}^{(1)}(\mu, \epsilon)  \,,\\
	\M_{(\pm s_t)}^{(2),(m_b)}(\mu, \epsilon) &=  \frac{1}{2}\biggl( Z_{[Q]}^{(2)} - \frac{1}{4}(Z_{[Q]}^{(1)})^2 \biggr)\,\M_{(\pm s_t)}^{(0)} 
	+ \frac{1}{2}Z_{[Q]}^{(1)}\,\M_{(\pm s_t)}^{(1)}(\mu, \epsilon)  + \M_{(\pm s_t)}^{(2)}(\mu, \epsilon)  \,.
	\label{top_decay_massifiedAmp}
\end{align}
\endgroup
We point out that, in order to obtain the correct IR poles and the finite remainder, $Z_{[Q]}^{(1)}$ and $\M_{(\pm s_t)}^{(1)}$ must be expanded up to $\mathcal{O}(\epsilon^2)$. 
\\

Till now we have considered $m_b$ effects associated with the external asymptotic state. However, additional logarithmically enhanced contributions in the bottom-quark mass arise from diagrams with an insertion of a heavy bottom-quark loop -- we will refer to them as ``$n_m$\footnote{Here we will use $n_m$ to denote the number of quarks with (small) mass $m_b$, in order to distinguish from $n_h$, which is the number of quarks with (heavy) mass $m_t$. In this specific case, we have \mbox{$n_m = 1$}.} terms". 
A priori, the impact of these terms can be numerically as large as the mass logarithms recovered by the massification procedure based on Eq.~\eqref{top_decay_massified}. Therefore, it is important to include them in order to retrieve all $\epsilon$ poles, logarithmically enhanced contributions in $m_b$ and mass-independent terms. 

The inclusion of $n_m$-dependent terms leads to a modification of the universal factor $Z_{[Q]}^{(m_b|0)}$, computed in Ref.~\cite{Mitov:2006xs}, and to a non-vanishing soft function $\mathcal{S}$ starting from two-loop order.
These contributions have been first calculated in Ref.~\cite{Engel:2018fsb}, where the effects due to two different masses $m$ and $M$ with hierarchy \mbox{$m \ll M \sim Q$} are taken into account. 

In order to include these $n_m$-dependent terms in our construction of the massified two-loop top-quark decay amplitudes, we need to restart from Eq.~\eqref{eq:top_decay_ampdecomp_decoupled} and replace \mbox{$\alphasNl \to \alphas^{(n_l + n_m)}$}, since the strong coupling gets renormalised in $\MSbar$ scheme with \mbox{$n_l + n_m$} flavours. Moreover, all (implicit) occurrences of $n_l$ in the massless two-loop amplitudes~\eqref{top_decay_masslessAmp_decoupled} must be replaced by \mbox{$n_l + n_m$}.

Following the notation of Ref.~\cite{Engel:2018fsb}, the mass factorisation formula in Eq.~\eqref{top_decay_massified} is modified as
\begin{equation}
	\M_{(\pm s_t)}^{(m_b)}(\mu,\epsilon) = \biggl( \sqrt{  Z_{[Q]}^{(m_b|0)} } + \sqrt{ Z_q } \bigl|_{n_m \mathrm{terms}}  \biggr) \mathcal{S} \M_{(\pm s_t)}(\mu,\epsilon) \,,
	\label{top_decay_massified_nmterms}
\end{equation}
where the additional $n_m$-dependent terms are entirely encoded in $\sqrt{ Z_q } \bigl|_{n_m \mathrm{terms}}$ and $\mathcal{S}$. 
The function $\sqrt{ Z_q } \bigl|_{n_m \mathrm{terms}}$ corresponds to the $n_m$-dependent part of the function $\sqrt{Z_q}$ in Ref.~\cite{Engel:2018fsb}, after renormalisation of the strong coupling. 
On the other hand, the soft function $\mathcal{S}$ receives contributions from the purely soft momentum region (\mbox{$k \sim (\lambda, \lambda, \lambda)$} with \mbox{$\lambda \sim m_b/m_t \ll 1$}) and, at bare level, it can be computed from first principles as
\begin{equation}
	\mathcal{S}^{0} = 1 + \frac{(\alphas^0)^2}{4\pi} C_F \left( \frac{\mu^2}{m_t^2} \right)^2 \!\Gamma(1-\epsilon)\! \int \frac{d^d k}{i\pi^{d/2}} \frac{(-2p_t^{\mu})(-2 p_{b-}^{\nu})}{(k^2)^2 (2 p_t \cdot k)(2 p_{b -} )} \Pi_{\mu\nu}^{(n_m)}(k) \,,
\end{equation}
where, in the rest frame of the decaying top quark, the external momenta scale like \mbox{$p_b = (p_{b+}, p_{b-}, p_{b \perp})$} $\sim (0, 1, \lambda)$ and \mbox{$p_t = (p_{t+}, p_{t-}, p_{t \perp}) \sim (1, 1,0)$}.%
\footnote{For the definition of the light-cone coordinates, we used the conventions of Ref.~\cite{Engel:2018fsb}.} The function $\Pi_{\mu\nu}^{(n_m)}$ is the contribution of $n_m$ massive fermions with mass $m_b$ to the usual tensorial vacuum polarisation. 
Explicit expressions for $\sqrt{ Z_q } \bigl|_{n_m \mathrm{terms}} $ and $\mathcal{S}$ are given in Appendix~\ref{app:Appendix1}.
The inclusion of $n_m$-dependent contributions leads to rapidity divergences that cancel only when $\sqrt{ Z_q } \bigl|_{n_m \mathrm{terms}}$ and $\mathcal{S}$ are added together. However, scale-factorisation breaking effects remain in the form of $\log(m_b m_t /s)$. 
The cancellation of the $n_m$-dependent $1/\epsilon$ poles, arising on the r.h.s.\ of Eq.~\eqref{top_decay_massified_nmterms}, is achieved after decoupling the bottom quark from the running of $\alphas$, i.e.\ by applying Eq.~\eqref{oneloop_decoupling} with the substitutions \mbox{$\{ n_h \to n_m, m \to m_b \}$}, 
and expanding in $\alphasNl$.
\\

As a first test of our construction of the massified top-quark decay amplitudes, we checked that their IR structure matches that predicted by the massive subtraction operator in the minimal-subtraction scheme of Refs.~\cite{Becher:2009qa,Becher:2009cu,Ferroglia:2009ii}, properly expanded in the small-$m_b$ limit. The explicit form of this subtraction operator, labelled as $\mathcal{Z}_{m_b \ll m_t}$, is given in Appendix~\ref{app:Appendix1}.

After subtracting the IR poles, we define the finite remainder at scale $\mu$ as
\begin{align}
	&\M_{(\pm s_t)}^{(m_b), \mathrm{fin}}(\mu) = \mathcal{Z}_{m_b \ll m_t}^{-1}(\mu, \epsilon) \M_{(\pm s_t)}^{(m_b)}(\mu, \epsilon) \,,
\label{top_decay_finite_remainder}
\end{align}
with $\M_{(\pm s_t)}^{(m_b)}(\mu, \epsilon)$ defined in Eq.~\eqref{top_decay_massified_nmterms}.
Analytic expressions of the massified one-loop and two-loop finite remainders, with the inclusion of $n_m$-dependent contributions, are reported in Appendix~\ref{app:Appendix1}.
We also verified that $\M_{(\pm s_t)}^{(m_b), \mathrm{fin}}$ has the correct scale dependence up to two-loop order or, in other words, that it satisfies the renormalisation-group evolution
\begin{equation}
	\frac{d}{d\log \mu} \M_{(\pm s_t)}^{(m_b), \mathrm{fin}}(\mu) = \left[ \frac{\alphasNl}{\pi} \Gamma_{0}  + \left( \frac{\alphasNl}{\pi} \right)^{\!2}\Gamma_{1} + \mathcal{O}(\alphas^3) \right] \M_{(\pm s_t)}^{(m_b), \mathrm{fin}}(\mu) \,,
\end{equation}
where 
\begingroup
\allowdisplaybreaks
\begin{align}
  \Gamma_{0} &= C_F \biggl(  -1 + \log(1-x) + \frac{1}{2} \log \frac{m_t^2}{m_b^2}  \biggr) + \mathcal{O}\left(\! \frac{m_b^2}{m_t^2} \!\right)  \quad\text{and}\\
	\Gamma_{1} &=  \frac{C_F}{8} \biggl\{  \biggl[ \biggl( \frac{134}{9} - \frac{2}{3}\pi^2 \biggr) C_A - \frac{20}{9} n_l  \biggr] \biggl(\! -1 + \log(1-x) + \frac{1}{2}\log \frac{m_t^2}{m_b^2} \biggr) 
	+ 4(1 - \zeta_3 ) C_A  \biggr\} + \mathcal{O}\left(\! \frac{m_b^2}{m_t^2} \!\right) \,
\end{align}
\endgroup
are the one-loop and two-loop anomalous dimensions controlling the evolution of $\mathcal{Z}_{m_b \ll m_t}$, respectively.

As a last check, we performed a pointwise comparison, at one-loop level, between the massified finite reminder and the exact, evaluated with \Recola or \OpenLoops. For a bottom quark of mass \mbox{$m_b = 4.75$\,GeV} and a top quark of mass \mbox{$m_t = 173$\,GeV}, pointwise differences are at the level of $0.3\%$, thus confirming that power corrections in the bottom-quark mass are numerically negligible.

\subsubsection[Polarised two-loop amplitudes for on-shell \texorpdfstring{$t \bar t$}{ttx} production]{Polarised two-loop amplitudes for on-shell \texorpdfstring{$\boldsymbol{t \bar t}$}{ttx} production}
\label{sec:onshell_twoloop_amplitudes}
A crucial ingredient for the construction of the contribution in Eq.~\eqref{eq:fact_twoloop_2x0} is given by the two-loop amplitudes for $t\bar{t}$ production where the top and anti-top quarks carry polarisations \mbox{$\lambda_t = \pm 1$} and \mbox{$\lambda_{\bar t} = \pm 1$}, respectively.
As already mentioned, we will rely on the results provided in Ref.~\cite{Chen:2017jvi} for the corresponding two-loop finite remainders at scale \mbox{$\mu=\muR = m_t$}, with $\mu$ being the scale at which the infrared poles have been subtracted.

In order to set the conventions, we consider the process 
\begin{equation}
	c(p_1) +\bar{c}(p_2) \to t(p_3,\lambda_t) + \bar{t}(p_4,\lambda_{\bar t}),\qquad c=q,g\,,
\end{equation}
in quark--antiquark annihilation or gluon fusion. 
Momentum conservation \mbox{$p_1 + p_2 = p_3 + p_4$} and on-shell conditions \mbox{$p_1^2 = p_2^2 = 0$}, \mbox{$p_3^2 = p_4^2 = m_t^2$} are satisfied.
The amplitudes are functions of the invariants 
\begin{equation}
	s \equiv (p_1+p_2)^2 ~,~~ t \equiv m_t^2 - (p_1-p_3)^2 = \frac{s}{2}(1-\beta\cos\theta) ~,~~ u \equiv m_t^2 - (p_2-p_3)^2 = \frac{s}{2}(1+\beta\cos\theta) \,,
\end{equation}
where \mbox{$\beta=\sqrt{1-4m_t^2/s}$} is the velocity of the top quark and $\theta$ is the top-quark scattering angle with respect to the beam axis in the $p_1$ direction.

To describe the polarisation state of the top and anti-top quarks, we introduce the spin four-vectors $s_t$ and $s_{\bar t}$, respectively. 
They are space-like four-vectors (\mbox{$s_t^2 = s_{\bar t}^2 = -1$}) which describe the spin of the top quarks in their respective rest frames. 
These vectors are defined as 
\begin{equation}
	s_t = \left( \frac{|\vec{p}_3|}{m_t}, \frac{\vec{p}_3}{|\vec{p}_3|} \frac{E_3}{m_t} \right) ~~~\text{and}~~~ s_{\bar t} = \left( \frac{|\vec{p}_4|}{m_t}, \frac{\vec{p}_4}{|\vec{p}_4|} \frac{E_4}{m_t} \right) \,,
\end{equation}
with \mbox{$s_t  \cdot p_3 = s_{\bar t} \cdot p_4 = 0$}, and they are used to construct the spin projectors~\eqref{eq:spin_projectors}.
To obtain the amplitude for a top quark with spin in the \mbox{$\pm s_t$} direction, the corresponding projector $P_{\Sigma}(\pm s_t)$ can be applied on the top-quark spinors.
When calculating matrix elements, this results in the insertion of the projector in the spin sum as
\begin{equation}
	u(p_3, \pm s_t) \bar{u}(p_3, \pm s_t) = (\slashed{p}_3 + m_t)P_{\Sigma}(\pm s_t) ~,~~~  v(p_4, \pm s_{\bar t} ) \bar{v}(p_4, \pm s_{\bar t} ) = (\slashed{p}_4 - m_t)P_{\Sigma}(\pm s_{\bar t}) \,.
\end{equation}

In Ref.~\cite{Chen:2017jvi} the polarised two-loop finite remainders have been calculated via two alternative procedures. 
The first approach involves a \textit{projection method}, which starts with an analysis of the functional dependence of the amplitude on the external quantities to identify linearly independent Lorentz structures that can be used to construct the amplitude itself.
These structures can be regarded as a basis in the spin space.
However, their number is process-dependent and grows dramatically with the number of external legs. 
In Ref.~\cite{Chen:2017jvi} the independent spin structures were found to be 8 in the $gg$ channel and 4 in the $q \bar q$ channel, respectively.

The second way to represent spin-dependent matrix elements is given by the \textit{spin-density matrix} defined in Ref.~\cite{Bernreuther:1993hq}.
This approach is more convenient for phenomenological applications.
In our construction we will mainly rely on this second method.

The two-loop finite remainder%
\footnote{As in Sec.~\ref{sec:heavy-to-light_FF}, the finite remainder is defined in the minimal subtraction scheme of Refs.~\cite{Becher:2009qa,Becher:2009cu,Ferroglia:2009ii}.} contribution to the spin-density matrix can be decomposed as (see Eq.~(2.31) of Ref.~\cite{Chen:2017jvi})
\begin{align}
	\hspace{-0.7cm}
	2 \mathrm{Re}\left\{ \M_{c \bar c}^{(2),\mathrm{fin}} \M_{c \bar c}^{(0) *} \right\}_{(\lambda_t s_t, \lambda_{\bar t} s_{\bar t})} \!\! &= 
        \frac{(4 \pi \alphas)^2}{N_{c \bar c} n_{c \bar c}} \biggl\{ 
	 A_{c \bar c}+\frac{m_t}{s^2} (B)_{c \bar c} \left( \lambda_t \epsilon^{\mu\nu\alpha\beta}p_{1\mu}p_{2\nu}p_{3\alpha}s_{t\beta} 
	+\lambda_{\bar t} \epsilon^{\mu\nu\alpha\beta}p_{1\mu}p_{2\nu}p_{3\alpha}s_{\bar{t} \beta}   \right) \notag \\
	&+ (C)_{c \bar c}  \lambda_t \lambda_{\bar t} \left(s_t \cdot s_{\bar t}  \right)
	+ \frac{1}{s}(D)_{c \bar c}  \lambda_t \lambda_{\bar t} \biggl( (p_1 \cdot s_t)(p_1 \cdot s_{\bar t}) + (p_2 \cdot s_t)(p_2 \cdot s_{\bar t}) \biggr) \notag \\
	& +  \frac{1}{s} (E_{12})_{c \bar c}  \lambda_t \lambda_{\bar t} \biggl( (p_1 \cdot s_t)(p_2 \cdot s_{\bar t}) \biggr)
	+ \frac{1}{s}(E_{21})_{c \bar c}  \lambda_t \lambda_{\bar t} \biggl( (p_2 \cdot s_t)(p_1 \cdot s_{\bar t})  \biggr)
	\biggr\} \,,
	\label{eq:SDM_ttx}
\end{align}
for a top (anti-top) quark with helicity $\lambda_t (\lambda_{\bar t})$ along the direction of $s_t (s_{\bar t})$. 
The normalisation factor
\begin{equation}
	n_{c \bar c} = \frac{\beta(1-\beta^2)}{ \pi}
\left\{
	\begin{array}{ccl}
		\frac{1}{4096} &\quad&\text{if}~ c = g\,,\\[1ex]
		\frac{1}{576}   &\quad&\text{if}~ c = q
	\end{array}
        \right.
\end{equation}
is introduced to damp the singular behaviour close to the production threshold \mbox{$(s \sim 4m_t^2)$} and in the high-energy limit \mbox{$(\beta \to 1)$}, while the factor $N_{c \bar c}$ defined in Eq.~\eqref{eq:Ncc_average-factor} takes into account the average over the IS spins and colours.
The coefficient functions $A_{c \bar c}, \dots, (E_{21})_{c \bar c}$ are given in the form of interpolation grids (in $\beta$ and $\cos\theta$) as well as analytic expansions near the production threshold (\mbox{$\beta \to 0$}).
In the threshold region, the coefficients provided in Ref.~\cite{Chen:2017jvi} have to be multiplied by a factor $n_{c \bar c}/4$ to be consistent with the overall normalisation.
Compared to Eq.~(2.31) of Ref.~\cite{Chen:2017jvi} additional factors are included in Eq.~\eqref{eq:SDM_ttx}, which account for a different convention in the definition of the spin four-vectors. 

It is worth mentioning that the coefficient $A_{c \bar c}$ is strictly related to the unpolarised result (except for an overall normalisation), while $(B)_{c \bar c}$ describes the effects due to the top-quark transverse polarisation with respect to the scattering plane spanned by $p_1$ and $p_3$. The remaining coefficient functions encode the spin correlations between the top and anti-top quarks.

We observe once more that the two-loop amplitudes in Eq.~\eqref{eq:SDM_ttx} depend on a single mass scale, $m_t$, thus they do not include effects from massive bottom-quark loops. 
In our implementation, we have evaluated the on-shell $t \bar t$ amplitudes with $\alphas$ running in the 4FS and \mbox{$n_l = 5$}, where $n_l$ denotes the number of massless flavours, and have restored $m_b$ effects up to power corrections of order $\mathcal{O}(m_b^2/m_t^2)$ via the massification procedure (see Sec.~\ref{sec:mb-effects_onshell_ttx_amplitudes} for more details on these additional contributions).

\subsubsection[Bottom-quark mass effects in the two-loop \texorpdfstring{$t \bar t$}{ttx} amplitudes]{Bottom-quark mass effects in the two-loop \texorpdfstring{$\boldsymbol{t \bar t}$}{ttx} amplitudes}
\label{sec:mb-effects_onshell_ttx_amplitudes}
In this section, we discuss how to include the bottom-quark mass dependence in the two-loop $t \bar t$ amplitudes of Ref.~\cite{Chen:2017jvi}.
The idea consists of extending the massification procedure outlined in Sec.~\ref{sec:heavy-to-light_FF} for the top-quark decay to the $t \bar t$ case. This is possible thanks to the mass factorisation formula derived in Ref.~\cite{Wang:2023qbf} for a general number of heavy flavours with different masses. In this specific case, we have \mbox{$n_t = 1$} and \mbox{$n_b =1$} massive flavours with mass $m_t$ and $m_b$, respectively. 
The factorisation formula of Ref.~\cite{Wang:2023qbf} allows us to restore the correct $\epsilon$-poles of the massive amplitude as well as the finite part in $\epsilon$ up to power corrections in the mass of the heavy quarks. Formally, it holds in the highly boosted regime \mbox{$m_t, m_b \ll \mu_h$}, where \mbox{$\mu_h \sim Q$} is any hard scale of the process, and for a generic hierarchy between the two masses. 

When applying the massification procedure starting from the massless di-jet amplitudes in the 6FS, we neglect power-suppressed terms of order $m_t/\mu_h$, in line with the power counting discussed above. However, these corrections are not negligible in the context of $t\bar{t}$ production. 
To restore the full top-mass dependence, we perform an intermediate matching of the massified amplitudes in the 5FS -- after decoupling the top quark from the running of $\alphas$ -- to the exact massive $t\bar{t}$ amplitudes of Ref.~\cite{Chen:2017jvi} evaluated with \mbox{$n_l = 5$} light flavours, i.e.\ including a massless bottom-quark contribution.
We finally apply the decoupling of the bottom quark and extract the new $n_b$-dependent terms, which have to be added to the (single mass-scale) two-loop $t \bar t$ finite remainder of Ref.~\cite{Chen:2017jvi} evaluated with \mbox{$n_l = 5$}. In this way, we restore the contributions from massive bottom-quark loops up to power corrections of order $m_b/\mu_h, m_b/m_t$, while retaining the full dependence on the top-quark mass. 

The analytic expression of these $n_b$ terms, at scale \mbox{$\mu = \muR = m_t$}, reads
\begingroup
\allowdisplaybreaks
\begin{align}
	\mathcal{M}_{q \bar q}^{(2), \text{fin}}\biggl|_{n_b} = &-\frac{1}{9} n_b^2 \log^2\left( \frac{m_t^2}{m_b^2}\right)  \mathcal{M}_{q \bar q}^{(0)} \notag \\
	&+ n_b \biggl\{ \frac{2}{3}  \log\frac{m_t^2}{m_b^2} \mathcal{M}_{q \bar q}^{(1), \text{fin}}(m_t, m_b)
	+ \biggl[ \frac{C_A}{18}  \left( -8 + 15 \log \frac{m_t^2}{m_b^2} \right) \notag \\
	&\hspace{1cm} + \frac{C_F}{432}\left(1909 + 60 \pi^2 - (508 + 24 \pi^2) \log \frac{m_t^2}{m_b^2} + 480 \log^2\left( \frac{m_t^2}{m_b^2}\right) - 360\zeta_3 \right) \notag \\
	&\hspace{1cm} + \frac{1}{108} \sum_{i > j} \mathbf{T}_i \cdot \mathbf{T}_j \left( 56 + 3 \pi^2 - 60 \log \frac{m_t^2}{m_b^2}  + 36 \log^2\left( \frac{m_t^2}{m_b^2}\right)  \right) \log \left(\frac{-s_{ij}}{m_t^2}\right)
	\biggr] \mathcal{M}_{q \bar q}^{(0)} \biggr\}\,
\end{align}
\endgroup
for the $q \bar q$ partonic channel, and
\begingroup
\allowdisplaybreaks
\begin{align}
	\mathcal{M}_{gg}^{(2), \text{fin}}\biggl|_{n_b} = &\, n_b \biggl\{ \frac{1}{3} \log\frac{m_t^2}{m_b^2} \mathcal{M}_{gg}^{(1), \text{fin}}(m_t, m_b)  \notag \\
	&\hspace{0.5cm}+ \biggl[ \frac{C_A}{108} \left( 131 + 3 \pi^2 - (94 - 6 \pi^2)\log \frac{m_t^2}{m_b^2} + 63\log^2\left( \frac{m_t^2}{m_b^2}\right) - 54 \zeta_3 \right) \notag \\
	&\hspace{0.7cm} + \frac{C_F}{108}\left( 56 + 3 \pi^2 - 60 \log \frac{m_t^2}{m_b^2}  + 36 \log^2\left( \frac{m_t^2}{m_b^2}\right)  \right) - \frac{n_l}{108}\left( \pi^2 + 6\log^2\left( \frac{m_t^2}{m_b^2}\right) \right) + \frac{n_t}{108}\pi^2 \notag \\
	&\hspace{0.7cm} + \frac{1}{108} \sum_{i > j} \mathbf{T}_i \cdot \mathbf{T}_j \left( 56 + 3 \pi^2 - 60 \log \frac{m_t^2}{m_b^2}  + 36 \log^2\left( \frac{m_t^2}{m_b^2}\right)  \right) \log \left(\frac{-s_{ij}}{m_t^2}\right)
	\biggr] \mathcal{M}_{gg}^{(0)} \biggr\} 
\end{align}
\endgroup
for the $gg$ channel.
Consistent with Sec.~\ref{sec:onshell_twoloop_amplitudes} and Ref.~\cite{Chen:2017jvi}, these expressions are given in an $\alphas/2\pi$ expansion, where $\alphas$ is our reference running coupling with \mbox{$n_l = 4$} massless flavours.
The one-loop $t \bar t$ finite remainder $\mathcal{M}_{c \bar c}^{(1), \text{fin}}(m_t, m_b)$ with \mbox{$c=q,g$} is evaluated at \mbox{$\mu = \muR = m_t$} and includes contributions from massive bottom-quark loops. 
The sum runs over the coloured external legs $i,j$ and the two-parton colour correlators $\mathbf{T}_i \cdot \mathbf{T}_j$ act on the Born amplitude $\mathcal{M}_{c \bar c}^{(0)}$. 
The kinematical invariants are defined as \mbox{$s_{ij} = 2 \sigma_{ij} p_i \cdot p_j$} with \mbox{$\sigma_{ij} = +1$} if both $i$ and $j$ are incoming (outgoing), \mbox{$\sigma_{ij} = -1$} otherwise.

We observe that, as expected, logarithmic corrections in $m_b$ appear up to the power of two. Moreover, $n_b$ terms do not affect the IR pole structure of the two-loop $t \bar t$ amplitude, but only the finite remainder. 

\section{Computational framework}\label{sec:framework}
All the computations considered in this paper are carried out within the \Matrix framework~\cite{Grazzini:2017mhc}, suitably extended to address the complexity and the multitude of topologies appearing in off-shell $t \bar t$ production. 
The stability of the numerical integration, even for \mbox{$2 \to 8$} partonic processes that start contributing at NNLO, is guaranteed by an efficient, multi-channel based phase-space generation.
The evaluation of all required tree-level and one-loop matrix elements is performed via the \OpenLoops~\cite{Cascioli:2011va,Buccioni:2017yxi,Buccioni:2019sur} and \Recola~\cite{Actis:2012qn,Actis:2016mpe,Denner:2017wsf,Denner:2016kdg} amplitude providers, where in particular the numerical stability of \OpenLoops is crucial for the reliable evaluation of the \mbox{$2 \to 7$} real--virtual amplitudes in the IR singular limits.

For the treatment of the IR singularities arising at NLO we rely on the Catani-Seymour (CS) dipole subtraction~\cite{Catani:1996jh,Catani:1996vz,Catani:2002hc} and on a process-independent implementation of the $q_T$-subtraction formalism \cite{Catani:2007vq}, suitably extended to heavy-quark production \cite{Bonciani:2015sha,Catani:2019iny,Catani:2019hip,Catani:2020kkl}.
The presence of a colourless system does not pose any additional complication from a conceptual viewpoint. However, it requires the computation of appropriate soft-parton contributions for generic kinematics of the heavy-quark pair~\cite{Catani:2023tby,inprep}. This extension has been successfully applied to the class of \mbox{$2 \to 3$} processes, up to NNLO in QCD \cite{Catani:2021cbl,Catani:2022mfv,Buonocore:2022pqq,Buonocore:2023ljm,Devoto:2024nhl}. 

Since $q_T$ subtraction is insensitive to the number of colourless particles in the final state and to the resonance structure of the process, the same procedure can be exploited to regularise all IR singularities appearing in the case of a kinematically complex process like off-shell $t \bar t$ production,
which is treated in practice as $b{\bar b}F$ production where $F$ is the four-lepton final state.
This holds as long as the bottom quarks are treated as \textit{massive}. The role of the bottom-quark mass is crucial to guarantee the regularisation of FS collinear singularities, which are not captured by the limit \mbox{$q_T \to 0$}.

As in all modern implementation of the $q_T$ subtraction formalism, a technical cut-off \mbox{$\rcut \equiv q_T^{\mathrm{cut}}/Q$} is introduced on the dimensionless
variable \mbox{$q_T/Q$}, where $q_T~(Q)$ is the transverse momentum (invariant mass) of the $b{\bar b}e^+\nu_e\mu^-{\bar \nu}_\mu$ system.
The final result, which corresponds to the limit \mbox{$\rcut\to 0$}, is extracted by simultaneously computing
the cross section at fixed values of $\rcut$ and then performing an \mbox{$\rcut\to 0$}
extrapolation. More details on the procedure and the ensuing uncertainties can be found in
Refs.~\cite{Grazzini:2017mhc,Catani:2021cbl}.

\subsection{The double-pole approximation in \Matrix}
\label{sec:dpa_matrix}
For the construction of the virtual DPA, as detailed in Sec.~\ref{sec:DPA}, we follow the approach of Ref.~\cite{Denner:2000bj}, where only the IR-finite virtual corrections are treated in DPA while all other contributions are computed exactly.

The $n$-loop virtual amplitude contributes to the N$^n$LO calculation (\mbox{$n=1,2$}) through its interference with the Born amplitude.%
\footnote{At NNLO we also have a contribution from the one-loop squared amplitude, which, however, does not need to be approximated.}
To isolate the part that requires an approximation, we define the hard-virtual coefficient
\begin{equation}
	H(\alphas(\muR); \muIR) = 1 + \frac{\alphas(\muR)}{2 \pi} H^{(1)}\left( \muIR \right) + \left(\frac{\alphas(\muR)}{2 \pi}\right)^2 H^{(2)}\left( \muIR \right) + \mathcal{O}(\alphas^3)\,,
	\label{eq:H}
\end{equation}
where
\begin{equation}
\label{eq:Hn_ttxoff-shell}
        H^{(n)}(\muIR)= \frac{2{\mathrm{Re}}\left(\M^{(n), \mathrm{fin}}(\muIR)\M^{(0)*}\right)}{|\M^{(0)}|^2} \, \,, ~~~~~~~~~~n=1,2,\dots,
\end{equation}
is computed through the interference of the Born amplitude ${\cal M}^{(0)}$ for \mbox{$c{\bar c}\to b{\bar b}e^+\nu_e\mu^-{\bar \nu}_\mu$} (\mbox{$c=q,g$}) with the $n$-loop virtual correction ${\cal M}^{(n),{\rm fin}}$ to the IR finite remainder
\begin{equation}
	\ket{\cal M^{{\rm fin}}(\muIR)} = \sum_{l=0}^{\infty}\left(\frac{\alphas(\muIR)}{2\pi}\right)^l \ket{{\cal M}^{(l),{\rm fin}}(\muIR)} \,.
\end{equation}
Unless stated otherwise, the IR subtraction is carried out in the scheme of Ref.~\cite{Ferroglia:2009ii} at scale $\muIR$, which is fixed to the invariant mass $Q$ of the Born-like system.

Instead of defining our approximation through the direct evaluation in the DPA of the interference between the tree-level and loop amplitudes, we adopt an approach inspired by Ref.~\cite{Buonocore:2021rxx}. 
In that work, the pole approximation has been used to estimate the mixed QCD--EW corrections to the charged-current Drell--Yan production. 
The prescription consists of evaluating the numerator and denominator of Eq.~\eqref{eq:Hn_ttxoff-shell} in the DPA, specifically
\begin{equation}
	H^{(n)}\left( \muIR \right) \bigl|_{\mathrm{DPA}} \equiv \frac{2{\mathrm{Re}}\left(\M^{(n), \mathrm{fin}}_{\rm DPA}(\muIR)\M_{\rm DPA}^{(0)*}\right)}{|\M_{\rm DPA}^{(0)}|^2} \, , ~~~~~~~~~~n=1,2,\dots.
	\label{eq:Hn_DPA_ttxoff-shell}
\end{equation}
Since $H^{(n)} \vert_{\mathrm{DPA}}$ multiplies the exact Born cross section $d\sigma_{\mathrm{LO}}$, this approach effectively reweights the loop--tree interference in DPA with the ratio between the full and approximated squared Born amplitudes.
As a consequence, tree-level off-shell and non-resonant effects, not captured by the standard DPA, are partially included by this reweighting.

The Born matrix element in DPA, appearing in Eq.~\eqref{eq:Hn_DPA_ttxoff-shell}, is constructed as in Eq.~\eqref{eq:LO_PA_ttx}, while, at the one-loop level, the finite remainder $\ket{ \M^{(1), \mathrm{fin}}_{\PA} }$ is obtained from the UV-renormalised one-loop amplitude 
\begin{equation}
	\ket{ \M^{(1)}_{\PA} } = \ket{ \M^{(1)}_{\fact} } + \ket{ \M^{(1)}_{\nonfact} } \,,
\end{equation}
after subtracting the IR singularities at scale $\muIR$.\footnote{Note that the DPA preserves the IR structure of the full virtual amplitude. More precisely, the coefficients of the $1/\epsilon^2$ and $1/\epsilon$ poles that multiply the Born matrix element in DPA are the same as in the exact amplitude, except for the evaluation of the kinematic invariants on the on-shell projected momenta.} 
The factorisable $\ket{ \M^{(1)}_{\fact} }$ and non-factorisable $\ket{ \M^{(1)}_{\nonfact} }$ one-loop corrections, interfered with the Born matrix element, are given in Eqs.~\eqref{eq:fact_M1M0_ttx} and~\eqref{eq:nonfact_M1M0_ttx}, respectively. 
At two-loop order, the factorisable virtual contribution $\ket{ \M^{(2)}_{\fact} }$ is well understood and schematically constructed in Eq.~\eqref{eq:fact_twoloop_ttx}.
In contrast, the corresponding non-factorisable corrections have not yet been computed.
A first determination of their impact, based on an on-shell matching procedure in the \mbox{$\Gamma_t\to 0$} limit, will be presented in Sec.~\ref{sec:ttxoffshell_NNLO_results}.

A final comment regards the scale $\muIR$ at which we define the hard-virtual contribution in DPA.
If no approximation is applied, the hard-virtual coefficient $H^{(n)}$ evaluated at two different scales will be exactly given by the running of the $n$-loop amplitudes, fully predicted by the lowest perturbative orders.
However, with an approximation, this is no longer the case, and differences can arise. The observed variation can indeed provide an estimate of the corresponding systematic uncertainties.
Therefore, at one-loop order, we will provide results at both the dynamic scale \mbox{$\muIR = \widetilde{Q}$}, where $\widetilde{Q}$ denotes the invariant mass of the off-shell $t\bar t$ event after applying the on-shell projection (see Appendix~\ref{app:Appendix_mapping}), and the fixed scale \mbox{$\muIR = m_t$}. 
This second choice is motivated by the fact that we want to explore how the approximation works at different subtraction scales, in view of its application at the next perturbative order. 
At NNLO, we will construct the DPA at scale $m_t$ since the polarised two-loop amplitudes for $t \bar t$ production, necessary for the construction of the factorisable two-loop corrections (see Sec.~\ref{sec:DPA_NNLO}), are provided in Ref.~\cite{Chen:2017jvi} only at that scale. 
The construction of the DPA at a scale \mbox{$\muIR \ne m_t$} would require the implementation of the running of the polarised two-loop amplitudes for $t \bar t$ production.
Even though this is in principle doable, its actual implementation is not a straightforward task, because it requires the imaginary parts of the polarised (and colour-correlated) one-loop $t \bar t $ amplitudes.  
Based on the validation we perform at NLO, we expect the performance of the DPA to be equally good at scales $Q$ and $m_t$.

\subsection{Numerical stability}
\label{sec:stability}
As already mentioned in Sec.~\ref{sec:framework}, the \Matrix framework provides an efficient implementation of the $q_T$-subtraction formula up to NNLO for $Q\bar{Q}F$ production, with $F$ being any colourless system. 
Therefore, in principle, the same machinery can be exploited to regularise all IR singularities appearing 
in off-shell top-quark pair production, which from the QCD viewpoint can be seen as a $b \bar b F$ process. 
In practice, however, the numerical complexity of the calculation is dramatically increased
compared to the \mbox{$2 \to 3$} processes studied so far. 
This is mainly related to the multi-scale nature of the problem and to the proliferation
of topologies, which carry a plethora of different resonance structures and thus challenge the convergence of the numerical integration.

To achieve sufficient numerical control over the power corrections
in the slicing parameter $\rcut$, a numerical integration deep into IR-singular phase-space regions is 
required. Therefore, we evaluate the cross section at fixed
$\rcut$ values in the range \mbox{$[0.01,1]\%$} to be able to reliably extrapolate to the
\mbox{$\rcut \to 0$} limit. Correspondingly, huge cancellations\footnote{These cancellations can be of about two orders of magnitude
for the lowest $\rcut$ values.} between the non-local $q_T$-subtraction
counterterm and the resolved contribution above the slicing cut take place, thus challenging in particular the numerical
integration of the double-real and real--virtual contributions.
Concerning the double-real contribution, only tree-level amplitudes --- notably with up to ten external legs --- are required.
The main challenge lies in a numerically stable integration of
the dipole-subtracted integrand in the presence of a cut on a second IR regulator, $\rcut$, which acts separately on each phase space, giving rise to miscancellations in particular close to the \mbox{$\rcut \to 0$} limit.

For the real--virtual contribution, a stable evaluation of one-loop amplitudes, which involve up to seven-point integrals, must be guaranteed
close to an IR-singular limit (\mbox{$\rcut \to 0$}) and moreover, as discussed below, in the small-width limit (\mbox{$\Gamma_t \to 0$}).
This is achieved through the one-loop provider \OpenLoops.
Indeed, in these extreme limits, large numerical cancellations occur within the coefficients of the real--virtual amplitudes, 
posing a significant challenge for numerical stability. The small top-quark width, which serves as a
regulator for soft radiation off the intermediate top quarks, yields a logarithmic
enhancement in $\Gamma_t/m_t$ that demands excellent numerical control for a reliable 
\mbox{$\Gamma_t \to 0$} extrapolation. 
Thanks to the sophisticated stability optimisations
in \OpenLoops~\cite{Buccioni:2019sur} based on the on-the-fly reduction techniques of
Ref.~\cite{Buccioni:2017yxi}, the numerical stability of the \mbox{$2 \to 7$} real--virtual amplitudes
demonstrates exceptional scaling in the combined \mbox{$\rcut \to 0$} and \mbox{$\Gamma_t \to 0$} limit. The stability optimisations of Ref.~\cite{Buccioni:2019sur} incorporate a hybrid-precision system that switches to quadruple-precision floating-point operations only where needed, with the option to optimise for IR singular regions. This system ensures manageable CPU evaluation times (an average of \mbox{$\sim 10$\,s} (\mbox{$\sim 1$\,s}) per phase-space point for the $gg$ ($q\bar q$) real--virtual subprocess in the vicinity of the IR singular limits), while ensuring the required numerical stability. In this way,  we retain an average accuracy of \mbox{$14.5 \pm 0.5$} digits for both the $gg$  and   $q\bar q$ channels at \mbox{$\rcut=0.1\%$} and \mbox{$\Gamma_t = \Gamma_t^{\rm phys}/100$} (measured against a full quadruple-precision benchmark). In contrast, full quadruple-precision reevaluation for unstable points would be prohibitive for Monte Carlo integration at a cost of \mbox{$\sim 180$\,s} (\mbox{$\sim 20$\,s}) per phase-space point for the $gg$ ($q\bar q$) real--virtual channel.

As already mentioned, we numerically determine the non-factorisable corrections at two-loop order by performing an 
extrapolation in the \mbox{$\Gamma_t \to 0$} limit. 
Obtaining these results for progressively smaller top-quark widths is far from trivial, as the increasingly peaked resonance structure poses further challenges to the local cancellations in the dipole-subtracted double-real contribution. 
In Sec.~\ref{sec:ttxoffshell_NNLO_results} we will, indeed, consider
technical top-quark widths as small as \mbox{$\Gamma_{t}^{\rm{phys}}/100$}.
While the adaptive multi-channel Monte Carlo approach applied in \Matrix is, in principle,
sufficiently flexible to deal with all these challenges, it turned out that a few refinements of the general methods 
lead to substantial improvements in performance, not
only reducing the required computational time, but also increasing the reliability of the produced results.

Among those developments, the biggest gain in the convergence of the Monte Carlo integration is achieved by
a pre-optimisation of the a-priori weights of the multi-channel approach.
These are, by default, distributed in a flat way. However, in the presence of many different resonance structures
and a huge number of integration channels,\footnote{The most complicated all-gluon subtracted double-real subprocess involves 23729 channels, 4397 from double-real topologies and 537 channels for each of the 36 dipoles.}
an adaptation of this flat starting configuration to the resonance structures, represented by the separate
integrations channels, turns out to significantly help the automated weight adaptation procedure in finding the
optimal weight configurations during the grid optimisation phase.
For the subtracted \linebreak(double-)real contributions 
where the optimisation phase is most challenging, this is further supported by a second multi-channelling stage to explicitly control that both (double-)real and all dipole topologies are sufficiently represented.

To further consolidate the phase-space generation, a new optional feature has been introduced that allows generating phase-space points using quadruple-precision arithmetic.
This approach essentially eliminates situations where the event generation fails due to bad numerics.
The need to properly handle such problematic points is crucial, as neglecting them could introduce a bias in the event generation, potentially impacting the integrated results if a substantial fraction of points falls into this category.
The quadruple-precision feature completely cures this potential issue,
and, if applied prudently (e.g.,\ by calculating the corresponding channel weights in the usual
double-precision arithmetic unless instabilities occur), does not induce a relevant increase in runtime.

Altogether, the above modifications result in a significant improvement in numerical convergence, most notably in the calculations involving the technical \mbox{$\Gamma_t \to 0$} extrapolation. For the smallest considered value of the top-quark width, \mbox{$\Gamma_t = \Gamma_{t}^{\rm{phys}}/100$}, the runtime required to reach the same relative precision as at the physical top-quark width is reduced to roughly a factor of three, 
whereas it would exceed an order of magnitude without these improvements.
Ultimately, only with these modifications the whole fitting procedure described in Sec.~\ref{sec:ttxoffshell_NNLO_results} can be carried out in a sustainable manner with a moderate computational cost.\footnote{To be more explicit, the NNLO calculation at \mbox{$\Gamma_t = \Gamma_{t}^{\rm{phys}}$} corresponding to the results presented in Sec.~\ref{sec:ttxoffshell_NNLO_results} takes about 400k CPU hours.}

\section{Validation and results at NLO}
\label{sec:ttxoffshell_NLO_results}

In this section, we present NLO QCD results for off-shell $t\bar t$ production and decays, i.e.\ for the complete process~\eqref{eq:off-shell_ttx_dilepton}.
This section has a dual purpose. 
On the one hand, we aim at validating the $q_T$-subtraction method for such a complex multi-leg off-shell process by comparison with the CS dipole subtraction implemented in \Matrix, and with NLO results published in the literature.
On the other hand, we want to study the performance of the virtual DPA construction, at both fiducial and differential cross-section levels, given its crucial role at NNLO where complete double-virtual amplitudes are currently out of reach.

The section is organised as follows.
In Sec.~\ref{sec:setup_nlo} we describe the setup used in our computation. 
In Secs.~\ref{sec:totalXS_ttx_off-shell} and \ref{sec:distributions_ttx_off-shell} we show LO and NLO QCD results for the total cross section and for a number of phenomenologically relevant distributions, respectively. 
The goal is to validate the implementation of the $q_T$ subtraction for this \mbox{$2 \to 6$} process against the results obtained with the CS method.
The sub-percent (few-percent) differences in the bulk (tails) of the distributions confirm that the missing power corrections in the slicing parameter $\rcut$ are under control, and that the \mbox{$\rcut \to 0$} extrapolation leads to the correct results. 
Section~\ref{sec:DPAvalidation_ttx_off-shell} focuses on the validation of the DPA at both fiducial and differential levels. These results demonstrate that by incorporating both factorisable and non-factorisable corrections, combined with the reweighting procedure described in Sec.~\ref{sec:dpa_matrix}, we achieve exceptional agreement with the full computation, even in phase space regions not dominated by resonant top quarks.
Differences between the exact results and those obtained via the DPA range from a few per mille up to $5\%$ in the tails of some kinematic distributions.  

\subsection{Setup}
\label{sec:setup_nlo}
In this section, we describe the setup used in our NLO computation. 
As for the EW parameters, we follow Ref.~\cite{Cascioli:2013wga}, where the $G_\mu$ input scheme is used with
\begingroup
\allowdisplaybreaks
\begin{align}
	m_W = 80.385 \,\mathrm{GeV} ~~&,~~~~\Gamma_W = 2.09530 \,\mathrm{GeV}  \,,\\
	m_Z = 91.1876 \,\mathrm{GeV} ~~&,~~~~\Gamma_Z = 2.50479 \,\mathrm{GeV}  \,,\\
	G_F = 1.16637 \cdot 10^{-5} \,\mathrm{GeV}^{-2}  ~~&,~~~~
	\alpha = \frac{\sqrt 2}{\pi} G_F m_W^2 \left( 1- \frac{m_W^2}{m_Z^2} \right) \,,
\end{align}
\endgroup
and the unstable top quarks, $Z$ and $W$ bosons are treated in the complex-mass scheme~\cite{Denner:2005fg}. 
We consider a diagonal CKM matrix.

The pole masses of the top and bottom quarks are set to \mbox{$m_t=173.2$\,GeV} and \mbox{$m_b=4.75$\,GeV}, respectively, while the top-quark width is \mbox{$\Gamma_t^{\rm NLO}=1.34264$\,GeV}. The Higgs boson mass and width are set to \mbox{$m_H=126$\,GeV} and \mbox{$\Gamma_H=4.21$\,MeV}, respectively.
Consistently with a finite bottom-quark mass, we work in the 4FS with \mbox{$n_f = 4$} active flavours, and we use the \verb|NNPDF23_nlo_FFN_NF4_as_0118| PDF set~\cite{Ball:2012cx} at both LO and NLO, via the LHAPDF interface~\cite{Buckley:2014ana}.

By default, unless stated otherwise, we set the central renormalisation ($\mu_R$) and factorisation ($\mu_F$) scales to the standard choice 
\begingroup
\allowdisplaybreaks
\begin{equation}
	\mu_0 = \mu_R = \mu_F = m_t \,.
\end{equation}
\endgroup
Theory uncertainties due to missing higher-order terms are estimated through the customary procedure of independently
varying $\mu_R$ and $\mu_F$ by a factor of two around their central value with
the constraint \mbox{$0.5 \leq \mu_R/\mu_F \leq 2$}. When performing scale variations, the NLO top-quark width is kept fixed. \\
\\
We have considered two fiducial setups. 
The first is inspired by the fully off-shell NLO calculation of Ref.~\cite{Cascioli:2013wga} and, for brevity, we will refer to it as ``CKMP".
This setup is chosen with the aim of reproducing, wherever feasible, results available in existing literature. 
The second setup, dubbed as ``CMP", implements the same fiducial cuts as those in Ref.~\cite{Czakon:2020qbd}. The main difference lies in the treatment of the bottom quarks: while in Ref.~\cite{Czakon:2020qbd} the calculation is performed in the 5FS with massless bottom quarks, in our framework we have to consider massive bottom quarks to avoid non-regularised FS collinear singularities. 
Due to the different nature of our 4FS setup, we cannot expect to directly reproduce the results of Ref.~\cite{Czakon:2020qbd}. However, we exploit this second setup to examine how well the DPA performs under different cuts applied to the decay products of the $t\bar t$ pair.
\begin{itemize}
\item In the CKMP setup we consider proton--proton collisions at a centre-of-mass energy of \mbox{$\sqrt{s}=8$\,TeV}.
We impose fiducial cuts on the transverse momenta and pseudo-rapidities of the charged leptons \mbox{$l \in \{ e^+, \mu^-\}$},
\begin{equation}
	p_{T,l} > 20 \,\mathrm{GeV} ~~,~~~~ |\eta_l| < 2.5 \,,
\end{equation}
and on the missing transverse momentum \mbox{$p_{T,\mathrm{miss}} > 20$\,GeV}, defined as the vectorial sum of the neutrinos' transverse momenta.
No requirement is imposed on the bottom quarks or any light-jet radiation.

\item In the CMP setup we consider proton--proton collisions at a centre-of-mass energy of \mbox{$\sqrt{s}=13$\,TeV}.
We require that the charged leptons pass the cuts
\begin{equation}
	p_{T,l} > 20 \,\mathrm{GeV} ~~,~~~~ |\eta_l| < 2.4 ~~,~~~~ m_{ll} > 20 \,\mathrm{GeV} \,,
\end{equation}
where $m_{ll}$ is the invariant mass of the two charged leptons.
We define bottom jets with a standard anti-$k_T$ algorithm~\cite{Cacciari:2008gp} of radius \mbox{$R=0.4$}, and classify any jet involving at least one bottom quark as a bottom jet, which also includes the case of collimated $b \bar b$ pairs resulting from the splitting of energetic gluons. This bottom jet definition is infrared and collinear safe due to the finite bottom-quark mass.
We require two bottom jets with transverse momentum and rapidity satisfying
\begin{equation}
	p_{T,b_{\mathrm{jet}}} > 30 \,\mathrm{GeV} ~~,~~~~ |y_{b_{\mathrm{jet}}}| < 2.4 \,,
\end{equation}
and demand that they are well separated from any lepton by imposing \mbox{$\Delta R_{l, b{\mathrm{jet}}} > 0.4$}.
This setup is inspired by a typical $t \bar t$-signal analysis, since the requirement of central bottom jets enhances contributions from double-resonant topologies and suppresses single-resonant and non-resonant contributions. 
\end{itemize}

\subsection{Fiducial cross sections}\label{sec:totalXS_ttx_off-shell}
In Tab.~\ref{tab:WWbb_LONLOxs}, we display the results for the integrated cross sections at LO and NLO for the two setups detailed in Sec.~\ref{sec:setup_nlo}. 
The reported uncertainties are obtained via a standard seven-point scale variation, while the numbers in parentheses take into account the statistical Monte Carlo uncertainty combined with the systematic error due to the \mbox{$\rcut \to 0$} extrapolation. 
The inclusion of the NLO corrections leads to an increase of the fiducial cross section of about $33\%$ in the CKMP setup and $22\%$ in the less inclusive CMP setup.
Scale uncertainties\footnote{Here and in the following, we conservatively quote the maximum between the upward and downward variations.}
decrease from $35\%$ ($30\%$) at LO to $11\%$ ($8\%$) at NLO in the CKMP (CMP) setup, respectively.
Numerical uncertainties are negligible with respect to the scale-variation bands. 

\begin{table}[t]
\centering  
\renewcommand{\arraystretch}{1.5}
\setlength{\tabcolsep}{0.3em}
\begin{tabular}{clll}
&
\multicolumn{1}{c}{CKMP}  &\multicolumn{1}{c}{CMP} \\
\toprule
$\sigma_{\mathrm{LO}}$ [fb] & 
 $\phantom{00}1364.99(6)\,^{+34.3\%}_{-23.7\%}$ & $\phantom{00}2811.1(1)\,^{+29.4\%}_{-21.2\%}$ 
\\\hline
$\sigma_{\mathrm{NLO}_{\mathrm{CS}}}$ [fb] & 
$\phantom{00}1816.9(4)\,^{+8.0\%}_{-11.0\%}$ & $\phantom{00}3424.9(7)\,^{+4.7\%}_{-8.0\%}$  
\\
$\sigma_{\mathrm{NLO}_{q_T}}$ [fb] & 
$\phantom{00}1817.7(1.2)\,^{+8.0\%}_{-11.0\%}$ & $\phantom{00}3416(4)\,^{+4.5\%}_{-8.0\%}$ 
\\ 
$\sigma_{\mathrm{NLO}_{\rm CKMP}}$ [fb] & 
$\phantom{00}1817\,^{+8.0\%}_{-11.0\%}$ & \multicolumn{1}{c}{-}
\\\hline
$\sigma_{\mathrm{NLO}_{q_T, \mathrm{DPA}}}$ [fb] & 
$\phantom{00}1822.9(1.0)\,^{+8.2\%}_{-11.2\%}$ & $\phantom{00}3408(4)\,^{+4.4\%}_{-8.0\%}$ 
\\
\bottomrule
\end{tabular}
\caption{\label{tab:WWbb_LONLOxs} LO and NLO cross sections at \mbox{$\sqrt{s}=8$\,TeV} (CKMP setup) and \mbox{$\sqrt{s}=13$\,TeV} (CMP setup), respectively. The renormalisation and factorisation scales are set to \mbox{$\mu_0 = m_t$}. The quoted uncertainties are obtained through scale variations as described in the main text. Numerical uncertainties on the last digit are stated in brackets: they include the Monte Carlo integration error combined with the systematic uncertainties from the \mbox{$\rcut \to 0$} extrapolation. The last row reports the results for the NLO cross section obtained within the $q_T$-subtraction framework, where the genuine one-loop contribution has been approximated in the DPA.}
\end{table}



%
In the second and third rows of Tab.~\ref{tab:WWbb_LONLOxs} we report the NLO results based on CS and $q_T$ subtraction, respectively. The difference is at the level of $0.04\%$ for the more inclusive CKMP setup and of $0.2\%$ for the more exclusive CMP setup. Such an agreement is comparable to the listed statistical \textit{plus} extrapolation uncertainties. Within these uncertainties, the CKMP result is compatible with the corresponding NLO cross section of Ref.~\cite{Cascioli:2013wga} listed in the fourth row of Tab.~\ref{tab:WWbb_LONLOxs}. We do not list a reference value for the CMP setup due to the different nature (4FS vs. 5FS) of our computation with respect to the reference.

At a technical level, the quality of the \mbox{$\rcut \to 0$} extrapolation within the $q_T$ subtraction has additionally been assessed by investigating the behaviour of the cross section at fixed values of the slicing parameter $\rcut$. 
As expected for heavy-quark production~\cite{Catani:2017tuc,Buonocore:2019puv}, the $\rcut$ dependence of the NLO cross section is found to be linear, due to the additional contribution of soft radiation emitted from the FS bottom quarks. This implies a significant sensitivity to the $\rcut$ parameter. At NLO, the only exception is represented by the $gq$ partonic channel, where the $\rcut$ dependence is quadratic, since soft wide-angle radiation is absent at this perturbative order.

The last row of Tab.~\ref{tab:WWbb_LONLOxs} shows the results obtained within the $q_T$-subtraction framework, but based on the DPA, as defined in Sec.~\ref{sec:dpa_matrix}.
At the level of fiducial cross sections, we find excellent agreement between $\sigma_{\mathrm{NLO_{q_T}}}$ and $\sigma_{\mathrm{NLO_{q_T, DPA}}}$.
In both setups, the error to be assigned to the DPA is at the per mille level relative to the NLO cross section, implying that the NLO correction is controlled to better than $2\%$. 
The quality of the DPA clearly depends on the numerical impact of the one-loop contribution. In this case, $\Delta\sigma_{\mathrm{NLO,H}}$ is particularly large, amounting to approximately $1.3~(1.6)$ times the NLO cross section in the CKMP (CMP) setup. Therefore, the level of agreement achieved by the DPA is even more remarkable. 
A similar level of compatibility between the DPA and the full NLO predictions is also observed when choosing different dynamical central scales.

Within the DPA, we observe a clear hierarchy between factorisable and non-factorisable corrections at the level of the considered fiducial cross sections, despite the fact that both are formally of the same order in the $\Gamma_t/m_t$ expansion. This hierarchy will be examined in more detail in Sec.~\ref{sec:DPAvalidation_ttx_off-shell}.
Here, we limit ourselves to stating that the pattern is rather different in the two setups. 
In the CKMP setup, the non-factorisable corrections represent \mbox{$\sim 0.5\%$} of the factorisable contribution and could be safely neglected at the integrated level without missing a significant portion of the NLO cross section. In contrast, within the CMP setup, the non-factorisable corrections amount to approximately $12\%$ of the factorisable contribution and account for nearly $95\%$ of the total NLO correction.
This is primarily due to large cancellations arising between different perturbative ingredients of the $q_T$-subtraction formula.
By carefully investigating this behaviour, we can affirm that the different pattern is mainly driven by the choice of the fiducial cuts applied on the bottom jets, while other aspects of the setup play only a minor role.

\subsection{Differential distributions}\label{sec:distributions_ttx_off-shell}
In Figs.~\ref{fig:LO_NLO_CKMP_setup} and \ref{fig:LO_NLO_CKMP_setup2} we show differential distributions within the CKMP setup at $8$\,TeV.
Each plot presents three panels. 
The upper panel displays the comparison between LO and NLO results, in both CS and $q_T$-subtraction schemes. 
In the central panel, we report the differential NLO over LO correction factors. 
In the bottom panel, we show the ratio between $\mathrm{NLO_{q_T}}$ and $\mathrm{NLO_{CS}}$ results.
The error bands in the first two panels are obtained from scale variations, while the error bars in the lower panel refer to the Monte Carlo integration uncertainties combined with \mbox{$q_T \to 0$} extrapolation uncertainties in the case of the $\mathrm{NLO_{q_T}}$ results.

As an upshot of these distributions, we observe that even at the differential level, the \mbox{$\rcut \to 0$} extrapolation is under control. Differences between the $\mathrm{NLO_{q_T}}$ and $\mathrm{NLO_{CS}}$ results are at most $(2-3)\%$, though covered by the combined statistical and extrapolation uncertainties. We point out that qualitatively equivalent results for the same differential distributions (and various others) have also been observed in the CMP setup.\\

\noindent In more detail, we first consider the invariant mass distribution, $m_{b\bar b}$, of the two bottom quarks (see  Fig.~\ref{fig:1_LO_NLO_CKMP_setup}).
 The inclusion of the NLO QCD corrections leads to large $K$-factors in the low invariant-mass region, i.e.\ for \mbox{$m_{b\bar b} < 100$\,GeV}.
The enhancement of the $K$-factor for low $m_{b\bar b}$ values can be explained by the fact that, in the CKMP setup, we do not impose any cut on the bottom quarks. 
Therefore, their invariant mass can take very small values (even though higher than the mass threshold \mbox{$m_{b\bar b} = 2 m_b$}), which are favoured by non-resonant configurations where the two bottom quarks originate from a \mbox{$g\to b\bar b$} splitting. 
Over the entire $m_{b\bar b}$ range we considered, the results based on the CS and $q_T$-subtraction schemes agree at below the per cent level.

Another interesting distribution is represented by the transverse momentum of the $b\bar b$ pair (shown in Fig.~\ref{fig:2_LO_NLO_CKMP_setup}).
This kinematic variable plays an important role in boosted-Higgs searches with a large $t \bar t$ background. The strategy proposed in the literature consists in the selection of boosted \mbox{$H \to b \bar b$} candidates with \mbox{$p_{T,b\bar b} > 200$\,GeV}, which reduces the $t \bar t$ contamination.
As shown in Ref.~\cite{SM:2012sed}, where off-shell results have been compared against those obtained in NWA, the suppression of $t \bar t$ events is indeed particularly strong for \mbox{$p_{T,b\bar b} > 150$\,GeV}. 
This is due to kinematic constraints that characterise LO and NWA. If the two top quarks are produced on-shell, in order to acquire a transverse momentum \mbox{$p_{T,b} > (m_t^2 - m_W^2)/(2 m_t) \simeq 65$\,GeV}, the bottom quarks need to be boosted via the $p_T$ of their parent (anti-top) top quarks. Since at LO the top quarks are produced back-to-back, it is difficult to generate a $b \bar b$ pair with high transverse momentum through resonant top-quark pairs. This implies that, at LO, non-resonant effects play a relevant role in populating this region in $p_{T,b \bar b}$. 
The reason for the dramatic NLO enhancement we observe at large $p_{T,b \bar b}$ values in Fig.~\ref{fig:2_LO_NLO_CKMP_setup} is represented by real configurations in which the $t \bar t$ system can acquire a large transverse momentum by recoiling against an extra jet.
In the high-$p_{T,b \bar b}$ region the NLO scale uncertainties blow up and can be even wider than the LO bands. 
We will further comment on the $p_{T, b\bar b}$ distribution in Sec.~\ref{sec:DPAvalidation_ttx_off-shell}, where we compare the exact results (as shown here) against those obtained with the DPA. In the tail of the considered $p_{T,b\bar b}$ range we observe deviations between the $\mathrm{NLO_{q_T}}$ and $\mathrm{NLO_{CS}}$ results of up to $2\%$, which are, however, covered by integration uncertainties.

\begin{figure}[tb]
  \subfloat[Invariant mass of the two bottom quarks.]{\label{fig:1_LO_NLO_CKMP_setup}{\includegraphics[width=0.47\textwidth]{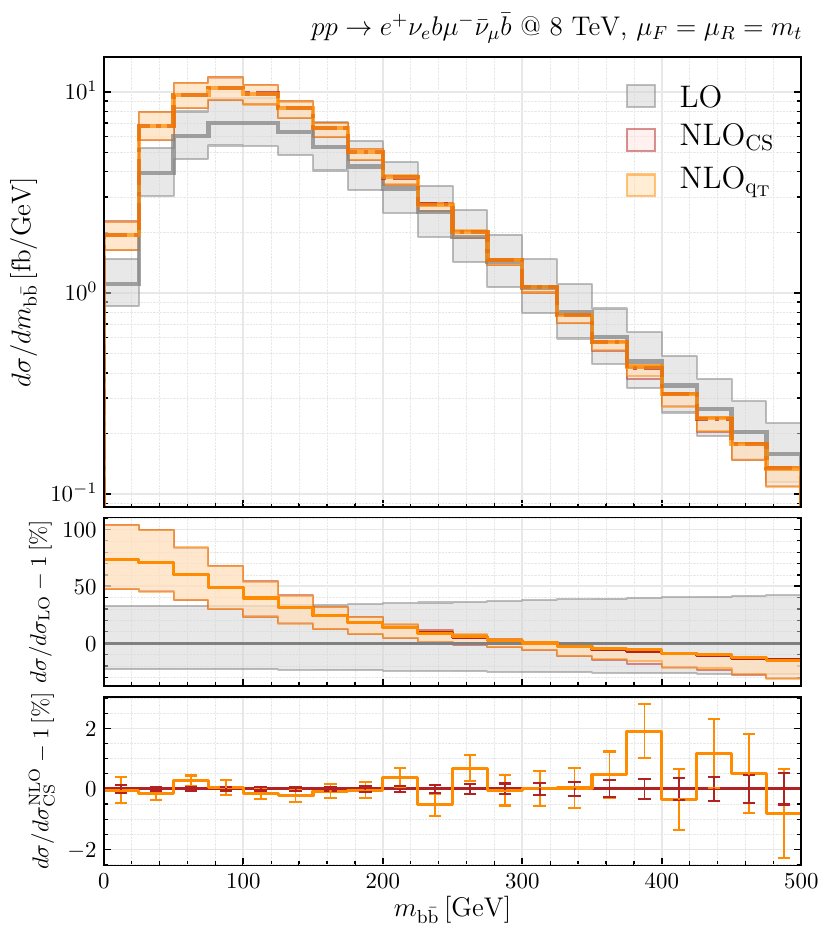} }} \hfill
  \subfloat[Transverse momentum of the two bottom quarks.]{\label{fig:2_LO_NLO_CKMP_setup}{\includegraphics[width=0.47\textwidth]{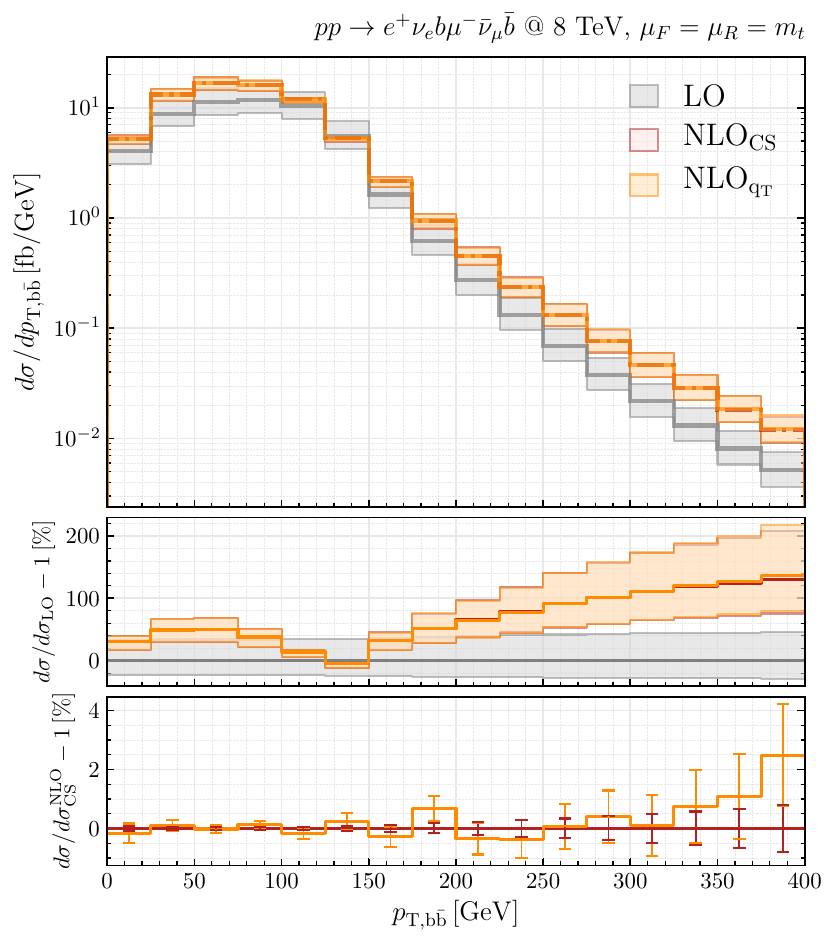} }} 
  \caption{LO and NLO differential results for the CKMP setup at \mbox{$\sqrt{s} = 8$\,TeV} in the invariant mass (a) and the transverse momentum (b) of the $b\bar b$ pair. The central scale is fixed at \mbox{$\mu_0 = m_t$}. Absolute predictions at LO (grey) are compared against NLO predictions based on CS (red) and $q_T$ (orange) subtraction. The bands in the first two panels are obtained from scale variations, while the error bars in the lower panel refer to Monte Carlo integration uncertainties, combined with \mbox{$q_T\to 0$} extrapolation uncertainties in the case of the $\mathrm{NLO_{q_T}}$ result.  }
    \label{fig:LO_NLO_CKMP_setup}
\end{figure}
\begin{figure}[tb]
  \subfloat[Invariant mass of the two charged leptons.]{\label{fig:3_LO_NLO_CKMP_setup}{\includegraphics[width=0.47\textwidth]{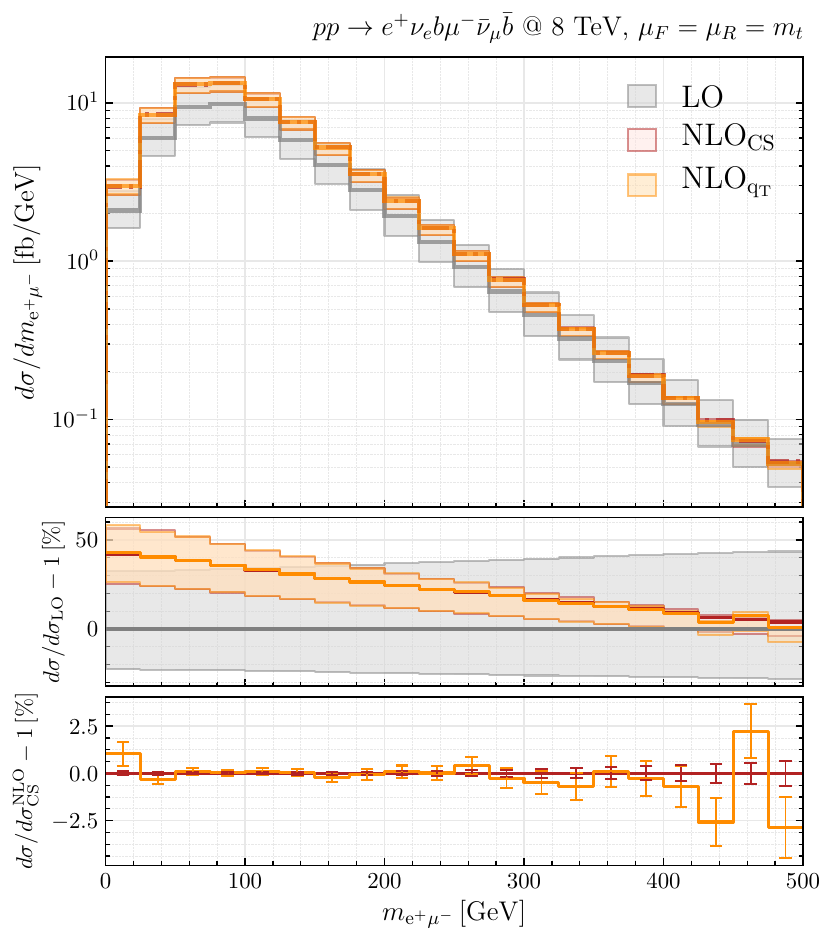} }}  \hfill
\subfloat[Azimuthal separation of the positron and the muon.]{\label{fig:4_LO_NLO_CKMP_setup}{\includegraphics[width=0.47\textwidth]{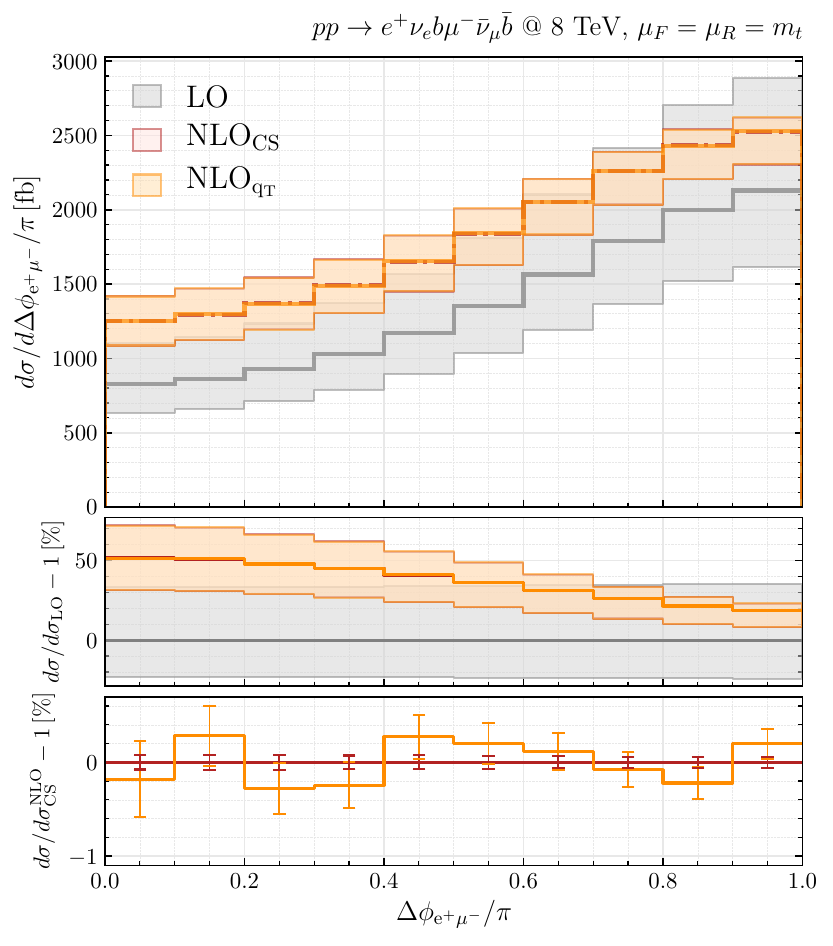} }} 
  \caption{LO and NLO differential results for the CKMP setup at \mbox{$\sqrt{s} = 8$\,TeV} for the invariant-mass distribution $m_{e^+\mu^-}$ of the two charged leptons (a) and the azimuthal-angle separation $\Delta\phi_{e^+\mu^-}$ (b). The central scale is fixed at \mbox{$\mu_0 = m_t$}. Curves and bands as in Fig.~\ref{fig:LO_NLO_CKMP_setup}. }
    \label{fig:LO_NLO_CKMP_setup2}
\end{figure}

Finally, we consider the invariant-mass distribution $m_{e^+\mu^-}$ and the azimuthal-angle separation $\Delta\phi_{e^+\mu^-}$ of the two charged leptons, shown in Figs.~\ref{fig:3_LO_NLO_CKMP_setup} and~\ref{fig:4_LO_NLO_CKMP_setup}, respectively. 
These distributions play, for example, a key role for background modelling in \mbox{$H \to W^+ W^-$} measurements, and are sensitive to $t \bar t$ spin correlations.
In Ref.~\cite{Cascioli:2013wga}, it was shown that finite-top-width effects can be relevant and lead to a shape distortion of the distribution due to their kinematic dependence. 
Both distributions are characterised by large $K$-factors, monotonically decreasing towards high $m_{e^+\mu^-}$ and $\Delta\phi_{e^+\mu^-}$ values. However, we note that the strong shape dependence of the NLO corrections is significantly reduced if a dynamic scale like $H_T/4$ is employed. 
For both distributions, results based on CS and $q_T$ subtraction agree at the sub-percent level.

\subsection{Validation of the double-pole approximation}\label{sec:DPAvalidation_ttx_off-shell}
In the following, we aim to explore the validity of the DPA at NLO, both at fiducial and differential levels. 
This validation is essential to assess the quality of the approximation and to understand the pattern of factorisable and non-factorisable corrections at one-loop order, in view of the application of the DPA to the double-virtual contribution at NNLO.

As a first test, we study the quality of the DPA for different choices of the scale $\muIR$, which appears in the definition of the hard-virtual coefficient in Eq.~\eqref{eq:Hn_DPA_ttxoff-shell}. 
We compute the hard-virtual contribution $\Delta\sigma_{\mathrm{NLO,H}}\vert_{\mathrm{DPA}}$.\footnote{We recall that the hard-virtual contribution to the NLO correction is driven by the coefficient $H^{(1)}$ in Eq.~(\ref{eq:H}).} at scale \mbox{$\muIR = \widetilde{Q}$} (as defined in Sec.~\ref{sec:dpa_matrix}), which is our default choice, and alternatively at the fixed scale \mbox{$\muIR = m_t$}. In this second prescription, we add the exact shift due to the running of the one-loop amplitudes, and then compare with the exact hard-virtual contribution $\Delta\sigma_{\mathrm{NLO,H}}$, conventionally defined at scale $Q$. 
This analysis provides a way to identify whether there is a preferred IR subtraction scale that improves the performance of the approximation, while also allowing for a quantitative assessment of the systematic uncertainty associated with the approximated one-loop result.
\begin{table}[t]
\centering  
\renewcommand{\arraystretch}{1.5}
\setlength{\tabcolsep}{0.5em}
\makebox[\textwidth][c]{
\begin{tabular}{clcccc}
& 
\multicolumn{2}{c}{CKMP} & \multicolumn{2}{c}{CMP}\\
\toprule
$\sigma$ [fb]  & \multicolumn{1}{c}{$gg$} & \multicolumn{1}{c}{$q \bar q$} & \multicolumn{1}{c}{$gg$} & \multicolumn{1}{c}{$q \bar q$}\\
\cmidrule(lr){1-1}\cmidrule(lr){2-3}\cmidrule(lr){4-5}
\hline
$\Delta\sigma_{\mathrm{NLO,H}} \,@\,Q$  & $1973.2(1.8)$ & $\phantom{00} 401.2(3)$  & $5004.5(6.0) $  & $\phantom{-0} 570.8(4)$\\
\hline
\hline
$\Delta\sigma_{\mathrm{NLO,H}}|_{\mathrm{fact}} \,@\,\widetilde{Q}$ & $ 1992.1(1.9)$ & $\phantom{00} 398.2(3)$ & $4479.2(2.6)$ & $\phantom{-0} 510.8(3)$   \\
$\Delta\sigma_{\mathrm{NLO,H}}|_{\mathrm{nonfact}} \,@\,\widetilde{Q}$   & $-16.20(8)$ & $\phantom{0000} 3.53(2)$ & $\phantom{0}522.8(3)$ & $\phantom{-00} 60.81(3)$ \\
\hline
$\Delta\sigma_{\mathrm{NLO,H}}|_{\mathrm{DPA}} \,@\,\widetilde{Q}$   & $1975.9(1.9)$ & $\phantom{00} 401.7(3)$ & $ 5002.0(2.6)$ & $\phantom{-0} 571.6(3)$\\
\hline
\hline
$\Delta\sigma_{\mathrm{NLO,H}}|_{\mathrm{fact}} \,@\,m_t$    & $ 1394.5(1.1)$ & $\phantom{00} 255.7(2)$ & $ 3066.3(1.7)$ & $\phantom{-0} 322.2(2)$ \\
$\Delta\sigma_{\mathrm{NLO,H}}|_{\mathrm{nonfact}} \,@\,m_t$  & $ -7.84(8)$ & $\phantom{0000} 4.37(2)$& $\phantom{0} 445.3(2)$ & $\phantom{-00} 52.69(3)$  \\
exact shift $m_t \to Q$    & $\phantom{0} 576.9(3)$ & $\phantom{00} 141.86(7)$& $ 1487.9(7)$ & $\phantom{-0} 196.96(9)$\\
\hline
$\Delta\sigma_{\mathrm{NLO,H}}|_{\mathrm{DPA}} \,@\,m_t$ + shift    & $1963.6(1.1)$ & $\phantom{00}401.9(2)$& $5001.5(1.8)$ & $\phantom{-0} 571.8(2)$\\
\bottomrule
\end{tabular}
}
\caption{\label{tab:WWbb_fact-nonfact} Comparison between the exact one-loop hard-virtual contribution $\Delta\sigma_{\mathrm{NLO,H}}$, defined at scale \mbox{$\muIR = \muR = Q$}, and the same quantity computed in DPA, $\Delta\sigma_{\mathrm{NLO,H}}|_{\mathrm{DPA}}$. We apply the approximation directly at scale \mbox{$\muIR = \muR = \widetilde{Q}$} or at scale \mbox{$\muIR = \muR = m_t$}. If the DPA is applied at scale $m_t$, we need to add the contribution of the one-loop running (next-to-last row), based on the exact $t \bar t$ off-shell Born amplitudes, in order to construct the virtual result at scale $Q$. We also show the numerical impact of the factorisable and non-factorisable one-loop corrections separately. The errors in parentheses refer to the statistical uncertainties from the Monte Carlo integration.}
\end{table}

In Tab.~\ref{tab:WWbb_fact-nonfact} we list the results for $\Delta\sigma_{\mathrm{NLO,H}}$ and $\Delta\sigma_{\mathrm{NLO,H}}\vert_{\mathrm{DPA}}$, in the two setups described in Sec.~\ref{sec:setup_nlo}, distinguishing the $gg$ and $q \bar q$ partonic channels. 
The main observations are:
\begin{itemize}
	\item the DPA works extremely well at both scales $\widetilde{Q}$ and $m_t$, leading to an impressive agreement with the exact results. 
	Indeed, differences between the exact and approximated $H^{(1)}$ contributions account for less than $0.5\%$ in both setups and for the two partonic channels.
	This implies that the systematic error due to the approximation is negligible at the cross-section level, being much smaller than the scale uncertainties;
	\item in both setups, the largest contribution to the virtual corrections originates from the factorisable one-loop corrections, but the overall numerical impact of the non-factorisable contribution on the NLO correction is rather different;
	\item in the CKMP setup, the non-factorisable corrections represent $0.8\%~(0.6\%)$ of the  corresponding factorisable corrections at scale $\widetilde{Q}~(m_t)$ in the $gg$ channel while, for the $q \bar q$ channel, the impact is $0.9\%~(1.7\%)$ at scale $\widetilde{Q}~(m_t)$;
	\item in contrast, in the CMP setup, where cuts are applied on the bottom jets, the non-factorisable corrections represent $12\%~(15\%)$ of the corresponding factorisable corrections at scale $\widetilde{Q}~(m_t)$ in both partonic channels.
	Even though numerically smaller, due to cancellations between the individual contributions, the sum of the non-factorisable virtual corrections in the two diagonal partonic channels is of the same size as the complete NLO correction in this setup.
\end{itemize}
\begin{figure}
  \subfloat[Azimuthal separation between the positron and the muon.]{\label{fig:1_fact-nonfact}{\includegraphics[width=0.48\textwidth]{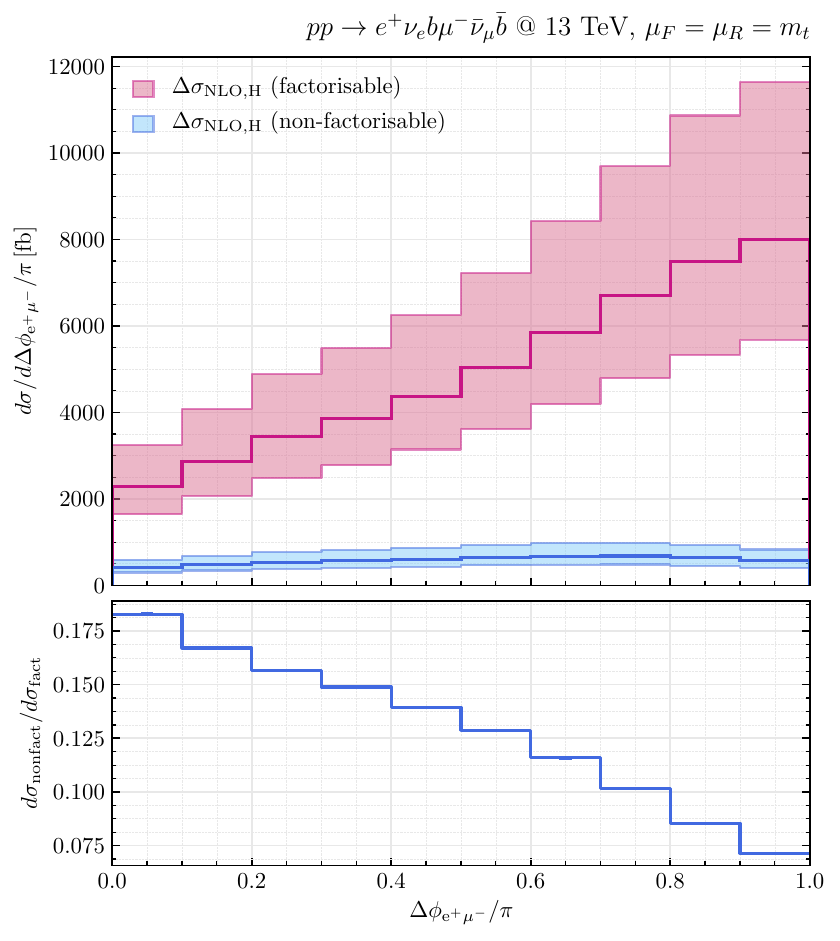}}} \hfill
  \subfloat[Invariant mass of the reconstructed top quark.]{\label{fig:2_fact-nonfact}{\includegraphics[width=0.48\textwidth]{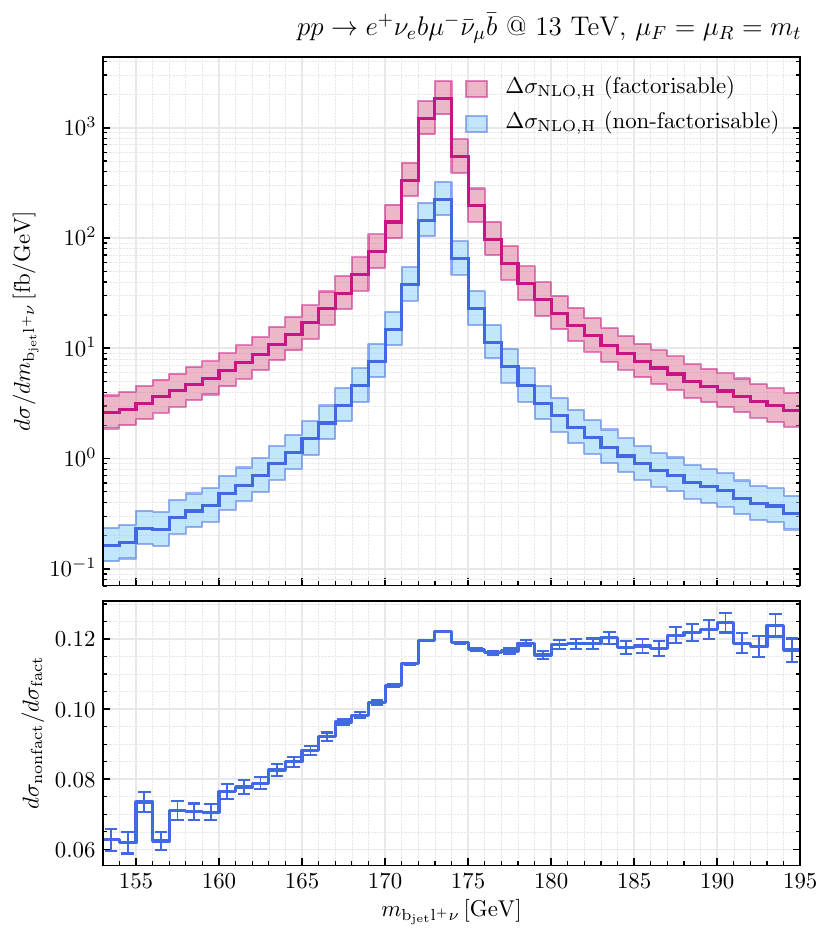}}} \\
  \subfloat[Transverse momentum of the charged lepton pair.]{\label{fig:3_fact-nonfact}{\includegraphics[width=0.48\textwidth]{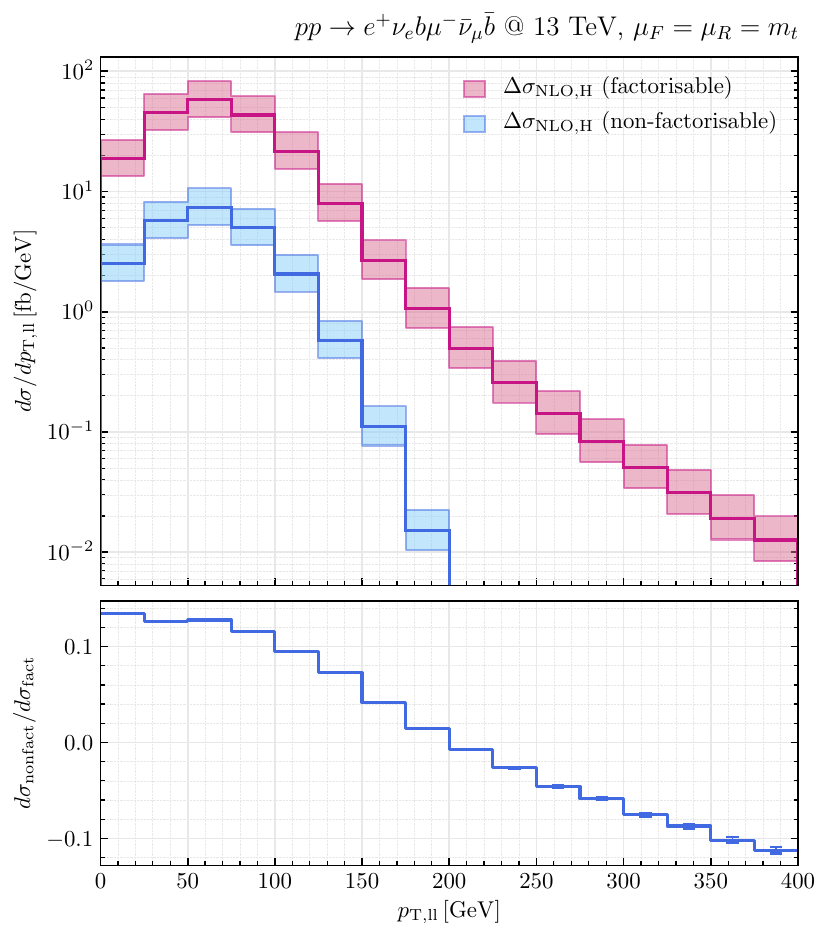}}} \hfill
  \subfloat[Invariant mass of the two bottom jets.]{\label{fig:4_fact-nonfact}{\includegraphics[width=0.48\textwidth]{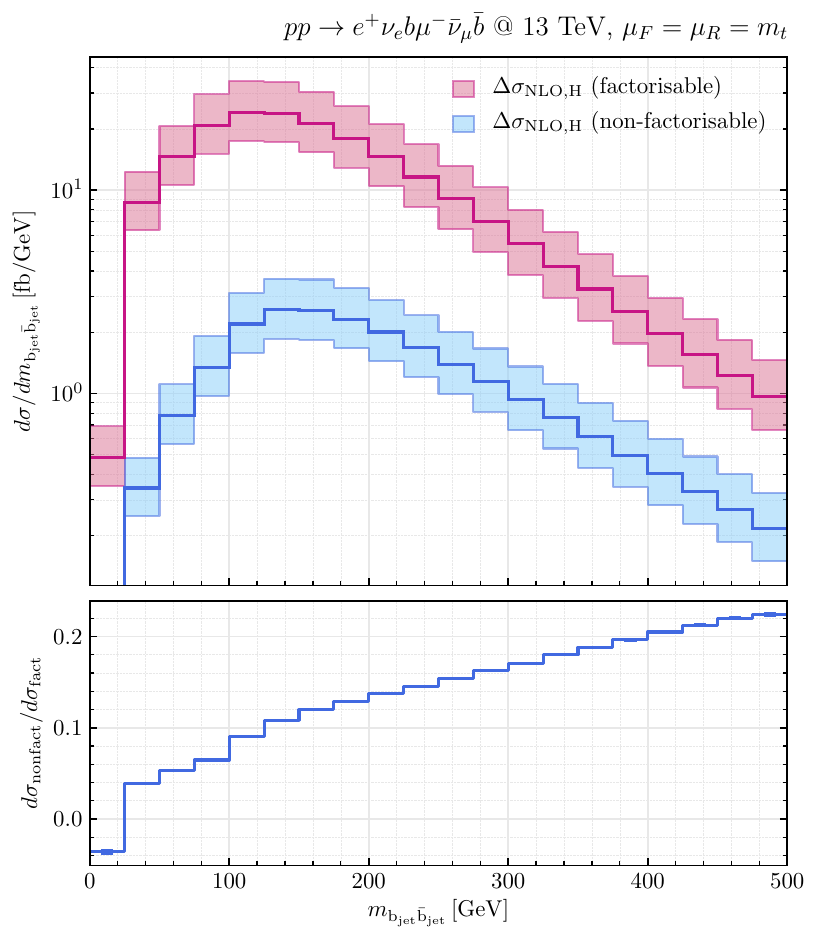}}}
    \caption{Comparison between factorisable (magenta curve) and non-factorisable (blue curve) one-loop corrections, computed at scale $\widetilde{Q}$ within the CMP setup at \mbox{$\sqrt{s} = 13$\,TeV} and central scale \mbox{$\mu_0 = m_t$}.}
    \label{fig:fact-nonfact_hierarchy}
\end{figure}
For the CMP setup, in Fig.~\ref{fig:fact-nonfact_hierarchy} we show differential distributions in the azimuthal-angle separation $\Delta\phi_{e^+\mu^-}$ and the transverse momentum $p_{T,ll}$ of the charged lepton pair, the invariant mass of the bottom-jet pair and of the \textit{reconstructed} top quark around the resonant peak (at Monte Carlo truth level). 
Despite the clear hierarchy between factorisable and non-factorisable one-loop corrections, the non-factorisable virtual contributions cannot be neglected. They account for up to $20\%$ of the corresponding factorisable corrections, thus representing a significant portion ($\mathcal{O}(15\%)$) of the NLO cross section. 
At the same time, the non-factorisable corrections show a strong kinematical dependence and can even change sign, as it happens for the $p_{T,ll}$ distribution. 

At this point, we want to highlight the well-known fact that the non-factorisable corrections are negligible at the integrated level if consistently removed from both real and virtual contributions~\cite{Melnikov:1993np,Fadin:1993kt}. This can be understood since the fully factorised construction of the NLO cross section is very close to the NWA, except for the inclusion of off-shell effects in the resonance propagators. 
We explicitly checked this statement by constructing a fully factorised version of the NLO cross section. 
This requires, beyond the factorisable one-loop corrections, the isolation of the factorisable part of the real-emission matrix elements as well as the construction of a fully factorised $q_T$-subtraction counterterm. Additionally, we need an extension of the on-shell projection, described in Appendix~\ref{app:Appendix_mapping} for a Born-like event, for the evaluation of the real-emission contribution.  

For this check, we considered the CMP setup where, as discussed in Sec.~\ref{sec:DPAvalidation_ttx_off-shell}, the net impact of the non-factorisable virtual corrections on the NLO correction is substantial.
At the integrated level, we observe that the fully factorised approach leads to an NLO correction which is \mbox{$\sim 3\%$} smaller than that obtained in the DPA. This implies that the non-factorisable corrections are indeed subleading if consistently removed from real and virtual contributions.
However, non-negligible effects can arise at the differential level, as illustrated in the following.

\begin{figure}
 \subfloat[Azimuthal separation between the positron and the muon.]{\label{fig:1_DPA-fullyfact}{\includegraphics[width=0.48\textwidth]{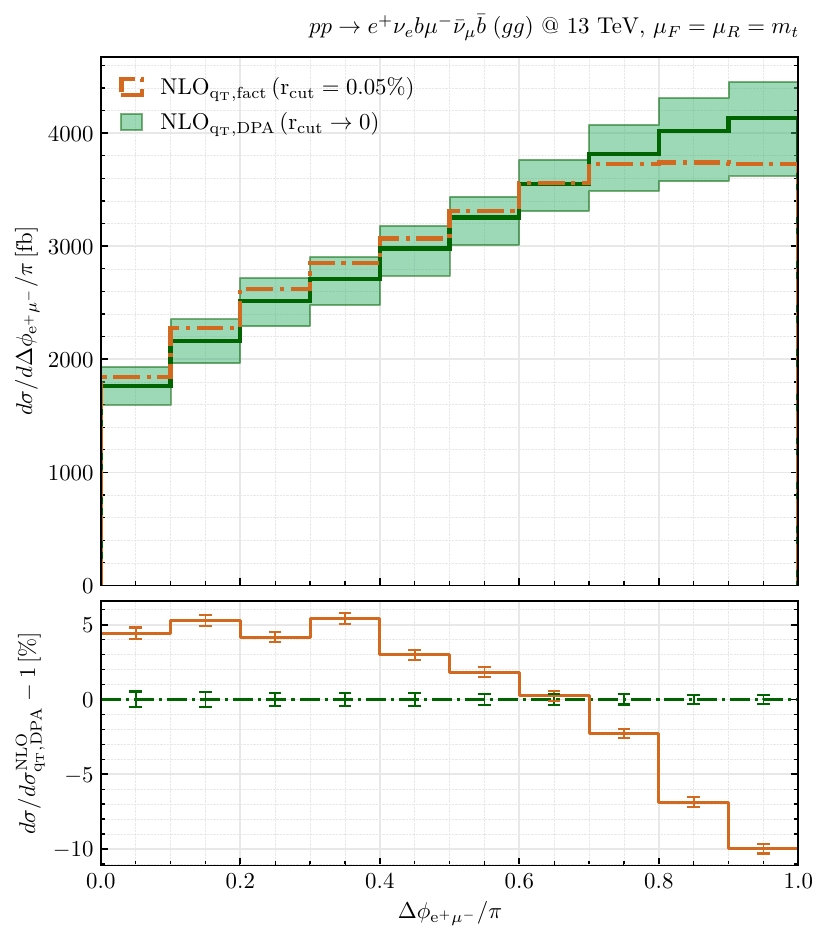}}} \hfill
 \subfloat[Invariant mass of the reconstructed top quark.]{\label{fig:2_DPA-fullyfact}{\includegraphics[width=0.48\textwidth]{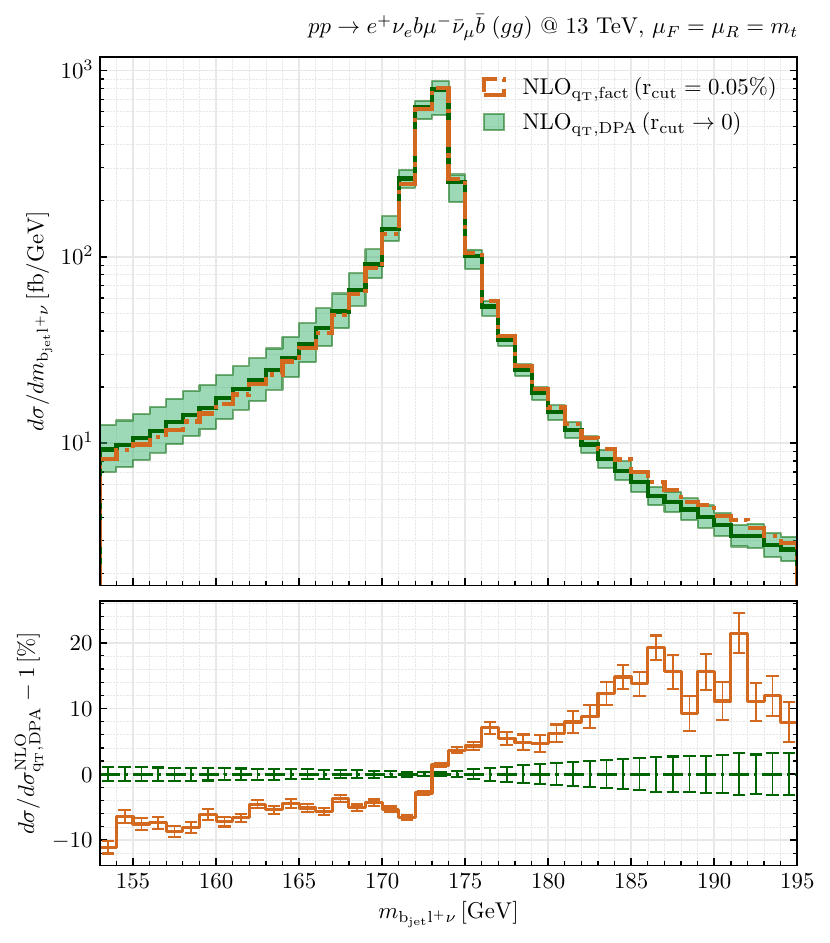}}} \\
 \subfloat[Transverse momentum of the charged lepton pair.]{\label{fig:3_DPA-fullyfact}{\includegraphics[width=0.48\textwidth]{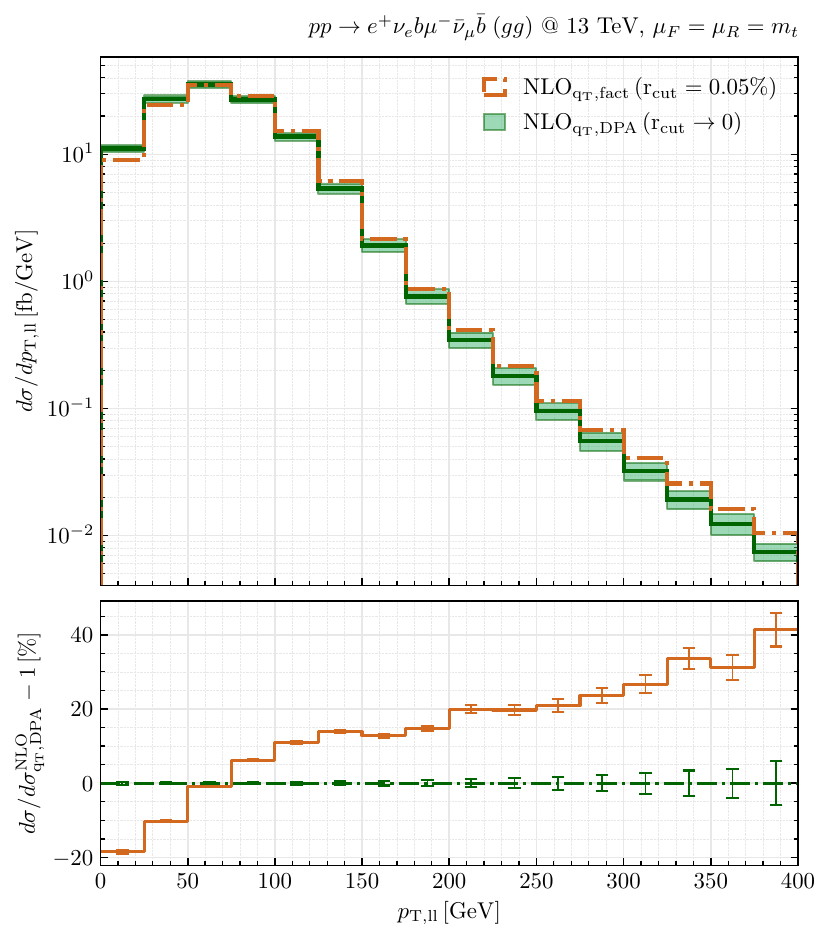}}} \hfill
   \subfloat[Invariant mass of the two bottom jets.]{\label{fig:4_DPA-fullyfact}{\includegraphics[width=0.48\textwidth]{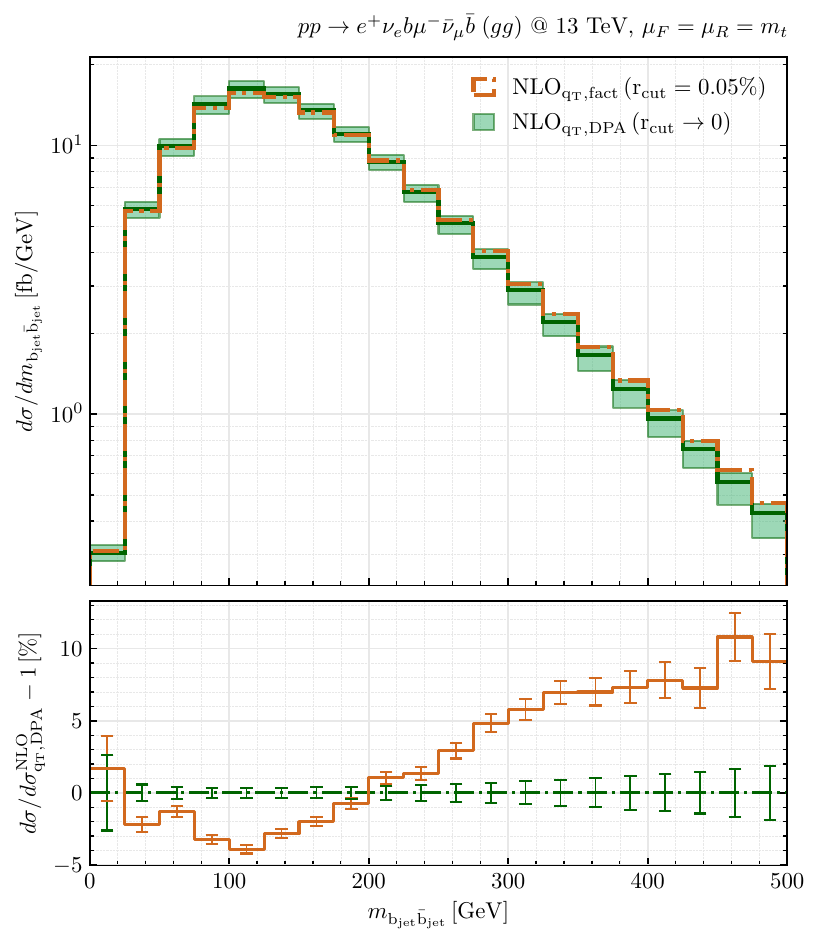}}}
   \caption{NLO differential results for the CMP setup at \mbox{$\sqrt{s} = 13$\,TeV} and central scale \mbox{$\mu_0 = m_t$}. 
  We consider the $gg$ partonic channel as a case study.
  We show a comparison between the NLO result in which only the virtual contribution at scale $Q$ has been approximated in DPA (green curve), and the NLO result obtained by consistently removing the non-factorisable corrections in both real and virtual contributions (brown curve). 
  The green bands in the upper panel of each plot are the perturbative scale-variation bands, while the error bars in the lower panels stand for the uncertainties from the Monte Carlo integration.}
    \label{fig:gg_DPA-fullyfact_comparison}
\end{figure}

In Fig.~\ref{fig:gg_DPA-fullyfact_comparison} we show the same kinematical distributions as in Fig.~\ref{fig:fact-nonfact_hierarchy}, but for the specific case of the gluon-fusion channel. We display a comparison between the NLO differential cross sections in the fully factorised approach denoted as NLO$_{\rm q_T,fact}$ (brown curves) and in the DPA labelled as NLO$_{\rm q_T,DPA}$ (green curves).
From the lower ratio plots, it becomes evident that the effects of the ``missing" non-factorisable corrections in the fully factorised approach have a clear kinematic dependence. In distributions particularly sensitive to off-shell effects, they can reach up to $40\%$ in the kinematic tails. 
Thus, a fully factorised construction cannot be seen as satisfactory for a reliable approximation at NNLO. 

Finally, we present a direct comparison, at the differential level, between the exact NLO results (labelled as $\mathrm{NLO_{q_T}}$) and those obtained by approximating the one-loop finite remainder in DPA (labelled as $\mathrm{NLO_{q_T, DPA}}$).
The impressive agreement observed in Tab.~\ref{tab:WWbb_LONLOxs} at the integrated level is confirmed for several relevant distributions. 
Here, we limit ourselves to commenting on the four observables already described in Sec.~\ref{sec:distributions_ttx_off-shell}, namely the invariant mass and the transverse momentum of the bottom-quark and the bottom-jet pair, respectively, as well as the invariant mass and the azimuthal-angle separation of the two charged leptons. 
Corresponding differential distributions are reported in Fig.~\ref{fig:NLOqT_NLOqTDPA_CKMP_setup} for the CKMP setup at 8 TeV, and in Fig.~\ref{fig:NLOqT_NLOqTDPA_CMP_setup} for the CMP setup at 13 TeV.
Each plot presents three panels. The upper panel displays the comparison between $\mathrm{NLO_{q_T}}$, $\mathrm{NLO_{q_T, DPA}}$ and $\mathrm{NLO_{q_T, DPA}^{fact}}$, where the last label refers to the DPA result in which we include only the factorisable one-loop corrections in the construction of the approximated finite remainder. 
In the central panel, we show the difference relative to the exact NLO results, expressed in per cent. 
In the bottom panel, we display the ratio between $\mathrm{NLO_{q_T, DPA}}$ and $\mathrm{NLO_{q_T}}$ to remark the excellent agreement between the exact results and those in DPA.
Error bands in the upper and central panels are obtained from the scale variations. 

\begin{figure}
  \subfloat[Invariant mass of the two bottom quarks.]{\label{fig:1_NLOqT_NLOqTDPA_CKMP_setup}{\includegraphics[width=0.47\textwidth]{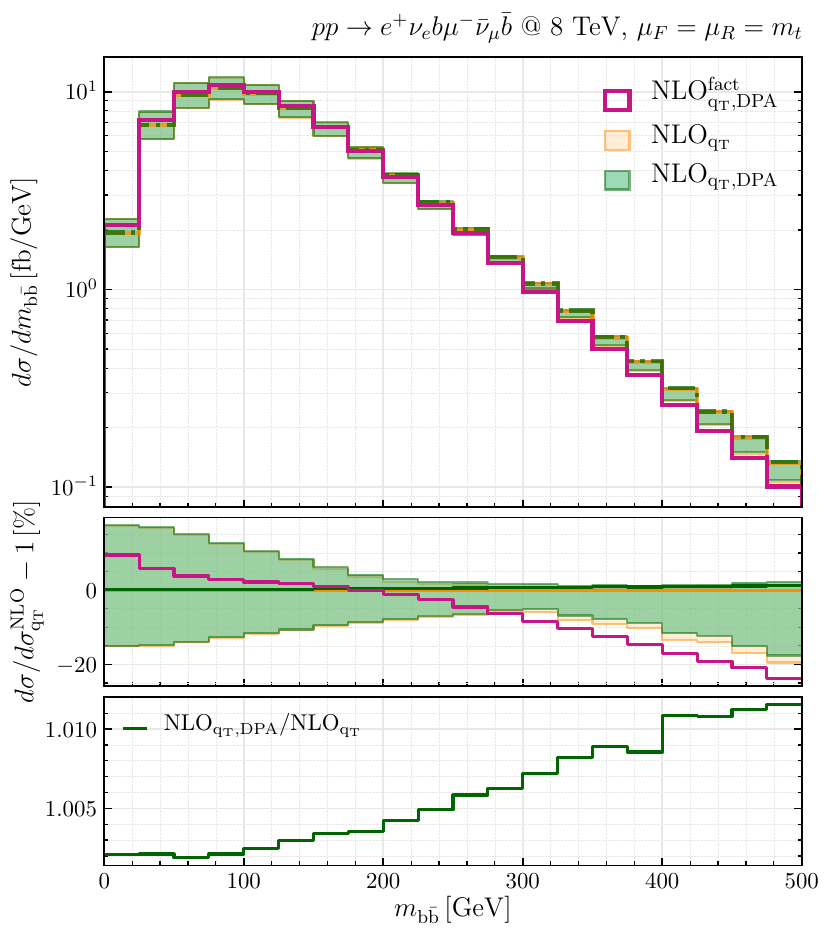} }} \hfill
  \subfloat[Transverse momentum of the two bottom quarks.]{\label{fig:2_NLOqT_NLOqTDPA_CKMP_setup}{\includegraphics[width=0.47\textwidth]{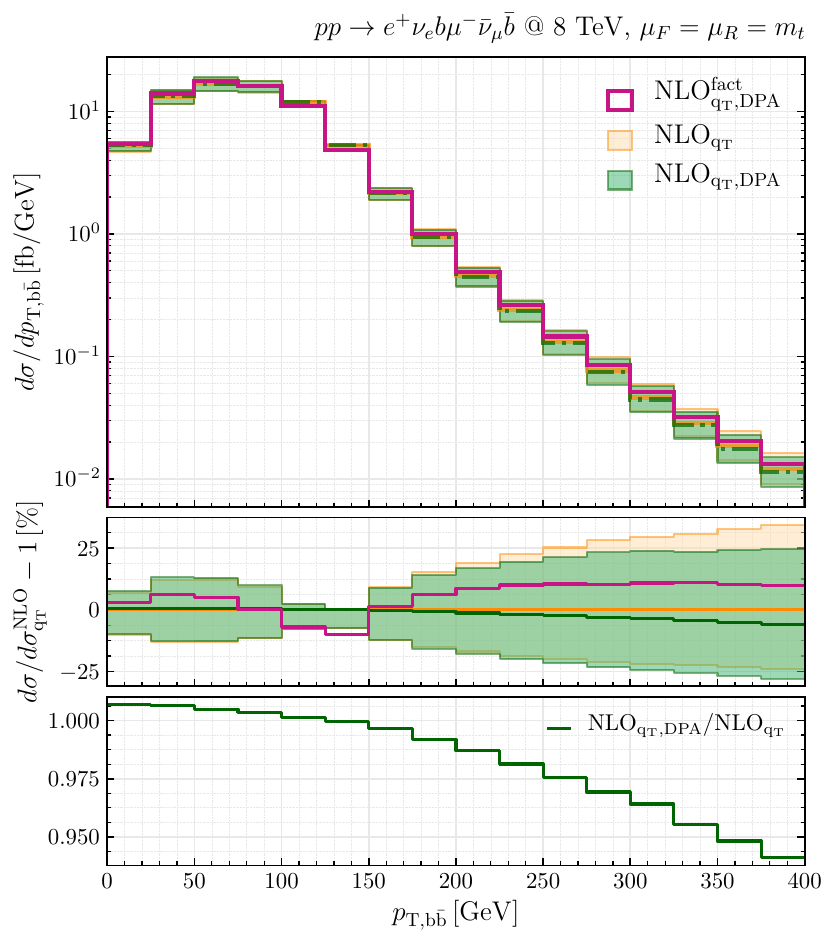} }} \\
  \subfloat[Invariant mass of the two charged leptons.]{\label{fig:3_NLOqT_NLOqTDPA_CKMP_setupp}{\includegraphics[width=0.47\textwidth]{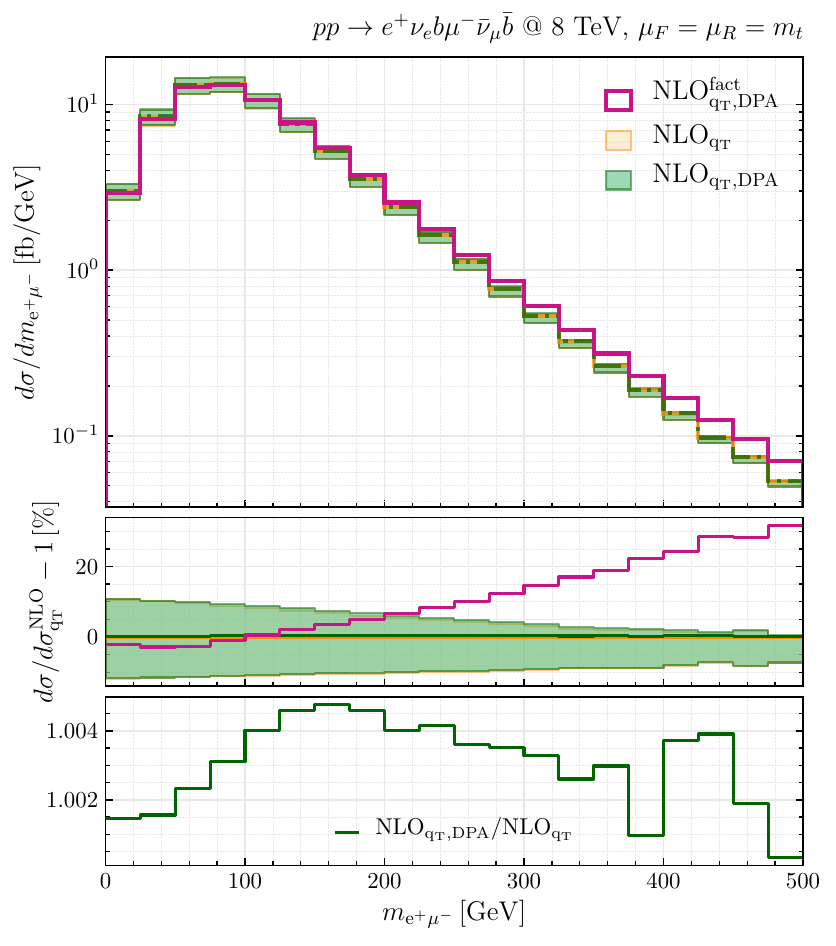} }}  \hfill
\subfloat[Azimuthal separation of the positron and the muon.]{\label{fig:4_NLOqT_NLOqTDPA_CKMP_setup}{\includegraphics[width=0.47\textwidth]{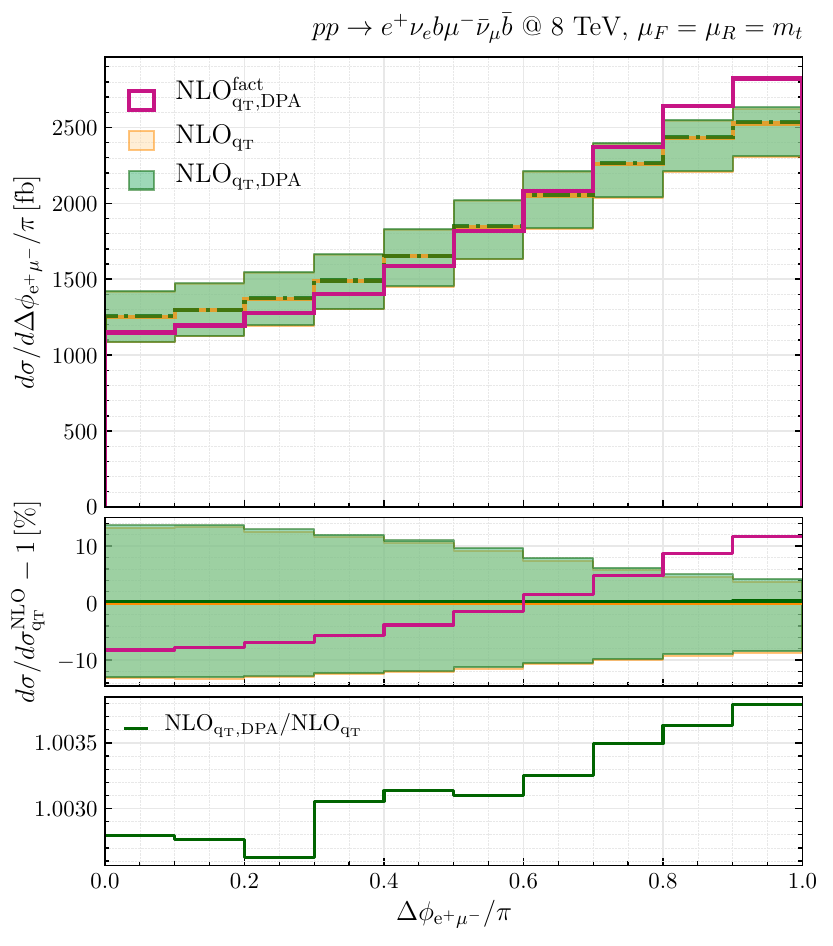} }} 
  \caption{NLO differential results for the CKMP setup at \mbox{$\sqrt{s} = 8$\,TeV}. The central scale is fixed at \mbox{$\mu_0 = m_t$}.}
    \label{fig:NLOqT_NLOqTDPA_CKMP_setup}
\end{figure}
\begin{figure}
  \subfloat[Invariant mass of the two bottom jets.]{\label{fig:1_NLOqT_NLOqTDPA_CMP_setup}{\includegraphics[width=0.47\textwidth]{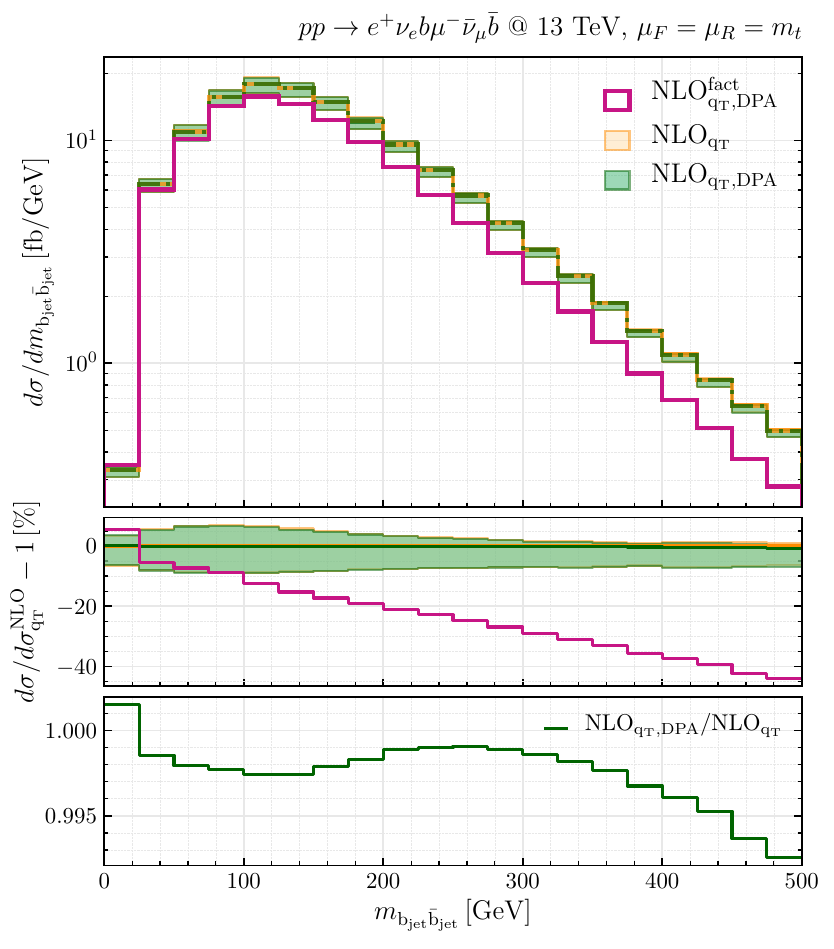} }} \hfill
  \subfloat[Transverse momentum of the two bottom jets.]{\label{fig:2_NLOqT_NLOqTDPA_CMP_setup}{\includegraphics[width=0.47\textwidth]{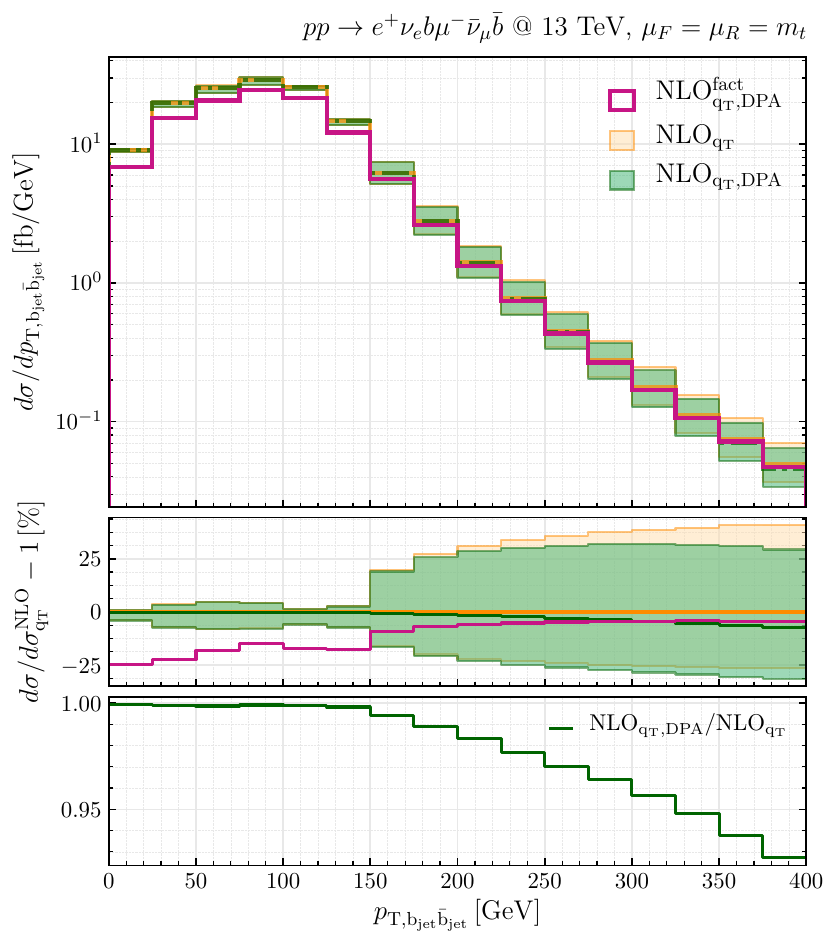} }} \\
  \subfloat[Invariant mass of the two charged leptons.]{\label{fig:3_NLOqT_NLOqTDPA_CMP_setup}{\includegraphics[width=0.47\textwidth]{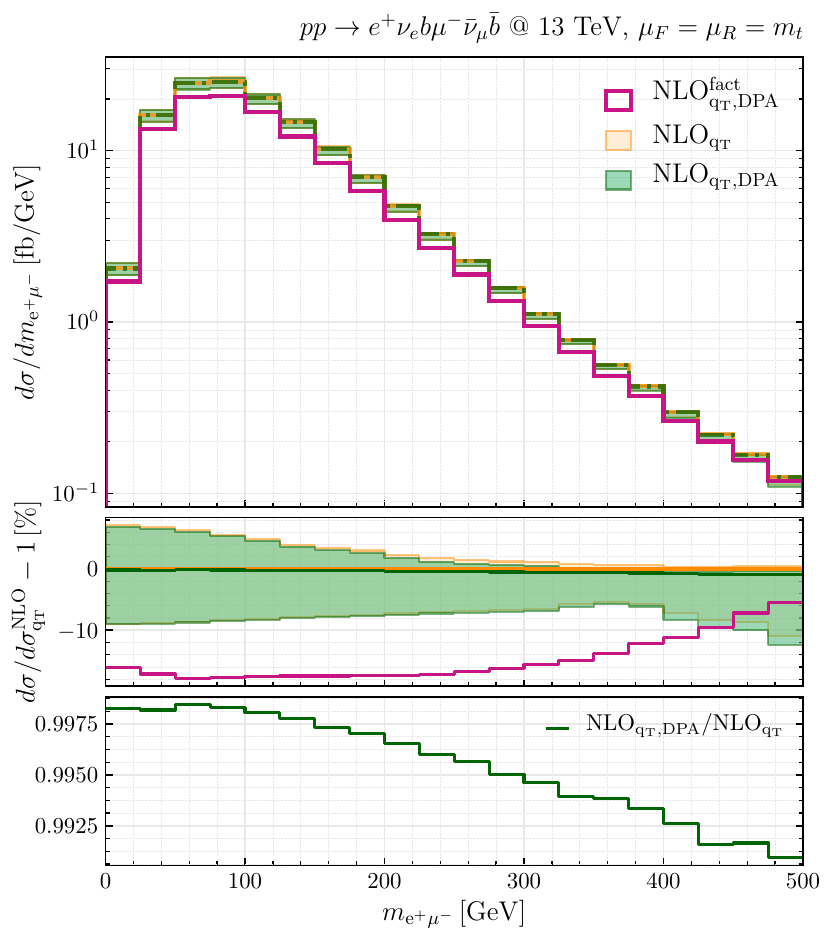} }}  \hfill
\subfloat[Azimuthal separation of the positron and the muon.]{\label{fig:4_NLOqT_NLOqTDPA_CMP_setup}{\includegraphics[width=0.47\textwidth]{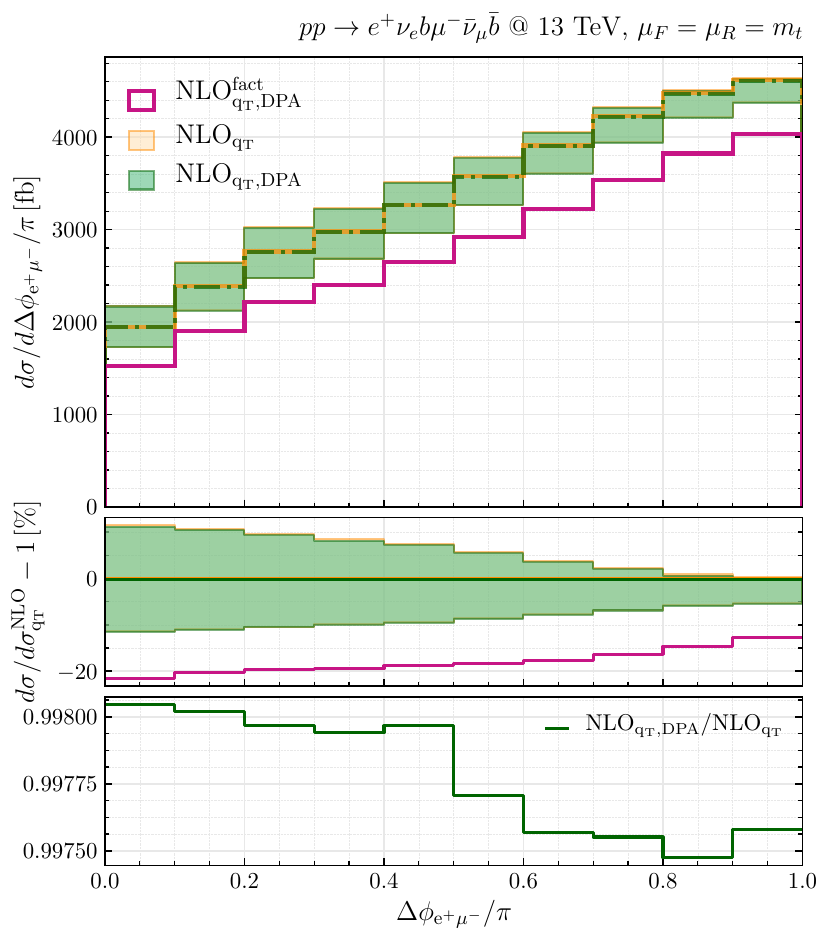} }}
  \caption{NLO differential results for the CMP setup at \mbox{$\sqrt{s} = 13$\,TeV}. The central scale is fixed at \mbox{$\mu_0 = m_t$}. }
    \label{fig:NLOqT_NLOqTDPA_CMP_setup}
\end{figure}
In both setups we observe that the difference between $\mathrm{NLO_{q_T}}$ and $\mathrm{NLO_{q_T, DPA}}$ is below $1\%$ in all phase space regions, except for the transverse momentum of the pair of bottom quarks (Fig.~\ref{fig:2_NLOqT_NLOqTDPA_CKMP_setup}) and bottom jets (Fig.~\ref{fig:2_NLOqT_NLOqTDPA_CMP_setup}), respectively, where differences can reach $5\%$ in the tails. 
Such discrepancies are expected since the high-$p_{T,b\bar b}$ tail is particularly sensitive to off-shell and non-resonant effects, as already commented in Sec.~\ref{sec:distributions_ttx_off-shell}, which are not captured by the DPA.

If we take into account only the contribution from the factorisable one-loop corrections (purple curves), effects due to the missing non-factorisable one-loop contribution could cause significant shape distortions.
The pattern is, however, different in the two setups.
In the CKMP setup (see Fig.~\ref{fig:NLOqT_NLOqTDPA_CKMP_setup}), the non-factorisable one-loop corrections are not uniform over the phase space and have the largest impact in the tails of the invariant-mass distributions. However, due to the strong hierarchy between factorisable and non-factorisable corrections (see Tab.~\ref{tab:WWbb_fact-nonfact}), the effects on the NLO cross section are only \mbox{$\sim 3\%$}.
On the contrary, in the CMP setup, the differential results for $\mathrm{NLO_{q_T, DPA}^{fact}}$ are systematically below the exact NLO results, and the neglected non-factorisable corrections can lead to differences of $15-20\%$ also in the bins that mostly contribute to the cross section.
Therefore, a large portion of the NLO cross section would be missed.
This is a consequence of the fact that, even though the non-factorisable virtual corrections represent on average \mbox{$\sim 10\%$} of the corresponding factorisable corrections (see Fig.~\ref{fig:fact-nonfact_hierarchy}), large numerical cancellations happening among different perturbative ingredients of the subtraction formula make the non-factorisable one-loop contribution relevant at the cross-section level.

In conclusion, we have shown that the construction of the NLO cross section based on the DPA leads to an impressive agreement with the exact computation, at both fiducial and differential levels, for two quite different setups. 
The non-factorisable one-loop corrections are found to be numerically smaller than the corresponding factorisable corrections, but this hierarchy does not justify neglecting those corrections a priori. Indeed, for a setup exclusive over the bottom jets, the non-factorisable virtual corrections, if not included, can lead to sizeable differences in the NLO cross section.
The impact of the non-factorisable corrections is small at the integrated level only if they are consistently excluded from both real and virtual contributions, achieved via a fully factorised approach similar to the NWA.
The suppression of these corrections has been proven in Refs.~\cite{Melnikov:1993np,Fadin:1993kt}.

\newpage
\section{Towards an NNLO accurate prediction}
\label{sec:ttxoffshell_NNLO_results}
%
At NNLO, all necessary building blocks for the construction of the factorisable virtual corrections are available and have been implemented in the \Matrix framework together with the complete real--virtual and double-real off-shell contributions, as discussed in Secs. \ref{sec:DPA} and \ref{sec:framework}, respectively.  

To complete the construction of the DPA at two-loop level for off-shell $t\bar t$ production and decays, the corresponding non-factorisable corrections are required. 
Although their full computation is not yet available, their functional dependence on the top-quark width is well understood. In particular, the non-factorisable virtual corrections exhibit a logarithmic sensitivity to $\Gamma_t$, arising from the exchange of soft gluons between the production and decay stages of the two resonances.
The top-quark width acts as a
regulator of the IR singularities associated with the emission (absorption) of
soft gluons from (by) a resonant top quark, and at most a single
$\log(\Gamma_t)$ can arise per loop order. Therefore, up to two-loop order, the
functional form of the non-factorisable virtual contribution to the cross section is given by
\begin{align}
\Delta\sigma_{\mathrm{NLO, H}}\vert_{\nonfact} &=  B^{(1)}_{\rm nf} \log\left( \frac{\Gamma_t}{Q_{h}} \right) + C^{(1)}_{\rm nf} + \mathcal{O}\left ( \frac{\Gamma_t}{Q_{h}}\right)\,,
	\label{eq:functional-form_oneloop-nonfact}\\
\Delta\sigma_{\mathrm{NNLO, H}}\vert_{\nonfact} &= A^{(2)}_{\rm nf} \log^2\left( \frac{\Gamma_t}{Q_{h}} \right) + B^{(2)}_{\rm nf} \log\left( \frac{\Gamma_t}{Q_{h}} \right) + C^{(2)}_{\rm nf} + \mathcal{O}\left ( \frac{\Gamma_t}{Q_{h}}\right)\,,
	\label{eq:functional-form_twoloop-nonfact}
\end{align}
where $Q_{h}$ refers to any relevant hard scale of $\mathcal{O}(m_{t})$ and ``nf" is a shortcut for ``non-factorisable".
Power-suppressed terms of $\mathcal{O}(\Gamma_t/Q_{h})$ contribute beyond the strict DPA,
which at NLO~(see Eq.~\eqref{eq:functional-form_oneloop-nonfact}) corresponds to the integrated version of the one-loop expression in Eq.~\eqref{eq:nonfact_M1M0_ttx}.
We notice that this functional form also holds at the
differential level, where the coefficients $A^{(i)}_{\rm nf}, B^{(i)}_{\rm nf}$ and $C^{(i)}_{\rm nf}$, with \mbox{$i=1,2$} for one- and two-loop
order, respectively, acquire a dependence on the Born kinematics. 
However, when combining factorisable and non-factorisable virtual corrections with real contributions,
the logarithmic dependence on $\Gamma_{t}$ must cancel at the level of the (N)NLO physical predictions.

Based on these considerations, a numerical strategy can be devised for extracting the non-factorisable corrections. 
Indeed, their logarithmic behaviour can be determined by evaluating the (N)NLO cross section at different
values of the top-quark width, with a rescaling ensuring that the
effective top-quark decay branching fractions remain constant. 
More precisely, we compute the total cross
section of the complete process~\eqref{eq:off-shell_ttx_dilepton} including only factorisable virtual corrections for progressively smaller
values of the top-quark width in the range \mbox{$0 < \Gamma_t < \Gatphys$}. We then rescale these cross sections by
the ratio $(\Gamma_t/\Gatphys)^2$ 
as
\begin{equation}
	\overline{\sigma}(\Gamma_t) = \sigma(\Gamma_t) \left( \frac{\Gamma_t}{\Gatphys} \right)^{\!2} \,,
	\label{eq:XS_width-rescaled}
\end{equation}
to adjust for the difference between $\Gamma_t$ and $\Gatphys$.
The residual logarithmic dependence of the (N)NLO
correction \textit{without} the non-factorisable virtual corrections, $\Delta\sigma_{\mathrm{(N)NLO}_{\mathrm{q_T, DPA}}^{\mathrm{fact}}}$, can be written as 
\begin{align}
\Delta\sigma_{\NLO_{\mathrm{q_T, DPA}}^{\mathrm{fact}}} &=  - B^{(1)} \log\left( \frac{\Gamma_t}{m_{t}} \right) - C^{(1)} - D^{(1)}  \left(\frac{\Gamma_t}{m_{t}}\right) +\mathcal{O}\left(\frac{\Gamma_t^2}{m_t^2} \right)\,,
\label{eq:functional-form_oneloop_real}\\
\Delta\sigma_{\NNLO_{\mathrm{q_T, DPA}}^{\mathrm{fact}}} &= - A^{(2)} \log^2\left( \frac{\Gamma_t}{m_{t}} \right) - B^{(2)} \log\left( \frac{\Gamma_t}{m_{t}} \right) - C^{(2)}  - D^{(2)}  \left(\frac{\Gamma_t}{m_{t}}\right) +\mathcal{O}\left( \frac{\Gamma_t^2}{m_t^2} \right)\,,
	\label{eq:functional-form_twoloop_real}
\end{align}
where we identify the hard scale $Q_{h}$ with the top-quark mass $m_{t}$.
The parameters $A^{(i)}$ and $B^{(i)}$ can be fitted to determine the logarithmic coefficients \mbox{$A^{(2)}_{\mathrm{nf}}=A^{(2)}$} and \mbox{$B^{(i)}_{\mathrm{nf}}=B^{(i)}$} of the non-factorisable corrections in Eqs.~\eqref{eq:functional-form_oneloop-nonfact}--\eqref{eq:functional-form_twoloop-nonfact}. 
However, the constant terms $C^{(i)}_{\mathrm{nf}}$ in Eqs.~\eqref{eq:functional-form_oneloop-nonfact}--\eqref{eq:functional-form_twoloop-nonfact} cannot be extracted solely through this approach. 
A key additional consideration is that the total off-shell (N)NLO cross section must match, in the \mbox{$\Gamma_{t} \to 0$} limit,
the result in NWA, where all necessary perturbative ingredients are available. 
This matching procedure enables the determination of the
missing constant in the non-factorisable corrections, ultimately yielding a complete
evaluation of the off-shell cross section at the physical top-quark width $\Gatphys$ in DPA.
In the parametrisation of Eqs.~\eqref{eq:functional-form_oneloop_real}--\eqref{eq:functional-form_twoloop_real}, which determines our fit model, we include explicit linear terms in $\Gamma_t/m_t$ with coefficients $D^{(i)}$ to account for potentially significant power-suppressed contributions to the cross section, which might otherwise jeopardise the accuracy of the fit for the non-factorisable virtual corrections.

In the following, we adopt the aforementioned strategy to predict the fully inclusive 
cross section for off-shell $t\bar{t}$ production and decays. In this specific case, 
a fully differential implementation of the NWA (as obtained in Ref.~\cite{Czakon:2020qbd}) is not required.
Indeed, our approach
consists of matching the inclusive $e^+\nu_e b \mu^-\bar{\nu}_{\mu} \bar{b}$ cross section,
in the limit \mbox{$\Gamma_t \to 0$}, to the on-shell $t \bar{t}$ cross section as obtained in Ref.~\cite{Catani:2019iny}, which has been validated against \TOPplusplus~\cite{Czakon:2011xx}. 
The precise matching of off-shell cross sections, in the small-width
limit, to corresponding predictions for the on-shell \mbox{$pp \to t\bar{t}$} process requires careful examination.
Indeed, spurious top-quark width effects arise when truncating the perturbative series. This issue will be discussed in detail in Sec.~\ref{sec:Delta-term} and in Appendix~\ref{app:Appendix_top_width}.

Although conceptually straightforward, the outlined numerical strategy for obtaining the non-factorisable two-loop corrections via a small-width expansion combined with an on-shell matching poses significant computational challenges for the
employed Monte Carlo framework. 
The numerical evaluation of the \mbox{$\Gamma_t \to 0$} limit within $q_T$ subtraction is particularly demanding due to the interplay of multiple singular behaviours induced by the widely separated scales: the top-quark width $\Gamma_t$, the $q_T^{\text{cut}}$ cutoff, and the hard scale \mbox{$Q \sim m_t$} of the process.
Achieving sufficiently accurate and numerically stable NNLO predictions becomes 
highly non-trivial and computationally intensive, as detailed in 
Sec.~\ref{sec:stability}.
The small-width analysis is further challenged by systematic uncertainties
related to the residual dependence on the technical cutoff $\rcut$. 
Corresponding studies of the behaviour of the cross section as a function of
$\rcut$ for different top-quark widths are presented in
Sec.~\ref{sec:num_extrapolation} for a numerical setup previously introduced in
Sec.~\ref{sec:nnlo_setup}. Subsequently, in Sec.~\ref{sec:num_extrapolation}, we
continue by validating the extrapolation procedure in the \mbox{$\Gamma_{t}\to 0$} limit
at lower orders (LO and NLO).
After performing these validations, this method is applied to extract the non-factorisable two-loop corrections in DPA.
Finally, in Sec.~\ref{sec:NNLO_results}, we present results for the off-shell NNLO inclusive
cross section at the physical top-quark width.

\subsection[Matching to the inclusive \texorpdfstring{$t \bar t$}{ttx} cross section]{Matching to the inclusive $\boldsymbol{t \bar t}$ cross section}\label{sec:Delta-term}
In the \mbox{$\Gamma_t \to 0$} limit, the cross section for the off-shell process~\eqref{eq:off-shell_ttx_dilepton} factorises into top-quark pair production
and decays, consistently with the NWA, with suppressed contributions from single-top
and non-resonant diagrams, as well as their interferences with double-resonant
topologies. In the NWA, this factorisation holds to all orders in perturbation
theory, with radiative corrections separating cleanly into contributions
associated with the production and decay subprocesses.
For a fully inclusive phase space, in the \mbox{$\Gamma_t \to 0$} limit, the
off-shell calculation should reproduce the on-shell $t \bar t$ cross
section multiplied by the $t$ and $\bar t$ decay branching fractions, namely
\begin{equation}
	\lim_{\Gamma_t \to 0} \int d\sigma_{\mathrm{off-shell}}  = \int d\sigma_{\proddec}
	 = \sigma_{t\bar{t}} \underbrace{ \frac{\Gamma_{t\to e^+ \nu_e b}}{\Gamma_t} }_{\text{BR}(t\to e^+ \nu_e b)} \underbrace{ \frac{\Gamma_{\bar{t}\to \mu^{-}\bar{\nu}_{\mu} \bar b}}{\Gamma_t}}_{\text{BR}(\bar t \to \mu^{-}\bar{\nu}_{\mu} \bar b)} \equiv \sigma_{\mathrm{on-shell}}  \,,
	 \label{eq:matching-to-onshell_ttx}
\end{equation}
where $d\sigma_{\proddec}$ denotes the all-order differential cross section in NWA and $\sigma_{t\bar{t}}$ is the inclusive cross section for on-shell $t \bar t$ production.
Since the (anti-)top quark decays almost exclusively into a $W$ boson and a (anti-)bottom quark, it is reasonable to approximate the partial (anti-)top width as
\begin{equation}
	\Gamma_{t\to e^+ \nu_e b} \approx \Gamma_t \times \text{BR}(W^{+} \to e^+ \nu_e) ~~~~\text{and}~~~~ \Gamma_{\bar{t}\to \mu^{-}\bar{\nu}_{\mu} \bar b} \approx \Gamma_t \times \text{BR}(W^{-} \to \mu^{-}\bar{\nu}_{\mu}) \,.
	\label{eq:approx_partial-top-width}
\end{equation}
Since we consider massless leptons in our off-shell calculation, we adopt for
both lepton species the branching ratio \mbox{$\text{BR}(W \to l \nu_l) = 10.8598\%$} computed from the
leading-order partial $W$ decay-width into leptons divided by the (measured)
total width $\Gamma_W$, using the EW parameters described below in Sec.~\ref{sec:nnlo_setup}.

Apart from Coulomb effects near the $t \bar t$ threshold,
Eq.~\eqref{eq:matching-to-onshell_ttx} is indeed valid to all orders in perturbation
theory~\cite{Fadin:1993kt}. However, when truncating the perturbative expansion, a careful treatment of 
$1/\Gamma_{t}$ terms is necessary to ensure that the NWA and off-shell results in the \mbox{$\Gamma_t \to 0$} limit exactly reproduce the inclusive
$t\bar{t}$ cross section order by order in the $\alphas$ expansion~\cite{Melnikov:2009dn}. In particular, setting
$\Gamma_{t}$ to the value computed at the same perturbative order as the off-shell calculation introduces spurious higher-order contributions.
Such terms are formally beyond the given fixed-order accuracy,
but they can lead to non-negligible effects, as pointed out in Ref.~\cite{Denner:2012yc} at NLO.
In order to ensure the perturbative equivalence between the NWA and the off-shell results in the small-width limit with the on-shell results,
we follow a procedure inspired by Ref.~\cite{Denner:2012yc}.%
\footnote{An alternative procedure consists
  in expanding and truncating the $1/\Gamma_{t}$ factors (see
  Refs.~\cite{Melnikov:2009dn,Campbell:2012uf,Czakon:2020qbd,Jezo:2023rht}).} Specifically, we compute an order-by-order correction term
$\Delta\sigma^{\mathrm{N}^n\LO}_{\mathrm{trunc}}$ as
\begingroup 
\allowdisplaybreaks
\begin{align}
	\Delta\sigma^{\mathrm{N}^n\LO}_{\mathrm{trunc}} &= \sigma_{\mathrm{on-shell}}^{\mathrm{N}^n\LO} - \int d\sigma^{\mathrm{N}^n\LO}_{\proddec} \notag \\
	&= \frac{1}{\left( \Gamma_t^{\mathrm{N}^n\LO} \right)^2} \sigma_{t \bar t}^{\mathrm{N}^n\LO}\, \Gamma_{t \to i}^{\mathrm{N}^n\LO}\, \Gamma_{\bar t \to j}^{\mathrm{N}^n\LO} - \int d\sigma^{\mathrm{N}^n\LO}_{\proddec} \,.
	\label{eq:delta-term_trunc}
\end{align}
\endgroup 
The expression of $\Delta\sigma^{\mathrm{N}^n\LO}_{\mathrm{trunc}}$
contains higher-order terms in the $\alphas$ expansion, and is obtained by
expanding the r.h.s.\ of Eq.~\eqref{eq:delta-term_trunc} in the strong coupling
while keeping the factor $(\Gamma_t^{\mathrm{N}^n\LO})^{-2}$ fixed. The relevant
expressions for $\Delta\sigma^{\mathrm{N}^n\LO}_{\mathrm{trunc}}$ up to \mbox{$n=2$} are
reported in Appendix~\ref{app:Appendix_top_width}. This subtraction term is used to match the off-shell computation in the \mbox{$\Gamma_t \to 0$} limit to the on-shell computation, thus removing the higher-order spurious width effects. 

\subsection{Numerical setup}\label{sec:nnlo_setup}
The numerical results shown in this section are obtained considering
a fully inclusive setup, i.e.\ no restrictions are imposed on the momenta of
charged leptons, bottom jets or missing energy. We consider proton--proton
collisions at a centre-of-mass energy of \mbox{$\sqrt{s}=13$\,TeV}. The
electroweak parameters are set to
\begin{align}
	m_W = 80.385 \,\mathrm{GeV} ~~&,~~~~\Gamma_W = 2.0928 \,\mathrm{GeV} \,, \\
	m_Z = 91.1876 \,\mathrm{GeV} ~~&,~~~~\Gamma_Z = 2.4952 \,\mathrm{GeV}  \,,\\
	G_F = 1.16637 \cdot 10^{-5} \,\mathrm{GeV}^{-2}  ~~&,~~~~
	\alpha = \frac{\sqrt 2}{\pi} G_F m_W^2 \left( 1- \frac{m_W^2}{m_Z^2} \right) \,,
	\label{eq:EW_parameters}
\end{align}
and unstable particles are treated in the complex-mass scheme. We consider a
diagonal CKM matrix. The pole masses of the top and bottom quarks are set to \mbox{$m_t=172.5$\,GeV} and \mbox{$m_b = 4.75$\,GeV},
while the Higgs boson mass and width are set to \mbox{$m_H = 125$\,GeV} and
\mbox{$\Gamma_H = 4.1$\,MeV}, respectively. The LO, NLO and NNLO top-quark widths are
\begin{align}
	\Gamma_{t}^{\mathrm{LO}} = 1.45364 \,\mathrm{GeV} ~~,~~~ \Gamma_{t}^{\mathrm{NLO}} &= 1.3365 \,\mathrm{GeV} ~~,~~~ \Gamma_{t}^{\mathrm{NNLO}} = 1.30913\,\mathrm{GeV} \,,
\end{align}
where the NLO and NNLO values have been computed at scale
\mbox{$\muR = m_t$}. For their calculation, we rely on the analytic results of
Ref.~\cite{Chen:2022wit}. The values of $\Gamma_t$ include bottom-mass effects
as well as finite $W$-width corrections, while NLO EW effects are not taken into
account. At LO we use the \verb|NNPDF30_lo_as_0118_nf_4| PDF
set~\cite{NNPDF:2014otw}, while we exploit the
\verb|NNPDF31_nlo_as_0118_nf_4|
(\verb|NNPDF31_nnlo_as_0118_nf_4|)
set~\cite{Ball:2017nwa} at NLO (NNLO).
The running QCD coupling $\alphas$ is evaluated at \mbox{$(n+1)$}-loop
accuracy within our N$^n$LO (\mbox{$n=0,1,2$}) predictions, as provided by the LHAPDF library~\cite{Buckley:2014ana}. In the
renormalisation of the strong coupling, only contributions from light-quark and
gluon loops are taken into account, while top and bottom quarks are decoupled.
We set the central renormalisation and factorisation scales to
\begin{equation}
	\mu_0 = \muR = \muF = m_t 
\end{equation}
and, as in Sec.~\ref{sec:ttxoffshell_NLO_results}, we estimate theory uncertainties due to missing higher-order terms via 
standard seven-point $\muF,\muR$ variations. 
When performing scale variations, the top-quark width is kept unchanged. This
implies that the matching with the on-shell $t \bar{t}$ prediction, in the 
\mbox{$\Gamma_t \to 0$} limit, is consistent only for the central scale choice $\mu_0$.

\subsection{Numerical extrapolations in the small width and slicing parameters}\label{sec:num_extrapolation}

In the following Sec.~\ref{sec:rcut_dependence} we present the behaviour of the cross section as a function of
$\rcut$ for different values of the top-quark width. 
Subsequently, in Sec.~\ref{sec:rcut_extrapolation} we present details of our $\rcut$ extrapolation procedure in the \mbox{$\Gamma_{t}\to 0$} limit. We continue in Sec.~\ref{sec:Gammat_extrapolation} by validating the \mbox{$\Gamma_{t}\to 0$} extrapolation at lower orders (LO and NLO), and for the non-diagonal NNLO partonic channels. 
In the latter case, our off-shell NNLO predictions are exact, and consistency with
the small-width limit serves as an additional validation of their correctness. 
These validations are performed prior to ultimately applying the method to
extract the non-factorisable two-loop corrections in DPA.

To perform the numerical extrapolation in the \mbox{$\Gamma_t \to 0$} limit, we compute
predictions for seven values of $\Gamma_t$ in the range $[0.01, 1] \times
\Gamma_t^{\text{phys}}$, namely $\{0.01, 0.02, 0.05, 0.1, 0.2, 0.5, 1\} \times
\Gamma_t^{\text{phys}}$. For the diagonal $q\bar{q}$ and $gg$ channels, we
include two additional larger values, $\{2, 5\} \times \Gamma_t^{\text{phys}}$,
to enhance sensitivity to power-suppressed corrections. Unless these additional
larger values provide significant additional insight, we restrict our
presentation to results within the $[0.01, 1] \times \Gamma_t^{\text{phys}}$
range. Here we emphasise again that achieving stable numerical results for
progressively smaller top-quark widths is highly non-trivial, as discussed in
Sec.~\ref{sec:stability}.

\subsubsection{Dependence on the slicing parameter in the small-width limit}\label{sec:rcut_dependence}
%
We begin the discussion of the numerical predictions by analysing the dependence
of the NLO and NNLO results on the slicing parameter \mbox{$r_{\text{cut}} \equiv q_T^{\mathrm{cut}}/Q$},
where $Q$ is the invariant mass of the two bottom quarks and the four leptons.
\begin{sidewaysfigure}
  \centering
  \includegraphics[width=\textwidth]{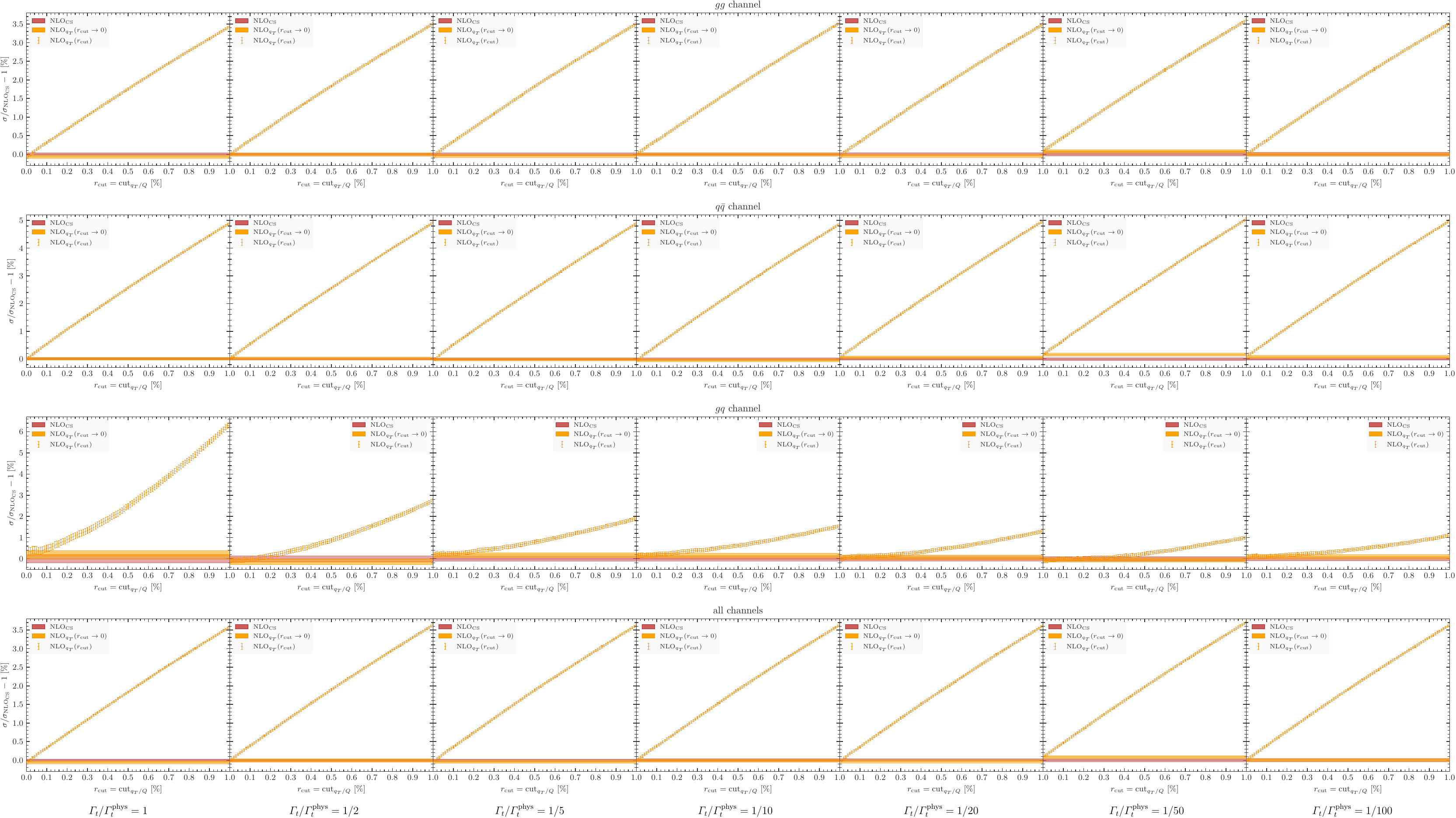}
  \caption{Study of the $\rcut$-dependence at NLO in QCD, for different partonic
	channels, $gg$, $q \bar q$, $gq$, as well as the combination of all
	channels. The seven columns refer to the different values of the top-quark
	width $\Gamma_t$ considered for the small-width extrapolation. The
	extrapolated result in the \mbox{$\rcut \to 0$} limit (orange band) is compared
	with the $\rcut$-independent cross section (red band) obtained by
	regularising the IR divergences with the CS dipole subtraction.}
	\label{fig:NLO_rcut-dependence}
\end{sidewaysfigure}
Figures~\ref{fig:NLO_rcut-dependence} and \ref{fig:NNLOqT_fact_rcut-dependence}
show the NLO and NNLO results, respectively. Each row corresponds to a specific
partonic channel, while columns represent progressively smaller top-quark width values
in the range $[0.01,1] \times \Gatphys$.

At NLO (see Fig.~\ref{fig:NLO_rcut-dependence}), we compare the extrapolated
result in the \mbox{$\rcut \to 0$} limit with the $\rcut$-independent cross section
obtained by regularising the IR divergences via CS dipole subtraction. For all partonic channels, we find perfect agreement with
the CS results even for the lowest $\Gamma_t$ value, thus showing an exquisite
numerical control and stability of the implementation. The power corrections in
$\rcut$ are linear for the $gg$/$q\bar{q}$ diagonal channels, and quadratic
for the $gq$ channel. This trend is expected since linear power corrections 
originate from soft wide-angle radiation emitted by the FS heavy
quarks~\cite{Buonocore:2019puv}. At NLO, soft corrections of this kind are absent
in the $gq$ channel. The dependence of the power corrections
on the top-quark width is rather mild and mostly visible in the $gq$ channel,
with relative variations of a few per cent in the range $\rcut \in [0.01, 1]\%$
at the cross-section level.

In Fig.~\ref{fig:NNLOqT_fact_rcut-dependence} we display corresponding NNLO 
results. For each partonic channel, the results are
normalised to the corresponding $\rcut$-independent cross sections obtained by
performing an extrapolation in the \mbox{$\rcut \to 0$} limit. Such extrapolation, together with the standard \Matrix extrapolation dubbed \textit{linquad} in Fig.~\ref{fig:NNLOqT_fact_rcut-dependence}, will be detailed in
Sec.~\ref{sec:rcut_extrapolation}. We highlight that the results for the off-diagonal channels are
complete and exact, while the double-virtual contribution in the diagonal $gg/q\bar{q}$ partonic channels is approximated
by considering only the factorisable
corrections in DPA. Being $\rcut$-independent, the double-virtual contribution
can only change the overall normalisation of the NNLO correction, but not the
behaviour in $\rcut$.
%
The off-diagonal partonic channels (third and fourth rows of
Fig.~\ref{fig:NNLOqT_fact_rcut-dependence}) display the expected $r_{\text{cut}}$
behaviour: missing power corrections are linear for the $gq$ channel and
quadratic for the \mbox{$qq + \bar{q} \bar{q} + q' \bar{q}$} channel (that we dub {\it rest} channel in the following) that opens
up at NNLO. In the $gq$ channel, relative power corrections are extremely large, 
reducing by a factor of 5 at \mbox{$r_{\text{cut}} \sim 1\%$} when $\Gamma_t$
decreases from $\Gatphys$ to $\Gatphys/100$.
However, here the normalisation is
to the full $gq$-initiated cross section, which is not necessarily positive
definite. When normalised to a stable quantity like the total NLO cross section,
power corrections are approximately \mbox{$-1.5\%$} at \mbox{$r_{\text{cut}} \sim 1\%$},
largely independent of $\Gamma_t$.

Power corrections in the diagonal $gg$ and $q\bar q$ channels (see first and
second rows of Fig.~\ref{fig:NNLOqT_fact_rcut-dependence}) exhibit a steeper
behaviour for \mbox{$\rcut \to 0$}, consistent with a linear scaling enhanced by
logarithmic terms. As we transition from $\Gatphys$ to progressively smaller
widths, we observe a smooth evolution in the shape of the power corrections. At
smaller widths, their behaviour qualitatively aligns with that seen in on-shell
$t\bar t$ production, although the relative impact of power corrections in $\rcut$
is more pronounced. The turning point in the shape of the power
corrections in $\rcut$, observed around $\Gatphys/10$, can be explained by a change of sign 
of the coefficients of the $\rcut \log(\rcut)$ and $\rcut \log^2(\rcut)$
power corrections. Separately, the various coefficients of the linear power
corrections in $\rcut$ can depend logarithmically on the top-quark width, but
this logarithmic dependence must cancel in the sum since the cross section is
IR-safe with respect to soft-gluon emissions from the resonant top-quark
propagators.

\subsubsection[\texorpdfstring{$\rcut \to 0$}{Gammat} extrapolation procedure]{$\boldsymbol{\rcut \to 0}$ extrapolation procedure}\label{sec:rcut_extrapolation}
%
The employed $r_{\text{cut}}$ extrapolation procedure warrants further discussion. The
standard \Matrix{} implementation of the $r_{\text{cut}}$ extrapolation uses a
simplified polynomial fit that ignores logarithmically enhanced $r_{\text{cut}}$
corrections and neglects power corrections beyond second order, while restricting the range of $\rcut$ values that enter the best fit.
We denote this ansatz as \textit{linquad} since the functional form of the fit
includes purely linear plus quadratic power corrections in $\rcut$. 
The advantage of this simplified ansatz lies in its process independence and, in particular, in its robustness when
  extrapolations are required for results with limited statistics and correspondingly large fluctuations between individual
  $\rcut$ values --- at the price of larger extrapolation uncertainties. However, this ansatz does not
capture the true functional form of power corrections in general, and it becomes sensitive
to the chosen fit range. When logarithmically enhanced power corrections are
present and create a steeper behaviour in $\rcut$, the fit range within the \textit{linquad} ansatz
should be as close as possible to the \mbox{$r_{\text{cut}} \to 0$} limit in order to minimise
extrapolation bias. This sensitivity can be exploited to optimise the range
selection and estimate uncertainties on the extrapolated result. For all
processes studied within the \Matrix{} framework based on $q_T$ subtraction, this
simplified quadratic ansatz with conservative uncertainty estimates provides
reliable results when the fit range is chosen as a reasonable subset of the full standard range $[0.01,1]\%$ (see Ref.~\cite{Grazzini:2017mhc}). 

In this work, we consider an alternative procedure which is tailored to the
expected behaviour of the power corrections in $\rcut$. We refine the model for
each individual partonic channel by also including logarithmically enhanced
contributions. More precisely, at NNLO we consider the following fit model for the different partonic channels:
\begin{itemize}
  \item  $\sigma^{(gg)}_{\rm NNLO}(\rcut) = \sigma^{(gg)}_{\rm NNLO}(0) + \rcut  \left( A_{2}^{(gg)}(\Gamma_t) \ln^2 \rcut + A_{1}^{(gg)}(\Gamma_t) \ln \rcut + A_{0}^{(gg)}(\Gamma_t) \right)$,
  \item  $\sigma^{(q\bar{q})}_{\rm NNLO}(\rcut) = \sigma^{(q\bar{q})}_{\rm NNLO}(0) + \rcut  \left( A_{2}^{(q\bar{q})}(\Gamma_t) \ln^2 \rcut + A_{1}^{(q\bar{q})}(\Gamma_t) \ln \rcut + A_{0}^{(q\bar{q})}(\Gamma_t) \right)$,
  \item  $\sigma^{(gq)}_{\rm NNLO}(\rcut) = \sigma^{(gq)}_{\rm NNLO}(0) + \rcut   A_{0}^{(gq)}(\Gamma_t)  + \rcut^{2} \left( B_{3}^{(gq)}(\Gamma_t) \ln^3 \rcut + B_{2}^{(gq)}(\Gamma_t) \ln^2 \rcut \right) $,
  \item  $\sigma^{(\rm rest)}_{\rm NNLO}(\rcut) = \sigma^{(\rm rest)}_{\rm NNLO}(0) +  \rcut^{2}\left( B_{1}^{(\rm rest)}(\Gamma_t) \ln \rcut + B_{0}^{(\rm rest)}(\Gamma_t) \right) $.
\end{itemize}
For the diagonal channels, we have checked that the inclusion of additional quadratic power
corrections leads to consistent results with larger uncertainties due to the
increased number of fitting parameters. For this reason, we do not include them
in the functional form of the fit used to produce our final results. Regarding
the off-diagonal channels, the $gq$ channel starts contributing at NLO, where
power corrections are quadratic with a single logarithmic enhancement~\cite{Ebert:2018gsn,Cieri:2019tfv}. In this channel, at NNLO a linear
behaviour emerges due to soft wide-angle radiation, similar to the NLO case for
the diagonal channels. Therefore, we include a purely linear term and quadratic
contributions with the highest powers of the logarithmically enhanced term
$\{\ln^{3}\rcut,\ln^{2}\rcut\}$. The {\it rest} channel opens up only at
NNLO, and the power corrections are expected to be quadratic with, at most, a
single logarithmic enhancement, informing the corresponding fit model.

\begin{sidewaysfigure}
  \centering
  \includegraphics[width=\textwidth]{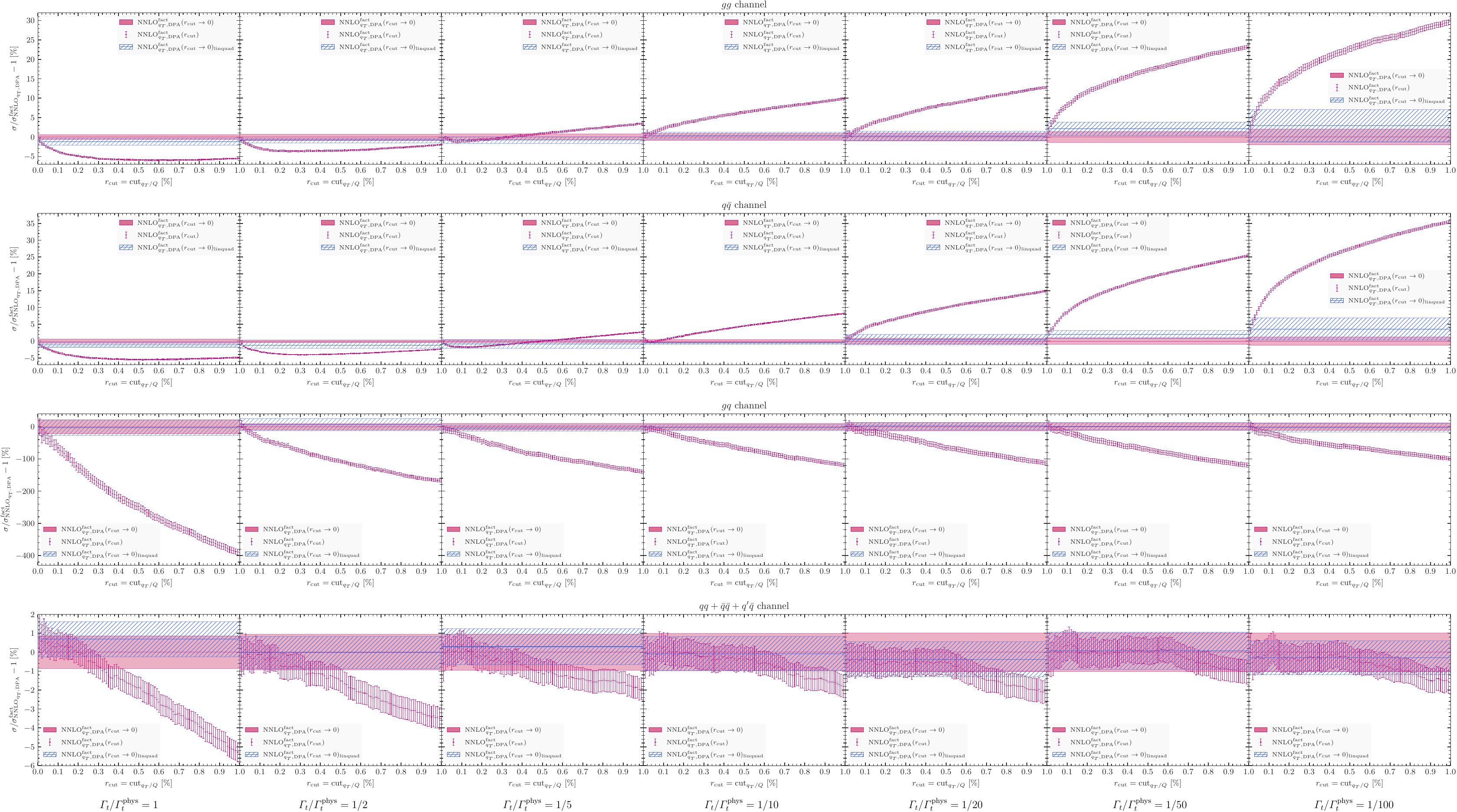}
  \caption{Study of the $\rcut$-dependence at NNLO in QCD, for different
	partonic channels: $gg$, $q \bar q$, $gq$ and
	\mbox{$qq + \bar{q} \bar{q} + q' \bar q$} from the first to the fourth row,
	respectively. The seven columns refer to the different values of the
	top-quark width $\Gamma_t$ considered for the small-width extrapolation. 
	For each partonic channel, the results at
	fixed $\rcut$ are normalised to the extrapolated cross section, in the limit
	\mbox{$\rcut \to 0$}, for that particular channel.}
	\label{fig:NNLOqT_fact_rcut-dependence}
\end{sidewaysfigure}

The cumulant distribution is obtained in a single run, thus the
  integration errors on the results at different $\rcut$ values are strongly
  correlated. In order to estimate the uncertainty associated with the
  extrapolated result in the \mbox{$\rcut \to 0$} limit, we apply a statistical
  procedure inspired by the replica method (see e.g.
  Ref.~\cite{Costantini:2024wby} and references therein). We do not make any
  assumption on correlations, and perform a Monte Carlo scan over all possible
  pseudo-datasets contained in (or even exceeding) the envelope of the original
  set. The procedure works as follows: for each partonic channel, we construct
$n$ replicas of the data points
\mbox{$(\rcuti, \sigma_{\rm NNLO}(\rcuti) \pm \delta\sigma_{\rm NNLO}(\rcuti))$} by
generating the pseudo-result $\sigma^{[j]}_{\rm NNLO}(\rcuti)$ for the $j$-th
replica at $\rcuti$ according to a Gaussian distribution centred at the original
central value $\sigma_{\rm NNLO}(\rcuti)$ with a variance given by the Monte
Carlo uncertainty $\delta\sigma_{\rm NNLO}(\rcuti)$. We generate \mbox{$n=1000$}
replicas and repeat the fit in $\rcut$ using the functional form discussed
above, thus obtaining $n$ values $\{\sigma^{[j]}_{\rm NNLO}(0)\}_{j=1}^{n}$ for
the extrapolated result at \mbox{$\rcut \to 0$}. We finally define our ``best"
extrapolated value as the arithmetic mean of the replicas and the corresponding
uncertainty as the standard deviation multiplied by a factor $\xi$,
\begin{equation}
  \begin{split}
  \sigma_{\rm NNLO}(0) &\approx  \bar{\sigma}_{\rm NNLO}(0) \pm \delta \sigma_{\rm NNLO}(0) \\
  &\equiv \frac{1}{n}\sum_{j=1}^{n}\sigma^{[j]}_{\rm NNLO}(0) \pm \frac{\xi}{\sqrt{n-1}} \sum_{j=1}^{n} \sqrt{\left(\sigma^{[j]}_{\rm NNLO}(0)-\bar{\sigma}_{\rm NNLO}(0)\right)^{2}} \,,
\end{split}
\end{equation}
separately for each partonic channel. We set \mbox{$\xi = 2$}, which in a strictly statistical interpretation of the uncertainty corresponds to the $95\%$ confidence level.

In Fig.~\ref{fig:NNLOqT_fact_rcut-dependence}, we display the results obtained
with the standard \textit{linquad} and improved extrapolation procedures.
Here, the \textit{linquad} fit was performed in the range \mbox{$\rcut \in [0.01\%-0.25\%]$}.
To assign a conservative estimate of the extrapolation error within this fit model, the upper bound of the $\rcut$ interval has been varied within the range \mbox{$[0.25\%-1\%]$}.
In Fig.~\ref{fig:NNLOqT_fact_rcut-dependence} the dashed blue band corresponds to the \textit{linquad} procedure,
while the violet bands correspond to the $\rcut$-extrapolated results obtained via the replica method described above.
As expected, the two procedures perform similarly for the off-diagonal channels,
with comparable error estimates across most of the considered width values. This
level of agreement can be explained by the fact that the logarithmically
enhanced contributions only contribute at the $\rcut^{2}$ level. In the diagonal
channels ($gg$ and $q \bar q$), the \textit{linquad} standard procedure
generally leads to an extrapolated result which
  is close to the cross section evaluated at the lowest $\rcut$ value in the
  considered fit range. This behaviour is a consequence of the fact that the
  \textit{linquad} ansatz fails to capture the steeper trend in $\rcut$ due to
  the logarithmically enhanced power correction. However, this is somewhat taken
  into account by the larger error assigned to the extrapolated result (dashed
  blue band), especially for the lowest width values. The replica procedure
  including logarithmically enhanced terms tends, instead, to follow the steeper
  behaviour, thus extrapolating further away from the last $\rcut$ point. In the
  following, we will consider the replica method based on the functional forms
  detailed above as our default input for the subsequent \mbox{$\Gamma_t \to 0$}
  extrapolation. As an alternative, we also consider the \textit{linquad} fit as
  input for the \mbox{$\Gamma_t \to 0$} extrapolation. The impact of this variation on
  the final NNLO cross section, reported in Sec.~\ref{sec:NNLO_results}, turns
  out to be well within the assigned uncertainties.

\subsubsection[\texorpdfstring{$\Gamma_t \to 0$}{Gammat} extrapolation procedure]{$\boldsymbol{\Gamma_t \to 0}$ extrapolation procedure}\label{sec:Gammat_extrapolation}
%
In this section, we discuss the extrapolation in the \mbox{$\Gamma_t \to 0$} limit. 
We start considering LO predictions, where the residual dependence on $\Gamma_{t}$ is due to finite-width effects in double-resonant diagrams as well as single-resonant and non-resonant topologies. 
In Fig.~\ref{fig:LO_Gammat_extrapolation}, we show the
behaviour of the LO cross section, in the $gg$
(upper panel) and $q\bar{q}$ (lower panel) partonic channels, as functions of the top-quark width. 
The error bars on the off-shell results at fixed $\Gamma_t$ correspond to Monte
Carlo uncertanties. The on-shell result is displayed in blue with an error band
corresponding to the numerical uncertainty of the on-shell $t \bar t$ calculation
multiplied by the BRs of the top and anti-top quarks. The extrapolated cross
section in the \mbox{$\Gamma_t/m_t \to 0$} limit is shown in grey and has been obtained by fitting the cross section results with a quadratic polynomial in $\Gamma_{t}$.

We observe a very different behaviour of power corrections in
$\Gamma_t/m_t$ for the two partonic channels, connected
to the impact of single-resonant ($tW$) topologies. For the $gg$ channel, the dominant
power corrections are linear, and their impact is quite sizeable at
\mbox{$\Gamma_t \sim \Gatphys$} with effects of $\mathcal{O}(6\%)$ with respect to the
extrapolated LO cross section in the vanishing-width limit. On the contrary, for
the $q\bar q$ channel, the effect is at the per mille level and the shape of the
LO cross section, as a function of $\Gamma_t/m_t$, is that of a quadratic curve.
These differing trends reflect the distinct origins of $\Gamma_t$ power
corrections in each channel. Linear power corrections can arise from genuine off-shell effects in double-resonant topologies, and from single-top contributions.
The linear behaviour occurs because the $(\Gamma_t/\Gatphys)^2$ factor in
Eq.~\eqref{eq:XS_width-rescaled} cancels only one $1/\Gamma_t$ factor from the
single top-quark propagator in $tW$ topologies, resulting in positive linear corrections from squared $tW$ diagrams.
In the $gg$ channel, linear corrections dominate because single-top
contributions are enhanced in the case of a quasi-collinear emission of a bottom quark from an incoming gluon. The $q\bar{q}$ channel lacks this enhancement, and $tW$
contributions only arise from \mbox{$q \bar{q} \to g^{*} \to b \bar{b}$} configurations
where one of the bottom quarks splits into a resonant top quark and a $W$ boson. Without the quasi-collinear enhancement, linear power corrections remain small in the $q\bar{q}$ channel, allowing quadratic corrections from non-resonant topologies to dominate numerically.

Lastly, we observe that the agreement with the on-shell result (blue curve) is
almost perfect, with a tiny mismatch at the sub-permille level, resolved mostly
in the $q \bar q$ channel. This subtle difference is likely due to the
approximation of the partial top-quark width~(see Eq.~\eqref{eq:approx_partial-top-width})
when computing the on-shell cross section (see Eq.~\eqref{eq:onshellXS} in Appendix ~\ref{app:Appendix_top_width}).
\begin{figure*}[tb] 
	\centering
	\includegraphics[width=0.5\textwidth]{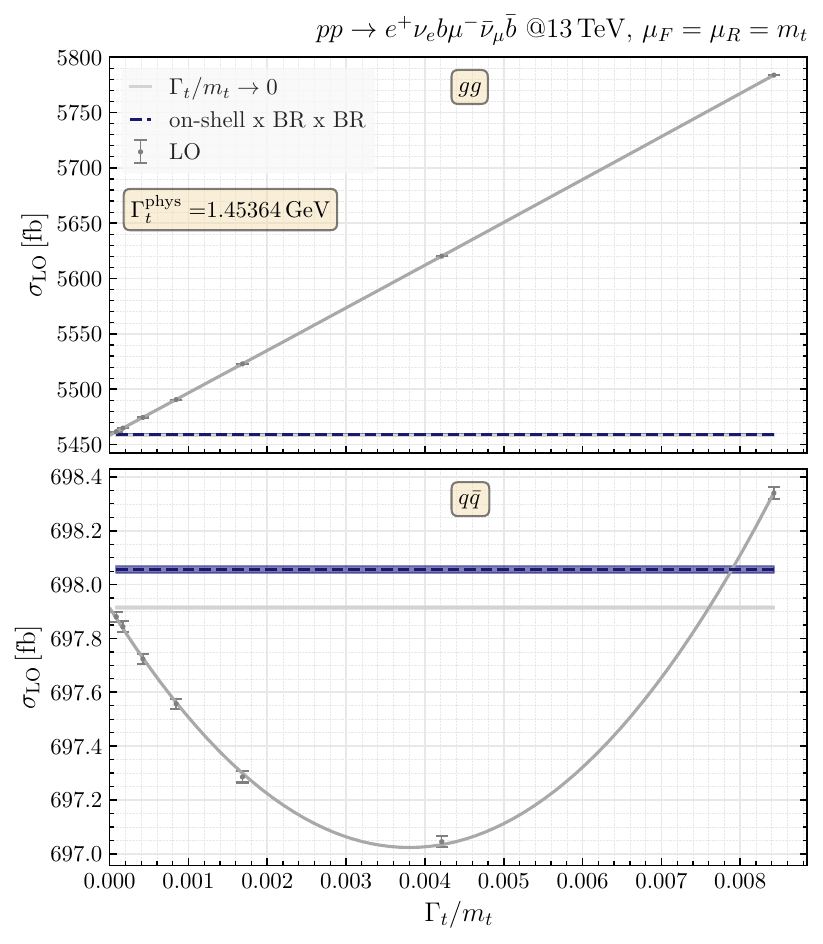} 
	\caption{\mbox{$\Gamma_t \to 0$} extrapolation at LO for the $gg$ (upper plot) and $q\bar q$ (lower plot) partonic channels.}
	\label{fig:LO_Gammat_extrapolation}
\end{figure*}
%

We now turn to the \mbox{$\Gamma_t \to 0$} extrapolation for the reconstruction of non-factorisable virtual corrections at higher orders. 
We first consider the NLO, where exact off-shell results are available and show remarkable agreement with predictions in the DPA for fiducial quantities (see Sec.~\ref{sec:ttxoffshell_NLO_results}). To mimic the NNLO situation, we use NLO predictions in DPA with non-factorisable corrections removed, denoted as $\NLO_{\mathrm{q_T, DPA}}^{\mathrm{fact}}$. By extrapolating these $\NLO_{\mathrm{q_T, DPA}}^{\mathrm{fact}}$ predictions to \mbox{$\Gamma_t \to 0$}, we reconstruct the non-factorisable one-loop corrections via a matching to the on-shell cross section. These reconstructed corrections can be compared against direct calculations, validating our procedure for the subsequent NNLO application.

At NLO, the functional form we use to fit $\NLO_{\mathrm{q_T, DPA}}^{\mathrm{fact}}$ as a function of $\Gamma_t$ is given by Eq.~\eqref{eq:functional-form_oneloop_real}.
Based on this functional form, we fit the
difference between on-shell and $\NLO_{\mathrm{q_T, DPA}}^{\mathrm{fact}}$
results (including a $\Delta\sigma^{\mathrm{N}\LO}_{\mathrm{trunc}}$ matching term, see Sec.~\ref{sec:Delta-term}) 
using a three-parameter ansatz in terms of the coefficients $B^{(1)},C^{(1)}$ and $D^{(1)}$, thus including a linear $\Gamma_t/m_t$ term. For the determination of systematic uncertainties, we employ a replica method similar to that introduced in Sec.~\ref{sec:rcut_dependence} for the \mbox{$r_{\text{cut}} \to 0$} extrapolation. 
This procedure allows us to infer the coefficients \mbox{$B^{(1)}_{\rm nf}=B^{(1)}$} and \mbox{$C^{(1)}_{\rm nf}=C^{(1)} + \Delta\sigma^{\NLO}_{\rm on-shell}-\Delta\sigma^{\mathrm{N}\LO}_{\mathrm{trunc}}$} in Eq.~\eqref{eq:functional-form_oneloop-nonfact}, thus numerically determining the non-factorisable one-loop corrections. Here, $\Delta\sigma^{\NLO}_{\rm on-shell}$ is the NLO correction to the on-shell cross section, as defined in Eq.~\eqref{eq:onshellXS} of Appendix~\ref{app:Appendix_top_width}. 

\begin{figure*}[t]
	\includegraphics[width=0.5\textwidth]{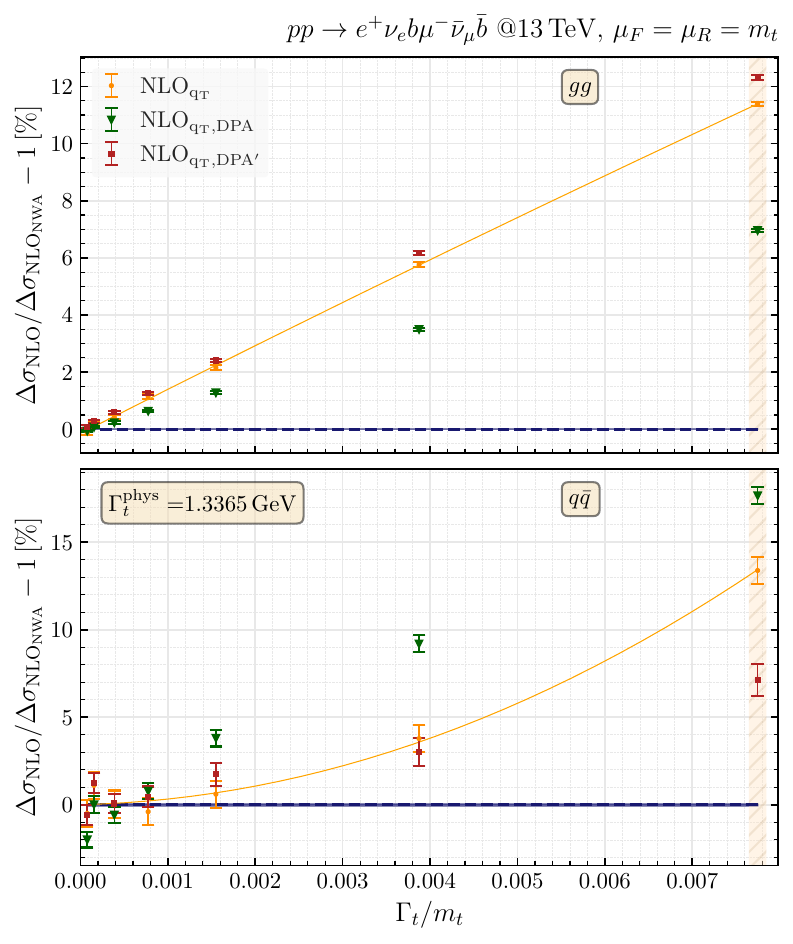}  \hfill
	\includegraphics[width=0.510\textwidth]{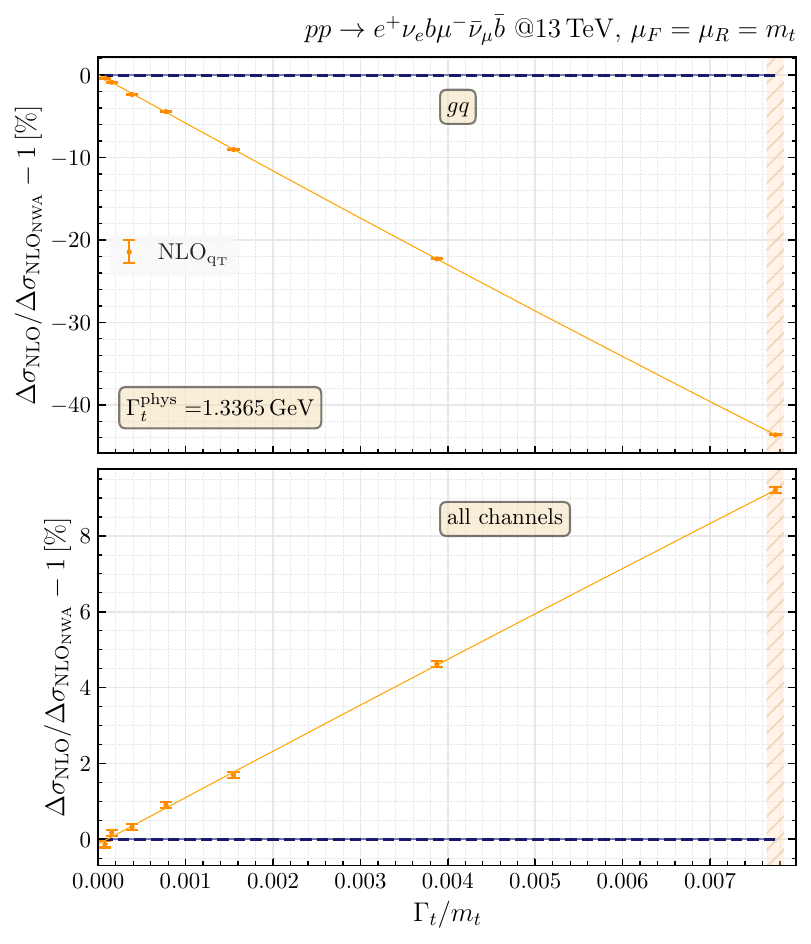}
	\caption{\mbox{$\Gamma_t \to 0$} extrapolation at NLO in $q_T$ subtraction. On the left, we show the comparison, for the diagonal partonic channels, between the exact NLO correction (orange markers) and the on-shell result (blue curve). We also display the NLO correction where the one-loop contribution has been computed in DPA (green markers) as well as the results obtained via the fitting procedure of the non-factorisable corrections (red markers). On the right, we compare the exact NLO correction and the on-shell result for the $gq$ channel (upper panel) and the combination of all partonic channels (lower panel). Results for the physical top-quark width are highlighted with an orange band.}
	\label{fig:NLO_Gammat_extrapolation}
\end{figure*}
In Fig.~\ref{fig:NLO_Gammat_extrapolation}, we assess the quality of this \mbox{$\Gamma_t \to 0$} extrapolation strategy at NLO. 
Specifically, for the diagonal channels (left plots), we compare the extrapolated results (red markers) with those from the exact off-shell calculation (orange markers) and from the complete DPA for the virtual contribution with explicit non-factorisable corrections (green markers), across various values of the top-quark width. For reference, the on-shell result (blue curve), which also includes the $\Delta\sigma^{\mathrm{N}\LO}_{\mathrm{trunc}}$ matching term, is shown.
In the right upper plot, we display the exact results for the $gq$ channel, where no approximation is required.
We observe a purely linear dependence on $\Gamma_t$, as discussed at LO for the $gg$ channel. These linear power corrections lead to a deviation from the on-shell result of approximately \mbox{$-40\%$} at the physical top-quark width, while their impact becomes negligible for the smallest width values.
For all three partonic channels as well as for their combination (shown in the lower right plot),
the numerical extrapolation of all the $\NLO_{\mathrm{q_T}}$ results in the \mbox{$\Gamma_t \to 0$} limit
is in excellent agreement with the NLO correction to the on-shell process.

Focusing on the diagonal channels, we observe that the $\NLO_{\mathrm{q_T, DPA}}$ results (green markers) closely reproduce the exact off-shell calculation (orange markers) in the small top-quark width limit. 
As the top-quark width increases, their slope deviates from that of the exact results, leading to a discrepancy of about $4\%$ at the physical width. 
This behaviour is, however, expected since power corrections responsible for such a difference are formally beyond the accuracy of the DPA.

A similar trend is seen in the $\NLO_{q_T, \mathrm{DPA'}}$ results obtained via the extrapolation of the
non-factorisable one-loop corrections (red markers): they match the on-shell
calculation in the small-width limit, but gradually depart from the
$\NLO_{\mathrm{q_T}}$ results as $\Gamma_t$ increases. Notably, their slope
differs from that of the $\NLO_{\mathrm{q_T, DPA}}$ predictions, which include a
direct calculation of the non-factorisable corrections. This difference
originates from the way the extrapolation is performed: power corrections
present in the input $\NLO_{\mathrm{q_T, DPA}}^{\mathrm{fact}}$ data for the extrapolation are
effectively fitted (as coefficient $D^{(1)}$), but removed in the final determination of the non-factorisable
corrections controlled by the coefficients $B^{(1)}_{\rm nf}$ and $C^{(1)}_{\rm nf}$ in Eq.~\eqref{eq:functional-form_oneloop-nonfact}.
In contrast, the direct calculation of the non-factorisable corrections includes power corrections from different origins.
Besides those associated with the use of the exact Born matrix elements,
additional spurious power corrections arise from the numerical evaluation of the
scalar integrals appearing in Eq.~\eqref{eq:deltanonfact_M1M0_ttx}. As such, the
observed difference is again formally beyond accuracy, and the better agreement
of the extrapolated results with the exact, mostly seen in the $gg$
channel, is partly accidental.

A last comment regards the relative impact of the different components of the non-factorisable one-loop corrections in the two diagonal partonic channels. In both cases, the largest contribution originates from the constant term $C^{(1)}_{\rm nf}$ in Eq.~\eqref{eq:functional-form_oneloop-nonfact}, while
the coefficient $B^{(1)}_{\rm nf}$ of the single logarithm is found to be numerically small.
This is compatible, within relatively large extrapolation
uncertainties,\footnote{ Given the smallness of the coefficient $B^{(1)}_{\rm nf}$ of the
  single logarithm, its numerical extraction is particularly challenging and
  sensitive to the exact functional form of the power corrections.} with the
direct analytical calculation provided by Eq.~\eqref{eq:coefficient_logGt}. For the $gg$
channel, the non-factorisable one-loop corrections at the physical top-quark width
represent only a few per cent of the entire NLO correction, while they account
for almost $85\%$ in the $q\bar q$ channel. This pattern is the consequence of
large cancellations between different contributions entering the
$q_T$-subtraction formula.

\begin{figure*}[t]
	\includegraphics[width=0.5\textwidth]{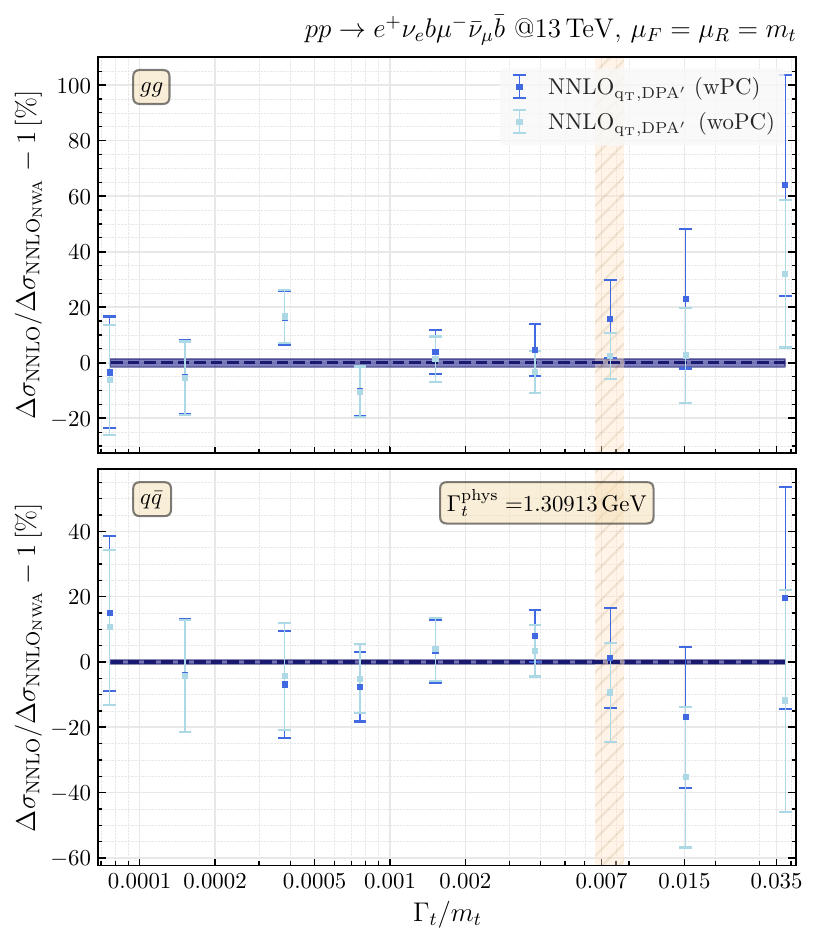}
	\hfill 
	\includegraphics[width=0.505\textwidth]{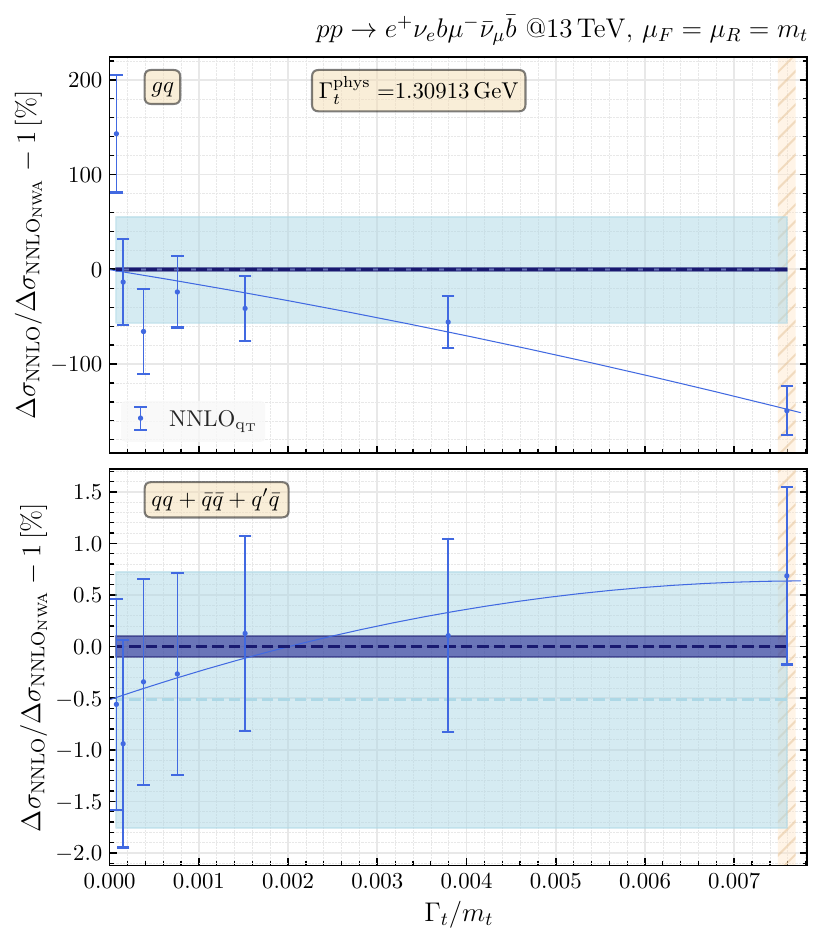} 
	\caption{\mbox{$\Gamma_t \to 0$} extrapolation at NNLO in $q_T$ subtraction. On the left, we present the results for the diagonal channels where the ``missing" non-factorisable two-loop corrections have been fitted according to the procedure described in the main text.	
On the right, we show the comparison, for the off-diagonal channels, between the exact NNLO correction (light-blue curve) and the on-shell result (dark-blue curve). Results for the physical top-quark width are highlighted with an orange band.}
	\label{fig:NNLO_Gammat_extrapolation}
\end{figure*}
After testing the feasibility of the extrapolation method used to determine the
non-factorisable virtual corrections at NLO, we now consider the same strategy at NNLO.
Corresponding numerical results are reported in Fig.~\ref{fig:NNLO_Gammat_extrapolation}, where we display the
off-shell results, obtained for seven different $\Gamma_t/m_t$ values in the range \mbox{$[0.01,1]\times \Gatphys/m_t$}, together with the on-shell result (blue curve) 
and the \mbox{$\Gamma_t \to 0$} extrapolated result equipped with its uncertainty (light-blue
band) obtained via the replica method.
We start our discussion considering the exact results for the off-diagonal $gq$ and \textit{rest} channels shown in the right plots. Since the non-factorisable corrections are absent in these channels, the agreement of the off-shell computation with the on-shell result, in the \mbox{$\Gamma_t \to 0$} limit, serves as an additional validation of the extrapolation procedure and thus of the correctness of our off-shell NNLO calculation.
Once again, the dependence on $\Gamma_t/m_t$ is expected to be power-suppressed
since the cross section is finite in the limit of vanishing top-quark width. 
For both partonic channels, we observe a fairly good agreement with the
on-shell result. We highlight that the large numerical errors seen in the $gq$ channel,
especially for the lower top-quark width values, are relative to the NNLO
correction in that channel. Their absolute impact on the total NNLO cross
section is well below the per cent level. Power corrections at the physical
top-quark width are numerically negligible in the \textit{rest} channel, while they are the
largest contribution for the $gq$ channel. 

We now turn to the dominant diagonal $gg$ and $q\bar{q}$ channels. In this case, the
on-shell result is used to numerically determine the non-factorisable
corrections as showcased at NLO. At NNLO, the functional form of the
integrated non-factorisable corrections is given by
Eq.~\eqref{eq:functional-form_twoloop-nonfact}. We perform a fit of the difference between
the on-shell calculation and the $\NNLO_{\mathrm{q_T, DPA}}^{\mathrm{fact}}$
results with the four-parameter functional form given in Eq.~\eqref{eq:functional-form_twoloop_real}, including a purely
linear term in $\Gamma_t/m_t$. In order to increase the sensitivity of the fit
to potential power corrections, we consider results for two additional width values,
\mbox{$\Gamma_t =\{2,5\} \times \Gatphys$}, larger than the physical top-quark width. Once
again, we rely on the replica method for the \mbox{$\Gamma_t \to 0$} extrapolation procedure. 
In this way, we numerically extract the double logarithm,
single logarithm and the constant of the non-factorisable two-loop corrections as the coefficients \mbox{$A^{(2)}_{\rm nf}=A^{(2)}$}, \mbox{$B^{(2)}_{\rm nf}=B^{(2)}$} and \mbox{$C^{(2)}_{\rm nf}=C^{(2)}+\Delta\sigma^{\NNLO}_{\rm on-shell}-\Delta\sigma^{\NNLO}_{\rm trunc}$} in Eq.~\eqref{eq:functional-form_twoloop-nonfact}.

As a test of the robustness of the above procedure, we considered an alternative
fit model in which power corrections are neglected. More precisely, we use a
three-parameter fit (\mbox{$D^{(2)}=0$}) and reduce the fit range to the original
seven-width set, only considering width values smaller than (or equal to) the
physical width. For ease of reference, we dub the default fitting procedure
``\textit{with power corrections}'' (fit-wPC) and the alternative variant
``\textit{without power corrections}" (fit-woPC). The final
$\NNLO_{q_T, \mathrm{DPA'}}$ results, which include the fitted 
non-factorisable two-loop corrections, are shown in
Fig.~\ref{fig:NNLO_Gammat_extrapolation} as blue (light-blue) markers for the
fit-wPC (fit-woPC) variant. These results clearly demonstrate the cancellation
of the logarithmic terms in the top-quark width. To highlight this cancellation
and the impact of the non-factorisable corrections at NNLO, in
Fig.~\ref{fig:NNLOfact-replica_Gammat_extrapolation} we show, for comparison,
the original $\Delta\sigma_{\NNLO_{\mathrm{q_T, DPA}}^{\mathrm{fact}}}$ results
for the diagonal channels together with the ensemble of the fitting replicas
(based on the fit-wPC model), where the logarithmic singular behaviour in the
small top-quark width is manifest.
\begin{figure*}[t]
\centering
	\includegraphics[width=0.5\textwidth]{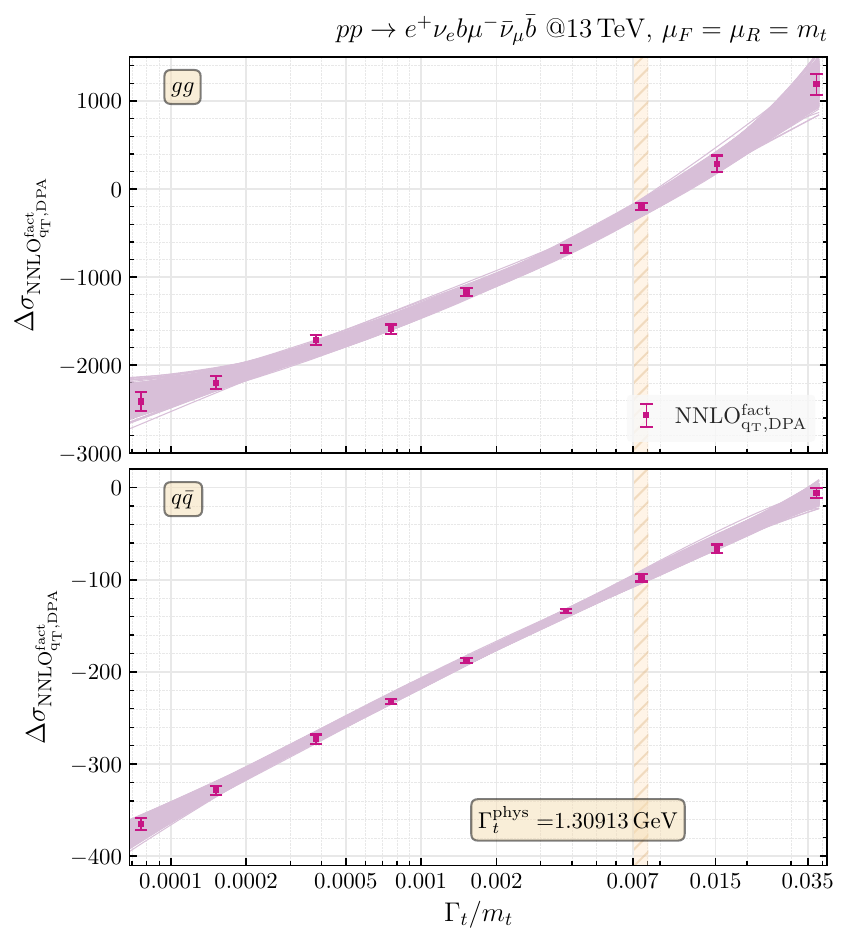} 
	\caption{Behaviour of the original $\Delta\sigma_{\NNLO_{\mathrm{q_T, DPA}}^{\mathrm{fact}}}$ results (purple markers) as a function of $\Gamma_t/m_t$. We also display the ensemble of 1000 fitting replicas (pink band), used in the numerical extrapolation of the non-factorisable corrections. Results for the physical top-quark width are highlighted with an orange band.}
	\label{fig:NNLOfact-replica_Gammat_extrapolation}
\end{figure*}

We observe a good agreement, within the estimated numerical uncertainties,
between the results obtained with the two considered fit models across all
available data points in the extended width range. Towards larger $\Gamma_t$
values, the fit-wPC model tends to give slightly larger results than those
obtained with fit-woPC, which in turn remain close to the result in the on-shell
limit. This is expected as the fit-wPC model is designed to remove the power
corrections from the extraction of the non-factorisable corrections, while the
fit-woPC may be affected by them. Overall, the impact of the power corrections
in $\Gamma_t/m_t$ is found to be mild and in line with what we already observed
at NLO. At the physical top-quark width, their contribution amounts to at
  most \mbox{$15-20\%$} of the on-shell correction, albeit being consistent with zero
  within the quoted extrapolation uncertainty, which has a similar size. The
  extrapolation error is large enough to account for intrinsic uncertainties
  associated with the application of the DPA.

In summary, at the physical top-quark width, we obtain
\begin{equation}\label{eq:nonfact-fit-wPC}
  \Delta\sigma_{\mathrm{NNLO, H}}\vert_{\nonfact}^{({\rm fit-wPC})}\biggl|_{gg} = 976 \pm 103 \,\text{fb}\;,
   \hspace{2cm}  \Delta\sigma_{\mathrm{NNLO, H}}\vert_{\nonfact}^{({\rm fit-wPC})}\biggl|_{q \bar q} =  131 \pm 6 \,\text{fb}\;,
\end{equation}
and
\begin{equation}\label{eq:nonfact-fit-noPC}
  \Delta\sigma_{\mathrm{NNLO, H}}\vert_{\nonfact}^{({\rm fit-woPC})}\biggl|_{gg}  = 880 \pm 49 \,\text{fb} \;,
 \hspace{2cm}   \Delta\sigma_{\mathrm{NNLO, H}}\vert_{\nonfact}^{({\rm fit-woPC})}\biggl|_{q \bar q}  = 129 \pm 4 \,\text{fb} \;,
\end{equation}
employing the fit-wPC and fit-woPC models, respectively. The two results agree within the estimated numerical uncertainties. As a final
determination of the integrated non-factorisable corrections, we adopt the result
obtained via the fit-wPC model,
\mbox{$ \Delta\sigma_{\mathrm{NNLO, H}}\vert_{\nonfact}= \Delta\sigma_{\mathrm{NNLO, H}}\vert_{\nonfact}^{({\rm fit-wPC})} $},
which provides a more flexible description in the presence of power-correction
effects. As a systematic uncertainty, to be conservative, we combine the errors 
from the two fit models linearly, separately for each partonic channel.
\begin{figure*}[t]
\centering
	\includegraphics[width=\textwidth]{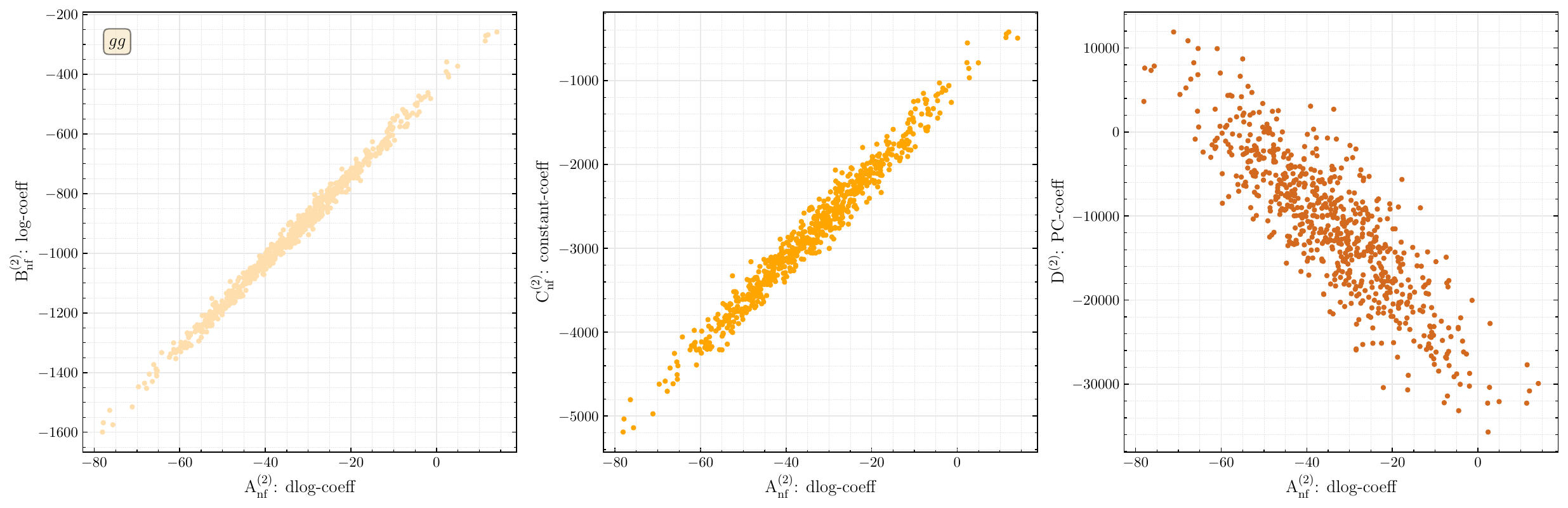} \\
	\includegraphics[width=\textwidth]{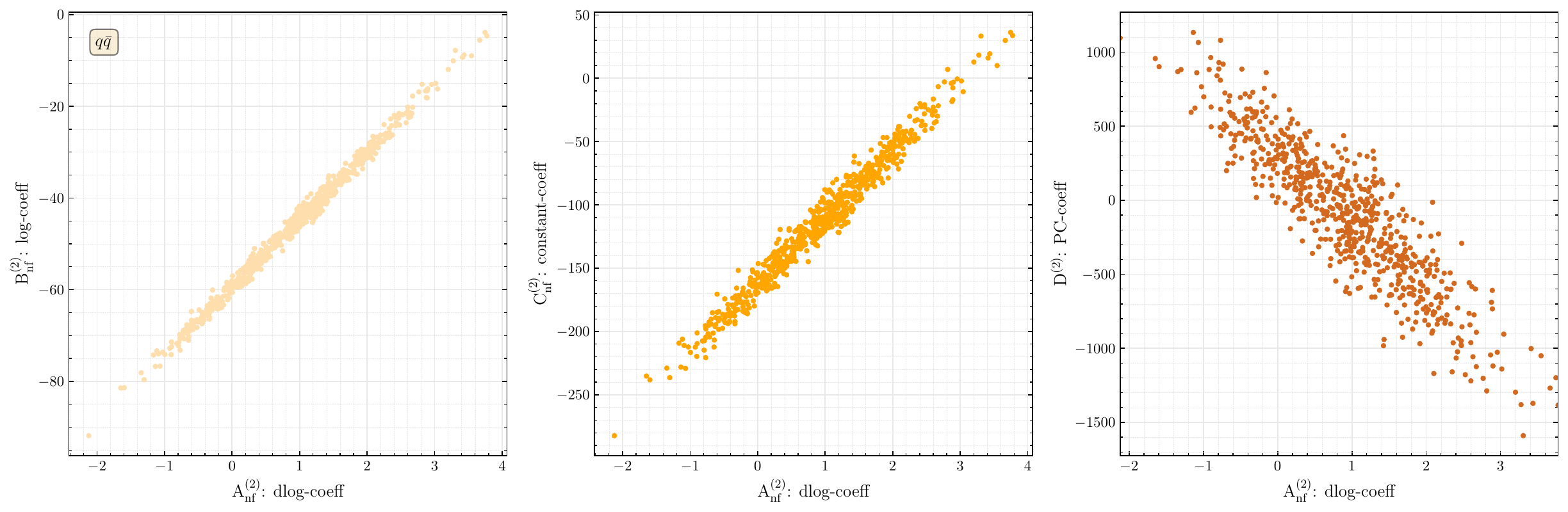}
	\caption{Two-dimensional correlation plots between pairs of fit parameters $A^{(2)}_{\rm nf}$, $B^{(2)}_{\rm nf}$, $C^{(2)}_{\rm nf}$, $D^{(2)}$ (fit-wPC model), corresponding to the $1\sigma$ variation of $\Delta\sigma_{\mathrm{NNLO, H}}\vert_{\nonfact}$. The first (second) row refers to $gg \,(q\bar q)$ channel.}
	\label{fig:correlations}
\end{figure*}

As a technical remark, we highlight that the chosen fitting procedure has been
designed to provide directly the quantity of interest, namely 
$\Delta\sigma_{\mathrm{NNLO, H}}\vert_{\nonfact}$ as a function of
$\Gamma_{t}$, with a consistent error propagation. Indeed, the
parameters of the fit are only loosely constrained given the available Monte Carlo statistics, but
are strongly correlated in the determination of the non-factorisable
corrections. This eventually leads to a moderate uncertainty for
$\Delta\sigma_{\mathrm{NNLO, H}}\vert_{\nonfact}$. 
In Fig.~\ref{fig:correlations}, we show two-dimensional correlation plots between
pairs of fit parameters corresponding to the $1\sigma$ variation of
$\Delta\sigma_{\mathrm{NNLO, H}}\vert_{\nonfact}$. We observe that the
coefficient of the double logarithm ($A^{(2)}_{\rm nf}$) is much smaller than the other
parameters\footnote{The smallness of the coefficient of the double logarithm might be
  expected on the basis of a naive non-abelian soft exponentiation of the single
  log coefficient at NLO.} and exhibits a strong linear correlation with the
coefficients of the single logarithm ($B^{(2)}_{\rm nf}$) and the constant term ($C^{(2)}_{\rm nf}$). Thus, we foresee a clear
improvement of the fitting procedure if one (or more) of the parameters becomes available through a direct calculation.

\subsection{NNLO results for the total cross section}\label{sec:NNLO_results}
%
\begin{table}[t]
\centering  
\renewcommand{\arraystretch}{1.5}
\setlength{\tabcolsep}{0.3em}
\begin{tabular}{cllllll}
$\sigma$ [fb] & \multicolumn{1}{c}{all} & \multicolumn{1}{c}{$gg$}  &\multicolumn{1}{c}{$q \bar q$} &\multicolumn{1}{c}{$gq$} &\multicolumn{1}{c}{$qq + \bar{q} \bar{q} + q' \bar q$}\\
\toprule
$\sigma_{\LO}$ & 
$\phantom{00}6482.21(6)\,^{+29.32\%}_{-21.24\%}$ &
$\phantom{00}5783.84(6)$ & $\phantom{00} 698.36(2)$ & $\phantom{00}--$ & $\phantom{00}--$
\\
\hline
\hline
$\sigma_{\NLO_{\mathrm{CS}}}$ &
$\phantom{00}9584(2)\,^{+8.18\%}_{-10.01\%}$ &
$\phantom{00}8631(2)$  & $\phantom{00}901.7(4)$ & $\phantom{00}50.4(1)$ & $\phantom{00}--$
\\
$\sigma_{\NLO_{q_T}}$ &
$\phantom{00}9589(6)\,^{+8.21\%}_{-10.02\%}$ &
$\phantom{00}8637(6)$ & $\phantom{00}901.8(1.3)$ & $\phantom{00}50.5(1)$ & $\phantom{00}--$
\\
$\sigma_{\NLO_{q_T, \mathrm{DPA}}}$ &
$\phantom{00}9498(6)\,^{+7.86\%}_{-9.84\%}$ &
$\phantom{00}8543(5)$ & $\phantom{00}903.9(1.3)$ & $\phantom{00}50.5(1)$ & $\phantom{00}--$
\\
\vspace{0.13cm}
$\sigma_{\NLO_{q_T, \mathrm{DPA'}}}$ &
$\phantom{00}9614(2) \pm 2$ &
$\phantom{00}8664(2) \pm 2$ & $\phantom{00}899.1(2) \pm 0.3$ & $\phantom{00}50.5(1)$ & $\phantom{00}--$
\\
\hline
\hline
$\sigma_{\NNLO_{q_T, \mathrm{DPA'}}}$ &
$\phantom{0}10623(55)\pm 152\,^{+3.2\%}_{-4.6\%}$ &
$\phantom{00}9599(54) \pm 152$ & $\phantom{00}979(5) \pm 6$ & $\phantom{00}67(10)$ & $\phantom{00}7.46(7)$
\\
\vspace{0.13cm}
$\sigma_{\NNLO_{q_T, \mathrm{DPA}}^{\fact}}$ &
$\phantom{00}9546(55)$ &
$\phantom{00}8623(54)$ & $\phantom{00}849(5)$ & $\phantom{00}67(10)$ & $\phantom{00}7.46(7)$
\\
\hline
$\Delta\sigma_{\NNLO, \mathrm{H}}\vert_{\fact}@m_t $ & 
$\phantom{00}5323.4(7)$ &
$\phantom{00}4726.7(7)$ & $\phantom{00}596.74(8)$ & $\phantom{00}--$ & $\phantom{00}--$
\\
$\Delta\sigma_{\NNLO, \mathrm{H}}\vert_{\nonfact}@m_t $ & 
$\phantom{000} 1107 \pm 152$ &
$\phantom{000} 976 \pm 152$ & $\phantom{000} 131 \pm 6$ & $\phantom{00}--$ & $\phantom{00}--$
\\
\bottomrule
\end{tabular}
\caption{\label{tab:WWbb_fully-inclusive_setup} Inclusive LO, NLO and NNLO cross sections at \mbox{$\sqrt{s}=13$\,TeV}, for the different contributing partonic channels. The quoted relative uncertainties are obtained through scale variations, while the errors in parentheses take into account the statistical errors from the Monte Carlo integration combined with the systematic errors from the \mbox{$\rcut \to 0$} extrapolation. The additional systematic error at NNLO originates from the \mbox{$\Gamma_t \to 0$} extrapolation procedure for the determination of the non-factorisable corrections.
In the last two rows, we report the factorisable two-loop corrections, computed at scale \mbox{$\mu = m_t$}, and the fitted non-factorisable corrections, respectively.}
\end{table}

This section presents the first NNLO results for off-shell $t\bar{t}$ production
and decays in the dilepton channel. Specifically, we consider the process~\eqref{eq:off-shell_ttx_dilepton} in a fully inclusive setup at
\mbox{$\sqrt{s} = 13$\,TeV}. The top-quark width is fixed to the physical value $\Gatphys$,
and all input parameters are chosen as defined in Sec.~\ref{sec:nnlo_setup}. In
Tab.~\ref{tab:WWbb_fully-inclusive_setup} we report our results for the LO, NLO and NNLO
cross sections with a breakdown into the different partonic channels.

At NLO, $\sigma_{\NLO_{q_T, \mathrm{DPA}}}$ includes both factorisable and non-factorisable one-loop corrections in DPA for the diagonal channels ($gg$ and $q \bar{q}$), while $\sigma_{\NLO_{\mathrm{CS}}}$ and $\sigma_{\NLO_{q_T}}$ provide the exact NLO result based on CS and $q_T$ subtraction, respectively. We also list the NLO result in DPA where the non-factorisable virtual corrections are inferred from a \mbox{$\Gamma_t \to 0$} fit (labelled $\sigma_{\NLO_{q_T, \mathrm{DPA'}}}$), as discussed in Sec.~\ref{sec:Gammat_extrapolation}. The latter agrees with the full NLO result to better than one per cent, with a permille-level uncertainty arising from the \mbox{$\Gamma_t \to 0$} extrapolation.

At NNLO, the off-diagonal channels ($gq$ and \mbox{$qq + \bar{q} \bar{q} + q' \bar{q}$}) are computed exactly, while for the
dominant diagonal channels we use the DPA for the two-loop finite remainder.
All other NNLO ingredients are based on exact off-shell matrix elements.
More specifically, the NNLO cross section $\sigma_{\NNLO_{q_T, \mathrm{DPA}}^{\fact}}$ approximates the double-virtual contribution using only the computed factorisable two-loop corrections (see Sec.~\ref{sec:DPA_NNLO}), while $\sigma_{\NNLO_{q_T, \mathrm{DPA'}}}$ additionally includes the non-factorisable corrections inferred via the \mbox{$\Gamma_t \to 0$} extrapolation.

The last two rows of Tab.~\ref{tab:WWbb_fully-inclusive_setup} show explicit results for the computed
factorisable and fitted non-factorisable two-loop corrections. The numerical hierarchy observed at NLO
(see Sec.~\ref{sec:ttxoffshell_NLO_results}) persists at two-loop order, with $\Delta\sigma_{\NNLO, \mathrm{H}}\vert_{\nonfact}$ being $\mathcal{O}(20\%)$ of $\Delta\sigma_{\NNLO, \mathrm{H}}\vert_{\fact}$. 
However, since various NNLO contributions largely cancel, here the non-factorisable corrections constitute 
an $\mathcal{O}(1)$ effect, thus making their inclusion essential to achieve meaningful NNLO predictions.

The total NNLO cross section for off-shell top-quark pair production and decays, obtained as the sum of 
$\sigma_{\NNLO_{q_T, \mathrm{DPA}}^{\fact}}$ and
$\Delta\sigma_{\NNLO, \mathrm{H}}\vert_{\nonfact}$, is
\begin{equation}
	\sigma_{\NNLO_{q_T, \mathrm{DPA'}}} = 10623(55) \pm 152\,^{+3.2\%}_{-4.6\%} \,\mathrm{fb}\,.
\end{equation}
The error in parentheses takes into account the Monte
Carlo statistical error and the $\rcut$-extrapolation error, while the second uncertainty represents the systematics
associated with the \mbox{$\Gamma_t \to 0$} fitting procedure. The latter has
been obtained by combining in quadrature the individual fit uncertainties of the
$gg$ and $q\bar{q}$ channels, given in
Eqs.~\eqref{eq:nonfact-fit-wPC}--\eqref{eq:nonfact-fit-noPC}. These fitting
systematics represent the dominant source of numerical uncertainties of our NNLO
prediction and correspond to \mbox{$\sim 1.5\%$} of the NNLO cross section. Such
systematic uncertainty is subdominant with respect to perturbative uncertainties
estimated through scale variations, shown as relative errors in sub- and
superscript. We also note that the perturbative uncertainties are reduced from
$\sim10\%$ at NLO to $\sim 5\%$ at NNLO.

Finally, we turn to the impact of the higher-order QCD corrections. The inclusive NLO and NNLO
$K$-factors are
\begin{align}
K_{\NLO}^{\mathrm{off-shell}}&=\sigma_{\NLO_{\mathrm{CS}}}/\sigma_{\LO}= 1.479\, \\
K_{\NNLO}^{\mathrm{off-shell}}&=\sigma_{\NNLO_{q_T, \mathrm{DPA'}}}/\sigma_{\NLO_{\mathrm{CS}}}= 1.1084 \, .
\end{align}
Due to the systematic errors of the NNLO cross section originating from the fit
of the non-factorisable corrections and the uncertainty of the
$\sigma_{\NNLO_{q_T, \mathrm{DPA}}^{\fact}}$ result at the physical top-quark width,
the NNLO $K$-factor is determined with a relative systematic error of \mbox{$\sim 1.6\%$}. 
For comparison, it is noteworthy that these off-shell results show remarkable
consistency with the $K$-factors for the on-shell cross
section~(see Eq.~\eqref{eq:onshellXS} in Appendix \ref{app:Appendix_top_width}), corrected with the required
$\Delta\sigma^{\mathrm{(N)N}\LO}_{\mathrm{trunc}}$ term of Eq.~\eqref{eq:delta-term_trunc}, namely
\mbox{$K_{\NLO}^{\mathrm{on-shell}}= 1.470$} and \mbox{$K_{\NNLO}^{\mathrm{on-shell}}= 1.1080$}.

Conversely, one can correct the off-shell results using the
$\Delta\sigma^{\mathrm{NN}\LO}_{\mathrm{trunc}}$ term~\eqref{eq:delta-term_trunc}, which removes
spurious higher-order contributions in the top-quark width.
In the limit of a small top-quark width, the resulting predictions correctly reproduce the on-shell $t\bar{t}$ cross section multiplied by the square of the branching ratio. 
This correspondence was explicitly verified at NLO, while at NNLO it follows from the methodology we employed for the construction of the non-factorisable corrections.
In this way, we obtain
\begin{align}
  \sigma_{\NLO_{\mathrm{CS}}}^{\Delta_{\text{trunc}}} &= 9113(2) \,^{+10.14\%}_{-10.06\%} \,\mathrm{fb}\,,\\
  \sigma_{\NNLO_{q_T, \mathrm{DPA'}}}^{\Delta_{\text{trunc}}} &= 10278(55) \pm 152 \,^{+4.5\%}_{-5.7\%} \,\mathrm{fb}\,,
\end{align}
for the NLO and NNLO off-shell cross sections, respectively. These cross sections agree with the corresponding results listed in Tab.~\ref{tab:WWbb_fully-inclusive_setup} within scale uncertainties. This is expected, since the respective results differ by terms that are formally beyond accuracy. The removal of these spurious contributions in the top-quark width is crucial when comparing the off-shell results against on-shell \mbox{$t\bar t + tW$} predictions in the NWA in order to identify remaining small differences due to genuine off-shell and interference effects. Such a comparison is beyond the scope of this paper.
In conclusion, this NNLO result currently represents the most precise prediction for the inclusive cross section of $W^+W^-b {\bar b}$ production with leptonic decays and massive bottom quarks. A conservative linear combination of the remaining perturbative uncertainties and the systematic errors leads to a residual theoretical uncertainty at NNLO of about $6.5\%$.

\newpage
\section{Summary}
\label{sec:summa}


An accurate description of top-quark pair production at hadron colliders
requires the inclusion of finite-width and off-shell effects. Focusing on the
cleanest leptonic signature, the process is characterised by a complex
multi-particle final state consisting of a bottom-quark pair in association with four leptons.
Beyond leading order, a strict theoretical separation between double-resonant $t\bar{t}$ 
and single-resonant $tW$ production mechanisms no longer exists. 
A consistent theoretical framework to unify the two processes including off-shell and interference effects is therefore given by a full calculation of the \mbox{$pp \to b\, \ell^{+}\, \nu_{\ell}\, \bar{b}\, \ell'^{-}\, \bar{\nu}_{\ell'} + X$} process
in the 4FS with massive bottom quarks. 
The finite bottom-quark mass serves as a regulator of collinear singularities and allows us to cover the full bottom-quark phase space, including single-top contributions that emerge from initial-state \mbox{$g \to b \bar b$} splittings. Furthermore, the finite bottom-quark mass
is essential for the $q_T$-subtraction formalism we employ, enabling the classification of the complete process as part of the $Q{\bar Q}F$ category (where \mbox{$Q=b$} and $F$ represents the remaining colourless final state).

Achieving NNLO accuracy in QCD for this process is a formidable challenge. The
required two-loop amplitudes for a \mbox{$2 \to 4$} process involving both internal and external masses
remain well beyond the capabilities of current multi-loop techniques. In addition, differential
NNLO calculations rely on suitable numerical subtraction schemes to regulate
intermediate IR divergences. These calculations are numerically demanding,
with computational costs that scale poorly with the complexity of the
final state.

In this work, we take the first foundational steps towards an NNLO-accurate
prediction for this process. Our main idea consists of constructing an
approximation of the two-loop amplitudes using the well-known \textit{double-pole approximation}.
This approach significantly simplifies the calculation while preserving 
the cancellation of IR singularities between real and virtual
contributions. The number of relevant Feynman diagrams is drastically reduced.
Moreover, the contributing diagrams can be organised into two classes: hard-virtual corrections acting
separately on the production or decay subamplitudes (\textit{factorisable} corrections), and
soft-gluon exchanges connecting different subprocesses (\textit{non-factorisable} corrections).
The implementation of the factorisable corrections relies on the same building blocks
as those used in the narrow-width approximation. The non-factorisable
corrections require the calculation of soft scalar integrals, which are
however considerably simpler than the exact virtual amplitude.

We have reviewed the construction of the virtual DPA at NLO and implemented it within
the $q_{T}$-subtraction framework. As a starting point, we conducted a thorough
study of the DPA at NLO across various setups. Our
findings confirm that the DPA provides an excellent approximation of the exact
one-loop amplitude. Notably, the quality of the DPA extends beyond the
double-resonant region, where it is formally expected to be valid, thanks to a
reweighting procedure using the exact LO matrix element. This reweighting
effectively captures the leading effects of the missing power-suppressed
corrections in the top-quark width, which include the single-resonant $tW$ subprocess and its interference with double-resonant top-quark pair production.

Beyond NLO, we have completed the construction of the
factorisable two-loop corrections for massive bottom quarks. Our work builds upon the
polarised two-loop amplitudes for top-quark pair production from
Ref.~\cite{Chen:2017jvi} and the heavy-to-light form factor from
Ref.~\cite{Bonciani:2008wf}, both already employed in the construction of
the NWA presented in Ref.~\cite{Czakon:2020qbd}. These amplitudes were
originally computed in the 5FS, assuming massless bottom quarks.
To recover the dominant contributions relevant in the 4FS, we
applied a \textit{massification} procedure that accounts for bottom-quark mass effects 
from both internal loops and external legs, while neglecting power-suppressed terms 
in the bottom-quark mass. The resulting factorisable corrections have been implemented
within the \Matrix framework.

The evaluation of the non-factorisable corrections requires up to two-loop
six-point scalar integrals with internal masses, which are currently not
available. However, certain simplifications are expected, as only soft-gluon
exchanges contribute to the double-pole expansion at leading power. 
In this work, we follow a different strategy and determine the missing non-factorisable two-loop corrections to the
inclusive cross section through a fully numerical procedure. Notably, in the limit of vanishing top-quark width, the off-shell
calculation should smoothly approach the result obtained in the NWA. 
This expected behaviour enables the extraction of the
non-factorisable corrections through a careful numerical extrapolation in the
\mbox{$\Gamma_{t} \to 0$} limit, where the total inclusive cross section factorises into the
on-shell $t\bar{t}$ production cross section multiplied by the branching ratios
of the top and anti-top quark decays.

Despite the inherent limitations of this numerical approach, it serves as a first proof of concept, 
demonstrating the feasibility of NNLO computations for such complex processes.
In particular, our subtraction framework has proven both efficient and flexible, successfully handling the calculation even at the small top-quark widths required for the extrapolation procedure. 
The significant reduction in computational cost and the improved reliability of the results have been made possible by the stability optimisation of the \mbox{$2 \to 7$} real--virtual amplitudes provided by \OpenLoops, along with a series of technical refinements in phase-space generation and Monte Carlo grid optimisation.
Furthermore, this work marks the first application of the DPA to estimate a two-loop QCD amplitude, 
providing valuable insight into the relevance of the different contributions required to build the DPA beyond NLO.

More specifically, we found that the non-factorisable virtual corrections,
although numerically smaller than their factorisable counterparts, are by no means
negligible. This observation does not contradict previous statements in the
literature concerning the smallness of the non-factorisable corrections. Indeed, our
NLO analysis confirms that their overall impact on the total cross section is
small when both real and virtual non-factorisable contributions are consistently
removed. However, the picture changes at the differential level, where significant
shape distortions emerge. 
At NNLO, for the specific inclusive setup we considered, we
observe a similar behaviour: the non-factorisable two-loop corrections amount to
approximately $20\%$ of the factorisable corrections, but due to cancellations among different
perturbative components they represent an $\mathcal{O}(1)$ effect to the NNLO correction.
Although this is not entirely unexpected -- as non-factorisable and factorisable
corrections are formally of the same order -- our analysis provides the first concrete
numerical assessment of their relevance at NNLO. This underlines the importance of
computing and consistently including non-factorisable contributions in the
construction of the DPA for the virtual amplitudes.

Based on the outlined construction of the DPA, we have presented inclusive NNLO predictions for $W^+W^-b {\bar b}$ production with leptonic decays at the LHC. The NNLO corrections increase the NLO result by approximately $11\%$ with a numerical uncertainty of $\pm 2\%$, due to the \mbox{$\Gamma_{t} \to 0$} extrapolation combined with Monte Carlo uncertainties, and about $\pm 5\%$ residual perturbative uncertainties due to missing higher-order QCD corrections beyond NNLO.

We conclude with some remarks on possible future directions. In this work, we have
defined the non-factorisable two-loop corrections as those contributions that cannot be
genuinely factorised. As an expansion in the top-quark width, their
functional form contains logarithmic terms in $\Gamma_{t}$, with up to
quadratic powers of the logarithm. Although a complete computation remains
challenging, it may be organised in such a way that manifestly gauge-invariant -- and
potentially simpler -- subsets of contributions become apparent. For instance, we expect 
a contribution arising from dressing the factorisable one-loop correction
with a soft-gluon exchange between production and decay stages, or between different
decay subamplitudes. Moreover, determining the coefficients of the logarithmic terms
in the top-quark width should be significantly simpler than computing the
constant term. These investigations, which will allow for phenomenological studies of fully differential NNLO predictions for $W^+W^-b {\bar b}$ production, are left for future work.

\section*{Acknowledgments}
We are grateful to Simone Devoto and Javier Mazzitelli for their contribution at the early stages of this work and for allowing us to use the results of Ref.~\cite{inprep} before publication.
This work is supported in part by the Swiss National Science Foundation (SNSF) under
contract 200020$\_$219367.
J.L. is supported by the Science and Technology Research Council (STFC) under the Consolidated Grant ST/X000796/1, and the STFC Ernest Rutherford Fellowship ST/S005048/1.
C.S. is supported in part by the Excellence Cluster ORIGINS funded by the Deutsche Forschungsgemeinschaft (DFG, German Research Foundation) under Germany's Excellence Strategy --EXC-2094-390783311.
The work of L.B. is funded by the European Union (ERC, grant agreement No.
101044599, JANUS). Views and opinions expressed are however those of the authors
only and do not necessarily reflect those of the European Union or the European
Research Council Executive Agency. Neither the European Union nor the granting
authority can be held responsible for them.

\appendix
\section[On-shell projection]{On-shell projection}
\label{app:Appendix_mapping}
We consider the process~\eqref{eq:off-shell_ttx_dilepton} which, at Born level, admits two partonic channels: gluon fusion ($gg$) and quark--antiquark ($q \bar q$) annihilation. 
We parametrise the Born-like kinematics as 
\begin{equation}
	c(p_1) + \bar{c}(p_2) \to e^{+}(p_3) + \nu_e(p_4) + b(p_5, m_b) + \mu^{-}(p_6) + \bar{\nu}_{\mu}(p_7) + \bar{b}(p_8, m_b) \,,\qquad c=q,g\,,	\label{eq:partonic_process_ttxoffshell}
\end{equation}
where the momenta $\{p_i\}_{i=1}^8$ satisfy momentum conservation 
\vspace{-0.2cm}
\begin{equation}
	p_1 + p_2 = \sum_{i=3}^{8} p_i
\end{equation}
and on-shell conditions 
\begin{equation}
	p_5^2 = p_8^2 = m_b^2 \ne 0\,, \quad p_1^2 = p_2^2 = p_3^2 = p_4^2 = p_6^2 = p_7^2 = 0 \,.
\end{equation}	
It is crucial in our framework to keep the bottom quarks massive to avoid unregularised final-state collinear singularities.
We introduce \mbox{$Q^2 = (p_1 + p_2)^2$} to refer to the total squared invariant mass of the event.
The reconstructed off-shell top ($t$) and anti-top ($\bar t$) quarks have momenta 
\begin{equation}
	p_t = p_3 + p_4 +p_5 ~~~\text{and}~~~ p_{\bar t} = p_6 + p_7 +p_8  \,,
	\label{eq:offshell_top_momenta}
\end{equation}
respectively. 

In order to apply the DPA, we need to project off-shell onto on-shell kinematics. 
In other words, for each off-shell event, we have to reconstruct the $t/\bar t$ momenta and project them on-shell simultaneously. 
This projection must ensure momentum conservation and on-shell mass conditions of all the external particles. 
To this end, we exploit the on-shell projection introduced in Refs.~\cite{Denner:2014wka,Denner:2016jyo}, extending it to include finite bottom-quark mass effects. 
Beyond momentum conservation and on-shell conditions, the key features of this projection are that 
\begin{itemize}
	\item it preserves the total invariant mass $Q$ of the off-shell event;
	\item it does not change the resonance of the $W^{\pm}$ propagators. 
\end{itemize}
Therefore, only events above the threshold, i.e with \mbox{$Q^2 > 4 m_t^2$}, can be projected while events not satisfying this condition have to be discarded. 
The starting point of the construction is the ansatz 
\begin{equation}
	\raisedhat{p}_t = (1-\xi)\,p_{\bar{t}} + (1-\eta)\,p_t ~~,   \hspace*{1cm}   \raisedhat{p}_{\bar{t}} = \xi\, p_{\bar{t}} + \eta\, p_t \,,
	\label{eq:onshell_top_momenta}
\end{equation}
where the projected top-quark momenta are defined as linear combinations of the off-shell skadd{ones}{ momenta}, satisfying the conditions 
\begin{equation}
	\raisedhat{p}_t + \raisedhat{p}_{\bar{t}} = p_t + p_{\bar{t}} \,,~~~~~ \raisedhat{p}_t^2 = \raisedhat{p}_{\bar{t}}^2 = m_t^2 \,.
\end{equation}
This ansatz leads to a system of equations for the parameters $\eta$ and $\xi$, which are solutions of the quadratic equation
\begin{align}
	0 &= \eta^2\, [ p_{\bar{t}}^2p_t - p_t^2p_{\bar{t}} + (p_t \cdot p_{\bar{t}})(p_t - p_{\bar{t}}) ]^2 + 
	\eta\, (p_t + p_{\bar{t}})^2[ (p_t \cdot p_{\bar{t}})^2 - p_t^2p_ {\bar{t}}^2 ]   \notag \\
	&\hspace{5.80cm}+ \frac{1}{4}[ (p_t + p_{\bar{t}})^2 ]^2p_ {\bar{t}}^2 - (p_ {\bar{t}}^2 + p_t \cdot p_{\bar{t}})^2m_t^2  \,,
	\label{eq:eta_equation}
\end{align}
and 
\begin{equation}
	\xi = \frac{ (p_t + p_{\bar{t}})^2 - 2\,\eta\, (p_t^2 + p_t \cdot p_{\bar{t}})  }{ 2(p_{\bar{t}}^2 + p_t \cdot p_{\bar{t}} ) } \,.
	\label{eq:xi_equation}
\end{equation}
In general, two pairs of solutions $(\eta_1,\xi_1)$ and $(\eta_2,\xi_2)$ are possible. 
For each event, we choose the pair $(\eta_i,\xi_i)$ that minimises the alteration of the top-quark momenta, or in other words, that minimises the distance \mbox{$\eta_i^2 + (\xi_i - 1)^2$} in the $(\eta,\xi)$ plane.

Clearly, this on-shell projection of the top and anti-top quarks alters the momenta of their decay products, which have to be adjusted accordingly to ensure on-shell conditions. 
Therefore, we compute the new momenta of the bottom and anti-bottom quarks as
\begin{equation}
	p_b' = \raisedhat{p}_t - p_{W^{+}}~~,    \hspace*{1cm}   p_{\bar{b}}' = \raisedhat{p}_{\bar{t}} - p_{W^{-}} \,,
\end{equation}
where \mbox{$p_{W^{+}} = p_3 + p_4$} and \mbox{$p_{W^{-}} = p_6 + p_7$}\,, respectively.
Then, we project the momenta on-shell by following a procedure similar to that described above. 
Specifically, for the top-quark decay products we define 
\begin{equation}
	\raisedhat{p}_{W^+} = \xi\, p_{W^+} + \eta\, p_b'~~,   \hspace*{1cm}   \raisedhat{p}_{b} = (1-\xi)\, p_{W^+} + (1-\eta)\, p_b' \,,
\label{eq:eq:onshell_w_momenta}
\end{equation}
which must satisfy the conditions \mbox{$\raisedhat{p}_{W^{+}} + \raisedhat{p}_b = p_{W^{+}} + p_b' $}
and \mbox{$\raisedhat{p}_{W^{+}}^2 = p_{W^{+}}^2$}, \mbox{$\raisedhat{p}_b^2 = m_b^2$}. 
The requirement \mbox{$\raisedhat{p}_{W^{+}}^2 = p_{W^{+}}^2$} ensures that the off-shell invariant mass of the $W^{+}$ boson is not modified.
A completely analogous procedure can be constructed for the $W^{-}$ boson and the anti-bottom quark.
With the ansatz in Eq.~\eqref{eq:eq:onshell_w_momenta} the equivalent of Eqs.~\eqref{eq:eta_equation} and~\eqref{eq:xi_equation} is given by
\begin{align}
	0 &= \eta^2\, [ q_1^2q_2 - q_2^2q_1 + (q_1 \cdot q_2)(q_2 - q_1) ]^2 + 
	\eta\, [(q_1 + q_2)^2 + q_1^2 - m_b^2] [ (q_1 \cdot q_2)^2 - q_1^2q_2^2 ]   \notag \\
	&\hspace{5.85cm}+ \frac{1}{4}[ (q_1 + q_2)^2 + q_1^2 -m_b^2 ]^2q_1^2 - (q_1^2 + q_1 \cdot q_2)^2q_1^2 \,, \\
	\xi &= \frac{ (q_1 + q_2)^2 + q_1^2 - m_b^2 - 2\,\eta\, (q_2^2 + q_1 \cdot q_2)  }{ 2(q_1^2 + q_1 \cdot q_2 ) }  \,,
\end{align}
where \mbox{$q_1 \in \{ p_{W^{+}}, p_{W^{-}} \}$} and \mbox{$q_2 \in \{p_b', p_{\bar{b}}' \}$}.
By solving these two systems of equations, we obtain the momenta $(\raisedhat{p}_{W^{-}})\, \raisedhat{p}_{W^{+}}$, $(\raisedhat{p}_{\bar b})\, \raisedhat{p}_b$ for the (anti-)top-quark decay products.

Finally, we need two more projections for the $W$-boson decays \mbox{$W^{+} \to e^{+}\nu_e$} and \mbox{$W^{-} \to \mu^{-} \bar{\nu}_\mu$}\,, respectively. We first compute the new momenta of the neutrinos,
\begin{equation}
	p_{\nu_e}' = \raisedhat{p}_{W^{+}} - p_{e^{+}}~~,    \hspace*{1cm}   p_{\nu_{\mu}}' = \raisedhat{p}_{W^{-}} - p_{\mu^{-}} \,,
\end{equation}
and then use the ansatz 
\begin{equation}
	\raisedhat{p}_l = \alpha\, p_l~~,  \hspace*{1cm} \raisedhat{p}_{\nu_l} = \raisedhat{p}_{W} - \raisedhat{p}_l \,,
\end{equation}
where the subscript ``$l$" refers to any charged lepton and ``$W$" to the corresponding $W$ boson. 
The on-shell condition \mbox{$\raisedhat{p}_l^2 = 0$} is automatically satisfied while the requirement \mbox{$\raisedhat{p}_{\nu_l}^2 = 0$} allows us to determine the parameter
\begin{equation}
	\alpha = \frac{ \raisedhat{p}_{W}^2 }{ 2\raisedhat{p}_{W}\cdot p_l} \,.
\end{equation}

In our implementation of the described on-shell projection within the \Matrix framework~\cite{Grazzini:2017mhc}, we introduce a modification to prevent events below threshold from being discarded.
This modification has an effect only on the initial step of the procedure, i.e.\ on the on-shell projection of the top quarks. The remaining steps are identical. 
More specifically, we relax the condition on the conservation of the invariant mass $Q$ of the event.
By so doing, we can impose that the on-shell projected momenta $\raisedhat{p}_t$ and $\raisedhat{p}_{\bar t}$ have the same spatial components as the original momenta $p_t$ and $p_{\bar t}$, while their energies get rescaled to ensure \mbox{$\raisedhat{p}_t^2 = \raisedhat{p}_{\bar t}^2 = m_t^2$}.
As in this way the invariant mass of the event is altered, the initial-state momenta $p_1$ and $p_2$ have to be adjusted accordingly to ensure momentum conservation. 
To avoid discontinuities close to the threshold, \mbox{$Q = 2 m_t$}, we use this modified version of the on-shell projection in the entire phase space. 

\section{Non-factorisable corrections at NLO}
\label{app:Appendix0}
%
In the following, we consider a representative example for each class of the non-factorisable one-loop corrections appearing in Eq.~\eqref{eq:deltanonfact_M1M0_ttx}. The remaining functions can be directly obtained upon substitution of labelling and momenta of the external particles and resonances.

\textit{Manifestly non-factorisable} corrections are expressed in terms of scalar boxes $\mathcal{D}_0$ and pentagons $\mathcal{E}_0$, and are given by
\begingroup
\allowdisplaybreaks
\begin{align}
 	\Delta_{if}(t,b;1) & = -(s_{1b}-m_b^2)K_t\,\mathcal{D}_0(m_b^2,\tilde{s}_{tb},s_{t1},0,p^2_t,s_{1b},0,m_b^2,\mu_t^2,0) \,,\\
	\Delta_{mf'}(t;\bar{t},\bar{b}) & = -(s_{t\bar{b}}-p_t^2-m_b^2) K_{\bar t}\,\mathcal{D}_0(p_t^2,s_{t\bar{t}},\tilde{s}_{\bar{t}\bar{b}},m_b^2,p_{\bar{t}}^2,s_{t\bar{b}},0,\mu_t^2,\mu_t^2,m_b^2) \notag \\
	& \sim -(s_{t\bar{b}}-m_t^2-m_b^2) K_{\bar t}\,\mathcal{D}_0(p_t^2,s_{t\bar{t}},\tilde{s}_{\bar{t}\bar{b}},m_b^2,p_{\bar{t}}^2,s_{t\bar{b}},0,\mu_t^2,\mu_t^2,m_b^2) \,,\\
	\Delta_{ff'}(t,b;\bar{t},\bar{b}) &= -(s_{b\bar{b}}-2m_b^2) K_t\,K_{\bar t}\,\mathcal{E}_0(p_b,p_t, -p_{\bar{t}}, -p_{\bar{b}},0,m_b^2,\mu_t^2,\mu_t^2,m_b^2) \,,
\end{align}
\endgroup
with \mbox{$K_t \equiv p_t^2 -\mu_t^2$} and \mbox{$K_{\bar t} \equiv (p_{\bar t}^2 -\mu_t^2)$}, and the $\sim$ sign implies that the on-shell limit \mbox{$p_t^2 \to m_t^2$}, \mbox{$p_{\bar t}^2 \to m_t^2$} is taken wherever possible. This means that all quantities are evaluated on the on-shell kinematics, i.e.\ on the set of projected momenta $\{ \raisedhat{p}_i \}$ defined in Appendix~\ref{app:Appendix_mapping}, while the momenta of the resonances are kept off-shell.

\textit{Non-manifestly non-factorisable} corrections can be entirely expressed in terms of scalar bubbles $\mathcal{B}_0$ and triangles $\mathcal{C}_0$,
\begingroup
\allowdisplaybreaks
\begin{align}
	\Delta_{mm'}(t;\bar{t}) &= - (s_{t\bar{t}} -p_t^2 -p_{\bar t}^2)\biggl\{ \mathcal{C}_0(p_t^2,s_{t\bar{t}},p_{\bar{t}}^2,0,\mu_t^2,\mu_t^2) -  \mathcal{C}_0(m_t^2,s_{t\bar{t}},m_t^2,0,m_t^2,m_t^2)   \biggr\} \notag \\
	&\sim - (s_{t\bar{t}} -2 m_t^2)\biggl\{ \mathcal{C}_0(p_t^2,s_{t\bar{t}},p_{\bar{t}}^2,0,\mu_t^2,\mu_t^2) -  \mathcal{C}_0(m_t^2,s_{t\bar{t}},m_t^2,0,m_t^2,m_t^2)   \biggr\} \,, \\
	\Delta_{im}(t;1) &= - (s_{t1}-p_t^2)\biggl\{ \mathcal{C}_0(p_t^2, s_{t1},0,0,\mu_t^2,0) -  \mathcal{C}_0(m_t^2, s_{t1},0,0,m_t^2,0)   \biggr\} \notag \\
	&\sim - (s_{t1}-m_t^2)\biggl\{ \mathcal{C}_0(p_t^2, s_{t1},0,0,\mu_t^2,0) -  \mathcal{C}_0(m_t^2, s_{t1},0,0,m_t^2,0)   \biggr\} \,, \\
	\Delta_{mm}(t) &= 2 p_t^2 \biggl\{ \frac{\mathcal{B}_0(p_t^2, 0, \mu_t^2) - \mathcal{B}_0(p_t^2, 0, p_t^2)}{K_t} - \mathcal{B}_0^{'}(p_t^2, 0, p_t^2) \biggr\}   \notag \\
	&\sim 2 m_t^2 \biggl\{ \frac{\mathcal{B}_0(p_t^2, 0, \mu_t^2) - \mathcal{B}_0(p_t^2, 0, p_t^2)}{K_t} - \mathcal{B}_0^{'}(p_t^2, 0, p_t^2) \biggr\}  \,,\\
	\Delta_{mf}(t,b) &= - (\tilde{s}_{tb}-p_t^2-m_b^2)\biggl\{ \mathcal{C}_0(p_t^2, \tilde{s}_{tb},m_b^2,0,\mu_t^2,m_b^2) -  \mathcal{C}_0(m_t^2, \tilde{s}_{tb},m_b^2,0,m_t^2,m_b^2)   \biggr\} \notag \\
	&\sim - (\tilde{s}_{tb}-m_t^2-m_b^2)\biggl\{ \mathcal{C}_0(p_t^2, \tilde{s}_{tb},m_b^2,0,\mu_t^2,m_b^2) -  \mathcal{C}_0(m_t^2, \tilde{s}_{tb},m_b^2,0,m_t^2,m_b^2)   \biggr\} \,.
\end{align}
\endgroup
The kinematical invariants appearing in the previous formulae are defined as 
\begin{equation}
	s_{ij} = (\raisedhat{p}_i + \sigma_{ij} \raisedhat{p}_j)^2 ~~, ~~~~ \tilde{s}_{ij} = (\raisedhat{p}_i - \sigma_{ij} \raisedhat{p}_j)^2 ~~~~~\forall ~i,j \in I \cup R_t \cup R_{\bar t} \cup \overline{R}
\end{equation}
where \mbox{$\sigma_{ij} = +1$} if the particles $i$ and $j$ are both incoming or outgoing, \mbox{$\sigma_{ij} = -1$} otherwise.
For the sake of clarity, we remind the reader that the index \mbox{$I = \{1,2\}$} collects the labels assigned to the IS particles, \mbox{$\overline{R} = \{ t, \bar t\}$}, and $R_{t}$ and $R_{\bar{t}}$, defined in Eq.~\eqref{eq:indices_topdecay_products}, are collective indices for the decay products of the two resonances.

\section{Two-loop amplitudes for top-quark decays}
\label{app:Appendix1}
%
In Sec.~\ref{sec:heavy-to-light_FF}, we have described the construction of the massified amplitudes associated with the decay of a polarised top quark into a massive bottom quark and a lepton pair. 
These amplitudes are an essential part for building the factorisable two-loop corrections to off-shell $t\bar t$ production, as discussed in Sec.~\ref{sec:DPA_NNLO}.

In the following, we present the analytic expressions for the finite remainders of the decay amplitudes up to two-loop order, decomposed into the relevant colour structures.
Without loss of generality, we consider a top quark with helicity \mbox{$\lambda_t = \pm 1$} along the direction of the spin four-vector $s_t$, defined in Eq.~\eqref{eq:top_spin-four-vector}.
We recall once more that $n_l$ denotes the number of massless quarks, $n_m$ the number of quarks with (small) mass $m_b$ and $n_h$ the number of heavy quarks with mass \mbox{$m_t \gg m_b$}. 
\\
\\
The employed \textit{massification} procedure requires explicit expressions for the $n_m$-dependent renormalised collinear and soft functions appearing in Eq.~\eqref{top_decay_massified_nmterms}. 
They are given by
\begingroup
\allowdisplaybreaks
\begin{align}
	 \hspace{-0.23cm} \sqrt{ Z_q } \bigl|_{n_m \mathrm{terms}} \,=\, \left( \frac{\alphas^{(n_l + n_m)} }{\pi} \right)^2&\! C_F n_m  \biggl\{  
	 \frac{1}{24 \epsilon^2} \left(1 + l_b \right)
	 - \frac{1}{ \epsilon} \left[ \frac{1}{16} l_b^2  
	 +\frac{1}{16} l_b +\frac{7}{576} + \frac{\zeta_2}{16}  \right] \notag \\
	 &+\frac{7}{144} l_b^3 +\frac{5}{96} l_b^2 + \frac{1}{288}  l_b \left( 31 + 42 \zeta_2 \right) +\frac{3403}{10368} -\frac{3}{32}\zeta_2 +\frac{\zeta_3}{18} \notag \\
	 &+ \frac{1}{16} \left( \frac{1}{\eta} + \log\left(\frac{\nu^2}{s} \right)\right) \biggl[ \frac{2}{3 \epsilon^2} -\frac{1}{\epsilon}\left( \frac{4}{3} l_b + \frac{10}{9} \right) 
	 + \frac{4}{3} l_b^2 +\frac{20}{9} l_b+\frac{56}{27}+\frac{2}{3}\zeta_2 \biggr] \biggr\}
	 \label{eq:Zq_Signer}
\end{align}
\endgroup
and 
\begingroup
\allowdisplaybreaks
\begin{align}
	  \mathcal{S} \,=\, 1 + \left( \frac{\alphas^{(n_l + n_m)} }{\pi} \right)^2 &\! C_F n_m \biggl\{ 
	 \frac{1}{32 \epsilon^3}
	 - \frac{1}{16 \epsilon^2}\left( l_b +\frac{1}{9}  \right) 
	 + \frac{1}{\epsilon}\biggl[ 
	 \frac{1}{16} l_b^2 + \frac{1}{72} l_b 
	 -\frac{13}{216}+\frac{\zeta_2}{32}  \biggr] \notag \\	 
	 &-\frac{1}{24} l_b^3 - \frac{1}{72} l_b^2 
	- l_b \left( -\frac{13}{108} + \frac{\zeta_2}{16} \right)  
	+\frac{11}{48} -\frac{\zeta_2}{144} -\frac{\zeta_3}{16} \notag \\
	  &- \frac{1}{16}  \left( \frac{1}{\eta} + \log\left(\frac{\nu^2}{m_t m_b} \right)\right) \biggl[ \frac{2}{3 \epsilon^2} - \frac{1}{\epsilon}\left( \frac{4}{3} l_b + \frac{10}{9} \right)
	 + \frac{4}{3} l_b^2 +\frac{20}{9} l_b+\frac{56}{27}+\frac{2}{3}\zeta_2 \biggr] \biggr\} \,,
	 \label{eq:Soft_Signer}
\end{align}
\endgroup
where \mbox{$l_b \equiv \log \left(m_b^2/\mu^2 \right)$} and \mbox{$s = 2p_t \cdot p_b = m_t^2(1-x) + m_b^2 \sim m_t^2(1-x)$}. 
In Eqs.~\eqref{eq:Zq_Signer} and \eqref{eq:Soft_Signer} $\eta$ is the rapidity regulator and $\nu$ the associated scale.

After subtracting the IR poles in the minimal subtraction scheme of Refs.~\cite{Becher:2009qa,Becher:2009cu,Ferroglia:2009ii}, we define the finite remainder at scale $\mu$.
The massified subtraction operator $\mathcal{Z}_{m_b \ll m_t}$ appearing in Eq.~\eqref{top_decay_finite_remainder} is defined as
\begin{equation}
	 \mathcal{Z}_{m_b \ll m_t}(\mu, \epsilon) = 1 + \frac{\alphasNl}{\pi} \mathcal{Z}^{(1)}_{m_b \ll m_t}(\mu, \epsilon) + \left( \frac{\alphasNl}{\pi}  \right)^2\! \mathcal{Z}^{(2)}_{m_b \ll m_t}(\mu, \epsilon)  + \mathcal{O}(\alphas^3) \,,
\end{equation}
where the expressions for the one-loop and two-loop coefficients read
\begingroup
\allowdisplaybreaks
\begin{align}
	\mathcal{Z}^{(1)}_{m_b \ll m_t}(\mu, \epsilon)  =& -\frac{C_F}{4 \epsilon} \left( 2 + l_b - L_{\mu} - 2\log(1-x) \right) \,,\\
	\mathcal{Z}^{(2)}_{m_b \ll m_t}(\mu, \epsilon)  =& \frac{1}{\epsilon^2} \biggl\{ C_A C_F \frac{11}{96}\left(2 + l_b -L_{\mu} -2 \log(1-x) \right) 
	+ \frac{C_F^2 }{32} \left(2 + l_b -L_{\mu} -2 \log(1-x)  \right)^2  \notag \\
	&\hspace{0.5cm} - \frac{C_F n_l}{48}\left( 2 + l_b - L_{\mu} - 2\log(1-x) \right) \biggr\}  \notag \\
	+&  \frac{1}{\epsilon} \biggl\{ C_A C_F \biggl[-\frac{49}{144} + \left(l_b - L_{\mu} -2 \log(1-x) \right) \left(-\frac{67}{288} +\frac{\pi^2}{96} \right)
	 + \frac{\pi^2}{48} -\frac{\zeta_3}{8} \biggr] \notag \\
	&\hspace{0.5cm} + C_F n_l \frac{5}{48}\left( 2 + l_b -L_{\mu} -2 \log(1-x) \right) \biggr\} \,.
\end{align}
\endgroup
Here, we introduce \mbox{$L_{\mu} = \log \left( m_t^2/\mu^2 \right)$} to denote the logarithmic terms in the large mass $m_t$. 
We remind the reader that the kinematic variable $x$ is defined as \mbox{$x = q^2/m_t^2$}, where $q^2$ is the invariant mass of the off-shell $W$ boson. 

The all-order massified finite remainder~\eqref{top_decay_finite_remainder}, after the decoupling of the massive quarks, can be decomposed as
\begin{align}
	&\M_{(\pm s_t)}^{(m_b), \mathrm{fin}}(\mu) =
	 g_W^2 V_{bt} \, \frac{i}{q^2 - \mu_W^2} \, \left[ \M_{(\pm s_t)}^{(0)} +  \frac{\alphasNl}{\pi} \M_{(\pm s_t)}^{(1), (m_b)}(\mu) \biggl|_{\mathrm{fin}}  +  \biggl( \frac{\alphasNl}{\pi} \biggr)^2 \M_{(\pm s_t)}^{(2), (m_b)}(\mu) \biggl|_{\mathrm{fin}} \right] + \mathcal{O}(\alphas^3) \,.
\end{align}
At one-loop order, we can write
\begin{align}
	\M_{(\pm s_t)}^{(1),(m_b)}(\mu) \biggl|_{\mathrm{fin}} &=  C_F \left( \mathcal{D}_{C_F}M_{(\pm s_t)}^{(G_1)} + \mathcal{E}_{C_F}M_{(\pm s_t)}^{(G_3)}  \right) \,,
\end{align}
with
\begin{align}
	 \mathcal{D}_{C_F} &=  \frac{ l_b^2}{8} -\frac{l_b}{8} -\frac{L_{\mu}^2}{8} + L_{\mu}\left( \frac{5}{8} -\frac{1}{2}\log(1-x) \right)  -1 
	 + \frac{\log(1-x)}{4 x} \left( 3x - 1\right) -\frac{\log(1-x)^2}{2} - \frac{{\rm Li}_2(x) }{2} \,,\\
	 \mathcal{E}_{C_F} &= \frac{\log(1-x)}{2x} \,.
\end{align}
Analogously, the corresponding two-loop finite remainder can be organised as
\begin{align}
	\M_{(\pm s_t)}^{(2),(m_b)}(\mu) \biggl|_{\mathrm{fin}} =&  \left( C_F^2 \mathcal{D}_{C_F^2} + C_A C_F \mathcal{D}_{C_A C_F} + C_F n_l \mathcal{D}_{C_F n_l}  +  C_F n_h \mathcal{D}_{C_F n_h}  +  C_F n_m \mathcal{D}_{C_F n_m} \right) M_{(\pm s_t)}^{(G_1)} \notag \\
	+& \left( C_F^2 \mathcal{E}_{C_F^2} + C_A C_F \mathcal{E}_{C_A C_F} + C_F n_l \mathcal{E}_{C_F n_l}  +  C_F n_h \mathcal{E}_{C_F n_h}  +  C_F n_m \mathcal{E}_{C_F n_m} \right) M_{(\pm s_t)}^{(G_3)} \,,
\end{align}
with 
\begingroup
\allowdisplaybreaks
\begin{align}
	\mathcal{D}_{C_F^2} =&  \frac{1}{128} l_b^4 -\frac{1}{64} l_b^3 
	+ \biggl(-\frac{1}{16} H(0,1,x)-\frac{L_{\mu}^2}{64}-\frac{1}{16} L_{\mu} \log (1-x)+\frac{5 L_{\mu}}{64}-\frac{1}{16} \log ^2(1-x) \notag \\
	&-\frac{\log (1-x)}{32 x}+\frac{3}{32} \log (1-x)-\frac{15}{128} \biggr) l_b^2 +  \biggl( \frac{1}{16} H(0,1,x)+\frac{L_{\mu}^2}{64} \notag \\
	&+ \frac{1}{16} L_{\mu} \left( \log (1-x)-\frac{5}{4} \right)+\frac{1}{16} \log ^2(1-x)+\frac{\log (1-x)}{32 x}\left(1-3x \right)-\frac{3 \zeta_3}{4}+\frac{\pi ^2}{16}+\frac{5}{64}\biggr)l_b \notag \\
	&+ \frac{1}{128}L_{\mu}^4 + \left(\frac{1}{16} \log (1-x)-\frac{5}{64} \right)L_{\mu}^3 \notag \\
	&+ \left(\frac{1}{16} H(0,1,x)+\frac{3}{16} \log ^2(1-x)+\frac{\log (1-x)}{32 x}(1- 13x)+\frac{41}{128} \right)L_{\mu}^2  \notag \\
	&+ \biggl( \frac{1}{4} \left( \log (1-x) -\frac{5}{4}\right)H(0,1,x)+\frac{1}{4} \log ^3(1-x)+\frac{\log ^2(1-x)}{16 x}\left( 2- 11 x\right) \notag \\
	&- \frac{\log (1-x)}{32 x}\left( 5 - 31 x\right) +\frac{3 \zeta_3}{4}-\frac{\pi ^2}{16}-\frac{37}{64} \biggr)L_{\mu} \notag \\
	& + \frac{1}{8} \log ^4(1-x)+\frac{1}{4} \log (x) \log ^3(1-x)+\frac{\log ^3(1-x)}{8 x}-\frac{3}{8} \log ^3(1-x) \notag \\
	&-\frac{1}{16} \log ^2(x) \log ^2(1-x)+\frac{1}{2} H(0,1,x) \log ^2(1-x)+\frac{63 \log (x) \log ^2(1-x)}{16 (x-1)} \notag \\
	&+\frac{7 \log (x) \log ^2(1-x)}{(x-1)^2}+\frac{3 \log (x) \log ^2(1-x)}{16 (x-1)^3}+\frac{1}{4} \log (x) \log ^2(1-x) \notag \\
	&+\frac{25 \log ^2(1-x)}{16-16 x}+\frac{\log ^2(1-x)}{8 x}-\frac{1}{24} \pi ^2 \log ^2(1-x)-\frac{3}{8} \log ^2(1-x) \notag \\
	&+\frac{1}{24} \log ^3(x) \log (1-x)+\frac{3 H(0,1,x) \log (1-x)}{2 (x-1)}+\frac{H(0,1,x) \log (1-x)}{8 x} \notag \\
	&+\frac{7 H(0,1,x) \log (1-x)}{2 (x-1)^2}-\frac{3}{4} H(0,1,x) \log (1-x)+\frac{3}{4} H(0,0,1,x) \log (1-x) \notag \\
	&-\frac{\log (x) \log (1-x)}{4 x}+\frac{1}{24} \pi ^2 \log (x) \log (1-x)+\frac{1}{2} \log (x) \log (1-x) \notag \\
	&+\frac{7 \pi ^2 \log (1-x)}{12 (x-1)}-\frac{\pi ^2 \log (1-x)}{12 x}+\frac{9 \log (1-x)}{32 x}+\frac{17 \pi ^2 \log (1-x)}{12 (x-1)^2}+\frac{3}{4} \zeta_3 \log (1-x) \notag \\
	&-\frac{1}{16} \pi ^2 \log (1-x)-\frac{21}{32} \log (1-x)-\frac{\log ^4(x)}{96}+\frac{H(0,1,x)^2}{4 (x-1)}+\frac{13 H(0,1,x)^2}{8 (x-1)^2} \notag \\
	&+\frac{7 H(0,1,x)^2}{4 (x-1)^3}+\frac{5}{8} H(0,1,x)^2-\frac{1}{48} \pi ^2 \log ^2(x)-\frac{\pi ^2 \left(x^3-3 x^2+5 x-1\right) H(0,-1,x)}{12 (x-1)^3} \notag \\
	&+\frac{45 H(0,1,x)}{8-8 x} +\frac{\pi ^2 H(0,1,x)}{4 (x-1)}-\frac{5 H(0,1,x)}{16 x}+\frac{31 \pi ^2 H(0,1,x)}{24 (x-1)^2}-\frac{25 H(0,1,x)}{8 (x-1)^2} \notag \\
	&+\frac{17 \pi ^2 H(0,1,x)}{12 (x-1)^3}-\frac{21}{16} H(0,1,x)+\frac{H(-1,0,0,1-x)}{8-8 x}+\frac{3 H(-1,0,0,1-x)}{8 (x-1)^3} \notag \\
	&+\frac{5 H(-1,0,1,x)}{x-1}+\frac{H(-1,0,1,x)}{2 x}+\frac{2 H(-1,0,1,x)}{(x-1)^2}+\frac{5}{2} H(-1,0,1,x)+\frac{63 H(0,0,1,1-x)}{8-8 x} \notag \\
	&+\frac{H(0,0,1,1-x)}{4 x}-\frac{14 H(0,0,1,1-x)}{(x-1)^2}-\frac{3 H(0,0,1,1-x)}{8 (x-1)^3}-\frac{5}{4} H(0,0,1,1-x) \notag \\
	&+\frac{3 H(0,0,1,x)}{4-4 x}+\frac{H(0,0,1,x)}{8 x}+\frac{3 H(0,0,1,x)}{2 (x-1)^2}-\frac{3 H(0,0,1,x)}{8 (x-1)^3} -\frac{5}{4} H(0,0,1,x) \notag \\
	&-\frac{2 H(0,-1,0,1,x)}{(x-1)^2}-\frac{2 H(0,-1,0,1,x)}{(x-1)^3}-H(0,-1,0,1,x)-\frac{1}{4} H(0,0,0,1,1-x) \notag \\
	&+\frac{H(0,0,0,1,x)}{2-2 x}-\frac{5 H(0,0,0,1,x)}{4 (x-1)^2}-\frac{3 H(0,0,0,1,x)}{2 (x-1)^3}-\frac{1}{4} H(0,0,0,1,x) \notag \\
	&-\frac{2 H(0,0,1,1,x)}{x-1}-\frac{13 H(0,0,1,1,x)}{(x-1)^2}-\frac{14 H(0,0,1,1,x)}{(x-1)^3}-\frac{3}{4} H(0,0,1,1,x) \notag \\
	&+\frac{H(0,1,1-x)}{8 (x-1)^3 x}  \biggl(2 x \log ^2(1-x) (x-1)^3+2 (2 x-1) (x-1)^3 \notag \\
	&+\left(7 x^4+41 x^3+10 x^2-56 x+1\right) \log (1-x)\biggr)  
	+\frac{\pi ^2 \log (2-x)}{16-16 x}+\frac{3 \pi ^2 \log (2-x)}{16 (x-1)^3} \notag \\
	&+\frac{5 \pi ^2 \log (x+1)}{12 (x-1)}+\frac{\pi ^2 \log (x+1)}{24 x}+\frac{\pi ^2 \log (x+1)}{6 (x-1)^2}+\frac{5}{24} \pi ^2 \log (x+1)-\frac{1}{4} Li_4\left(\frac{x-1}{x}\right) \notag \\
	&+\frac{3 \pi ^4}{80-80 x}+\frac{7 \pi ^2}{12-12 x}+\frac{\pi ^2}{24 x}-\frac{277 \pi ^4}{1440 (x-1)^2} 
	-\frac{59 \pi ^2}{48 (x-1)^2}-\frac{19 \pi ^4}{90 (x-1)^3}+\frac{39 \zeta_3}{4 (x-1)} \notag \\
	&-\frac{\zeta_3}{4 x}+\frac{14 \zeta_3}{(x-1)^2}+\frac{3 \zeta_3}{8 (x-1)^3}+\frac{3 \zeta_3}{2}+\frac{\pi ^2 \log (2)}{1-x}-\frac{1}{2} \pi ^2 \log (2)-\frac{K \pi ^4}{45}+\frac{17 \pi ^4}{960}+\frac{\pi ^2}{16}+\frac{15}{8} \,,
\end{align}
\endgroup
\vspace{-0.5cm}
\begingroup
\allowdisplaybreaks
\begin{align}
	\mathcal{D}_{C_A C_F} =& -\frac{11}{288} l_b^3 + \left( \frac{167}{576}-\frac{\pi ^2}{96} \right) l_b^2 
	+ \left( \frac{15 \zeta_3}{16}-\frac{1165}{1728}-\frac{7 \pi ^2}{72} \right) l_b \notag \\
	&+ \frac{11}{288} L_{\mu}^3 +  \left( \frac{11}{48} \log (1-x)+\frac{\pi ^2}{96}-\frac{299}{576} \right) L_{\mu}^2 \notag \\
	&+\frac{1}{1728 x} \biggl( 792 x H(0,1,x)+x \left(-1188 \zeta_3+3925+96 \pi ^2\right)+792 x \log ^2(1-x) \notag \\
	&+12 \left(6 \pi ^2 x-233 x+33\right) \log (1-x)\biggr)L_{\mu} \notag \\
	&+ \frac{11}{36} \log ^3(1-x)+\frac{37 \log (x) \log ^2(1-x)}{32 (x-1)}+\frac{7 \log (x) \log ^2(1-x)}{4 (x-1)^2} \notag \\
	&-\frac{3 \log (x) \log ^2(1-x)}{32 (x-1)^3}+\frac{13}{48} \log (x) \log ^2(1-x)-\frac{17 \log ^2(1-x)}{32 (x-1)} \notag \\
	&+\frac{13 \log ^2(1-x)}{96 x}+\frac{1}{24} \pi ^2 \log ^2(1-x)-\frac{349}{288} \log ^2(1-x) \notag \\
	&+\frac{37 H(0,1,1-x) \log (1-x)}{16 (x-1)}+\frac{7 H(0,1,1-x) \log (1-x)}{2 (x-1)^2} \notag \\
	&-\frac{3 H(0,1,1-x) \log (1-x)}{16 (x-1)^3}+\frac{13}{24} H(0,1,1-x) \log (1-x)-\frac{1}{2} H(0,0,1,x) \log (1-x)\notag \\
	&+\frac{47 \pi ^2 \log (1-x)}{96 (x-1)}+\frac{\pi ^2 \log (1-x)}{24 x}-\frac{269 \log (1-x)}{288 x} \notag \\
	&+\frac{29 \pi ^2 \log (1-x)}{48 (x-1)^2}-\frac{7}{8} \zeta_3 \log (1-x)+\frac{7}{36} \pi ^2 \log (1-x) \notag \\
	&+\frac{2545}{864} \log (1-x)-\frac{\left(4 x^3-16 x^2+10 x-5\right) H(0,1,x)^2}{16 (x-1)^3} \notag \\
	&+\frac{\pi ^2 \left(x^3-3 x^2+5 x-1\right) H(0,-1,x)}{24 (x-1)^3}+\frac{H(-1,0,0,1-x)}{16 (x-1)} \notag \\
	&-\frac{3 H(-1,0,0,1-x)}{16 (x-1)^3}+\frac{5 H(-1,0,1,x)}{2-2 x}-\frac{H(-1,0,1,x)}{4 x} \notag \\
	&-\frac{H(-1,0,1,x)}{(x-1)^2}-\frac{5}{4} H(-1,0,1,x)-\frac{37 H(0,0,1,1-x)}{16 (x-1)}-\frac{7 H(0,0,1,1-x)}{2 (x-1)^2} \notag \\
	&+\frac{3 H(0,0,1,1-x)}{16 (x-1)^3}-\frac{13}{24} H(0,0,1,1-x)+\frac{31 H(0,0,1,x)}{16 (x-1)} \notag \\
	&+\frac{15 H(0,0,1,x)}{8 (x-1)^2}+\frac{3 H(0,0,1,x)}{16 (x-1)^3}+\frac{5}{12} H(0,0,1,x) \notag \\
	&+\frac{H(0,-1,0,1,x)}{(x-1)^2}+\frac{H(0,-1,0,1,x)}{(x-1)^3}+\frac{1}{2} H(0,-1,0,1,x) \notag \\
	&+\frac{H(0,0,0,1,x)}{2-2 x}-\frac{9 H(0,0,0,1,x)}{4 (x-1)^2}-\frac{15 H(0,0,0,1,x)}{8 (x-1)^3} \notag \\
	&-\frac{2 H(0,0,1,1,x)}{x-1}-\frac{5 H(0,0,1,1,x)}{(x-1)^2}-\frac{7 H(0,0,1,1,x)}{2 (x-1)^3} \notag \\
	&+\frac{1}{144 (x-1)^3 x} \biggl( H(0,1,x) \biggl(4 \left(-29+3 \pi ^2\right) x^4+273 x^3+6 \left(-64+13 \pi ^2\right) x^2 \notag \\
	&+3 \left(44 x^3-93 x^2+96 x-47\right) \log (1-x) x+\left(260-3 \pi ^2\right) x-33\biggr) \biggr) \notag \\
	&+\frac{\pi ^2 \log (2-x)}{32 (x-1)}-\frac{3 \pi ^2 \log (2-x)}{32 (x-1)^3}-\frac{5 \pi ^2 \log (x+1)}{24 (x-1)} \notag \\
	&-\frac{\pi ^2 \log (x+1)}{48 x}-\frac{\pi ^2 \log (x+1)}{12 (x-1)^2}-\frac{5}{48} \pi ^2 \log (x+1) \notag \\
	&+\frac{3 \pi ^4}{80-80 x}-\frac{13 \pi ^2}{16 (x-1)}-\frac{43 \pi ^4}{360 (x-1)^2} 
	-\frac{67 \pi ^2}{96 (x-1)^2}-\frac{263 \pi ^4}{2880 (x-1)^3}+\frac{11 \zeta_3}{8 (x-1)} \notag \\
	&+\frac{7 \zeta_3}{2 (x-1)^2}-\frac{3 \zeta_3}{16 (x-1)^3}+\frac{11 \zeta_3}{12}
	+\frac{ \pi ^2 \log (2)}{4 (x-1)}\left( x + 1 \right)+\frac{ \pi ^4}{90}\left(K -\frac{167}{64}\right)-\frac{41 \pi ^2}{288}-\frac{1595}{432} \,,
\end{align}
\endgroup
\vspace{-0.5cm}
\begingroup
\allowdisplaybreaks
\begin{align}
	\mathcal{D}_{C_Fn_l} =& \frac{1}{144}  l_b^3 -\frac{13}{288}  l_b^2 + \left( \frac{77}{864}+\frac{\pi ^2}{72} \right) l_b \notag \\
	& -\frac{1}{144} L_{\mu}^3 + \left( \frac{25}{288}-\frac{1}{24} \log (1-x) \right) L_{\mu}^2  \notag \\
	& + \left( -\frac{1}{12} H(0,1,x)-\frac{1}{12} \log ^2(1-x)-\frac{\log (1-x)}{24 x}+\frac{19}{72} \log (1-x)-\frac{\pi ^2}{72}-\frac{341}{864} \right) L_{\mu} \notag \\
	& -\frac{H(0,1,x)}{24 x}+\frac{19}{72} H(0,1,x)+\frac{1}{6} H(0,0,1,1-x)+\frac{1}{12} H(0,0,1,x)-\frac{1}{6} \log (1-x) H(0,1,1-x) \notag \\
	&-\frac{1}{6} \log (1-x) H(0,1,x)-\frac{1}{18} \log ^3(1-x)-\frac{1}{12} \log (x) \log ^2(1-x)-\frac{\log ^2(1-x)}{24 x}  \notag \\
	& +\frac{19}{72} \log ^2(1-x) + \log (1-x) \left( \frac{19}{144 x}-\frac{1}{36} \pi ^2 -\frac{209}{432} \right)-\frac{\zeta_3}{6}+\frac{\pi ^2}{36}+\frac{53}{108} \,,
\end{align}
\endgroup
\vspace{-0.5cm}
\begingroup
\allowdisplaybreaks
\begin{align}
	\hspace{-0.5cm}
	\mathcal{D}_{C_Fn_h} =& \frac{1}{5184 (x-1)^3 x} \biggl( -72 \left(19 x^4+24 x^3-8 x-3\right) H(0,1,x) \notag \\
	&+432 \left(x^3-3 x^2-3 x-1\right) x H(0,0,1,x)-432 x^4 \zeta_3-168 \pi ^2 x^4+7951 x^4-3180 x^4 \log (1-x) \notag \\
	&+1296 x^3 \zeta_3+1656 \pi ^2 x^3-14709 x^3+2304 x^3 \log (1-x)+1296 x^2 \zeta_3 \notag \\
	&-1080 \pi ^2 x^2+7869 x^2+1944 x^2 \log (1-x)+432 x \zeta_3-24 \pi ^2 x-1111 x \notag \\
	&-384 x \log (1-x)-684 \log (1-x) \biggr) \,,
\end{align}
\endgroup
\vspace{-0.5cm}
\begingroup
\allowdisplaybreaks
\begin{align}
	\hspace{-0.5cm}
	\mathcal{D}_{C_Fn_m} =& \frac{l_b^3}{144} + \left( -\frac{ L_{\mu}}{48} -\frac{1}{24}\log (1-x) +\frac{25}{288} \right) l_b^2 \notag \\
	&+ \biggl( \frac{L_{\mu}^2}{48} +\frac{1}{12} \left( \log (1-x) -\frac{25}{12} \right) L_{\mu} + \frac{1}{12} H(0,1,x) +\frac{1}{12} \log ^2(1-x) \notag \\
	&+\frac{ \log (1-x)}{24 x} -\frac{19}{72} \log (1-x)+\frac{\pi ^2}{72}+\frac{397}{864}  \biggr) l_b  \notag \\
	&-\frac{L_{\mu}^3}{144} + \left( -\frac{1}{24} \log (1-x)+\frac{25}{288}\right) L_{\mu}^2  \notag \\
	&+ \left( -\frac{1}{12} H(0,1,x) -\frac{1}{12}\log ^2(1-x)+\frac{19}{72} \log (1-x)-\frac{\log (1-x)}{24 x}-\frac{397 }{864}-\frac{\pi ^2}{72} \right)  L_{\mu} \notag \\
	&-\frac{H(0,1,x)}{24 x}+\frac{19}{72} H(0,1,x)+\frac{1}{6} H(0,0,1,1-x)+\frac{1}{12} H(0,0,1,x)-\frac{1}{6} \log (1-x) H(0,1,1-x) \notag \\
	&-\frac{1}{6} \log (1-x) H(0,1,x) -\frac{1}{18} \log ^3(1-x) + \left( -\frac{1}{24 x}+\frac{19}{72} -\frac{1}{12} \log (x) \right) \log ^2(1-x) \notag \\
	&+\frac{19 \log (1-x)}{144 x}-\frac{1}{36} \pi ^2 \log (1-x)-\frac{265}{432} \log (1-x)-\frac{\zeta_3}{12}+\frac{11 \pi ^2}{432}+\frac{517}{432} \,,
\end{align}
\endgroup
\vspace{-0.5cm}
\begingroup
\allowdisplaybreaks
\begin{align}
	\mathcal{E}_{C_F^2} =& \frac{\log (1-x)}{16 x} l_b^2 -\frac{\log (1-x)}{16 x} l_b -\frac{\log (1-x)}{16 x} L_{\mu}^2 \notag \\
	& + \frac{(5-4 \log (1-x)) \log (1-x)}{16 x}  L_{\mu} \notag \\
	& -\frac{1}{240 (x-1)^4 x} \biggl(\! 60 \log ^3(1-x) x^4+450 \log ^2(1-x) x^4+750 H(0,1,x) x^4+60 H(-1,0,0,1-x) x^4 \notag \\
	&-240 H(-1,0,1,x) x^4 -180 H(0,0,1,1-x) x^4-60 H(0,0,1,x) x^4+300 H(0,1,x) \log (1-x) x^4 \notag \\
	&-80 \pi ^2 \log (1-x) x^4 +225 \log (1-x) x^4+30 \pi ^2 \log (2-x) x^4 +150 \log ^2(1-x) \log (x) x^4 \notag \\
	&-120 \log (1-x) \log (x) x^4-20 \pi ^2 \log (x+1) x^4-480 \zeta_3 x^4+240 \pi ^2 \log (2) x^4  \notag \\
	&+80 \pi ^2 x^4 -240 \log ^3(1-x) x^3-480 H(0,1,x)^2 x^3+840 \log ^2(1-x) x^3  \notag \\
	&-480 \pi ^2 H(0,1,x) x^3+3600 H(0,1,x) x^3-120 H(-1,0,0,1-x) x^3 -960 H(-1,0,1,x) x^3  \notag \\
	&+12360 H(0,0,1,1-x) x^3 -840 H(0,0,1,x) x^3+960 H(0,0,0,1,x) x^3  \notag \\
	&+3840 H(0,0,1,1,x) x^3-3960 H(0,1,x) \log (1-x) x^3 -980 \pi ^2 \log (1-x) x^3  \notag \\
	&-840 \log (1-x) x^3-60 \pi ^2 \log (2-x) x^3-6420 \log ^2(1-x) \log (x) x^3  \notag \\
	&+480 \log (1-x) \log (x) x^3-80 \pi ^2 \log (x+1) x^3 -11040 \zeta_3 x^3-720 \pi ^2 \log (2) x^3  \notag \\
	&+72 \pi ^4 x^3+1570 \pi ^2 x^3+360 \log ^3(1-x) x^2-1800 H(0,1,x)^2 x^2-2970 \log ^2(1-x) x^2 \notag \\
	&+240 \pi ^2 H(0,-1,x) x^2-1320 \pi ^2 H(0,1,x) x^2-5040 H(0,1,x) x^2  \notag \\
	&+360 H(-1,0,0,1-x) x^2-4200 H(0,0,1,1-x) x^2+420 H(0,0,1,x) x^2 \notag \\
	&+2880 H(0,-1,0,1,x) x^2+720 H(0,0,0,1,x) x^2+14400 H(0,0,1,1,x) x^2  \notag \\
	&+2040 H(0,1,x) \log (1-x) x^2+120 \pi ^2 \log (1-x) x^2+1170 \log (1-x) x^2 \notag \\
	&+180 \pi ^2 \log (2-x) x^2+2460 \log ^2(1-x) \log (x) x^2-720 \log (1-x) \log (x) x^2  \notag \\
	&+2580 \zeta_3 x^2+720 \pi ^2 \log (2) x^2 +196 \pi ^4 x^2 \notag \\
	&-1590 \pi ^2 x^2-240 \log ^3(1-x) x-240 H(0,1,x)^2 x+1620 \log ^2(1-x) x  \notag \\
	&-240 \pi ^2 H(0,1,x) x+780 H(0,1,x) x-840 H(-1,0,0,1-x) x \notag \\
	&+960 H(-1,0,1,x) x-7560 H(0,0,1,1-x) x+960 H(0,0,1,x) x+480 H(0,0,0,1,x) x  \notag \\
	&+1920 H(0,0,1,1,x) x+1560 H(0,1,x) \log (1-x) x \notag \\
	&+980 \pi ^2 \log (1-x) x-720 \log (1-x) x-420 \pi ^2 \log (2-x) x+3540 \log ^2(1-x) \log (x) x  \notag \\
	&+480 \log (1-x) \log (x) x+80 \pi ^2 \log (x+1) x \notag \\
	&+8520 \zeta_3 x-240 \pi ^2 \log (2) x+36 \pi ^4 x-80 \pi ^2 x+60 \log ^3(1-x)+60 \log ^2(1-x)  \notag \\
	&-90 H(0,1,x)+240 H(-1,0,1,x)+120 H(0,0,1,1-x) \notag \\
	&+60 H(0,0,1,x)+60 H(0,1,x) \log (1-x)-40 \pi ^2 \log (1-x)+165 \log (1-x) \notag \\
	&+60 H(0,1,1-x) \left(\left(4 x^4-210 x^3+76 x^2 
	+122 x-1\right) \log (1-x)-2 (x-1)^4\right)  \notag \\
	&-120 \log (1-x) \log (x)+20 \pi ^2 \log (x+1) 
	-120 \zeta_3+20 \pi ^2 \biggr) \,,
\end{align}
\endgroup
\vspace{-0.5cm}
\begingroup
\allowdisplaybreaks
\begin{align}
	\hspace{-0.7cm}
	\mathcal{E}_{C_A C_F} =& -\frac{11 \log (1-x)}{24 x} L_{\mu} \notag \\
	&-\frac{1}{1440 (x-1)^4 x} \biggl( 1740 \log ^2(1-x) x^4-180 H(-1,0,0,1-x) x^4+720 H(-1,0,1,x) x^4 \notag \\
	&+4500 H(0,0,1,1-x) x^4-540 H(0,0,1,x) x^4-4500 H(0,1,1-x) \log (1-x) x^4 \notag \\
	&-210 \pi ^2 \log (1-x) x^4 -1730 \log (1-x) x^4-90 \pi ^2 \log (2-x) x^4-2250 \log ^2(1-x) \log (x) x^4 \notag \\
	&+60 \pi ^2 \log (x+1) x^4 -2520 \zeta_3 x^4-720 \pi ^2 \log (2) x^4+1200 \pi ^2 x^4-540 x^4-1020 \log ^2(1-x) x^3 \notag \\
	&+360 H(-1,0,0,1-x) x^3+2880 H(-1,0,1,x) x^3+16200 H(0,0,1,1-x) x^3 \notag \\
	&-11880 H(0,0,1,x) x^3+4320 H(0,0,0,1,x) x^3+17280 H(0,0,1,1,x) x^3 \notag \\
	&-16200 H(0,1,1-x) \log (1-x) x^3-3900 \pi ^2 \log (1-x) x^3+8540 \log (1-x) x^3  \notag \\
	&+180 \pi ^2 \log (2-x) x^3-8100 \log ^2(1-x) \log (x) x^3+240 \pi ^2 \log (x+1) x^3 \notag \\
	&-20160 \zeta_3 x^3+2160 \pi ^2 \log (2) x^3+324 \pi ^4 x^3+1950 \pi ^2 x^3+1620 x^3-2790 \log ^2(1-x) x^2 \notag \\
	&-720 \pi ^2 H(0,-1,x) x^2-1080 H(-1,0,0,1-x) x^2-14760 H(0,0,1,1-x) x^2 \notag \\
	&+7200 H(0,0,1,x) x^2 -8640 H(0,-1,0,1,x) x^2+11160 H(0,0,0,1,x) x^2 \notag \\
	&+10080 H(0,0,1,1,x) x^2+14760 H(0,1,1-x) \log (1-x) x^2+3330 \pi ^2 \log (1-x) x^2  \notag \\
	&-15240 \log (1-x) x^2-540 \pi ^2 \log (2-x) x^2+7380 \log ^2(1-x) \log (x) x^2+19620 \zeta_3 x^2  \notag \\
	&-2160 \pi ^2 \log (2) x^2+411 \pi ^4 x^2-1470 \pi ^2 x^2-1620 x^2  \notag \\
	&-180 \left(12 x^2+7 x+2\right) H(0,1,x)^2 x+1680 \log ^2(1-x) x+2520 H(-1,0,0,1-x) x  \notag \\
	&-2880 H(-1,0,1,x) x-7560 H(0,0,1,1-x) x+3600 H(0,0,1,x) x+720 H(0,0,0,1,x) x  \notag \\
	&+2880 H(0,0,1,1,x) x+7560 H(0,1,1-x) \log (1-x) x+660 \pi ^2 \log (1-x) x  \notag \\
	&+11780 \log (1-x) x+1260 \pi ^2 \log (2-x) x+3780 \log ^2(1-x) \log (x) x-240 \pi ^2 \log (x+1) x  \notag \\
	&+4680 \zeta_3 x+720 \pi ^2 \log (2) x+54 \pi ^4 x-1680 \pi ^2 x+540 x+390 \log ^2(1-x)  \notag \\
	&-720 H(-1,0,1,x)+120 \pi ^2 \log (1-x)-3350 \log (1-x)  \notag \\
	&-60 H(0,1,x) \biggl(-59 x^4+\left(-79+36 \pi ^2\right) x^3+9 \left(19+5 \pi ^2\right) x^2 \notag \\
	&+3 \left(9 x^3+14 x^2-13 x-10\right) \log (1-x) x+\left(-22+6 \pi ^2\right) x-11\biggr)-60 \pi ^2 \log (x+1) \biggr) \,,
\end{align}
\endgroup
\vspace{-0.5cm}
\begingroup
\allowdisplaybreaks
\begin{align}
	\hspace{-3cm}
	\mathcal{E}_{C_Fn_l} =& \frac{1}{12 x} \left( \log (1-x) L_{\mu} + H(0,1,x) + \log ^2(1-x) -\frac{25}{6} \log (1-x) \right) \,,
\end{align}
\endgroup
\vspace{-0.5cm}
\begingroup
\allowdisplaybreaks
\begin{align}
	\mathcal{E}_{C_Fn_h} =& -\frac{1}{72 (x-1)^4 x} \biggl( 216 x^2 H(0,0,1,x)+6 \left(x^4+20 x^3-20 x-1\right) H(0,1,x) \notag \\
	&+4 \pi ^2 x^4-102 x^4+25 x^4 \log (1-x)-52 \pi ^2 x^3-120 x^3+272 x^3 \log (1-x)-216 x^2 \zeta_3 \notag \\
	&+12 \pi ^2 x^2+546 x^2-594 x^2 \log (1-x)+36 \pi ^2 x-324 x+272 x \log (1-x)+25 \log (1-x) \biggr) \,,
\end{align}
\endgroup
\vspace{-0.5cm}
\begingroup
\allowdisplaybreaks
\begin{align}
	\hspace{-2.2cm}
	\mathcal{E}_{C_Fn_m} = \frac{1}{12x} \left( - \log (1-x)l_b  + \log (1-x)L_{\mu} + H(0,1,x) + \log ^2(1-x) -\frac{25}{6} \log (1-x)  \right) \,,
\end{align}
\endgroup
and \mbox{$K = 3.32812$}~\cite{Bonciani:2008wf}.

All these two-loop form factors are expressed in terms of logarithms, one classical polylogarithm of transcendental weight 4 and eleven harmonic polylogarithms 
$H(0, -1, x)$, $H(0, 1, 1 - x)$, $H(0, 1, x)$, $H(-1, 0, 0, 1 - x)$, $H(-1, 0, 1, x)$,  $H(0, 0, 1, 1 - x)$, $H(0, 0, 1, x)$, $H(0, -1, 0, 1, x)$, 
$H(0, 0, 0, 1, 1 - x)$, $H(0, 0, 0, 1, x)$, $H(0, 0, 1, 1, x)$ up to weight four and with letters~$\{-1,0,1\}$.

We stress that these results hold at the amplitude level. Since we are interested in the double-virtual contribution, we need to compute the interference between the two-loop finite remainder and the Born amplitude, namely
\begin{equation}
	\frac{g_W^4 |V_{bt}|^2}{(q^2-m_W^2)^2 + \Gamma_W^2 m_W^2} \,2 \mathrm{Re}{ \left(  \M_{(\pm s_t)}^{(0) *} \M_{(\pm s_t)}^{(2),(m_b)}(\mu) \biggl|_{\mathrm{fin}} \right) } \,,
\end{equation}
with \mbox{$\M_{(\pm s_t)}^{(0)} = M_{(\pm s_t)}^{(G_1)}$} since \mbox{$G_3^{(0)} = 0$} in Eq.~\eqref{eq:heavy-to-light_QCDvertex}. For completeness, we also give the analytic expressions of $|M_{(\pm s_t)}^{(G_1)}|^2$ and $M_{(\pm s_t)}^{(G_1) *} M_{(\pm s_t)}^{(G_3)}$:
\vspace{0.4cm}
\begin{equation}
	\hspace{-2.7cm}
	|M_{(\pm s_t)}^{(G_1)}|^2 =  2 (p_3 \cdot p_b) \left[ (p_t \cdot p_4) \mp m_t (s_t \cdot p_4)  \right] 
	\label{eq:top_decay_polarised_born1} \,,
\end{equation}
\begin{align}
	M_{(\pm s_t)}^{(G_1) *} M_{(\pm s_t)}^{(G_3)} &=  2 (p_t \cdot p_3) (p_t \cdot p_4) - (p_3 \cdot p_4) m_t^2 
	\,\pm \,2 (p_3 \cdot p_4) (s_t \cdot p_4) m_t \notag \\
	&\mp  (p_t \cdot p_3) (s_t \cdot p_4) m_t \left( 1 +  \frac{2 (p_t \cdot p_4)}{m_t^2} \right)
	\notag \\
	&\mp \left[ (p_t \cdot p_4) (s_t \cdot p_3) + i \varepsilon_{\mu \nu \rho \sigma} p_t^{\mu} p_3^{\nu} p_4^{\rho} s_t^{\sigma} \right] m_t \left( 1-  \frac{2 (p_t \cdot p_4)}{m_t^2} \right) \,,
	\label{eq:top_decay_polarised_born2}
\end{align}
where we label the four-momentum of the neutrino (charged lepton) with $p_3 (p_4)$.

For the decay of an anti-top quark with helicity \mbox{$\lambda_{\bar t} = \pm 1$} along the direction of the spin four-vector $s_{\bar t}$, all results discussed in Sec.~\ref{sec:heavy-to-light_FF} and in this appendix remain valid with the substitution \mbox{$s_t \to - s_{\bar t}$} on the r.h.s.\ of Eqs.~\eqref{eq:top_decay_polarised_born1} and~\eqref{eq:top_decay_polarised_born2}.

\section{Top-quark width treatment and on-shell matching}
\label{app:Appendix_top_width}
In Sec.~\ref{sec:Delta-term}, we have discussed how to consistently perform the matching of an off-shell calculation, in the \mbox{$\Gamma_t \to 0$} limit, to the on-shell $t \bar t$ cross section. 
We recall that Eq.~\eqref{eq:matching-to-onshell_ttx} is valid at all orders in perturbation theory.
Nevertheless, when truncating the perturbative expansion at a fixed order in $\alphas$, spurious width contributions are generated if $1/\Gamma_{t}$ terms are not carefully treated, and the agreement with the on-shell calculation can be spoiled.
Before entering into the details of how these spurious terms can be removed, we set the notation that will be used in this appendix.

We systematically expand the total top-quark width in powers of $\alphas$ as
\begin{equation}
	\Gamma_t = \Gamma_t^{(0)} + \alphas \Gamma_t^{(1)} + \alphas^2 \Gamma_t^{(2)} + \mathcal{O}(\alphas^3) \,,
	\label{eq:top-width_expanded}
\end{equation}
where the superscript $(n)$ refers to the N$^n$LO correction.
Analogously, we expand the differential cross section for $t\bar{t}$ production,
\begin{align}
	d\sigma_{t\bar{t}} = d\sigma_{t\bar{t}}^{(0)} + \alphas d\sigma_{t\bar{t}}^{(1)} +  \alphas^2  d\sigma_{t\bar{t}}^{(2)} + \mathcal{O}(\alphas^3) \,,
	\label{eq:ttxXS_expanded}
\end{align}
and for the (anti)top-quark decay into leptons,
\begingroup
\allowdisplaybreaks
\begin{align}
	d\Gamma_{t \to i} &= d\Gamma_{t \to i}^{(0)} + \alphas d\Gamma_{t \to i}^{(1)} +  \alphas^2 d\Gamma_{t \to i}^{(2)} + \mathcal{O}(\alphas^3) \,,\\
	d\Gamma_{\bar{t} \to j} &= d\Gamma_{\bar{t} \to j}^{(0)} + \alphas d\Gamma_{\bar{t} \to j}^{(1)} +  \alphas^2 d\Gamma_{\bar{t} \to j}^{(2)} + \mathcal{O}(\alphas^3) \,.
\end{align}
\endgroup
For brevity, we introduced the subscripts $(j) i$ to refer to the final state of the (anti)top-quark decay \mbox{$(\mu^{-}\bar{\nu}_{\mu} \bar b) e^+ \nu_e b$}.
The all-order cross section in the NWA is given by
\begingroup
\allowdisplaybreaks
\begin{equation}
	d\sigma_{\proddec} = d\sigma_{t\bar{t}}\frac{d\Gamma_{t \to i}}{\Gamma_t} \frac{d\Gamma_{\bar{t} \to j}}{\Gamma_t} \,,
	\label{eq:proddecXS}
\end{equation}
\endgroup
where top-quark spin correlations are implicitly understood.
If the factors $1/\Gamma_t$ in the previous formula are not expanded, the perturbative coefficients of $d\sigma_{\proddec}$, up to $\mathcal{O}(\alphas^2)$ corrections, would be
\begingroup
\allowdisplaybreaks
\begin{align}
	d\sigma^{(0)}_{\proddec} &= \frac{1}{(\Gamma_t^{\NNLO})^2}  d\sigma_{t\bar{t}}^{(0)} d\Gamma_{t \to i}^{(0)} d\Gamma_{\bar{t} \to j}^{(0)}  \,,  \\
	d\sigma^{(1)}_{\proddec} &= \frac{1}{(\Gamma_t^{\NNLO})^2}\left [ d\sigma_{t\bar{t}}^{(0)}\left( d\Gamma_{t \to i}^{(1)}d\Gamma_{\bar{t} \to j}^{(0)}   + d\Gamma_{t \to i}^{(0)} d\Gamma_{\bar{t} \to j}^{(1)}\right )
	+ d\sigma_{t\bar{t}}^{(1)} d\Gamma_{t \to i}^{(0)} d\Gamma_{\bar{t} \to j}^{(0)}  \right] \,,  \\
	d\sigma^{(2)}_{\proddec} &= \frac{1}{(\Gamma_t^{\NNLO})^2}\biggl[ d\sigma_{t\bar{t}}^{(0)}\left( d\Gamma_{t \to i}^{(2)}d\Gamma_{\bar{t} \to j}^{(0)}   + d\Gamma_{t \to i}^{(0)} d\Gamma_{\bar{t} \to j}^{(2)}  + d\Gamma_{t \to i}^{(1)} d\Gamma_{\bar{t} \to j}^{(1)} \right )
	   + d\sigma_{t\bar{t}}^{(2)} d\Gamma_{t \to i}^{(0)} d\Gamma_{\bar{t} \to j}^{(0)} \notag \\
	&\hspace{2cm} + d\sigma_{t\bar{t}}^{(1)}\left( d\Gamma_{t \to i}^{(1)}d\Gamma_{\bar{t} \to j}^{(0)}   + d\Gamma_{t \to i}^{(0)} d\Gamma_{\bar{t} \to j}^{(1)}\right )  \biggr] \,,
	\label{eq:proddecXS_coefficients}
\end{align}
\endgroup
and the corresponding LO, NLO and NNLO cross sections could be schematically written as
\begingroup
\allowdisplaybreaks
\begin{align}
	d\sigma^{\LO}_{\proddec} &= \left(\frac{\Gamma_t^{\NNLO}}{\Gamma_t^{\LO}}\right)^{\!2} d\sigma^{(0)}_{\proddec}  \,, \\
	d\sigma^{\NLO}_{\proddec} &= \left(\frac{\Gamma_t^{\NNLO}}{\Gamma_t^{\NLO}}\right)^{\!2} \left(d\sigma^{(0)}_{\proddec} + \alphas d\sigma^{(1)}_{\proddec} \right) \,, \\
	d\sigma^{\NNLO}_{\proddec} &= d\sigma^{(0)}_{\proddec} + \alphas d\sigma^{(1)}_{\proddec} + \alphas^2 d\sigma^{(2)}_{\proddec} \,,
\end{align}
\endgroup
where all coefficients $d\sigma^{(k)}_{\proddec}$ and widths $\Gamma_t^{\mathrm{N}^n\LO}$ appearing in the definition of $d\sigma^{\mathrm{N}^n\LO}_{\proddec}$ (\mbox{$k \le n$}) are evaluated with $\alphas$  and PDFs at N$^n$LO.

In this na\"ive approach, the integral of $d\sigma^{\mathrm{N}^n\LO}_{\proddec}$ over the FS phase space does not reproduce the on-shell result
\begin{equation}
	\sigma_{\mathrm{on-shell}}^{\mathrm{N}^n\LO} \approx \sigma_{t \bar t}^{\mathrm{N}^n\LO} \times \text{BR}(W^{+} \to e^+ \nu_e) \times \text{BR}(W^{-} \to \mu^{-}\bar{\nu}_{\mu})
	\label{eq:onshellXS}
\end{equation}
beyond the leading order, as a consequence of the truncation of the perturbative expansion.
The difference consists in terms that are formally beyond the order of the calculation.

In order to remove these spurious effects, two approaches have been proposed in
the literature, namely
\begin{enumerate}
  \item the ``inverse-width expansion", which consists in expanding at fixed
		order the inverse of the total decay width of the resonant top quarks.
		The method was proposed in Ref.~\cite{Melnikov:2009dn}, and then applied
		in the NWA computations at NLO in Ref.~\cite{Campbell:2012uf}, and at NNLO in
		Ref.~\cite{Czakon:2020qbd}. It has also been applied in the off-shell NLOPS computation of Ref.~\cite{Jezo:2023rht};
  \item the ``$\Delta$-term" approach, which accounts for the mismatch terms
		arising from the truncation of the perturbative series in the off-shell
		or narrow-width calculation. This was introduced at NLO in the off-shell computation of
		Ref.~\cite{Denner:2012yc}.
\end{enumerate}
In the following two subsections, we summarise both methods up to NNLO. In our numerical results we have adopted the $\Delta$-term method as our default to perform the matching between the calculation of off-shell $t \bar t$ production and decays and the on-shell cross section.
This method is easier to construct, but limited to inclusive cross sections. In contrast, the inverse-width expansion can also be applied at the differential level.
%
%
%
%
%
\subsection{Inverse-width expansion}
In order to get a consistent perturbative expansion of Eq.~\eqref{eq:proddecXS}, we can individually expand the inverse width factors $1/\Gamma_t$ as
\begin{equation}
	\frac{1}{\Gamma_t} = \frac{1}{\Gamma_t^{(0)}} \left[ 1 - \alphas\frac{\Gamma_t^{(1)}}{\Gamma_t^{(0)}} + \alphas^2 \frac{\left(\Gamma_t^{(1)} \right)^{\!2} -\Gamma_t^{(0)}\Gamma_t^{(2)} }{\left(\Gamma_t^{(0)}\right)^{\!2}} \right] + \mathcal{O}(\alphas^3) \,.
	\label{eq:inverse_Gammat_expansion}
\end{equation}
Based on this, only the terms that contribute to a specific perturbative order are retained.

We first consider the application of the inverse-width expansion to the production$\times$decay cross section in the NWA.
Then, we will extend it to the off-shell calculation.
Having Eq.~\eqref{eq:inverse_Gammat_expansion} in mind, we can connect the expanded coefficients to the unexpanded, given in Eq.~\eqref{eq:proddecXS_coefficients}, by
\begingroup
\allowdisplaybreaks
\begin{align}
	d\sigma^{(0)\, \mathrm{exp}}_{\proddec} &=  \left(\frac{\Gamma_t^{\NNLO}}{\Gamma_t^{(0)}}\right)^{\!2} d\sigma^{(0)}_{\proddec} \,, \\
	d\sigma^{(1)\, \mathrm{exp}}_{\proddec} &=  \left(\frac{\Gamma_t^{\NNLO}}{\Gamma_t^{(0)}}\right)^{\!2} \!\left\{ d\sigma^{(1)}_{\proddec} - \frac{2 \Gamma_t^{(1)} }{ \Gamma_t^{(0)} }d\sigma^{(0)}_{\proddec} \right\} \,, \\
	d\sigma^{(2)\, \mathrm{exp}}_{\proddec} &= \left(\frac{\Gamma_t^{\NNLO}}{\Gamma_t^{(0)}}\right)^{\!2} \!\biggl\{ d\sigma^{(2)}_{\proddec}
	- \frac{2 \Gamma_t^{(1)} }{ \Gamma_t^{(0)} } d\sigma^{(1)}_{\proddec}
	+ \left[ 3 \!\left( \frac{\Gamma_t^{(1)} }{ \Gamma_t^{(0)} }\right)^{\!2} -\frac{2 \Gamma_t^{(2)} }{ \Gamma_t^{(0)} } \right]\! d\sigma^{(0)}_{\proddec} \biggr\} \,.
\end{align}
\endgroup

\vspace{-0.2cm}
\noindent
As a result, the expanded versions of the LO, NLO and NNLO cross sections in NWA read
\begingroup
\allowdisplaybreaks
\begin{align}
	d\sigma^{\LO,\, \mathrm{exp}}_{\proddec} &= d\sigma^{\LO}_{\proddec} \,, \\
	d\sigma^{\NLO,\, \mathrm{exp}}_{\proddec} &= \left(\frac{\Gamma_t^{\NLO}}{\Gamma_t^{(0)}}\right)^{\!2} \!\left\{ d\sigma^{\NLO}_{\proddec} - 2\alphas \frac{ \Gamma_t^{(1)} }{ \Gamma_t^{(0)} } \left(\frac{\Gamma_t^{\NNLO}}{\Gamma_t^{\NLO}}\right)^{\!2} \!d\sigma^{(0)}_{\proddec} \right\} \,, \\
	d\sigma^{\NNLO,\, \mathrm{exp}}_{\proddec} &= \left(\frac{\Gamma_t^{\NNLO}}{\Gamma_t^{(0)}}\right)^{\!2}
	\!\biggl\{ d\sigma^{\NNLO}_{\proddec}  - 2\alphas \frac{\Gamma_t^{(1)} }{ \Gamma_t^{(0)} } \left( d\sigma^{(0)}_{\proddec} + \alphas d\sigma^{(1)}_{\proddec} \right)   \notag \\
	&\hspace{4.3cm} +\alphas^2 \left[ 3 \!\left( \frac{\Gamma_t^{(1)} }{ \Gamma_t^{(0)} }\right)^{\!2} -\frac{2 \Gamma_t^{(2)} }{ \Gamma_t^{(0)} } \right]\! d\sigma^{(0)}_{\proddec}  \biggr\} \,,
\end{align}
\endgroup
where all coefficients $d\sigma^{(k)}_{\proddec}$ appearing in the definition of $d\sigma^{\mathrm{N}^n\LO, \,\mathrm{exp}}_{\proddec}$ (\mbox{$k < n$}) are evaluated with PDFs and $\alphas$ at N$^n$LO.

We can now consider the corresponding off-shell calculations.
In the \mbox{$\Gamma_t \to 0$} limit, the off-shell N$^n$LO cross section tends to the respective NWA result, i.e.\
\begin{equation}
	d\sigma^{\mathrm{N}^n\LO}_{\mathrm{off-shell}} \xrightarrow[\Gamma_t \to 0]{} d\sigma^{\mathrm{N}^n\LO}_{\proddec} \,.
\end{equation}
Thus, the inverse-width expansion can be easily extended to the off-shell case by replacing the NWA cross sections with their off-shell counterparts, i.e.,
\begingroup
\allowdisplaybreaks
\begin{align}
	d\sigma^{\LO,\, \mathrm{exp}}_{\offshell} &= d\sigma^{\LO}_{\offshell} \,, \label{eq:expanded_offshellXS_lo}\\
	d\sigma^{\NLO,\, \mathrm{exp}}_{\offshell} &= \left(\frac{\Gamma_t^{\NLO}}{\Gamma_t^{(0)}}\right)^{\!2} \!\left\{ d\sigma^{\NLO}_{\offshell} - 2\alphas \frac{ \Gamma_t^{(1)} }{ \Gamma_t^{(0)} } d\sigma^{\LO, \,\Gamma_t^{\NLO}}_{\offshell} \right\}  \,, \label{eq:expanded_offshellXS_nlo}\\
	d\sigma^{\NNLO,\, \mathrm{exp}}_{\offshell} &= \left(\frac{\Gamma_t^{\NNLO}}{\Gamma_t^{(0)}}\right)^{\!2}
	\!\biggl\{ d\sigma^{\NNLO}_{\offshell}  - 2\alphas \frac{\Gamma_t^{(1)} }{ \Gamma_t^{(0)} } d\sigma^{\NLO, \,\Gamma_t^{\NNLO}}_{\offshell}
	+\alphas^2 \left[ 3 \!\left( \frac{\Gamma_t^{(1)} }{ \Gamma_t^{(0)} }\right)^{\!2} -\frac{2 \Gamma_t^{(2)} }{ \Gamma_t^{(0)} } \right] d\sigma^{\LO, \,\Gamma_t^{\NNLO}}_{\offshell} \biggr\} \,,
	\label{eq:expanded_offshellXS}
\end{align}
\endgroup

\vspace{-0.2cm}
\noindent where the notation $d\sigma^{\mathrm{N}^k\LO, \,\Gamma_t^{\mathrm{N}^n\LO}}_{\offshell}$ \mbox{($k < n$)} refers to the $\mathrm{N}^k\LO$ cross section computed with the $\mathrm{N}^n\LO$ value of the top-quark width, $\alphas$ and PDFs.
The relations in Eqs.~\eqref{eq:expanded_offshellXS_lo}--\eqref{eq:expanded_offshellXS} ensure that, by construction, the integral over the full phase space reproduces, in the vanishing width limit, the correct on-shell cross section~\eqref{eq:matching-to-onshell_ttx} at fixed perturbative order.
%
%
\subsection{Alternative approach: the \texorpdfstring{$\Delta$}{Delta}-term}
%
An alternative approach to obtain a consistent perturbative expansion is inspired by the procedure introduced at NLO in
Ref.~\cite{Denner:2012yc}. This will be our default method to perform the
on-shell matching, as discussed in Sec.~\ref{sec:Delta-term}. 
Instead of removing higher-order contributions by expanding
and truncating the factor $1/\Gamma_t$ in Eq.~\eqref{eq:proddecXS}, the spurious
term at $\mathrm{N}^n\LO$ is explicitly computed as follows \begingroup \allowdisplaybreaks
\begin{align}
	\Delta\sigma^{\mathrm{N}^n\LO}_{\mathrm{trunc}} &= \sigma_{\mathrm{on-shell}}^{\mathrm{N}^n\LO} - \int d\sigma^{\mathrm{N}^n\LO}_{\proddec} \notag \\
	&= \frac{1}{\left( \Gamma_t^{\mathrm{N}^n\LO} \right)^2} \sigma_{t \bar t}^{\mathrm{N}^n\LO}\, \Gamma_{t \to i}^{\mathrm{N}^n\LO}\, \Gamma_{\bar t \to j}^{\mathrm{N}^n\LO} - \int d\sigma^{\mathrm{N}^n\LO}_{\proddec} \,.
\end{align}
\endgroup
The expression of $\Delta\sigma^{\mathrm{N}^n\LO}_{\mathrm{trunc}}$ contains higher-order terms in $\alphas$, and it is obtained by expanding the r.h.s.\ of the previous equation in the strong coupling while keeping the factor $(\Gamma_t^{\mathrm{N}^n\LO})^{-2}$ fixed.
Up to NNLO, we have
\begingroup
\allowdisplaybreaks
\begin{align}
	\hspace{-0.8cm} \Delta\sigma^{\LO}_{\mathrm{trunc}} &= 0 \,, \\
	\hspace{-0.8cm} \Delta\sigma^{\NLO}_{\mathrm{trunc}} &= \alphas^2 \left( \frac{1}{\Gamma_t^{\NLO}}\right)^{\!2} \!
	\biggl[ \sigma_{t\bar{t}}^{(1)}\left( \Gamma_{t\to i}^{(1)}\Gamma_{\bar{t}\to j}^{(0)}   + \Gamma_{t \to i}^{(0)} \Gamma_{\bar{t} \to j}^{(1)}\right )
	+ \sigma_{t\bar{t}}^{(0)} \Gamma_{t\to i}^{(1)}\Gamma_{\bar{t} \to j}^{(1)}
	+ \alphas \sigma_{t\bar{t}}^{(1)} \Gamma_{t\to i}^{(1)}\Gamma_{\bar{t} \to j}^{(1)}  \biggr] \notag \\
	&\approx \bigl(\text{BR}(W \to l \nu_l)\bigr)^2 \left( 1 - \delta^{\NLO}\right)
	\biggl[  ( \sigma_{t\bar{t}}^{(0)} + \alphas \sigma_{t\bar{t}}^{(1)}) \left( 1- \delta^{\NLO} \right) + 2 \alphas \sigma_{t\bar{t}}^{(1)} \delta^{\NLO} \biggr] \notag \\
	&= \bigl(\text{BR}(W \to l \nu_l)\bigr)^2 \left( 1 - \delta^{\NLO}\right)
	\biggl[  \sigma_{t\bar{t}}^{\NLO} \left( 1 + \delta^{\NLO} \right) - 2 \sigma_{t\bar{t}}^{\LO_{\NLO}}\delta^{\NLO} \biggr] \,, \\
	\hspace{-0.8cm} \Delta\sigma^{\NNLO}_{\mathrm{trunc}} &= \alphas^3 \left( \frac{1}{\Gamma_t^{\NNLO}}\right)^{\!2}
	\biggl[ \sigma_{t\bar{t}}^{(2)} \left( \Gamma_{t\to i}^{(1)}\Gamma_{\bar{t}\to j}^{(0)}  + \Gamma_{t \to i}^{(0)} \Gamma_{\bar{t} \to j}^{(1)}\right ) \notag \\
	&\hspace{3cm}+ \left( \sigma_{t\bar{t}}^{(1)} + \alphas \sigma_{t\bar{t}}^{(2)}  \right) \left( \Gamma_{t\to i}^{(1)}\Gamma_{\bar{t}\to j}^{(1)}   + \Gamma_{t \to i}^{(2)} \Gamma_{\bar{t} \to j}^{(0)} + \Gamma_{t \to i}^{(0)} \Gamma_{\bar{t} \to j}^{(2)}\right )  \notag \\
	&\hspace{3cm}+ \left( \sigma_{t\bar{t}}^{(0)} + \alphas\sigma_{t\bar{t}}^{(1)} + \alphas^2 \sigma_{t\bar{t}}^{(2)}  \right) \left( \Gamma_{t\to i}^{(2)}\Gamma_{\bar{t}\to j}^{(1)}   + \Gamma_{t \to i}^{(1)} \Gamma_{\bar{t} \to j}^{(2)} +\alphas  \Gamma_{t \to i}^{(2)} \Gamma_{\bar{t} \to j}^{(2)}\right )    \biggr] \notag \\
	&\approx \bigl(\text{BR}(W \to l \nu_l)\bigr)^2 \left( 1- \delta^{\NNLO} \right)
	\biggl\{   \left( \sigma_{t\bar{t}}^{(0)} + \alphas \sigma_{t\bar{t}}^{(1)} + \alphas^2 \sigma_{t\bar{t}}^{(2)}  \right)
	\left( 1- \delta^{\NNLO} \right) \notag \\
	&\quad + 2 \!\left( \alphas \sigma_{t\bar{t}}^{(1)}  + \alphas^2 \sigma_{t\bar{t}}^{(2)}\right)\delta^{\NNLO}
	-  \frac{ \delta^{\NNLO} }{\delta^{\NLO}} \left( \frac{1 - \delta^{\NLO}}{1-\delta^{\NNLO} } \right)  \!\left[ 2 \alphas \sigma_{t\bar{t}}^{(1)} \delta^{\NNLO}
	+ \sigma_{t\bar{t}}^{(0)} \frac{ \delta^{\NNLO} }{\delta^{\NLO}}  \left( 1- \delta^{\NLO} \right)  \right]
	\biggr\} \notag \\
	&= \bigl(\text{BR}(W \to l \nu_l)\bigr)^2 \left( 1- \delta^{\NNLO} \right)
	\biggl\{  \sigma_{t\bar{t}}^{\NNLO}
	\left( 1 + \delta^{\NNLO} \right)
	- 2 \sigma_{t\bar{t}}^{\LO_{\NNLO}} \delta^{\NNLO} \notag \\
	&\hspace{2cm} -  \left(\frac{ \delta^{\NNLO} }{\delta^{\NLO}}\right)^{\!2} \left( \frac{1 - \delta^{\NLO}}{1-\delta^{\NNLO} } \right)  \!\left[ 2 \sigma_{t\bar{t}}^{\NLO_{\NNLO}} \delta^{\NLO}
	+ \sigma_{t\bar{t}}^{\LO_{\NNLO}} \left( 1- 3 \delta^{\NLO} \right)  \right]
	\biggr\} \,,
\end{align}
\endgroup
where \mbox{$\delta^{\NLO} \equiv \Gamma_{t}^{(0)} / \Gamma_t^{\NLO} $} and \mbox{$\delta^{\NNLO} \equiv \Gamma_{t}^{(0)} /\Gamma_t^{\NNLO}$}, respectively. 
The notation $\sigma_{t\bar{t}}^{\mathrm{N}^k\LO_{\mathrm{N}^n\LO}}$ \mbox{($k < n$)} stands for the $\mathrm{N}^k\LO$ $t \bar t$ on-shell production cross section computed with the $\mathrm{N}^n\LO$ value of $\alphas$ and PDFs.
Under the assumption of massless leptons, we set \mbox{$\text{BR}(W^{+} \to e^+ \nu_e) = \text{BR}(W^{-} \to \mu^{-}\bar{\nu}_{\mu}) = \text{BR}(W \to l \nu_l)$}.

It is important to recall that, in the evaluation of these spurious terms, the top-quark width has to be computed at the same renormalisation scale $\muR$ used in the off-shell calculation, to avoid additional spurious terms.

\bibliography{biblio}

\end{document}